\documentclass[%
 reprint,
 superscriptaddress,
preprintnumbers,
nofootinbib,
 amsmath,amssymb,
 aps,
]{revtex4-1}

\usepackage{tikz}
\usepackage{graphicx}
\usepackage{dcolumn}
\usepackage{bm}
\usepackage{hyperref}
\usepackage{color}

\usepackage{multirow, array}

\hypersetup{
    colorlinks = true
}
\usepackage{aas_macros}

\newcommand{\iMpc}{\ensuremath{h/\mathrm{Mpc}}}

\newcommand{\Mpc}{\ensuremath{\mathrm{Mpc}/h}}

\renewcommand{\vec}{\bm}

\def\la{\mathrel{\mathpalette\fun <}}
\def\ga{\mathrel{\mathpalette\fun >}}
\def\fun#1#2{\lower3.6pt\vbox{\baselineskip0pt\lineskip.9pt
        \ialign{$\mathsurround=0pt#1\hfill##\hfil$\crcr#2\crcr\sim\crcr}}}
\newcommand{\lexp}{\mathop{\langle}}    
\newcommand{\rexp}{\mathop{\rangle}}    

\newcommand{\beq}{\begin{equation}}
\newcommand{\eeq}{\end{equation}}
\newcommand{\beqa}{\begin{eqnarray}}
\newcommand{\eeqa}{\end{eqnarray}}
\newcommand{\be}{\begin{equation}}
\newcommand{\ee}{\end{equation}}
\newcommand{\bea}{\begin{eqnarray}}
\newcommand{\eea}{\end{eqnarray}}
\newcommand{\nn}{\nonumber}

\newcommand{\vpt}{\mbox{\sf VPT}}

\begin{document}

\preprint{TUM-HEP 1424/22}

\title{Perturbation theory with dispersion and higher cumulants: non-linear regime}

\author{Mathias Garny}
\email{mathias.garny@tum.de}
\affiliation{Physik Department T31, Technische Universit\"at M\"unchen, James-Franck-Stra\ss{}e 1, D-85748 Garching, Germany
}%

\author{Dominik Laxhuber}
\email{dominik.laxhuber@tum.de}
\affiliation{Physik Department T31, Technische Universit\"at M\"unchen, James-Franck-Stra\ss{}e 1, D-85748 Garching, Germany
}%

\author{Rom\'an Scoccimarro}
\email{rs123@nyu.edu}
\affiliation{
 Center for Cosmology and Particle Physics, Department of Physics, New York University, NY 10003, New York, USA
}%

\date{October 14, 2022}

\begin{abstract} 
We present non-linear solutions of Vlasov Perturbation Theory (\vpt), describing gravitational clustering of collisionless dark matter with dispersion and higher cumulants induced by orbit crossing. We show that {\vpt} can be cast into a form that is formally analogous to standard perturbation theory (SPT), but including additional perturbation variables, non-linear interactions, and a more complex propagation. {\vpt} non-linear kernels have a crucial decoupling property: for fixed total momentum, the kernels becomes strongly suppressed when any of the individual momenta cross the dispersion scale into the non-linear regime. This screening of UV modes allows us to compute non-linear corrections to power spectra even for cosmologies with very blue power-law input spectra, for which SPT diverges. We compare predictions for the density and velocity divergence power spectra as well as the bispectrum at one-loop order to N-body results in a scaling universe with spectral indices  $-1\leq n_s\leq +2$. We find a good agreement up to the non-linear scale for all cases, with a reach that increases with the spectral index $n_s$.  We discuss the generation of vorticity as well as vector and tensor modes of the velocity dispersion, showing that neglecting vorticity when including dispersion would lead to a violation of momentum conservation. We verify momentum conservation when including vorticity, and compute the vorticity power spectrum at two-loop order, necessary to recover the correct large-scale limit with slope $n_w=2$. Comparing to our N-body measurements confirms the cross-over from $k^4$ to $k^2$ scaling on large scales.  Our results provide a proof-of-principle that perturbative techniques for dark matter clustering can be systematically improved based on the known underlying collisionless dynamics.
\end{abstract}

\maketitle

\tableofcontents

\section{Introduction}
\label{sec:introduction}

Perturbative techniques play a central role in understanding the formation and evolution of large-scale structures in the universe. The foundation  for this understanding rests on  the  gravitational clustering of collisionless cold dark matter (CDM). As such it is fully described by the Vlasov equation,  which follows the evolution of the phase-space distribution function (DF) coupled to gravity via the Poisson equation at scales below the Hubble radius. The conventional approach of standard perturbation theory (SPT) is a truncation of this description, in which only density (zero-th moment of the DF) and velocity (first moment of the DF) fields are taken into account (see~\cite{BerColGaz02} for a review). 

This truncation is motivated by the fact that CDM has negligible primordial velocity dispersion (second cumulant of the DF). However, even if absent initially, second and higher cumulants of the DF are generated by orbit crossing, and they should play a role in the transition to the non-linear regime. This is particularly so for initial conditions with significant small-scale power, i.e. blue spectral indices, for which SPT loop corrections describing this transition are UV divergent~\cite{MakSasSut92,ScoFri9612}. These UV divergences stand in sharp contrast to what is seen in N-body simulations~\cite{ColBouHer96},  showing that for such initial conditions fluctuations in the non-linear regime are {\em most suppressed} compared to  linear evolution in SPT. 

This motivates to systematically improve the perturbative description by taking second and higher cumulants into account, which obey a coupled hierarchy of equations of motion~\cite{PueSco0908}. Linearized solutions to this hierarchy were developed in paper~I~\cite{cumPT}, showing a structure that is much richer than in SPT as a result of the backreaction of orbit crossing on linear modes. This leads to  a suppression on small scales that saturates (for a given fixed wavenumber) when truncating the cumulant expansion at sufficiently high order.   This damping of linear UV modes sets in at a new scale, {\em the dispersion scale}, determined by the background value of the second cumulant, i.e. the velocity dispersion tensor. Physically, this damping corresponds to the backreaction of halo formation at small scales on linear modes.

The purpose of the present paper is to present non-linear solutions obtained by expanding around this new linear theory. We call this new perturbative expansion Vlasov Perturbation Theory or \vpt.   We show that \vpt~can be cast into a form that is formally analogous to SPT, but including additional perturbation variables, non-linear interactions, and a more complex propagation. This all follows from the Vlasov-Poisson equation in a straightforward way, without having to adhere to any ad hoc assumptions. 

The final outcome is to replace the well known non-linear kernels of SPT describing the perturbative solutions by new kernels that have a crucial decoupling property: for fixed total momentum ${\vec k}$, the kernels become strongly suppressed when any of the individual momenta cross the dispersion scale into the non-linear regime. As we shall see, this is precisely what is needed to bring the UV divergencies of SPT for blue spectra under control. This screening of UV modes is also what is expected physically to happen in the non-linear regime, where the predominant  structures are fairly stable dark matter halos. For CDM spectra, this regime is characterized by the ``virial turnover", where the non-linear power grows with wavenumber  less than in the weakly non-linear regime. The physical reason for this turnover can, at least approximately, be understood by the stable clustering picture~\cite{DavPee7708,Pee80}, where pairwise velocities cancel the Hubble flow.

It may be convenient to illustrate here the main ideas behind how \vpt~accounts for the screening mechanism of  UV modes for the simplest case, the density power spectrum (analogous results hold for other variables, e.g. velocity divergence, and other statistics, e.g. the bispectrum).  Consider scaling universes, i.e. scale-free Gaussian initial conditions with initial power spectra $P_0\propto k^{n_s}$ and $\Omega_m=1$. Then the power spectrum in \vpt~follows a perturbative expansion of the form,
\be
{P(k,\eta) \over e^{2\eta} P_0}= \sum_{L=0}^\infty \Big({k\over k_{\rm nl}(\eta)}\Big)^{(n_s+3)L}\ H_L(k/k_\sigma(\eta), \bar{\omega}, \ldots)\,,
\label{VPTep}
\ee
where $e^\eta$ is the scale factor, $L$ is the number of loops, $k_{\rm nl}(\eta)$ is the usual non-linear scale, $k_\sigma(\eta) \equiv 1/\sqrt{\epsilon}$ is the dispersion scale obtained from the background value of the velocity dispersion tensor $\epsilon_{ij}$ (suitably rescaled, see Eq.~\ref{eq:rescaledcumulants} below), $\bar{\omega}\equiv \omega/\epsilon^2$ is a constant that characterizes the background value of the fourth cumulant $\omega$, and the dots denote other (even) background values of the  cumulants, describing the DF non-Gaussianity\footnote{These are the normalized cumulants $\bar{\cal E}_{2n}$ discussed in paper~I~\cite{cumPT}.}. By self-similarity $k_\sigma/k_{\rm nl}$ is a constant that depends only on $n_s$. The dispersion scale is a new scale induced by orbit crossing, and as such is absent in SPT. 

\begin{figure*}[t]
  \begin{center}
  \includegraphics[width=\textwidth]{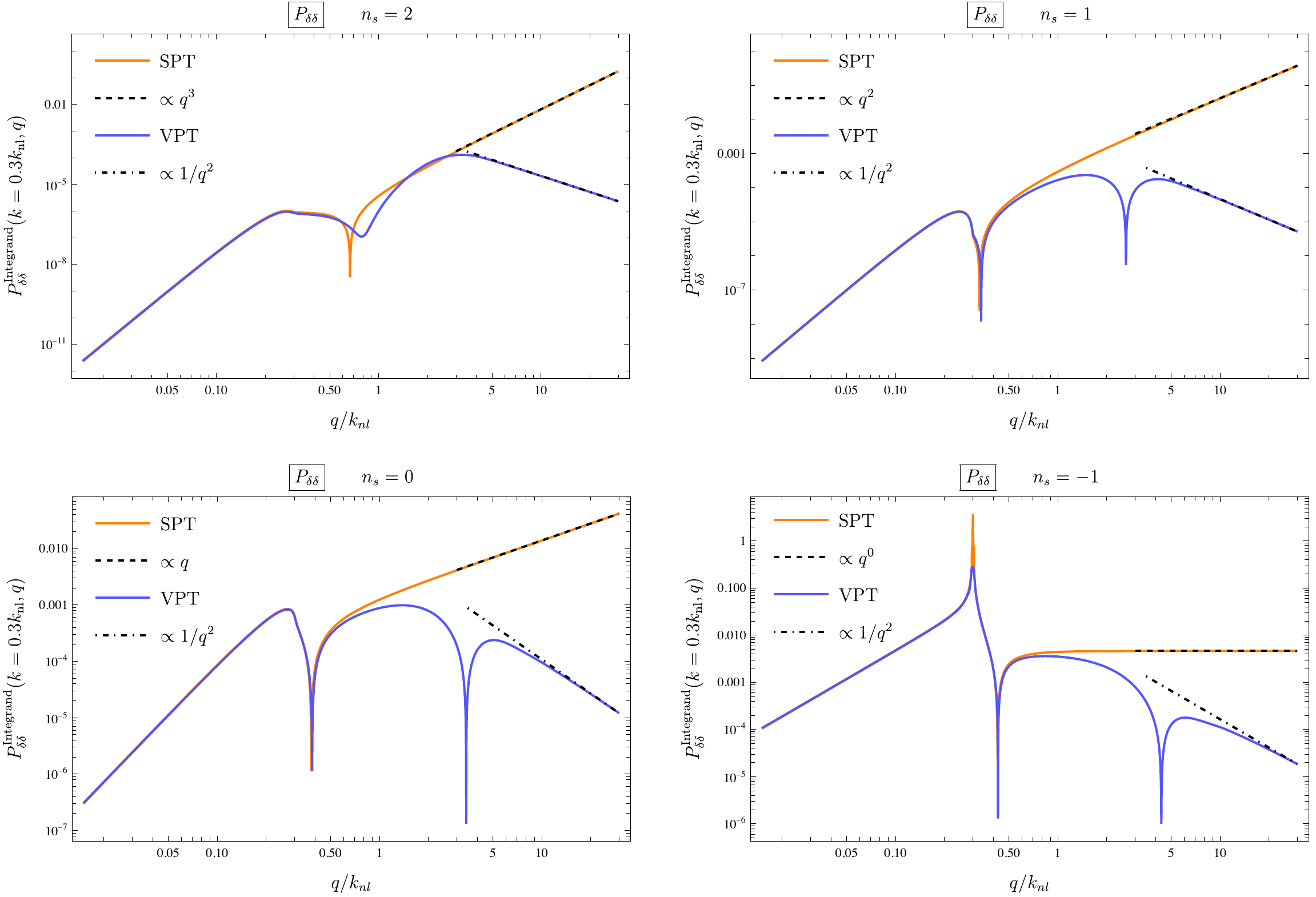}
  \end{center}
  \caption{\label{fig:integrand}
  Integrand of the one-loop contribution to the density power spectrum in \vpt~and SPT, for a scaling universe with power-law input spectrum $P_0\propto k^{n_s}$ with spectral indices $n_s=2,1,0,-1$ shown in the four panels.
  In SPT the integrand grows as $F_3(\vec k,\vec q,-\vec q)P_0(q)d^3q\sim q^{n_s+1}d\ln q$ for large loop wavenumber $q$, i.e. the well-known suppression $F_3\propto k^2/q^2$ due to momentum conservation is not sufficient to render the
  one-loop integral UV finite for all $n_s\geq-1$. In contrast, the physically expected screening of UV modes within the full collisionless Vlasov-Poisson dynamics is captured by \vpt, compensating the increase for large $q$ for blue spectra with high $n_s$ and leading to a finite integral with asymptotic decay of the integrand as $F_{1,\delta}(k,\eta)F_{3,\delta}(\vec k,\vec q,-\vec q,\eta)P_0(q)d^3q\sim q^{-2}d\ln q$ independently of the spectral index $n_s$. The sum of linear and one-loop contributions within \vpt~matches well with N-body results (see Fig.\,\ref{fig:Pdd_cumPT_Nbody}, including details on the cumulant hierarchy truncation used for this figure). Note that the $1/q^2$ scaling within \vpt~is universal and occurs for all truncations considered in this work, see Fig.~\ref{fig:F23_delta} and Eq.~\eqref{eq:kernelscalingvpt} for details.
  }
\end{figure*}

The prefactors of $(k/k_{\rm nl})$ in Eq.~(\ref{VPTep}) are the standard loop factors that appear in SPT, while the functions $H_L$ describe the scale-dependence due to  dispersion and higher cumulants that give rise to the screening of UV modes. For example, $H_0$ is the square of the linear kernel which is highly suppressed when $k\gg k_\sigma$. As stressed in paper~I~\cite{cumPT}, this suppression is physically different from a Jeans scale due to pressure, or viscosity in a normal fluid. In fact, the linear theory of collisionless matter is much richer than that of such fluids, with an effective description in terms of density and velocity divergence that is non-local in time, which is key to satisfy the cosmic energy equation~\cite{cumPT}. The dispersion scale can in principle be measured independently from simulations, and as we show in this paper it can be estimated from basic properties of dark matter halos. Intuitively we know the dispersion scale is close to the non-linear scale from the ``fingers of god" features seen in the redshift-space matter distribution.

The SPT limit of Eq.~(\ref{VPTep}) is somewhat delicate. Naively, it is reached by taking the $k_\sigma\to\infty$ limit (zero dispersion, including vanishing higher cumulants). In that case, however, the functions $H_L$ for $L\geq 1$ only have a finite limit (giving $k$-independent finite functions of $n_s$)  when $n_s<-1$\footnote{For example, for the one-loop density power spectrum $H_1$ becomes the function $\alpha(n_s)$ given by Eq.~(58) and Fig.~9 in ~\cite{Sco97}.}. This means that the limit of zero dispersion only exists for spectra which are sufficiently red-tilted. For such spectra though, there is still orbit crossing, $k_\sigma$ is still finite and corrections to SPT are predicted. 

In this paper we focus on spectra with $n_s\geq -1$, in order to comprehensively assess the applicability of \vpt~in a situation where the limitations of SPT are most severe\footnote{For an effective field theory description of these spectra at the one-loop level see~\cite{PajZal1308}.}. We compare one-loop predictions for the density power spectrum and bispectrum, the velocity divergence power spectrum and the density-velocity divergence power spectrum against N-body simulations. In addition, we  discuss the generation of vorticity, presenting solutions up to two-loop in order to capture the correct scaling of the vorticity power spectrum on large scales. We also highlight the importance of backreaction of vorticity modes on the density contrast, both from theoretical considerations as well as numerical results.

To illustrate the screening of UV modes captured by \vpt~(as opposed to SPT) we show in Fig.~\ref{fig:integrand} the \emph{integrand} of the one-loop density power spectrum for blue power-law input spectra $P_0\propto k^{n_s}$ with $n_s=2,1,0,-1$. The normalization is such that $P_{\delta\delta}^{1L}(k,\eta)=\int d\ln q P_{\delta\delta}^\text{Integrand}(k,q,\eta)$. 
This integrand is closely related to the one-loop contribution to the response function studied in~\cite{NisBerTar1611,NisBerTar1712}\footnote{$P_{\delta\delta}^\text{Integrand}(k,q,\eta)=T^{1L}(k,q)qP_0(q,\eta)P_0(k,\eta)$ with response $T(k,q)$ in the notation of~\cite{NisBerTar1611}.}.
In SPT one would have 
\be
  P_{\delta\delta}^\text{Integrand}(k,q,\eta)\big|_\text{SPT} \propto q^{n_s+1}\,,
\ee
such that the loop integration is UV divergent for all $n_s\geq -1$. In contrast, the prediction from the Vlasov theory is that the integrand drops as
\be\label{eq:vptintegrandscaling}
  P_{\delta\delta}^\text{Integrand}(k,q,\eta)\big|_\text{\vpt} \propto q^{-2}\,,
\ee
\emph{independently of $n_s$}, rendering the integral well-behaved and implying that perturbation modes with wavenumber far above the dispersion scale (which is of order of the non-linear scale, as will be discussed in detail below) give a negligible contribution, in accordance with physical expectations. We emphasize that also the linear theory in \vpt~is changed compared to SPT, and we find that the sum of linear and integrated one-loop contributions matches well with N-body results.

Previous work in the literature on corrections to SPT implied by the Vlasov equation spans a number of different fronts. In~\cite{PueSco0908}  the equation of motion for the Vlasov cumulant generating function hierarchy was written down, explaining the conditions under which it matches solution of the Vlasov equation after shell-crossing. They used measurements of the stress tensor in simulations to close the hierarchy and estimate the backreaction of shell-crossing on large-scale density and velocity divergence power spectra, including a study on the growth of vorticity.  In~\cite{McD1104} a low-$k$ expansion of the linearized hierarchy truncated at the second cumulant is used to obtain the impact of velocity dispersion on density and velocity divergence. In addition, an approximate treatment of non-linear evolution is presented to show that  the growth of the velocity dispersion background  is sustained by mode coupling. Still within the second-cumulant truncation,~\cite{Avi1603} and~\cite{ErsFlo1906} discuss the growth of the velocity dispersion tensor, with the latter work featuring the evolution of the background dispersion at the non-linear level while keeping the fluctuations in the dispersion linear (see also~\cite{Erschfeld:2021kem}). Different strategies for closing the Vlasov hierarchy are presented in~\cite{Uhl1810}. Complementary  approaches from the Lagrangian picture include~\cite{McDVla1801} and~\cite{CusTanDur1703} that computes  the growth of vorticity driven by velocity dispersion.

This work is structured as follows. In Section~\ref{sec:kernels} we discuss the basics of the Vlasov cumulant hierarchy, splitting cumulants of the DF into their expectation values and fluctuations and presenting their equations of motion. We also introduce the non-linear kernels, and study analytically their scaling in the limit of small wavenumber compared to the dispersion scale. In addition, we study numerically the case when wavenumbers cross the dispersion scale highlighting the screening of UV modes. Along the way we discuss the properties of these kernels in detail for various truncations of the cumulant hierarchy.  Section~\ref{sec:limits} discusses the symmetry properties of the non-linear kernels, including Galilean invariance and mass and momentum conservation. In Section~\ref{sec:vorticity} we discuss the generation of vorticity, and vector and tensor modes of the velocity dispersion tensor. Section~\ref{sec:powerlaw} computes the dispersion scale in two different, complementary ways. First, using perturbation theory, and second, using knowledge of dark matter profiles together with the halo mass function measured in our N-body simulations. In Section~\ref{sec:simulation} we present the predictions of \vpt~against measurements in our simulations, including the density power spectrum and bispectrum, velocity divergence power spectrum, the cross-spectrum between density and velocity divergence, and the vorticity power spectrum. Section~\ref{sec:conclusions} presents our conclusions. 

A  number of appendices contain supplementary results that support the discussion in the main text. Appendix~\ref{app:tadpoles} discusses our approach to tadpoles, App.~\ref{app:timeintegrals} presents analytic results for time integrations over the background dispersion, and App.~\ref{app:galilei} proves the scaling of non-linear kernels in the squeezed limit due to Galilean invariance. Appendix~\ref{app:vorticity} discusses our implementation of vorticity, vector and tensor modes in the numerical scheme, and App.~\ref{app:rescaling} shows how to rescale solutions to different values of the dispersion scale. Appendix~\ref{app:Nbody} presents our N-body simulations,  App.~\ref{app:HMF} discusses the determination of the halo mass function and details about the calculation of the dispersion scale from halos, and App.~\ref{app:DivVort} presents our algorithm for measuring velocity divergence and vorticity in the simulations.

\section{Non-linear kernels with dispersion and higher cumulants}
\label{sec:kernels}

We follow the formalism for perturbation theory introduced in paper~I~\cite{cumPT}, which presents linear solutions for the density and velocity  perturbations in the presence of a non-zero velocity dispersion tensor and higher cumulants. We start from the fact that the phase-space distribution function $f(\tau,\vec x,\vec p)$  of non-relativistic matter can be characterized by its cumulants, obtained by expanding the cumulant generating function in the auxiliary vector $\vec l$~\cite{PueSco0908,cumPT},
\bea\label{eq:genfunc}
  {\cal C}(\tau,\bm{x},\bm{l})  &=& \ln \int d^3p\, e^{ \bm{l}\cdot\bm{p}/a }\, f(\tau,\bm{x},\bm{p})\nn\\
  &=& \ln(1+\delta) + l_iv_i + \frac12 l_il_j \sigma_{ij} + \frac16 l_il_jl_k {\cal C}_{ijk} \nn\\
  && {} + \frac{1}{24} l_il_jl_kl_m{\cal C}_{ijkm}+\dots\,,
\eea
where $\delta(\tau,\vec x)$ is the density contrast, $v_i(\tau,\vec x)$ the peculiar velocity, 
$\sigma_{ij}(\tau,\vec x)$ the velocity dispersion tensor, and ${\cal C}_{ijk} ({\cal C}_{ijkm})$ the third (fourth) cumulant.
Furthermore, $a$ is the scale-factor, $\tau$ the conformal time and $\bm{p}$ the comoving momentum per unit particle mass.
In this section we first review the relevant equations and notation, and then describe how to recursively obtain non-linear kernels. These characterize the non-linear solutions, generalizing the well known non-linear kernels of standard perturbation theory (SPT)~\cite{Fry84,GorGriRey86,BerColGaz02}.

\subsection{Setup}

The Vlasov or collisionless Boltzmann equation for the phase-space distribution in the non-relativistic, sub-horizon limit
yields the following equation of motion for the cumulant generating function~\cite{PueSco0908}
\be\label{eq:genfunceom}
 \partial_\tau{\cal C}+{\cal H}(\bm{l}\cdot\nabla_l){\cal C}+(\nabla{\cal C})\cdot(\nabla_l{\cal C})+(\nabla\cdot\nabla_l){\cal C}=-\bm{l}\cdot\nabla\Phi\,,
\ee
where $\Phi$ is the gravitational potential given by $\nabla^2\Phi=\frac32{\cal H}^2\Omega_m \delta$, with conformal Hubble rate ${\cal H}=\partial_\tau\ln a$
and time-dependent matter density parameter $\Omega_m$.

The cumulants satisfy a coupled, non-linear hierarchy of differential equations, obtained by Taylor
expanding Eq.~\eqref{eq:genfunceom} in $\vec l$, given by the continuity and momentum conservation equation
at zeroth and first order,
\bea\label{eq:fluid}
  \partial_\tau\delta+\nabla_i[(1+\delta) v_i] &=& 0 \,, \\
  \partial_\tau v_i +{\cal H}v_i+v_j\nabla_j v_i + \nabla_i\Phi 
  &=& -\nabla_j \sigma_{ij} -\sigma_{ij}\nabla_j\ln(1+\delta), \nn
\eea
with the latter containing the velocity dispersion tensor on the right-hand side. Its equation of motion
is obtained when expanding Eq.~\eqref{eq:genfunceom} to second order in $\vec l$, and it is in turn coupled
to the third cumulant, whose equation follows from expanding to third order. They are given by
\bea\label{eq:eomsigmaij}
  \partial_\tau \sigma_{ij} &+& 2{\cal H}\sigma_{ij}+v_k\nabla_k\sigma_{ij}+\sigma_{jk}\nabla_k v_i+\sigma_{ik}\nabla_k v_j   \nn\\
 &=& - \nabla_k{\cal C}_{ijk}  - {\cal C}_{ijk}\nabla_k\ln(1+\delta) \,,\nn\\
  \partial_\tau {\cal C}_{ijk} &+& 3{\cal H}{\cal C}_{ijk} +v_m\nabla_m{\cal C}_{ijk} \nn\\
  &+& \sigma_{km}\nabla_m \sigma_{ij} +\sigma_{im}\nabla_m \sigma_{kj}+\sigma_{jm}\nabla_m \sigma_{ik}   \nn\\
  &+& {\cal C}_{jkm}\nabla_m v_i  +{\cal C}_{ikm}\nabla_m v_j  +{\cal C}_{ijm}\nabla_m v_k    \nn\\
 &=& - \nabla_m{\cal C}_{ijkm}  - {\cal C}_{ijkm}\nabla_m\ln(1+\delta) \,.
\eea
The framework of SPT is based on neglecting the velocity dispersion in the momentum conservation equation, which then becomes the standard Euler equation for a perfect fluid with zero pressure. Here we include the dispersion tensor and also higher cumulants, systematically taking non-linear terms in their equations of motion into
account within perturbation theory.

In order to be able to deal with the occurrence of the log-density field
\be
  A  \equiv \ln(1+\delta)\,,
\ee 
on the right-hand side of the evolution equations, we follow the hybrid scheme introduced in paper~I~\cite{cumPT} and perturbatively
solve for both $\delta$ as well as $A$, complementing the set of evolutions equations by
\be
  \partial_\tau A+\nabla_iv_i + v_i\nabla_iA = 0\,,
\ee
which follows directly from the continuity equation. It is convenient to consider the rescaled cumulants,
\bea\label{eq:rescaledcumulants}
  u_i &=& \frac{v_i}{-{\cal H}f}, \quad
  \epsilon_{ij} = \frac{\sigma_{ij}}{({\cal H}f)^2}\,,\nn\\
  \pi_{ijk} &=& \frac{{\cal C}_{ijk}}{(-{\cal H}f)^3},\quad
  \Lambda_{ijkm} = \frac{{\cal C}_{ijkm}}{({\cal H}f)^4}\,,
\eea
where $f=d\ln D/d\ln a$ is the usual growth rate, with $D(a)$ being the conventional growth factor.
In addition, we switch from conformal time $\tau$ to $\eta=\ln(D)$ using $\partial_\tau={\cal H}f\partial_\eta$
and for any quantity $X$ we have,
\bea
  && (\partial_\tau+n{\cal H})[(-{\cal H}f)^nX]\nn\\
  &=& -(-{\cal H}f)^{n+1}\left(\partial_\eta+n\left(\frac32\frac{\Omega_m}{f^2}-1\right)\right)X\,.
\eea

Rotational invariance allows all even cumulants to possess a homogeneous background value,
that we parametrize for the second and fourth cumulant by $\epsilon(\eta)$ and $\omega(\eta)$ via
\bea
  \langle\epsilon_{ij}(\eta,\vec x)\rangle &=& \delta_{ij}\ \epsilon(\eta)\,,\nn\\
  \langle\Lambda_{ijkm}(\eta,\vec x)\rangle &=& \left(\delta_{ij}\delta_{km}+2\,{\rm cyc.}\right)\, \frac{\omega(\eta)}{5}\,.
\eea
Equations of motion for the background values can be obtained by taking an ensemble average of the evolution equations
of the second and fourth cumulant, respectively (see paper~I~\cite{cumPT}). 
For the second cumulant, it is given by
\be\label{eq:epseom}
  \partial_\eta\epsilon(\eta) + 2\left( \frac32 \frac{\Omega_m}{f^2}-1 \right) \epsilon(\eta) = Q(\eta)\,,
\ee
with source term
\be\label{eq:Qdef}
  Q(\eta) = 
  \frac{1}{3}\langle u_l\nabla_l \epsilon_{ii}\rangle
  + \frac{2}{3}\langle \epsilon_{il}\nabla_l u_i \rangle 
  + \frac13\langle \pi_{iil}\nabla_l\ln(1+\delta)\rangle\,.
\ee
Note that while $\langle\delta\rangle=0$ is ensured by mass conservation,
the zeroth cumulant  $A  = \ln(1+\delta)$ does also have a background value ${\cal A}=\langle A\rangle$.
Nevertheless, since $A$ enters only via $\nabla_i\ln(1+\delta)$
in the equations of motion, the background value is not needed explicitly~\cite{cumPT}.

To obtain equations for the perturbations we expand the cumulants around their background values, in particular
\be
  \epsilon_{ij}(\eta,\vec x) = \delta_{ij}\, \epsilon(\eta) + \delta\epsilon_{ij}(\eta,\vec x)\,.
\ee
While $\delta,u_i,{\cal C}_{ijk}$ have no background contribution, a similar split is necessary for $A$ and analogously for $\Lambda_{ijkm}$.
Following the previous discussion, it is understood in the following that $A$ denotes the perturbation part only, with ${\cal A}$ subtracted.
The equation of motion for $\delta\epsilon_{ij}$ is obtained by subtracting from Eq.~\eqref{eq:eomsigmaij} its ensemble average, and similarly for $A$
and higher cumulants. 

The velocity can be split into its divergence $\theta=\nabla_iu_i$ and vorticity $w_i=\varepsilon_{ijk}\nabla_j u_k$ contributions via
\be\label{eq:uSV}
  u_i = u_i^S+u_i^V=\frac{\nabla_i}{\nabla^2}\theta-\frac{\varepsilon_{ijk}\nabla_j }{\nabla^2}w_k\,,
\ee
corresponding to one scalar and two vector modes,
while $\delta\epsilon_{ij}$ consists of two scalar modes $g$ and $\delta\epsilon$, two vector modes contained in the divergence-free vector $\nu_i$, and two tensor modes represented by the symmetric, transverse-traceless matrix $t_{ij}$,
\bea\label{eq:epsSVT}
  \delta\epsilon_{ij} &=& \delta\epsilon_{ij}^S + \delta\epsilon_{ij}^V + \delta\epsilon_{ij}^T\,,
\eea
where
\bea
  \delta\epsilon_{ij}^S &=& \delta_{ij}\,\delta\epsilon+\frac{\nabla_i\nabla_j}{\nabla^2}g \,,\nn\\
  \delta\epsilon_{ij}^V &=& -\frac{\varepsilon_{ilk}\nabla_l \nabla_j}{\nabla^2}\nu_k 
    -\frac{\varepsilon_{jlk}\nabla_l \nabla_i}{\nabla^2}\nu_k \,,\\
 \delta\epsilon_{ij}^T &=& t_{ij}\equiv P_{ij,ls}^T\,\delta\epsilon_{ls}\,,\nn
\eea
and the tensor projector $P_{ij,ls}^T$ is given in~\cite{cumPT}.

The impact of third, fourth and higher cumulants in the linear approximation has been analyzed in detail in paper~I~\cite{cumPT}.
In this work, we assess their impact at non-linear level by including the two scalar perturbations $\pi$ and $\chi$ of the third cumulant,
\be
  \pi_{ijk}^S = -\left(\delta_{ij}\frac{\nabla_k}{\nabla^2}+2\,{\rm cyc.}\right)\frac{\chi}{5}-\frac{\nabla_i\nabla_j\nabla_k}{\nabla^4}(\pi-\chi)\,,
\ee
as well as the background value $\omega(\eta)$ of the fourth cumulant. Note that the latter enters the evolution equations of $\pi_{ijk}$ via
the last term in the last line of Eq.~\eqref{eq:eomsigmaij}. We neglect non-scalar contributions to $\pi_{ijk}$ as well as perturbation modes of
the fourth cumulant, in order to obtain a tractable approximation scheme. The impact of higher cumulants will be quantified by comparing solutions
with and without taking $\pi_{ijk}^S$ into account.

Apart from the complete set of perturbation modes for the zeroth, first and second cumulant, it is important to note that 
our present setup includes all higher cumulant perturbation modes relevant for the source term of the background dispersion, given by
\be\label{eq:Q3rd}
  Q(\eta) = \frac{1}{3}\int d^3k \left( P_{\theta \tilde g}(k,\eta) + 2 P_{w_i\nu_i}(k,\eta) + P_{A\pi}(k,\eta)\right)\,,
\ee
where $\tilde g\equiv g-\delta\epsilon$.

The scalar, vector and tensor perturbation modes can be collected in a component vector
\be\label{eq:psidef}
  \psi=(\psi^S,\psi^V,\psi^T)\,,
\ee
with
\bea
  \psi^S &=& (\delta, \theta, g, \delta\epsilon, A, \pi, \chi)\,,\nn\\
  \psi^V &=& (w_i,\nu_i)\,,\nn\\
  \psi^T &=& (t_{ij})\,.
\eea
In Fourier space, the evolution equations can be written in the form
\be\label{eq:eom}
  \psi_{k,a}'(\eta)+\Omega_{ab}(k,\eta)\psi_{k,b}(\eta) = \int_{pq} \gamma_{abc}(\vec p,\vec q)\psi_{p,b}(\eta)\psi_{q,c}(\eta)\,,
\ee
where $\psi_{k,a}$ stands for the Fourier mode $\vec k$ of $\psi$, and $a$ runs over all perturbation modes, including their
spatial indices in case of vectors and tensors, with summation over repeated indices implied.
Furthermore, $'=d/d\eta$ and $\int_{pq}=\int d^3p\,d^3q\,\delta^{(3)}(\vec k-\vec p-\vec q)$. The linear
evolution is characterized by the scale- and time-dependent matrix $\Omega_{ab}(k,\eta)$. Due to rotational invariance
it has a block-diagonal structure for scalar, vector and tensor modes, with the blocks given by
\bea\label{eq:Omega}
  \Omega^S &=& \left(\begin{array}{ccccccc}
  & -1 \\
  -\frac32 d & \frac12 e & k^2 & k^2 & k^2\epsilon \\
  & -2\epsilon & e & & & 1 & -\frac35\\
  & & & e & & &  \frac15\\
  & -1 \\
  & & -3k^2\epsilon & -5 k^2\epsilon & -k^2\omega & \frac32 e \\
  & &  & -5 k^2\epsilon & -k^2\omega & & \frac32 e \\
  \end{array}\right)\,,\nn\\
  \Omega^V &=& \left(\begin{array}{cc}
  \frac12 e & k^2 \\
  -\epsilon & e\\
  \end{array}\right)\,,
  \quad
  \Omega^T = e\,,
\eea
with $d(\eta)\equiv \Omega_m/f^2$ and $e(\eta)\equiv 3d(\eta)-2$, with $d,e\mapsto 1$ in the EdS approximation. Note that the upper two-by-two part of $\Omega^S$ corresponds to the well-known form within SPT. In addition, each entry in $\Omega^V$ is understood to act on all three vector components of $w_i$ and $\nu_i$, respectively, with the off-diagonal entries describing a mixing of vorticity and vector modes of the dispersion tensor. Similarly, $\Omega^T$ acts equally on all components of $t_{ij}$. 

It is worth emphasizing that background values of the second and fourth cumulant, $\epsilon(\eta)$ and $\omega(\eta)$, are not considered to be a priori of any particular order in the perturbative expansion as they are sourced by one-point moments (see Eq.~\ref{eq:Qdef}) that correspond to integrals of power spectra over {\em all} momenta (see Eq.~\ref{eq:Q3rd}). That is, these quantities are not proportional to the fluctuation modes $\psi_{k,a}(\eta)$, and we treat them as any function of time (e.g. $\Omega_m$), as we are explicitly agnostic about the size of velocity dispersion effects in our approach. This is crucial and allows us, as discussed in paper~I~\cite{cumPT}, to obtain the expected decoupling of large-scale modes from halo formation at small scales. As we shall see below, this is described by the suppression of the non-linear kernels at high momenta, generalizing the results for the linear kernel described in detail in paper~I~\cite{cumPT}. This reduces the sensitivity of \vpt~to UV modes when compared to SPT,  helping with the convergence of perturbative calculations.  
 The decoupling effect appears for large wavenumbers $k\gtrsim k_\sigma$ above the scale set by the background dispersion
\be
  k_\sigma(\eta) \equiv 1/\sqrt{\epsilon(\eta)}\,.
\ee
As shown in paper~I~\cite{cumPT}, the absence of exponential instabilities of the linear system for $k\gg k_\sigma$ yields a restriction on the size of the fourth
relative to the second cumulant expectation value, particularly for the dimensionless ratio
\be
  \bar\omega\equiv\frac{\omega}{\epsilon^2}\,.
\ee
For our setup considered here, the stability condition is given by
\be
  -10\leq \bar\omega\leq 5\,,
\ee
provided that $\alpha\equiv\partial_\eta\ln\epsilon$ and $\bar\omega$ are constant or slowly varying. This is the case
in the limit of a scaling universe, where $\Omega_m=1$ and the initial perturbations are scale-free, $P_0(k) \propto k^{n_s}$. 

All non-linearities are captured by the \emph{vertices} $\gamma_{abc}(\vec p,\vec q)$ in Eq.~\eqref{eq:eom}. Their explicit form is
given in paper~I~\cite{cumPT}, satisfying
\be\label{eq:gammasymm}
  \gamma_{abc}(\vec p,\vec q) = \gamma_{acb}(\vec q,\vec p)\,,
\ee
and we take all of them into account in our analysis\footnote{Up to the second cumulant, all vertices obtained from the equations of motion are included.
For vertices involving at least one third cumulant perturbation $\pi$ or $\chi$, we only include contributions where all three fields are scalar modes. We checked that this is consistent with the symmetry constraints discussed in Sec.~\ref{sec:limits}.}.
Note that the equations of motion include at most quadratic terms when including both $\delta$ and $A$ in the set of perturbation variables. They include the familiar SPT vertices
$\gamma_{\delta\theta\delta}$, $\gamma_{\delta\delta\theta}$ and $\gamma_{\theta\theta\theta}$, as well as a large set of extra vertices
involving second and third cumulant perturbations and $A$. It is crucial to take the full set of vertices into account along with the
matrix $\Omega_{ab}$ to ensure that symmetry constraints on the resulting perturbative solutions from Galilean symmetry as well
as mass and momentum conservation hold, as we explain in detail in Sec.~\ref{sec:limits}.

A final comment regarding Eq.~\eqref{eq:eom} is worth pointing out. When deriving the equation for the perturbation part by subtracting the full from the ensemble averaged equations of motion, one obtains an equation that contains an additional
contribution compared to Eq.~\eqref{eq:eom}. It is proportional to a three-dimensional Dirac delta supported at $\vec k=0$, and contributes only to the components
that possess also a background value, being $\delta\epsilon$ and $A$ in the setup adopted here.
We refer to App.~\ref{app:tadpoles} for a discussion of these terms, that ensure the cancellation of so-called ``tadpole'' diagrams in the perturbative solution.
As shown there, these terms can be traded for a modification of the vertices, that consists in setting $\gamma_{abc}(\vec p,\vec q)$ to zero for $\vec p=-\vec q$.
We adopt this choice in the following, and use Eq.~\eqref{eq:eom} to generate perturbative solutions.

\subsection{Non-linear kernels}

We consider perturbative solutions of the equation of motion, Eq.~\eqref{eq:eom}, assuming that at some ``initial'' time $\eta_\text{ini}$ long before the onset of non-linearity, but long after recombination,
all perturbation modes are proportional to the initial density field $\delta_{k0}$. This is the case for adiabatic initial conditions and when
assuming an initially negligibly small (but non-zero, see below) velocity dispersion, as appropriate for cold dark matter. The vector $\psi$ of perturbations
may then be formally  Taylor expanded in powers of $\delta_{k0}$ as
\be\label{eq:psiexpansion}
  \psi_{k,a}(\eta) = \sum_{n\geq 1} \int_{k_i} F_{n,a}(\vec k_1,\dots,\vec k_n,\eta)\,e^{n\eta}\,\delta_{k_10}\cdots\delta_{k_n0}\,,
\ee
where $F_{n,a}$ is the $n$th order non-linear kernel for component $a$, and $\int_{k_i}\equiv \int d^3k_1\cdots d^3k_n\,\delta^{(3)}(\vec k-\sum_i\vec k_i)$, with the
constraint on the sum over all wavevectors arising from spatial translation symmetry of the equations of motion. Even if initially all higher cumulant \emph{perturbations} are taken to be zero, they are generated by time-evolution in presence of a background dispersion (and background values of higher cumulants in general). As discussed in paper~I~\cite{cumPT}, the perturbations in turn source the background values, and we come back to the possibility of determining self-consistent solutions of this system in Sec.~\ref{sec:selfconsistent}. For the moment, we consider the background values $\epsilon(\eta)$ and $\omega(\eta)$ entering via Eq.~\eqref{eq:Omega} as given and discuss the implications for the perturbations $\psi_{k,a}(\eta)$.

As usual, the $F_{n,a}$ can be taken to be symmetric
under arbitrary exchanges of wavenumbers in their arguments, which we assume in the following. The normalization is chosen such that, when ignoring second and higher
cumulants and using the EdS approximation, $F_{n,\delta}$ and $F_{n,\theta}$ would become time-independent and coincide with the conventional EdS-SPT kernels $F_n$
and $G_n$, with $F_1=G_1=1$. When taking second and higher cumulants into account, the kernels do depend on time, and even the linear solutions described by $F_{1,a}(k,\eta)$ become non-trivial, featuring a power-like suppression for $k\gg k_\sigma$, see paper~I~\cite{cumPT}.

The expansion in Eq.~\eqref{eq:psiexpansion} can be used to perturbatively compute the (cross-)power spectra
\be\label{eq:Pab}
\langle \psi_{k,a}(\eta)\psi_{k',b}(\eta)\rangle = \delta^{(3)}(\vec k+\vec k')P_{ab}(k,\eta)\,,
\ee
where the subscript $ab$ labels the various modes, with for example $a=b=\delta$ corresponding to the density power spectrum,
and $a=\delta, b=\theta$ the density/velocity divergence cross spectrum.
Following the standard scenario, the initial density field $\delta_{k0}$ can be taken to be drawn from a Gaussian random field
fully characterized by its initial power spectrum $\langle \delta_{k0}\,\delta_{k'0}\rangle= \delta^{(3)}(\vec k+\vec k')P_0(k)$.
By inserting Eq.~\eqref{eq:psiexpansion} into Eq.~\eqref{eq:Pab} and using the Wick theorem, we obtain a loop expansion in close analogy to SPT,
\be
  P_{ab}(k,\eta) = P_{ab}^\text{lin}(k,\eta) + P_{ab}^{1L}(k,\eta) + P_{ab}^{2L}(k,\eta)+\dots\,,
\ee
but with modified non-linear kernels. For example, the linear and one-loop contributions are given by
\bea\label{eq:P1Ldef}
  P_{ab}^\text{lin}(k,\eta) &=& e^{2\eta}F_{1,a}(k,\eta)F_{1,b}(k,\eta)P_0(k) \,,\nn\\
  P_{ab}^{1L}(k,\eta) &=& e^{4\eta}\int d^3q \Big\{ 2F_{2,a}(\vec k-\vec q,\vec q,\eta)F_{2,b}(\vec k-\vec q,\vec q,\eta)\nn\\
  && {} \times P_0(|\vec k-\vec q|)P_0(q) \nn\\
  &+& 3F_{1,a}(k,\eta)F_{3,b}(\vec k,\vec q,-\vec q,\eta)P_0(k)P_0(q) \nn\\
  &+& 3F_{3,a}(\vec k,\vec q,-\vec q,\eta)F_{1,b}(k,\eta)P_0(k)P_0(q) \Big\}\,.
\eea
The expressions can be used to numerically compute loop corrections, with $P_0(k)$ being the usual linear matter power spectrum obtained from
Boltzmann solvers such as CLASS~\cite{Blas:2011rf} or CAMB~\cite{Lewis:1999bs}. As in SPT, the integration region with $|\vec q|\ll k\equiv |\vec k |$ yields large contributions to
the individual summands, that cancel in their sum (see Sec.\,\ref{sec:galilei}). It is therefore advantageous to ensure their cancellation at the integrand level by multiplying the first summand
in the integrand of $P_{ab}^{1L}$ with $2\Theta(|\vec k-\vec q|-|\vec q|)$ and then symmetrizing it with respect to the substitution $\vec q\to -\vec q$~\cite{Blas:2013bpa}.
Here $\Theta(x)$ is the Heaviside function. This operation leaves the integral unchanged, but is preferable for numerical Monte Carlo integration.
The two-loop contribution to the power spectrum can be computed analogously (see Sec.\,\ref{sec:Pww} and~\cite{Blas:2013bpa,BlaGarKon1309,Garny:2020ilv}).
Similarly, the bispectrum can be computed using the modified non-linear kernels and employing the same algorithm as in SPT to ensure the cancellation of
large contributions for low loop wavenumber at the integrand level (see Sec.\,\ref{sec:Pddd} and~\cite{Floerchinger:2019eoj}).

The difference in \vpt~compared to SPT enters via the non-linear kernels $F_{n,a}$.
Equations of motion for the kernels can be obtained by inserting Eq.~\eqref{eq:psiexpansion} into Eq.~\eqref{eq:eom} and collecting all terms with a given power of initial
density fields. This yields a set of coupled, ordinary differential equations given by
\bea\label{eq:kernelODE}
 \lefteqn{ (\partial_{\eta}\delta_{ab}+n\delta_{ab}+\Omega_{ab}(k,\eta))F_{n,b}(\vec k_1,\dots ,\vec k_n,\eta) }\nn\\
  &=& \sum_{m=1}^{n-1} \Big\{\gamma_{abc}(\vec q_1+\cdots+\vec q_m,\vec q_{m+1}+\cdots+\vec q_n) \nn\\
  && \times F_{m,b}(\vec q_1,\dots,\vec q_m,\eta)F_{n-m,c}(\vec q_{m+1},\dots,\vec q_n,\eta) \Big\}^s\,, \nn\\
\eea
where $\{\cdots\}^s=\sum_\text{perm}\{\cdots\}/|\text{perm}|$ denotes an average over all $|\text{perm}|=n!/m!/(n-m)!$ possibilities to
choose the subset of wavevectors $\{\vec q_1,\dots,\vec q_m\}$ from $\{\vec k_1,\dots,\vec k_n\}$, and $k\equiv |\sum_i \vec k_i|$.
When neglecting second and higher cumulants (i.e. restricting to $a,b=\delta,\theta$) and in the EdS approximation these equations have
time-independent solutions for $F_{n,\delta}$ and $F_{n,\theta}$. Indeed, by combining the equations for these two kernels in that limit
and using time-independence one recovers the well-known algebraic recursion relations~\cite{GorGriRey86}.

When taking second and higher cumulants into account, one obtains a coupled system for all components of $F_{n,a}$, with
$\Omega_{ab}(k,\eta)$ being scale- and time-dependent. In this case, solutions can in general only be found numerically by integrating the set of
coupled equations, and for this purpose we follow the strategy developed in~\cite{Blas:2015tla,Garny:2020ilv}. Nevertheless, it is instructive to consider analytical
results in the limit when the impact of velocity dispersion and higher cumulants is small (but non-zero), which we discuss before presenting our numerical results.
In the following we set $\Omega_m/f^2\mapsto 1$ for simplicity.

\subsection{Analytical results in the limit $\epsilon\to 0$}
\label{sec:analytical_results}

For wavenumbers that satisfy $\epsilon k_i^2\ll 1$ it is possible to find approximate analytical solutions by
solving for the non-linear kernels perturbatively in powers of the background dispersion $\epsilon$. Using the counting $\omega\propto \epsilon^2$
the kernels $F_{n,a}$ for $a=\delta,\theta,A$ start at order $\epsilon^0$, the kernels for $a=w_i,g,\delta\epsilon,\nu_i,t_{ij}$ at order $\epsilon^1$,
and for $a=\pi,\chi$ at order $\epsilon^2$, in accordance with the scaling derived in paper~I~\cite{cumPT}.
The kernels $F_{n,\delta}|_{\epsilon^0}=F_n$ and $F_{n,\theta}|_{\epsilon^0}=G_n$ coincide with the EdS-SPT kernels at lowest order in $\epsilon$, and for all $n$.

\subsubsection{First order kernels}

Let us first derive the expansion of the linear kernels $F_{1,a}$ in $\epsilon$, for which the right-hand side of Eq.~\eqref{eq:kernelODE} is zero.
At linear order one has $F_{1,A}=F_{1,\delta}$ and $F_{1,w_i}=F_{1,\nu_i}=F_{1,t_{ij}}=0$, see Sec.~\ref{sec:vorticity} for more details on vorticity, vector and tensor modes. At lowest order in the expansion in $\epsilon$ one recovers the EdS-SPT result $F_{1,\delta}|_{\epsilon^0}=F_{1,\theta}|_{\epsilon^0}=1$. The evolution equation for $g$ related to the second line of Eq.~\eqref{eq:Omega} yields
\bea
 (\partial_\eta+2)F_{1,g}(k,\eta) &=& 2\epsilon(\eta)F_{1,\theta}(k,\eta)-F_{1,\pi}(k,\eta)\nn\\
 && {} +\frac35 F_{1,\chi}(k,\eta)\,.
\eea
Using that the third cumulant modes $\pi,\chi$ contribute only starting at order $\epsilon^2$ and inserting the lowest order expression for $F_{1,\theta}$ yields
\be\label{eq:Tgsmalleps}
  F_{1,g}(k,\eta) = 2E_2(\eta)+{\cal O}(\epsilon^2)\,,
\ee
where we define the weighted time integral of $\epsilon(\eta)$,
\be\label{eq:epsweight}
  E_m(\eta)\equiv \int^\eta d\eta'\,e^{m(\eta'-\eta)} \epsilon(\eta')\,,
\ee
which e.g. for $\epsilon=\epsilon_0\, e^{\alpha\eta}$ gives $E_m(\eta)=\epsilon(\eta)/(m+\alpha)$.

Similarly one finds $F_{1,\delta\epsilon}={\cal O}(\epsilon^2)$, i.e. the linear kernel for $\delta\epsilon$ starts only at order $\epsilon^2$.
Note that the $\epsilon^2$ contribution is generated only by the third cumulant. Indeed, when neglecting third and higher cumulants $F_{1,\delta\epsilon}$ would be zero
at all orders in $\epsilon$, because the evolution equation for $\delta\epsilon$ becomes trivial in that limit and corresponds to a decaying mode. However, 
below we shall see that starting from second order, the kernels for $\delta\epsilon$ have contributions at order $\epsilon^1$, as expected from the general scaling discussed above.

The evolution equations for $\delta$ and $\theta$ can be written as
\be
  \left(\partial_\eta+{\bf 1}+\Omega_\text{SPT}\right)\left(F_{1,\delta}\atop F_{1,\theta}\right) = -k^2\left( 0 \atop F_{1,g}+F_{1,\delta\epsilon}+\epsilon F_{1,A}\right)\,,
\ee
where
\be
  \Omega_\text{SPT}=\left(\begin{array}{cc} 0 &-1\\ -\frac32 &\frac12\end{array}\right)\,,
\ee
is the standard EdS-SPT evolution matrix, ${\bf 1}$ denotes the unit matrix, and the right-hand side contains the impact of velocity dispersion.
We can integrate out this equation to write the formal solution as
\be
  \left(F_{1,\delta}\atop F_{1,\theta}\right) = \left(1\atop 1\right)+\int^\eta d\eta' \,e^{\eta'-\eta}g_\text{SPT}(\eta-\eta')\left( 0 \atop {\cal S}(k,\eta')\right)\,,
\ee
where we used SPT initial conditions in the far past (which specifies the lower limit of integration),  
and $g_\text{SPT}(\eta-\eta')$ is the standard EdS-SPT linear Green function~\cite{Scoccimarro:1998b} with ${\cal S}(k,\eta)\equiv -k^2(F_{1,g}+F_{1,\delta\epsilon}+\epsilon F_{1,A})$.
We obtain the ${\cal O}(\epsilon)$ corrections to $F_{1,\delta}$ and $F_{1,\theta}$ by inserting the lowest order results for ${\cal S}$, giving
\bea\label{eq:F1correction}
  F_{1,\delta}(k,\eta) &=& 1 - k^2I_\delta(\eta)+{\cal O}(\epsilon^2)\,,\nn\\
  F_{1,\theta}(k,\eta) &=& 1 - k^2I_\theta(\eta)+{\cal O}(\epsilon^2)\,,
\eea
where
\bea
  I_\delta(\eta) &\equiv& \frac25 \int^\eta d\eta'\,\left(1-e^{5(\eta'-\eta)/2}\right)\nn\\
  && {} \left(\epsilon(\eta')+2\int^{\eta'} d\eta''\,e^{2(\eta''-\eta')}\epsilon(\eta'')\right)\,,\nn\\
  I_\theta(\eta) &\equiv& \frac25 \int^\eta d\eta'\,\left(1+\frac32 e^{5(\eta'-\eta)/2}\right)\nn\\
  && {} \left(\epsilon(\eta')+2\int^{\eta'} d\eta''\,e^{2(\eta''-\eta')}\epsilon(\eta'')\right)\,.
\eea
Since $\epsilon\geq 0$ and $\eta'\leq \eta$ we observe that the correction term has a negative sign irrespective of the precise time-dependence of $\epsilon(\eta)$, such that
velocity dispersion necessarily leads to a \emph{suppression} relative to SPT at first order in $\epsilon$. The integrals can be simplified using the relation
\bea
  &&\int^\eta d\eta'\,e^{a(\eta'-\eta)}\int^{\eta'} d\eta''\,e^{b(\eta''-\eta')}\,f(\eta'')\nn\\
  &&= -\frac{1}{a-b} \int^\eta d\eta'\,(e^{a(\eta'-\eta)} - e^{b(\eta'-\eta)})\,f(\eta')\,,
\eea
for $a\not= b$. Using the definition of Eq.~\eqref{eq:epsweight} of the time-weighted background dispersion one finds,
\bea
  I_\delta(\eta) &=& \frac45 E_0(\eta) -2 E_2(\eta) +\frac65 E_{5/2}(\eta)\,,\nn\\
  I_\theta(\eta) &=& \frac45 E_0(\eta) +2 E_2(\eta) -\frac95 E_{5/2}(\eta)\,.
\eea
Proceeding similarly for the third cumulant perturbations gives
\bea
  F_{1,\pi} &=& F_{1,\chi} + 3k^2\int^\eta d\eta'\,e^{5(\eta'-\eta)/2}\epsilon(\eta')F_{1,g}(k,\eta')\,,\nn\\
  F_{1,\chi} &=& k^2\int^\eta d\eta'\,e^{5(\eta'-\eta)/2}(5\epsilon(\eta')F_{1,\delta\epsilon}(k,\eta')\nn\\
  && {} +\omega(\eta')F_{1,A}(k,\eta'))\,,
\eea
which readily yields the leading ${\cal O}(\epsilon^2)$ contribution when inserting the lowest order expressions $F_{1,A}|_{\epsilon^0}=1$, $F_{1,\delta\epsilon}|_{\epsilon^1}=0$ and $F_{1,g}|_{\epsilon^1}$
from Eq.~\eqref{eq:Tgsmalleps} on the right-hand side
and recalling the power counting $\omega\propto\epsilon^2$.

\subsubsection{Second order kernels}

At second order in perturbation theory we can proceed analogously, taking the vertices on the right-hand side of Eq.~\eqref{eq:kernelODE} into account.
For example, the differential equation for the $g$ mode reads
\bea
 (\partial_\eta+3)F_{2,g}(\vec p,\vec q,\eta) = 2\epsilon(\eta)F_{2,\theta}(\vec p,\vec q,\eta) -F_{2,\pi}(\vec p,\vec q,\eta)\nn\\
  + \frac35 F_{2,\chi}(\vec p,\vec q,\eta) 
  + \gamma_{gbc}(\vec p,\vec q)F_{1,b}(p,\eta)F_{1,c}(q,\eta)\,,\nn\\
\eea
where summation over all $b,c$ is implied, with contributions coming in general
from $\gamma_{g\theta g}$, $\gamma_{g\theta\epsilon}$, $\gamma_{gA \pi}$, $\gamma_{gA\chi}$, $\gamma_{gw_i g}$, $\gamma_{gw_i\epsilon}$, $\gamma_{g\theta\nu_i}$, $\gamma_{g\theta t_{ij}}$, $\gamma_{gw_i\nu_j}$, $\gamma_{gw_i t_{jk}}$,
as well as corresponding contributions with the second two entries in the subscript flipped. All of these vertices are elementary functions of their arguments and are
given explicitly in paper~I~\cite{cumPT}. Note that $\epsilon$ in the subscript stands for the mode $\delta\epsilon$. At second order, none of the vertices involving vorticity, vector or tensor modes contribute since $F_{1,a}=0$ for them.
In addition, for the leading (linear) contribution in the expansion in $\epsilon$ we can omit third cumulant contributions and also those involving $\delta\epsilon$, since $F_{1,\delta\epsilon}$ starts at $\epsilon^2$.
This means only 
\be
  \gamma_{g \theta g}(\vec p,\vec q) = \frac12\frac{\vec p\cdot \vec q}{q^2}\left(\frac{(\vec p+\vec q)^2}{p^2}+\frac12-\frac32\frac{((\vec p+\vec q)\cdot \vec p)^2}{(\vec p+\vec q)^2p^2}\right)\,,
\ee
and $\gamma_{gg\theta}(\vec p,\vec q)=\gamma_{g\theta g}(\vec q,\vec p)$ need to be kept at ${\cal O}(\epsilon)$.
Using in addition that $F_{2,\theta}(\vec p,\vec q,\eta)|_{\epsilon^0}=G_2(\vec p,\vec q)$ is equal to the usual EdS-SPT second order velocity kernel
and $F_{1,g}|_{\epsilon^1}$ from~\eqref{eq:Tgsmalleps} yields
\bea
  F_{2,g}(\vec p,\vec q,\eta) &=& 2G_2(\vec p,\vec q)E_3(\eta) + 2(\gamma_{g \theta g}(\vec p,\vec q) \nn\\
  && {} +\gamma_{g \theta g}(\vec q,\vec p))(E_2(\eta)-E_3(\eta))\nn\\
  && {} +{\cal O}(\epsilon^2)\,.
\eea
Analogously one finds
\bea
  F_{2,\delta\epsilon}(\vec p,\vec q,\eta) &=&  2(\gamma_{\epsilon \theta g}(\vec p,\vec q)+\gamma_{\epsilon \theta g}(\vec q,\vec p))(E_2(\eta)-E_3(\eta))\nn\\
  && {} +{\cal O}(\epsilon^2)\,,
\eea
where~\cite{cumPT}
\be
  \gamma_{\epsilon \theta g}(\vec p,\vec q) = \frac12\frac{\vec p\cdot\vec q}{2p^2q^2(\vec p+\vec q)^2}((\vec p\cdot \vec q)^2-p^2q^2)\,.
\ee
Note that, while $F_{1,\delta\epsilon}$ starts only at order $\epsilon^2$, the second-order kernel $F_{2,\delta\epsilon}$ has a non-zero contribution
already at (leading) order $\epsilon^1$. The same is true for higher-order kernels of $\delta\epsilon$.

For the density and velocity divergence kernels we use
\be
  \left(\partial_\eta+2\cdot{\bf 1}+\Omega_\text{SPT}\right)\left(F_{2,\delta}\atop F_{2,\theta}\right) = \left( \gamma_{\delta bc}F_{1,b}F_{1,c} \atop \gamma_{\theta bc}F_{1,b}F_{1,c}+{\cal S}_2\right)\,,
\ee
where we omit the arguments for brevity and set ${\cal S}_2\equiv -(\vec p+\vec q)^2(F_{2,g}+F_{2,\delta\epsilon}+\epsilon F_{2,A})$.
Possible vertices are $\gamma_{\delta\theta\delta}, \gamma_{\delta w_i\delta}$ and $\gamma_{\theta\theta\theta}$, $\gamma_{\theta A g}$, $\gamma_{\theta A \epsilon}$, $\gamma_{\theta w_i\theta}$, $\gamma_{\theta w_iw_i}$, $\gamma_{\theta A\nu_i}$, $\gamma_{\theta At_{ij}}$  given in paper~I~\cite{cumPT} along with the flipped contributions. As before, only scalar vertices contribute to $F_{2,\delta}$ and $F_{2,\theta}$, while contributions involving $F_{1,\delta\epsilon}$ matter only at order $\epsilon^2$. Up to first order in $\epsilon$ the relevant vertices therefore amount to the two SPT vertices $\gamma_{\delta\theta\delta}(\vec p,\vec q)=\alpha(\vec p,\vec q)/2=(\vec p+\vec q)\cdot\vec p/(2p^2)$
and $\gamma_{\theta\theta\theta}(\vec p,\vec q)=\beta(\vec p,\vec q)=(\vec p+\vec q)^2\vec p\cdot\vec q/(2p^2q^2)$ as well as 
\be
  \gamma_{\theta A g}=-((\vec p+\vec q)\cdot \vec q) (\vec p\cdot \vec q)/(2q^2)\,.
\ee
Writing the equation in integral form
\bea\label{eq:F2analytIntegralForm}
  \left(F_{2,\delta}\atop F_{2,\theta}\right) &=& \int^\eta d\eta' \,e^{2(\eta'-\eta)}g_\text{SPT}(\eta-\eta')\nn\\
  && {} \times \left( \gamma_{\delta bc}F_{1,b}F_{1,c} \atop \gamma_{\theta bc}F_{1,b}F_{1,c}+{\cal S}_2\right)\Bigg|_{\eta'}\,,
\eea
and expanding the right-hand side up to linear order in $\epsilon$ yields
\bea\label{eq:F2analytRaw}
  F_{2,\delta}(\vec p,\vec q,\eta) &=& F_2(\vec p,\vec q) 
   - (\vec p+\vec q)^2\sum_{j=1}^7 \Gamma_j(\vec p,\vec q)J^\delta_j(\eta)\nn\\
  && {} + {\cal O}(\epsilon^2)\,,\nn\\
  F_{2,\theta}(\vec p,\vec q,\eta) &=& G_2(\vec p,\vec q) 
   - (\vec p+\vec q)^2\sum_{j=1}^7 \Gamma_j(\vec p,\vec q)J^\theta_j(\eta)\nn\\
  && {} + {\cal O}(\epsilon^2)\,,
\eea
where the time-dependence at linear order in $\epsilon$ is given by the integrals $J_j^{\delta/\theta}(\eta)$, with explicit form given in App.~\ref{app:timeintegrals},
and the dependence on wavenumber arising from the various non-linear vertices can be written as
\bea
  \Gamma_1(\vec p,\vec q)&\equiv& G_2(\vec p,\vec q)\,,\nn\\
  \Gamma_2(\vec p,\vec q)&\equiv& \gamma_{g \theta g}(\vec p,\vec q)+\gamma_{g \theta g}(\vec q,\vec p)+\gamma_{\epsilon \theta g}(\vec p,\vec q)\nn\\
  && {} +\gamma_{\epsilon \theta g}(\vec q,\vec p)\,,\nn\\
  \Gamma_3(\vec p,\vec q)&\equiv& F_{2,A}(\vec p,\vec q,\eta)|_{\epsilon^0}\nn\\
  &=& \left[ G_2(\vec p,\vec q) + \gamma_{A\theta A}(\vec p,\vec q)+ \gamma_{A\theta A}(\vec q,\vec p)\right]/2\,,\nn\\
  \Gamma_4(\vec p,\vec q)&\equiv& (p^2+q^2)\gamma_{\theta\theta\theta}(\vec p,\vec q)/(\vec p+\vec q)^2\,,\nn\\
  \Gamma_5(\vec p,\vec q)&\equiv& -(\gamma_{\theta Ag}(\vec p,\vec q)+\gamma_{\theta Ag}(\vec q,\vec p))/(\vec p+\vec q)^2\,,\nn\\
  \Gamma_6(\vec p,\vec q)&\equiv& (p^2\gamma_{\delta\theta\delta}(\vec p,\vec q)+q^2\gamma_{\delta\theta\delta}(\vec q,\vec p))/(\vec p+\vec q)^2\nn\\
  &=& 1/2\,,\nn\\
  \Gamma_7(\vec p,\vec q)&\equiv& (q^2\gamma_{\delta\theta\delta}(\vec p,\vec q)+p^2\gamma_{\delta\theta\delta}(\vec q,\vec p))/(\vec p+\vec q)^2\,,
\eea
with $\gamma_{A\theta A}(\vec p,\vec q)=\vec p\cdot\vec q/(2q^2)$.
Each of the $\Gamma_j$ approaches a constant when the sum of wavenumbers goes to zero, leading to a scaling of $F_{2,\delta}$ in accordance with the requirement from mass and momentum conservation (see Sec.\,\ref{sec:ktozerolimit}).

\begin{table}
  \centering
  \caption{Expansion coefficients in Eq.~\eqref{eq:Gammajexpansion} of the functions $\Gamma_j(\vec p,\vec q)$ in terms of the four basis functions given by Eq.~\eqref{eq:abcd}.}
  \begin{ruledtabular}
    \begin{tabular}{c|ccccccc} 
   &&&&&&&\\[-2ex]
   $j$ & $1$ & $2$ & $3$ & $4$ & $5$ & $6$ & $7$  \\[1.5ex] \hline &&&&&&&\\[-1.ex]
   $\beta_{j,a}$ & $0$ & $-1$ & $\frac{1}{2}$ & $1$ & $-\frac{1}{2}$ & $-\frac{1}{2}$ & $\frac{3}{2}$ \\[1.5ex]
   $\beta_{j,b}$ &  $1$ & $2$ & $\frac{1}{2}$ & $0$ & $\frac{1}{2}$ & $\frac{1}{2}$ & $-\frac{1}{2}$ \\[1.5ex]
   $\beta_{j,c}$ &  $0$ & $-\frac{3}{14}$ & $\frac{5}{14}$ & $\frac{5}{7}$ & $\frac{1}{7}$ & $-\frac{5}{14}$ & $\frac{1}{14}$ \\[1.5ex]
   $\beta_{j,d}$ & $\frac{2}{7}$ & $\frac{13}{14}$ & $\frac{1}{7}$ & $0$ & $-\frac{1}{7}$ & $-\frac{1}{7}$ & $\frac{1}{7}$ \\
    \end{tabular}
  \end{ruledtabular}
  \label{tab:betajabcd}
\end{table}

Using the explicit expressions for the vertices, one can check that each
\be\label{eq:Gammajexpansion}
\Gamma_j = \beta_{j,a}\Delta_a+\beta_{j,b}\Delta_b+\beta_{j,c}\Delta_c+\beta_{j,d}\Delta_d\,,
\ee
can be expressed in terms the four basis functions
\bea\label{eq:abcd}
  \Delta_a(\vec p,\vec q) &\equiv& (p^2+q^2)F_2(\vec p,\vec q)/(\vec p+\vec q)^2\,,\nn\\
  \Delta_b(\vec p,\vec q) &\equiv& F_2(\vec p,\vec q)\,,\nn\\
  \Delta_c(\vec p,\vec q) &\equiv& (p^2+q^2)K(\vec p,\vec q)/(\vec p+\vec q)^2\,,\nn\\
  \Delta_d(\vec p,\vec q) &\equiv& K(\vec p,\vec q) \equiv (\vec p\cdot\vec q)^2/(p^2q^2)-1\,,\
\eea
with coefficients $\beta_{j,k}$ given in Table~\ref{tab:betajabcd}. This basis has been introduced in~\cite{Eggemeier:2018qae}, and the present result proves equation (109) therein. Here $K$ is the kernel corresponding to the second order Galileon operator ${\cal G}_2$ in~\cite{Eggemeier:2018qae}, which physically represents the tidal field.

The decomposition in Eq.~\eqref{eq:Gammajexpansion} implies that one can write the first-order correction in $\epsilon$ to the second-order density kernel $F_{2,\delta}$ also as
\bea\label{eq:F2analyt_abcd}
  F_{2,\delta}(\vec p,\vec q,\eta) &=& F_2(\vec p,\vec q) 
   - (\vec p+\vec q)^2\sum_{k=a,b,c,d}\Delta_k(\vec p,\vec q)\beta^\delta_k(\eta)\nn\\
  && {}  + {\cal O}(\epsilon^2)\,,\nn\\
  F_{2,\theta}(\vec p,\vec q,\eta) &=& G_2(\vec p,\vec q) 
   - (\vec p+\vec q)^2\sum_{k=a,b,c,d}\Delta_k(\vec p,\vec q)\beta^\theta_k(\eta)\nn\\
  && {}  + {\cal O}(\epsilon^2)\,,
\eea
with coefficients $\beta^{\delta/\theta}_k(\eta)$ for $k=a,b,c,d$ given by linear combinations of the $J^{\delta/\theta}_j(\eta)$ with coefficients $\beta_{j,k}$ from Table~\ref{tab:betajabcd}.
Using the results from App.~\ref{app:timeintegrals} we find
\bea\label{eq:betaa}
  \beta^\delta_a(\eta) &=& \frac25 (2 E_{0}(\eta) - 6 E_{1}(\eta) + 5 E_{2}(\eta) + 3 E_{5/2}(\eta) \nn\\
  && {} -    5 E_3(\eta) + E_{7/2}(\eta))\,,\nn\\
  \beta^\delta_b(\eta) &=& \frac25 (6 E_{1}(\eta) - 10 E_{2}(\eta) + 5 E_3(\eta) - E_{7/2}(\eta)) \,,\nn\\
  \beta^\delta_c(\eta) &=& \frac{1}{35} (3 E_{1}(\eta) - 20 E_{2}(\eta) + 30 E_{5/2}(\eta) - 15 E_3(\eta) \nn\\
  && {} +    2 E_{7/2}(\eta))\,,\nn\\
  \beta^\delta_d(\eta) &=& \frac{3}{35} (3 E_{1}(\eta) - 10 E_{2}(\eta) + 15 E_3(\eta) - 8 E_{7/2}(\eta)) \,,\nn\\
\eea
and
\bea
  \beta^\theta_a(\eta) &=& \frac{1}{10} (8 E_{0}(\eta) - 24 E_{1}(\eta) + 40 E_{2}(\eta) - 33 E_{5/2}(\eta) \nn\\
  && {} +    20 E_3(\eta) - 6 E_{7/2}(\eta))\,,\nn\\
  \beta^\theta_b(\eta) &=& \frac{1}{10} (24 E_{1}(\eta) - 20 E_{2}(\eta) + 15 E_{5/2}(\eta) - 20 E_3(\eta) \nn\\
  && {} +    6 E_{7/2}(\eta))\,,\nn\\
  \beta^\theta_c(\eta) &=& \frac{1}{70} (16 E_{0}(\eta) + 6 E_{1}(\eta) + 60 E_{2}(\eta) - 81 E_{5/2}(\eta)\nn\\
  && {} +    30 E_3(\eta) - 6 E_{7/2}(\eta))\,,\nn\\
  \beta^\theta_d(\eta) &=& \frac{1}{35} (9 E_{1}(\eta) + 20 E_{2}(\eta) - 15 E_{5/2}(\eta) - 45 E_3(\eta) \nn\\
  && {} +    36 E_{7/2}(\eta))\,.
\eea

The kernels for the third cumulant perturbations $\pi$ and $\chi$ start at order $\epsilon^2$, and can be obtained by inserting the results up to first order from above
in the right-hand side of their equations of motion
\bea
  \left(\partial_\eta+\frac72\right)F_{2,\pi} &=& 3k^2\epsilon F_{2,g}+5k^2\epsilon F_{2,\delta\epsilon}+k^2\omega F_{2,A}\nn\\
  && {} +\gamma_{\pi bc}F_{1,b}F_{1,c}\,,\nn\\
  \left(\partial_\eta+\frac72\right)F_{2,\chi} &=& 5k^2\epsilon F_{2,\delta\epsilon}+k^2\omega F_{2,A} +\gamma_{\chi bc}F_{1,b}F_{1,c}\,,\nn\\
\eea
and using the respective vertices.

In addition, at second order, also vorticity, vector and tensor modes are generated. The corresponding kernels are discussed in Sec.\,\ref{sec:vorticity} and Sec.\,\ref{sec:tensor}, respectively.

In principle, with the same strategy one can obtain the correction terms to the third-order kernels $F_{3,a}$ at linear order in $\epsilon$. Similarly, it is possible to systematically expand to higher powers in $\epsilon$.
Nevertheless, the expressions become too lengthy to be shown, and we revert to a numerical treatment, that does \emph{not} require any expansion in $\epsilon$ and therefore remains valid when
$\epsilon k_i^2$ is sizeable.

\subsubsection{Wilson coefficients}
\label{sec:Wilson}

Note that the four basis functions in Eq.~\eqref{eq:abcd} appearing in the order $\epsilon$ correction to the
density kernel $F_{2,\delta}(\vec p,\vec q,\eta)$, Eq.~\eqref{eq:F2analyt_abcd}, are equivalent to the shape functions introduced in the context of the effective field theory (EFT) treatment of the
bispectrum~\cite{Baldauf:2015,Angulo:2015}\footnote{Using the basis functions $E_{1,2,3}$ and $\Gamma$ in the convention of~\cite{Baldauf:2021zlt} one has
$E_1=k^2(-\Delta_a+\Delta_b-\frac57\Delta_c-\frac27\Delta_d)$,
$E_2=k^2(-\frac23\Delta_a+\frac23\Delta_b-\frac{10}{21}\Delta_c+\frac{17}{21}\Delta_d)$,
$E_3=k^2(-\frac13\Delta_a+\frac13\Delta_b+\frac{11}{42}\Delta_c-\frac{2}{21}\Delta_d)$,
$\Gamma=k^2(\frac{2}{11}\Delta_a+\frac{9}{11}\Delta_b-\frac{1}{7}\Delta_c-\frac{6}{77}\Delta_d)$,
where $k^2\equiv (\vec p+\vec q)^2$.}. Indeed, within the effective field theory language the coefficients $\beta^\delta_a,\beta^\delta_b,\beta^\delta_c,\beta^\delta_d$
are the set of four counter-terms (or rather Wilson coefficients) at second order in perturbation theory and at leading order in the derivative expansion.
The result obtained here can be viewed as a first-principle \emph{determination} of those Wilson coefficients by matching the EFT to the exact UV theory, being the Vlasov equation (at least when considering cold dark matter only). 
Nevertheless, this does not mean that their values are identical to the corresponding counter-terms obtained when complementing an SPT loop computation with correction terms
of the form of Eq.~\eqref{eq:abcd} and fitting them to simulation results, since these counter-terms have to absorb also the unphysical contributions from the UV region of the SPT loop integration (or alternatively account for the missing modes when imposing a cutoff). Instead, the Wilson coefficients $\beta^\delta_a,\beta^\delta_b,\beta^\delta_c,\beta^\delta_d$ obtained here capture the actual impact of non-linearly induced dark matter velocity dispersion on the second-order kernel. Similarly, $I^\delta$ appearing in the $\epsilon$ correction to $F_{1,\delta}$, see Eq.~\eqref{eq:F1correction}, may be viewed as the result of matching the EFT correction at first order in perturbations and leading order in gradients to the Vlasov theory. 

Apart from that, we stress that we do not expand in powers of $\epsilon$ in our numerical
analysis (see below), and therefore our results go beyond the four correction terms in Eq.~\eqref{eq:abcd}.
For example, formally, the contributions to $F_{2,\delta}$ at order $\epsilon^2$ would correspond to additional higher-derivative operators in the EFT with two additional spatial gradients compared to those captured by Eq.~\eqref{eq:abcd}.
At order $\epsilon^3$, the EFT matching would require to include operators with four additional gradients, and so on. Furthermore, the correction to the third-order kernel $F_{3,\delta}$ would correspond to EFT operators at third order in perturbation theory, etc. All of these contributions are implicitly contained in the kernels $F_{n,a}$ considered here.

\subsection{Numerical results}\label{sec:numericalkernels}

\begin{table}
  \centering
  \caption{Background values and perturbation modes taken into account in various approximation schemes of \vpt.}
  \begin{ruledtabular}
    \begin{tabular}{l|ccccc} 
   &&&&&\\[-2ex]
              &\multicolumn{2}{c}{\bf s} & {\bf sw} & {\bf sv} & {\bf svt} \\[1.5ex] \hline &&&&&\\[-1.ex]
  {\bf cum2}  &$\epsilon(\eta)$ & $\delta,\theta,g,\delta\epsilon,A$ & $+w_i$ & $+\nu_i$ & $+t_{ij}$ 
  \\[1.5ex]
  {\bf cum3+} &$\epsilon(\eta), \omega(\eta)$ & $\delta,\theta,g,\delta\epsilon,A,\pi,\chi$ & $+w_i$ & $+\nu_i$ & $+t_{ij}$ 
  \\
    \end{tabular}
  \end{ruledtabular}
  \label{tab:approxschemes}
\end{table}

\begin{figure*}[t]
  \begin{center}
  \raisebox{0mm}{\includegraphics[width=\columnwidth]{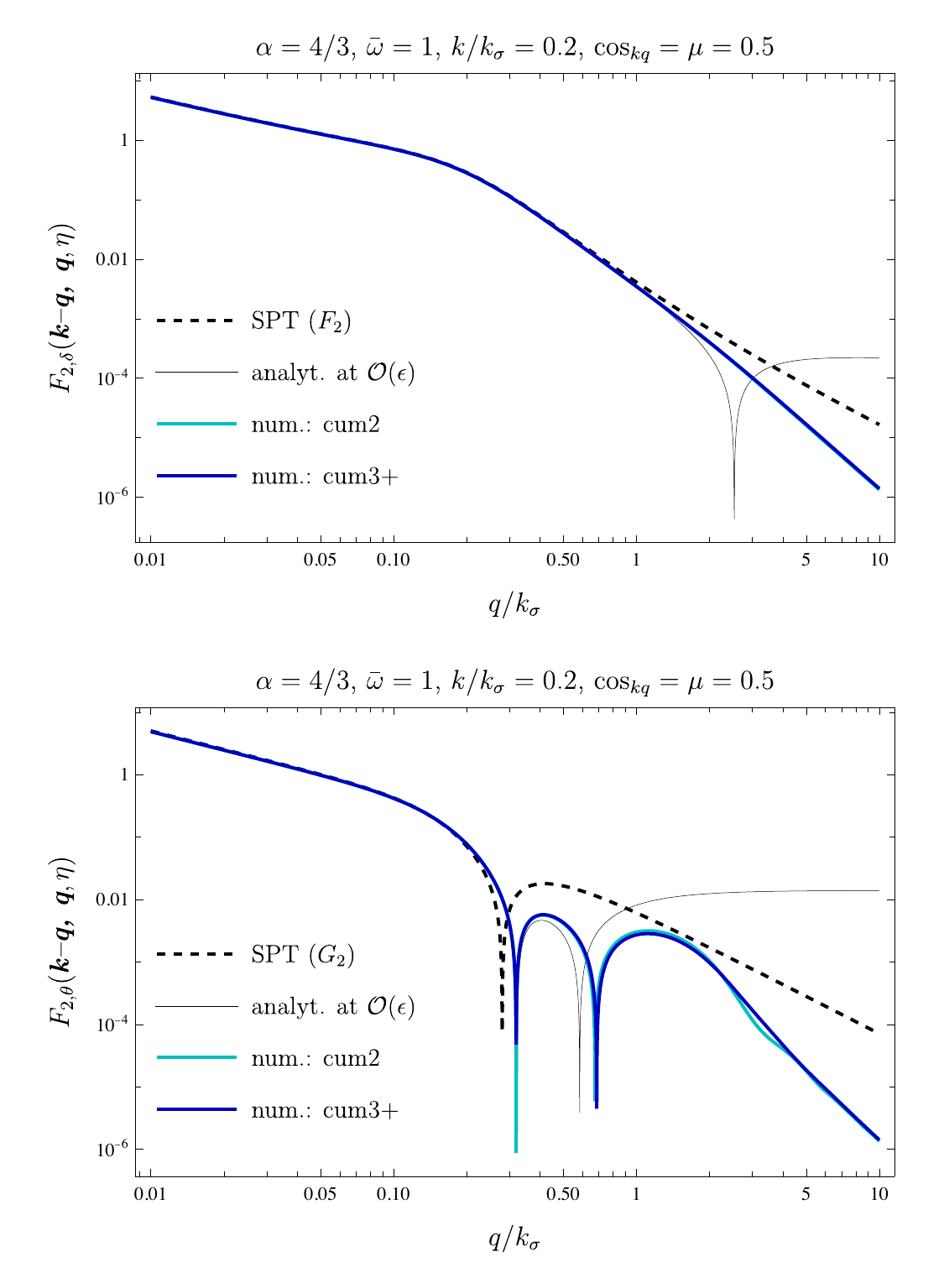}}
  \includegraphics[width=\columnwidth]{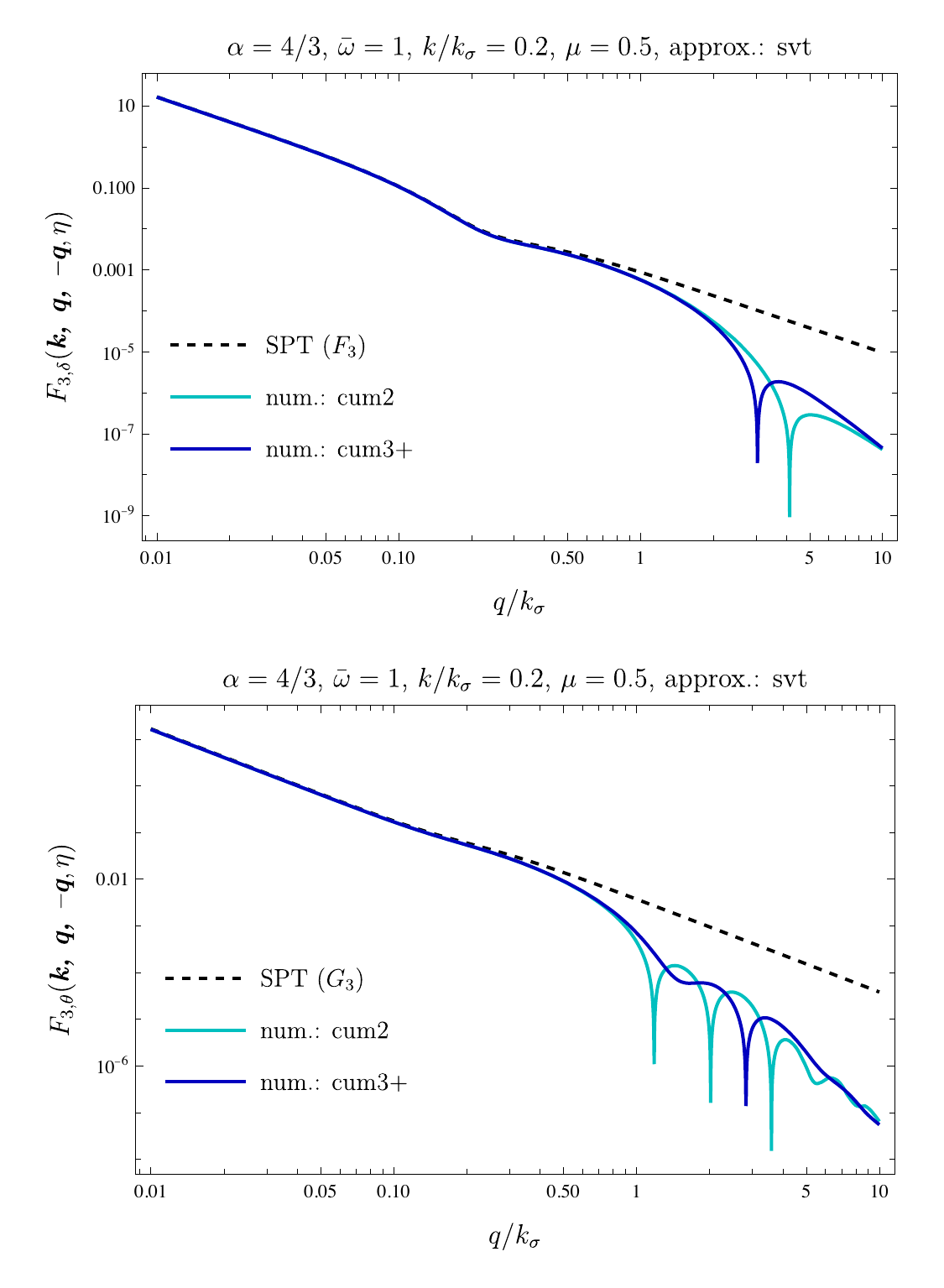}
  \end{center}
  \caption{\label{fig:F23_delta} 
  Non-linear kernels $F_{2,a}(\vec k-\vec q,\vec q,\eta)$ (left column) and $F_{3,a}(\vec k,\vec q,-\vec q,\eta)$ (right column)
  versus $q=|\vec q|$ in \vpt~(blue lines) compared to SPT (black dashed) for the density contrast $a=\delta$ (top row) and velocity divergence $a=\theta$ (bottom row).
  For the background dispersion we chose $\epsilon(\eta)=\epsilon_0\,e^{\alpha\eta}$ as well as $\omega(\eta)=\bar\omega\times\epsilon(\eta)^2$,
  with parameters as indicated in the figures, and $\eta=0$. 
  We show the dependence on $q$ relative to the dispersion scale $k_\sigma$, such that the result applies to any value of $\epsilon_0=1/k_\sigma^2$.
  The blue lines show second and third cumulant approximations (see Table~\ref{tab:approxschemes}), which are both suppressed compared to the EdS-SPT kernels for $q\gtrsim k_\sigma$ (shown in black dashed lines). The thin black solid line shows the analytical result of Eq.~\eqref{eq:F2analyt_abcd} at first order in $\epsilon$.
  }
\end{figure*}

\begin{figure}[t]
  \begin{center}
  \includegraphics[width=\columnwidth]{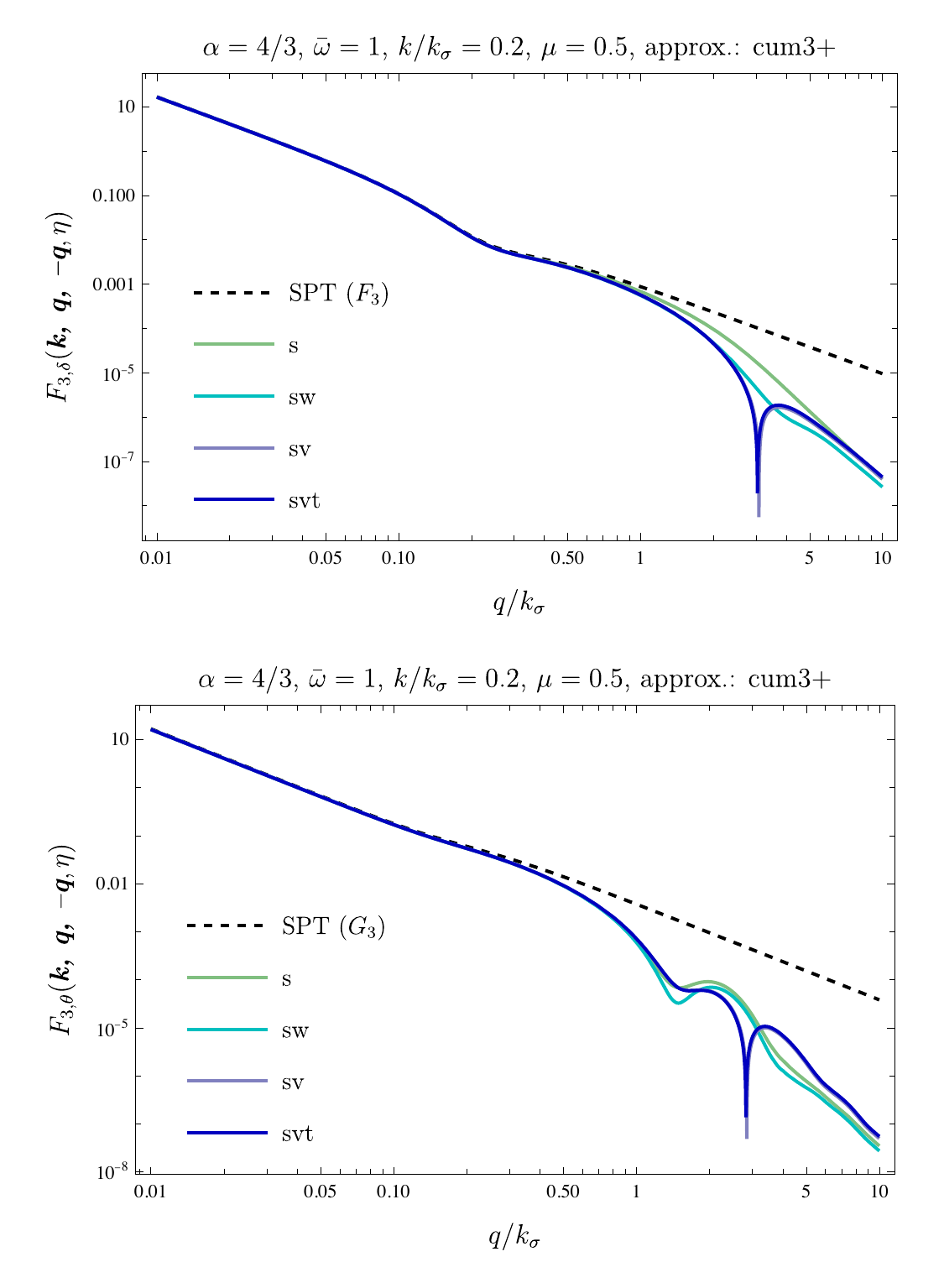}
  \end{center}
  \caption{\label{fig:F3_delta_svt} 
  Non-linear kernels $F_{3,a}(\vec k,\vec q,-\vec q,\eta)$ versus $q=|\vec q|$ for $a=\delta,\theta$ as in Fig.\,\ref{fig:F23_delta}, but comparing the impact of various approximations related to vorticity, as well as vector and
  tensor modes of the velocity dispersion tensor (see Table~\ref{tab:approxschemes}). Including vorticity backreaction has a significant impact on $F_{3,\delta}$ within the most relevant regime $k_\sigma \lesssim q\lesssim 2k_\sigma$ where the suppression of \vpt~relative to SPT sets in (({\bf sw}) vs ({\bf s}) in upper panel).
  }
\end{figure}

In our analysis we solve the system of coupled differential equations, Eq.~\eqref{eq:kernelODE}, for the non-linear kernels numerically
using a recursive algorithm~\cite{Blas:2015tla,Garny:2020ilv}. 

We consider the approximation schemes given in Table~\ref{tab:approxschemes}.
The most inclusive scheme is ({\bf cum3+/svt}), for which apart from $\delta$ and $\theta$ also
all scalar, vector and tensor perturbations of the dispersion tensor $\epsilon_{ij}$, its background expectation value $\epsilon(\eta)$, as well
as the scalar modes $\pi$ and $\chi$ of the third cumulant are taken into account. Furthermore, their
evolution equations contain the expectation value $\omega(\eta)$ of the fourth cumulant, that we also include.
For comparison, we consider approximations ({\bf cum2}) that omit the third cumulant perturbations. In that case,
$\omega(\eta)$ does not enter in the perturbation equations. In addition, we consider various approximations related to scalar, vector and tensor modes.
For approximation scheme ({\bf s}), only scalar modes are included. For ({\bf sw}) we include in addition vorticity. Within the schemes ({\bf sv}) and ({\bf svt}) also
the vector and tensor modes of $\epsilon_{ij}$ are added successively.

Since the second and higher cumulants are expected to play a role only at low redshift, we initialize the kernels at some finite time $\eta_\text{ini}$, using EdS-SPT kernels
for the density and velocity divergence, and setting all other kernels to zero (except $A$, which is initialized as $\delta$),
\bea \label{VlasovHics}
  F_{n,\delta}(\vec k_1,\dots ,\vec k_n,\eta_\text{ini}) &\equiv& F_n(\vec k_1,\dots ,\vec k_n)\,,\nn\\
  F_{n,\theta}(\vec k_1,\dots ,\vec k_n,\eta_\text{ini}) &\equiv& G_n(\vec k_1,\dots ,\vec k_n)\,,\nn\\
  F_{n,A}(\vec k_1,\dots ,\vec k_n,\eta_\text{ini}) &\equiv& F_n(\vec k_1,\dots ,\vec k_n)\,,\nn\\
  F_{n,a}(\vec k_1,\dots ,\vec k_n,\eta_\text{ini}) &\equiv& 0, \qquad a=w_i,g,\delta\epsilon,\nu_i,t_{ij},\pi,\chi\,.\nn\\
\eea
The most important part of the initial condition are the first two lines, and in particular for the linear kernels $n=1$. These conditions must be supplemented by an initial condition on the background dispersion $\epsilon$ of infinitesimal value, which makes the Vlasov cumulant hierarchy evolution consistent with the solution of the Vlasov equation\footnote{See Sec.~III.B in~\cite{PueSco0908} for more discussion on this.}, otherwise no dispersion and higher cumulants would be generated by time evolution out of Eqs.~(\ref{VlasovHics}). 
As long a $\eta_\text{ini}$ is chosen early enough, the results at low redshift are insensitive to the choice of the initial condition for the higher kernels with $n>1$, as well as for the perturbations of $A$ and the second and higher cumulants, since the dominant contribution is generated by the time-evolution at late times. In practice we use $\eta_\text{ini}=-20$, and checked that our results are insensitive to the precise choice of the initial time, as expected since transients will be down by at least $\exp(-20)$. Our implementation for the vector and tensor modes is described in App.\,\ref{app:vorticity}.

\begin{figure}[t]
  \begin{center}
  \includegraphics[width=\columnwidth]{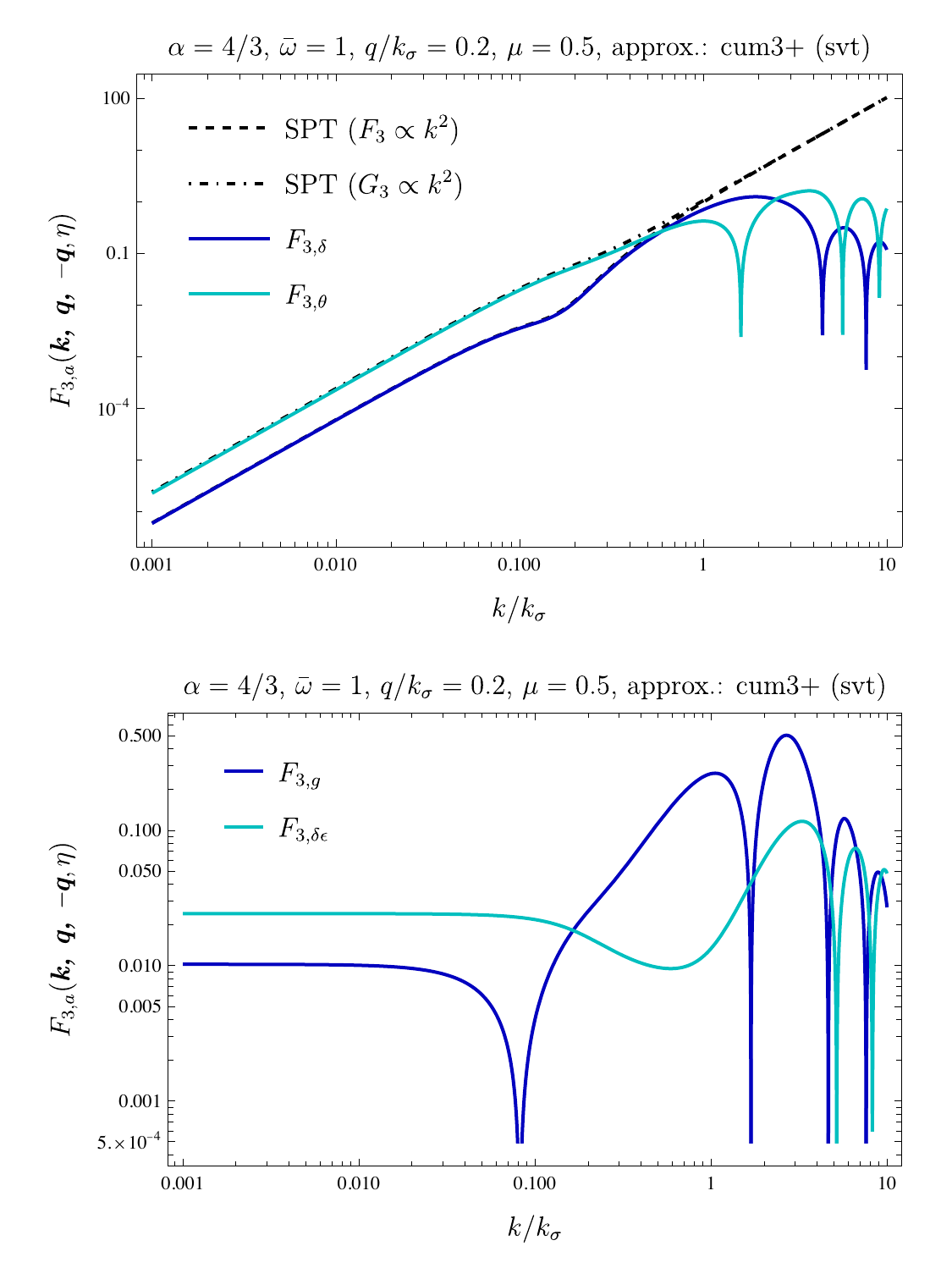}
  \end{center}
  \caption{\label{fig:F3_dteg} 
  Non-linear kernels $F_{3,a}(\vec k,\vec q,-\vec q,\eta)$ versus $k=|\vec k|$
  for $a=\delta,\theta$ (top) and the scalar perturbations $a=g,\delta\epsilon$
  of the stress tensor (bottom). $F_{3,\delta}\propto k^2$ for $k\to 0$ is ensured by mass and momentum conservation, while for
  $F_{3,\theta}$ it is a consequence of the symmetry $\vec q\to-\vec q$, see Sec.\,\ref{sec:ktozerolimit}. In contrast, the kernels of the velocity dispersion
  modes approach a constant at low $k$.
  }
\end{figure}

The numerical result for the kernels $F_{2,a}(\vec k-\vec q,\vec q,\eta)$ and $F_{3,a}(\vec k,\vec q,-\vec q,\eta)$, that enter in the one-loop density ($a=\delta$)
and velocity divergence ($a=\theta$) power spectra, Eq.~\eqref{eq:P1Ldef}, are shown in Fig.\,\ref{fig:F23_delta} (see also Fig.\,\ref{fig:F3_delta_svt}).
We show the dependence on the magnitude of the loop wavenumber $q=|\vec q|$ for fixed $k=|\vec k|$
and $\cos_{kq}=\mu=\vec k\cdot\vec q/(k q)$, as well as assuming a power-law time-dependence $\epsilon(\eta)=\epsilon_0\,e^{\alpha\eta}$ and $\omega(\eta)=\bar\omega\times\epsilon(\eta)^2$, with parameters as stated in the figure. Note that the kernels are dimensionless, such that the result is independent of $\epsilon_0=1/k_\sigma^2$ when plotted against the ratio $q/k_\sigma$.
The numerical results for the density kernels (in blue) approach the corresponding EdS-SPT kernels (black dashed) in the limit when both $q\ll k_\sigma$ and $k\ll k_\sigma$. Due to our choice $k/k_\sigma=0.2$, the blue and black dashed curves therefore lie (almost) on top of each other for $q/k_\sigma\ll 1$, as expected. Furthermore, we show the analytical result at order $\epsilon$ (Eq.~\ref{eq:F2analyt_abcd}) for $F_{2,\delta}$ and $F_{2,\theta}$. It agrees with the numerical result for $q/k_\sigma\lesssim 1$, and captures the onset of the deviation of \vpt~from SPT.

In the limit of large loop wavenumber $q$ we observe a substantial \emph{suppression} of the kernels relative to EdS-SPT. This implies that, when integrating over the loop wavenumber $q$, the one-loop correction to the density power spectrum is less sensitive to the UV regime as compared to SPT. This result is in line with theoretical expectations~\cite{Pee80} and in qualitative agreement with numerical findings on the response of the non-linear power spectrum to small-scale initial fluctuations~\cite{NisBerTar1611,NisBerTar1712}, which reflects the decoupling of small-scale modes.  {\em The inclusion of velocity dispersion and higher cumulants in \vpt~therefore addresses one of the major shortcomings of {\rm SPT}, namely the sensitivity to UV modes, by taking the physically expected screening of those perturbation modes into account.}

The asymptotic slope of the kernels $F_{n,\delta/\theta}(\vec k_1,\dots,\vec k_n,\eta)$ in the limit $k\equiv|\sum\vec k_i|\ll k_\sigma\ll |\vec k_i|$ can be understood analytically. Within SPT momentum conservation implies $F_n/G_n\propto k^2/q^2$ in the corresponding limit $k\ll |\vec k_i|$, where $q\equiv\text{max}_i|\vec k_i|$, and the same factor arises in \vpt. In addition, within \vpt~the dispersion scale $k_\sigma$ is relevant. For all wavenumbers with $|\vec k_i| \gg k_\sigma$, the corresponding modes contributing to the kernel have already entered the ``dispersion horizon''. The moment of entry corresponds to time $\eta=\eta_{k_i}$ satisfying the condition $\epsilon(\eta_{k_i})k_i^2=1$. For $\epsilon(\eta)=\epsilon_0e^{\alpha\eta}$ this corresponds to $e^{\eta_{k_i}}=(k_\sigma^2/k_i^2)^{1/\alpha}$. Now, the most important effect is that the linear modes effectively stop growing in the usual way (i.e.~$\propto e^\eta$) once entering the dispersion horizon. This implies that
\bea\label{eq:kernelscalingvpt}
F_{n,\delta/\theta}(\vec k_1,\dots,\vec k_n,\eta)\sim \frac{k^2}{q^2}\, e^{\eta_{k_1}}\, e^{\eta_{k_2}} \cdots e^{\eta_{k_n}}\nn\\
\text{for}\quad k\ll k_\sigma\ll q\,,
\eea
within \vpt. Using $e^{\eta_{k_i}}=(k_\sigma^2/k_i^2)^{1/\alpha}$ this yields a power-law screening of UV modes. Indeed, applying this argument to $F_{2,a}(\vec k-\vec q,\vec q,\eta)\propto q^{-2-4/\alpha}$ and $F_{3,a}(\vec k,\vec q,-\vec q,\eta)\propto q^{-2-4/\alpha}$ yields the slope by which these kernels decay at very large $q$. Using that for a power-law universe $\alpha=4/(n_s+3)$ this furthermore explains the universal $1/q^2$ decay stated in Eq.~\eqref{eq:vptintegrandscaling} and shown in Fig.~\ref{fig:integrand}. Moreover, the argument leading to this scaling is independent of the precise way on how the linear growth stalls when entering the dispersion horizon, and therefore shows that the asymptotic scaling of the \vpt~kernels is universal, independent of the precise approximation scheme. We shall also confirm this with our numerical solutions in the following.

In Fig.~\ref{fig:F23_delta} we compare the (\textbf{cum2}) and (\textbf{cum3+}) approximation schemes (see Table~\ref{tab:approxschemes}). For $F_{2,\delta(\theta)}$ the difference is very small (with both lines being almost on top of each other), while for $F_{3,\delta(\theta)}$ a more noticeable difference arises. But overall, the impact of third cumulant perturbations is small within the regime $q \sim {\cal O}(k_\sigma)$ where the suppression relative to SPT sets in.

In Fig.~\ref{fig:F3_delta_svt}, we show the dependence of the kernel $F_{3,\delta(\theta)}(\vec k,\vec q,-\vec q,\eta)$  on the approximation scheme for vorticity, vector and tensor modes (see Table~\ref{tab:approxschemes}; note that only scalar modes contribute to $F_{2,\delta(\theta)}$). Within the regime $q\lesssim 2k_\sigma$, where the suppression relative to SPT becomes important, the (\textbf{sw}), (\textbf{sv}) and (\textbf{svt}) schemes are all close to each other for $F_{3,\delta}$, while the (\textbf{s}) approximation differs significantly. This implies that the backreaction of vorticity on the third-order density kernel is sizeable, and leads to a substantial suppression of the $F_{3,\delta}$ kernel. The difference between (\textbf{s}) and (\textbf{sw}) captures the impact of vorticity, while including the vector mode of the stress tensor in addition within the (\textbf{sv}) scheme leads to further differences at large $q\gtrsim 2k_\sigma$. On the other hand, we find that the impact of the tensor perturbation of the stress tensor on both $F_{3,\delta}$ and $F_{3,\theta}$ is very minor, as quantified by the difference between (\textbf{sv}) and (\textbf{svt}). We find a qualitatively similar behaviour, including in particular the suppression at large $q$ relative to SPT, independently of the precise choice of the parameters $k$, $\cos_{kq}$, $\alpha$ and $\bar\omega$.

The kernels $F_{3,a}(\vec k,\vec q,-\vec q,\eta)$ of the two scalar perturbation modes $a=g,\delta\epsilon$ of the velocity dispersion tensor are shown in Fig.~\ref{fig:F3_dteg}, as compared to those for $a=\delta,\theta$. Note that we show the dependence on $k$ for fixed $q$ here. For $a=\delta,\theta$ the kernels scale as $k^2$ for small $k$, as in SPT, and guaranteed by mass and momentum conservation for the density, as well as by symmetry $\vec k\to -\vec k$ for $\theta$ (see Sec.\,\ref{sec:limits}). For large $k\gtrsim k_\sigma$ the kernels are suppressed compared to their SPT values, similarly as for the dependence on $q$.

The behaviour for the stress tensor perturbations $a=g,\delta\epsilon$ is qualitatively different for small $k$, where their kernels approach a constant value. This implies that their linear theory values are modified also in the limit $k\to 0$, as quantified e.g. by a ``bias" $b_a\equiv P_{a\delta}/P_{\delta\delta}|_{k\to 0}$. In the limit
$k\to 0$, the bias is given by
\bea
  b_a &=& F_{1,a}(k=0,\eta) + 3\int d^3q \, F_{3,a}(\vec k,\vec q,-\vec q,\eta)\, P_0(q)|_{k\to 0} \nn\\
  && {} + \dots\,,
\eea
where the ellipsis stand for two and higher loop contributions. For $a=g$ the linear contribution
is given by $b_g^\text{lin}=F_{1,g}(k=0,\eta)=2E_2(\eta)$ (see Eq.~\ref{eq:Tgsmalleps}), while for the case $a=\delta\epsilon$ we have $b_{\delta\epsilon}^\text{lin}=0$\footnote{Note that the ${\cal O}(\epsilon^2)$ contributions to $F_{1,a}(k,\eta)$
have to vanish for $a=g,\delta\epsilon$ in the limit $k\to 0$ for dimensional reasons, being proportional to some time integral of either $\epsilon(\eta')\epsilon(\eta'') k^2$ or $\omega(\eta') k^2$.}.

Overall, we conclude that vorticity as well as scalar and vector modes of the velocity dispersion tensor should be taken into account when computing the one-loop correction to the density power spectrum, while third cumulant perturbations and especially the tensor mode are less relevant. Taking vorticity into account is also important to guarantee momentum conservation, as we shall see in the next section.

\section{Symmetry constraints on non-linear kernels}
\label{sec:limits}

The symmetries underlying the Vlasov-Poisson system lead to constraints on the behaviour of the non-linear kernels $F_{n,a}(\vec k_1,\dots,\vec k_n,\eta)$ in particular limiting cases.
In the following we discuss the implications of
\begin{itemize}
\item Galilean invariance in the limit where one of the arguments $\vec k_i\to 0$, corresponding to the impact of a very large scale mode on the propagation on smaller scales, and
\item mass and momentum conservation in the limit when the total wavenumber $\vec k\equiv\sum_i\vec k_i\to 0$, while the individual $\vec k_i$ remain finite, corresponding to the impact of small-scale modes onto perturbation modes on larger scales.
\end{itemize}
The corresponding constraints are well-known within EdS-SPT~\cite{Zel6501,Pee80,GorGriRey86}, and here we discuss and show how they are
generalized when including velocity dispersion and higher cumulants. Since these symmetry constraints guarantee delicate cancellations
among various terms at a given order in perturbation theory, it is important to ensure their validity for a given approximation scheme.
For example, as we will see, and unlike what happens in SPT, momentum conservation requires to take vorticity into account. Finally, the symmetry constraints
also provide a rather non-trivial check of the formalism and implementation.

\subsection{Galilean invariance}\label{sec:galilei}

The Vlasov-Poisson system within the non-relativistic limit features a shift symmetry
\be
  \vec x\to \vec x'=\vec x+\vec n(\tau),\quad \tau\to \tau'=\tau\,,
\ee
with an arbitrary time-dependent shift function $\vec n(\tau)$, as  pointed out in~\cite{Kehagias:2013yd, Peloso:2013zw} generalizing the earlier work in~\cite{ScoFri9607}. 
More broadly, this symmetry is a remnant of those induced by general relativistic coordinate invariance, being the only one that has a non-trivial Newtonian limit~\cite{CreNorSim1312,HorHuiXia1409}, corresponding to the equivalence principle. It  can be viewed as a generalized Galilean invariance (in the absence of gravity and expansion of the universe, ${\vec n}$ becomes linear in time), with gravitational potential transforming as $\Phi(\tau,\vec x)\to \Phi(\tau',\vec x')=\Phi'(\tau,\vec x)-(d^2{\vec n}/d\tau^2+{\cal H}d{\vec n}/d\tau)\cdot\vec x$. While the density contrast transforms like a scalar under this symmetry, $\delta(\tau,\vec x)\to\delta(\tau',\vec x')=\delta'(\tau,\vec x)$, the velocity field is shifted by a time-dependent, but spatially constant vector, $\vec v(\tau,\vec x)\to\vec v(\tau',\vec x')=\vec v'(\tau,\vec x)+d\vec n/d\tau$. 

In order to determine the transformation properties of higher cumulants, we use that the phase-space distribution function itself is a scalar under Galilean transformations,
$f(\tau,\vec x,\vec p)\to f(\tau',\vec x',\vec p')=f'(\tau,\vec x,\vec p)$, where $\vec p'=\vec p+a(\tau)d\vec n/d\tau$.
From this, one finds that the cumulant generating function, Eq.~\eqref{eq:genfunc}, transforms as
\be
  {\cal C}(\tau,\vec x,\vec l) \to {\cal C}(\tau',\vec x',\vec l) = {\cal C}'(\tau,\vec x,\vec l)+\vec l\cdot d\vec n/d\tau\,,
\ee
which implies that all but the first cumulant, i.e. the velocity, are scalars under Galilean transformations.
In particular,
\bea
  \sigma_{ij}(\tau,\vec x) &\to& \sigma_{ij}(\tau',\vec x')=\sigma_{ij}'(\tau,\vec x)\,,\nn\\
  {\cal C}_{ijk\cdots }(\tau,\vec x) &\to& {\cal C}_{ijk\cdots }(\tau',\vec x')={\cal C}_{ijk\cdots }'(\tau,\vec x)\,,
\eea
for the velocity dispersion tensor, the third and all higher cumulants. Note that also the divergence of the velocity $\theta$ as well
as the vorticity $w_i$ transform as scalars, or in general any gradient $\nabla_i v_j$ of the velocity field, as well as
second gradients $\nabla_i\nabla_j\Phi$ of the gravitational potential.

Galilean invariance of the equations of motion of the density contrast, the velocity divergence and vorticity, as well
as the dispersion tensor and higher cumulants therefore implies that only covariant quantities may enter these equations.
This requires in particular that only gradients of the velocity field, and second gradients of the gravitational potential
may appear, as can readily be checked to be the case for the continuity and Euler equations~\eqref{eq:fluid}, and also for the equations of motion~\eqref{eq:eomsigmaij} of the dispersion tensor and the higher cumulants (note that the equations for $\theta$ and $w_i$ contain an extra gradient compared to the Euler equation for $\vec v$).
The only exception is the time-derivative. Galilean invariance
requires that it appears in the covariant combination $\partial_\tau+{\bf v}\cdot\nabla$. This is indeed the case in Eqs.~\eqref{eq:fluid} and~\eqref{eq:eomsigmaij}.
In terms of the rescaled time variable $\eta$ and velocity $\vec u$, the covariant derivative reads
\be
  D_\eta \equiv \partial_\eta - \vec u\cdot\nabla = \partial_\eta - \left(\frac{\nabla\theta}{\nabla^2}-\frac{\nabla\times\vec w}{\nabla^2}\right)\cdot\nabla\,.
\ee
Galilean invariance therefore implies that the covariant time derivative is the only place where the velocity field may appear \emph{without} a gradient acting on it, when
considering equations of motion for quantities that are scalars under Galilean transformations. As shown above, this is the case for all perturbation variables considered here.
When transforming the equations of motion into Fourier space as in Eq.~\eqref{eq:eom}, this implies that the non-linear vertices satisfy the property
\be\label{eq:gammasoft}
  \gamma_{abc}(\vec p,\vec q) = \delta_{ac}\left(\delta_{b\theta}\frac{\vec q\cdot\vec p}{2p^2}-\delta_{b w_i}\frac{(\vec q\times \vec p)_i}{2p^2}\right)+ {\cal O}(p^0)\,,
\ee
in the limit $\vec p\to 0$, generalizing the result in~\cite{CroSco0603b} to include vorticity. 
As shown in App.~\ref{app:galilei}, this implies the relation
\bea\label{eq:Fnsoft}
  F_{n+1,a}(\vec p,\vec k_1,\dots,\vec k_n,\eta) &=& \frac{1}{n+1}\frac{\vec k\cdot\vec p}{p^2}F_{n,a}(\vec k_1,\dots,\vec k_n,\eta) \nn\\
  && {} + {\cal O}(p^0)\ \ \ \text{for}\ \vec p\to 0\,,
\eea
for the squeezed limit of the kernels, where $\vec k=\sum_i\vec k_i$. This property is well-known for the EdS-SPT kernels, see e.g.~\cite{Sugiyama:2013pwa}. Galilean invariance ensures its validity also for the
\vpt~non-linear kernels, including those of higher cumulants. The limit is approached when $|\vec p|\ll |\vec k_i|$ \emph{and} $|\vec p|\ll k_\sigma$, i.e. the
size $\sim 1/\vec p$ of the large-scale mode has to be larger than the dispersion length-scale $\sqrt{\epsilon}$. We checked that the relation above is satisfied for the analytical results of the second-order kernels at order $\epsilon$. In addition we checked its validity analytically
for the third-order kernels $F_{3,a}$ for $a=\delta,\theta,g,\delta\epsilon,A$ expanded to order $\epsilon$. 

Equation~\eqref{eq:Fnsoft} ensures that, as in SPT~\cite{ScoFri9607,JaiBer9601}, all contributions to the equal-time power spectra and bispectra for which the kernels become singular in the limit of small loop wavenumber cancel. This cancellation occurs only after adding all contributions at a given order in perturbation theory, i.e. for example among the various lines in the one-loop expression, Eq.~\eqref{eq:P1Ldef}.
It is easy to see that Eq.~\eqref{eq:Fnsoft} also implies that the cancellation can be made manifest on the level of the loop integrand by the same strategy as in SPT, for the power spectrum~\cite{Blas:2013bpa,BlaGarKon1309}
and bispectrum~\cite{Floerchinger:2019eoj}, respectively. We already described its implementation at one-loop above, and use it throughout this work.

For unequal times, one can check that Eq.~\eqref{eq:Fnsoft} gives rise to the standard $(1/p)$ pole present in the so-called consistency relations between the squeezed $(N+1)$-point functions of density fluctuations and its regular $N$-point function~\cite{Kehagias:2013yd, Peloso:2013zw}. 

\subsection{Mass and momentum conservation}\label{sec:ktozerolimit}

Within SPT, it is well-known that $F_n(\vec k_1,\dots,\vec k_n)\propto k^2$ and $G_n(\vec k_1,\dots,\vec k_n)\propto k^2$ in the large-scale limit when the sum $\vec k\equiv\sum_i\vec k_i$ of wavevectors goes to zero (with the magnitude of at least two of the individual wavevectors remaining constant). Here we show that, as expected from mass and momentum conservation, this property is preserved for the kernels $F_{n,\delta}$ of the density contrast also in \vpt~{\em provided vorticity is taken into account}, as hinted at already above. However, as we show below, the scaling of the velocity divergence kernels $F_{n,\theta}$ in \vpt~is in general different compared to SPT, but still conforms with all symmetry requirements.

\subsubsection{Density contrast}

\begin{figure}[t]
  \begin{center}
  \includegraphics[width=\columnwidth]{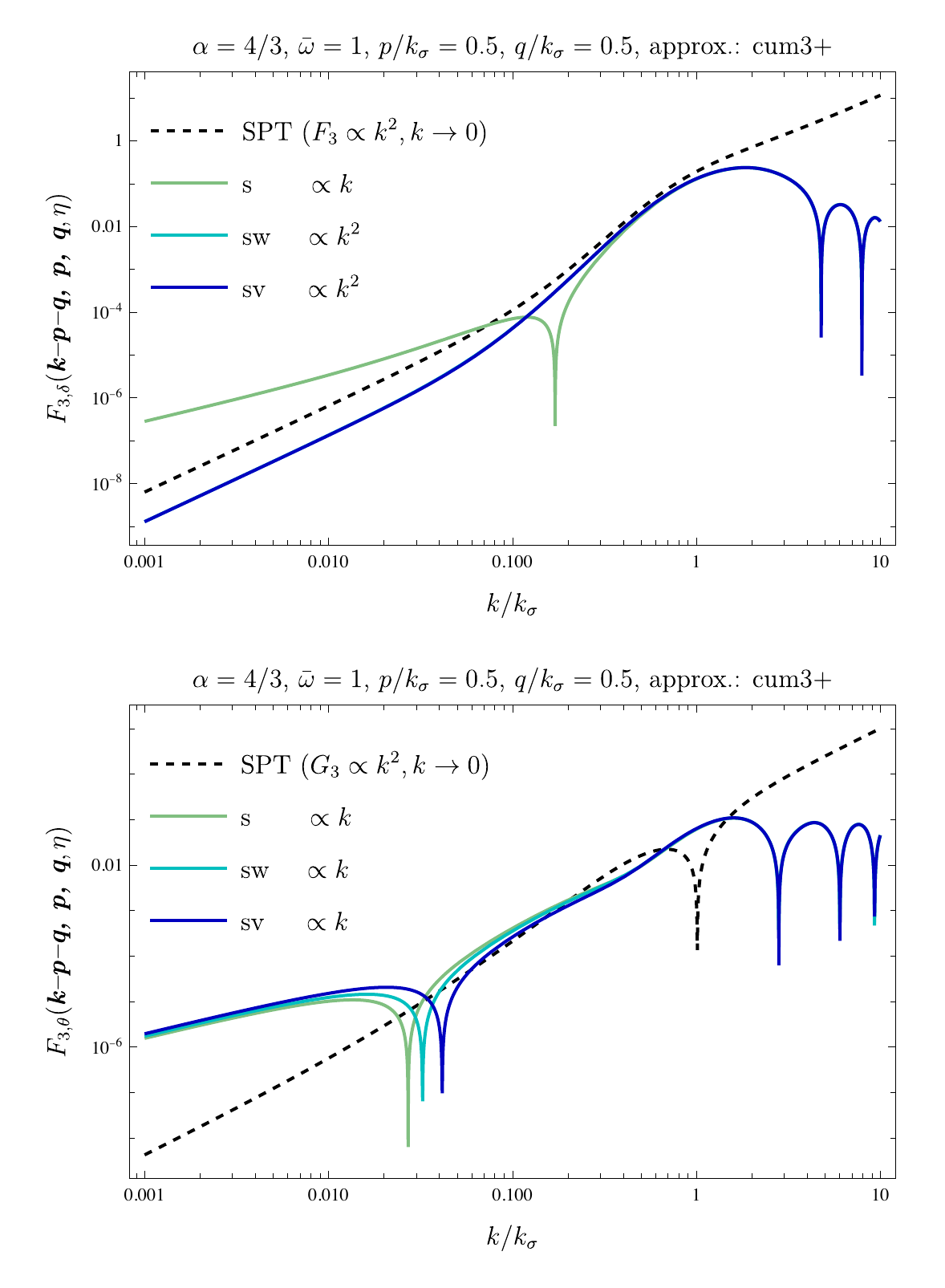}
  \end{center}
  \caption{\label{fig:F3_d_t_kmpmq_vs_k}
  Non-linear kernels $F_{3,a}(\vec k-\vec p-\vec q,\vec p,\vec q,\eta)$ for $a=\delta$ (upper panel) and $a=\theta$ (lower panel) versus $k=|\vec k|$, for fixed $\vec p$ and $\vec q$ at $\eta=0$,
  and assuming $\epsilon=\epsilon_0\,e^{\alpha\eta}$ with $\alpha=4/3$. The blue lines show the ({\bf s}), ({\bf sw}) and ({\bf sv}) approximation schemes, respectively, for ({\bf cum3+}) with $\bar\omega=1$. The black dashed lines show the EdS-SPT kernels $F_3$ and $G_3$ for comparison. For $F_{3,\delta}$, a spurious linear scaling with $k$ occurs in the limit $k\to 0$ when vorticity is neglected ({\bf s}), while the $k^2$ scaling ensured by mass and momentum conservation is realized for all other approximation schemes that include vorticity. For $F_{3,\theta}$, the linear scaling with $k$ is a physical effect, and occurs independently of the approximation scheme as long as $\vec p+\vec q\not=0$. For the figure we chose cosines $c_{kp}=0.3, c_{kq}=0.5, c_{pq}=0.875, p=0.5k_\sigma, q=0.5k_\sigma$ where $k_\sigma=1/\sqrt{\epsilon_0}$. 
  } 
\end{figure}

Let us start with the kernels $F_{n,\delta}$ of the density contrast. To study their asymptotic behaviour it is convenient to write the continuity equation in the form
\be \label{eq:massevolution}
  \partial_\tau \delta + \nabla_i P_i = 0\,,
\ee
where
\be
  P_i(\tau,\vec x) \equiv (1+\delta)v_i\,,
\ee
is the momentum field. Using the Euler equation, its equation of motion can be written as
\be\label{eq:momentumevolution}
  \partial_\tau P_i +{\cal H}P_i +\nabla_j T_{ij}=0\,,
\ee
which takes the form of a conservation equation for the comoving momentum $a P_i$, with
\bea
  T_{ij} &\equiv& (1+\delta)(\sigma_{ij}+v_iv_j)+\Phi\delta_{ij}\nn\\
  &+& \frac{1}{3{\cal H}^2\Omega_m}(\nabla_i\Phi\nabla_j\Phi-\Phi\nabla_i\nabla_j\Phi+\Phi\delta_{ij}\nabla^2\Phi)\,.\nn\\
\eea
This is of the expected  form coming from momentum conservation for free particles described by a distribution function in general relativity, where the stress-energy-momentum tensor ${\cal T}^{\mu \nu} = \int dV_p f({\vec p}) p^\mu p^\nu$, and $dV_p\equiv d^3p/\sqrt{-g}\,p^0$ is the invariant 3-volume in phase-space and $g$ the metric's determinant~\cite{MaBer9512}. Indeed, in the weak-field limit for non-relativistic velocities and scalar-only metric perturbations described by $\Phi$, the covariant derivative in momentum conservation induces the $\Phi$-dependent terms inside the Newtonian limit $T_{ij}$ that satisfy $\nabla_jT^\Phi_{ij}=(1+\delta) \nabla_i \Phi$, as required.  Combining Eqs.~(\ref{eq:massevolution}) and (\ref{eq:momentumevolution}) gives
\be\label{eq:densityksqscaling}
  \partial_\tau^2 \delta + {\cal H}\partial_\tau \delta = \nabla_i\nabla_j T_{ij}\,,
\ee
which is valid non-linearly and including the stress tensor contribution $\Sigma_{ij}\equiv (1+\delta)\sigma_{ij}$. Its form also does
not change when allowing for third or higher cumulants. Inserting the perturbative expansion of the density
contrast readily yields the claimed scaling due to the two overall derivatives in front of $T_{ij}$,
\be
  F_{n,\delta}(\vec k_1,\dots,\vec k_n,\eta)\propto k^2,\qquad \vec k=\sum_i\vec k_i\to 0\,.
\ee
This proves that the quadratic scaling of the kernels $F_{n,\delta}$ due to mass and momentum conservation is a general property, as expected.

Let us now discuss in how far this constraint is compatible with the various approximation schemes considered in this work.
One can check that Eq.~\eqref{eq:momentumevolution} holds independently of whether or not third and higher cumulants
are taken into account, i.e. in particular for both the ({\bf cum2}) and ({\bf cum3+)} approximations.
Furthermore, it also holds when including only a subset of the velocity dispersion tensor perturbations, as is
the case in the ({\bf sw}) or ({\bf sv}) approximations. Therefore also the scaling property is preserved for all of these approximation
schemes. 

However, we find that \emph{neglecting vorticity breaks momentum conservation}. 
The reason for this is that the velocity dispersion tensor contribution to the Euler equation sources vorticity. Neglecting it discards part of the
velocity field, and therefore also some part of the momentum, leading to artificial violation of momentum conservation.
Therefore, the ({\bf s}) approximation scheme should not be used in general, and we do indeed always include vorticity in this work.

Nevertheless, it's worth pointing out that when neglecting vorticity, the spurious linear terms do not affect the one-loop density power spectrum.
Technically, this can be seen by applying the scalar projector $P^s_{ij}=\nabla_i\nabla_j/\nabla^2$ to the Euler equation, and using this projected equation
to derive an equation for the momentum. This would lead to a contribution to the momentum equation~\eqref{eq:momentumevolution} of the form
\be
  (1+\delta)\left[\frac{1}{1+\delta}\partial_j[(1+\delta) \epsilon_{ij}]\right]^s\,,
\ee
where the superscript $s$ stands for applying the scalar projector, $[X_i]^s\equiv P^s_{ik}X_k$. This implies that the two factors of $1+\delta$
in front do not cancel in that case, and therefore do not yield an expression with an overall derivative in front. This, in turn, would lead to a violation
of the $k^2$ scaling when neglecting vorticity.
When taking vorticity into account, one effectively adds the corresponding contribution with a vector instead of scalar projection, $[X_i]^v\equiv P^v_{ik}X_k$. Due to
$P^v_{ij}+P^s_{ij}=\delta_{ij}$, the projectors drop out and one obtains again Eq.~\eqref{eq:momentumevolution}, i.e. a $k^2$ scaling of the density kernels.
However, when neglecting vorticity, one obtains an extra term on the right-hand side of Eq.~\eqref{eq:densityksqscaling} given by
\be
  \nabla_i\left[ \delta\left( [\nabla_j\sigma_{ij}]^s-\nabla_j\sigma_{ij} + \left[\sigma_{ij}\nabla_j A\right]^s - \sigma_{ij}\nabla_j A \right)\right]\,,
\ee
which only contains a single overall derivative and would therefore potentially lead to terms linear in $\vec k$ in the large scale limit of $F_{n,\delta}$
when neglecting vorticity. The first two summands cancel for the scalar contributions to the stress tensor, and are at least of second order for the
vector and tensor contributions. The last two summands are at least of second order as well. Therefore, taking the additional factor $\delta$ in front of the parenthesis into account, the spurious linear scaling in $\vec k$ can only appear
starting at third order in perturbation theory, i.e. for $F_{3,\delta}$. Furthermore, due to the property
$F_{n,\delta}(\vec k_1,\dots,\vec k_n,\eta)=F_{n,\delta}(-\vec k_1,\dots,-\vec k_n,\eta)$, the kernel $F_{3,\delta}(\vec k,\vec q,-\vec q,\eta)$ that appears
in the one-loop power spectrum is odd under $\vec k\to-\vec k$, and therefore any linear term in $\vec k$ has to cancel.

Thus, the spurious linear scaling that occurs when neglecting vorticity
can only affect the power spectrum starting at two-loop order, e.g. via the kernel $F_{3,\delta}(\vec k-\vec p-\vec q,\vec p,\vec q,\eta)$.
For this reason we include the ({\bf s}) approximation scheme in our analysis below at one-loop order for illustration and in order to compare to the
approximation schemes ({\bf sw}), ({\bf sv}) and ({\bf svt}), that are all compatible with momentum conservation.
However, we do not use the ({\bf s}) scheme when going beyond one-loop.

For illustration, we show the kernel $F_{3,\delta}(\vec k-\vec p-\vec q,\vec p,\vec q,\eta)$ in the upper panel of Fig.\,\ref{fig:F3_d_t_kmpmq_vs_k}.
For the ({\bf sw}) and ({\bf sv}) approximations, that take vorticity into account, the kernel scales as $k^2$ for small $k$. However, in the ({\bf s})
approximation scheme, a spurious linear scaling with $\propto k$ appears. Note that, similarly as before, the kernels are strongly suppressed for large
$k$ as compared to SPT, and also differ for small $k$ due to the finite, fixed size of $\vec p$ and $\vec q$. The behaviour is qualitatively similar
for the ({\bf cum2}) and ({\bf cum3+}) approximations, respectively, and we therefore only show the latter.

Finally, we comment on why neglecting vorticity within the SPT approximation does not lead to spurious linear terms.
Within SPT, there is no vorticity source term in the Euler equation.
The remaining non-linear term in the Euler equation has the property $[v^s_j\partial_j v^s_i ]^s=v^s_j\partial_j v^s_i $,
which implies the well-known property that no vorticity is generated in the perfect fluid approximation.
This has the consequence that, when neglecting the velocity dispersion tensor altogether, the omission of vorticity does not spoil momentum conservation. Therefore, the $k^2$ scaling of the density kernels
holds in SPT even in absence of vorticity. However, this is not true any longer as soon as taking the velocity dispersion tensor into account, in which case it is mandatory to include also vorticity in order to guarantee that momentum conservation holds.

\subsubsection{Velocity divergence}

In contrast to the density, the scaling of the velocity divergence kernels is indeed modified when
including the velocity dispersion tensor and higher cumulants along with vorticity,
\be
  F_{n,\theta}(\vec k_1,\dots,\vec k_n,\eta)\propto k,\qquad \vec k=\sum_i\vec k_i\to 0\,.
\ee
This scaling is in accordance with mass and momentum conservation, i.e. allowed by the
symmetries of the system. However, we observe that for $n=2$ it does not occur, i.e. $F_{2,\theta}\propto k^2$,
independently of the approximation scheme. Furthermore, due to
$F_{n,\theta}(\vec k_1,\dots,\vec k_n,\eta)=F_{n,\theta}(-\vec k_1,\dots,-\vec k_n,\eta)$
the linear term cancels in the one-loop kernel $F_{3,\theta}(\vec k,\vec q,-\vec q,\eta)\propto k^2$.
Therefore, the linear scaling does not affect the one-loop power spectrum of the velocity divergence,
or its cross spectrum with the density contrast. The simplest kernel that displays linear scaling is
$F_{3,\theta}(\vec k-\vec p-\vec q,\vec p,\vec q,\eta)$, for $\vec p+\vec q\not=0$. This kernel
is shown in the lower panel of Fig.\,\ref{fig:F3_d_t_kmpmq_vs_k}. In contrast to the density kernel,
we find linear scaling $\propto k$ in the limit $k\to 0$ also when taking vorticity into account, i.e.
for the ({\bf sw}) and ({\bf sv}) approximation schemes. Furthermore, the linear scaling occurs both
in the ({\bf cum2}) and ({\bf cum3+}) approximation schemes (only the latter is shown in Fig.\,\ref{fig:F3_d_t_kmpmq_vs_k}).

Proceeding analogously as in Sec.\,\ref{sec:analytical_results} we find the following analytical result for the
${\cal O}(\epsilon)$ contribution to the third-order velocity divergence kernel, when keeping the leading term
in a Taylor expansion in $k$,
\bea\label{eq:F3thetakpq}
  \lefteqn{ F_{3,\theta}(\vec k-\vec p-\vec q,\vec p,\vec q,\eta) = G_3(\vec k-\vec p-\vec q,\vec p,\vec q) } \nn\\
  &-&  \left[ \frac{(\vec k\cdot \vec p)(\vec q\cdot(\vec q+2\vec p))s_{pq}^2}{(\vec p+\vec q)^2} +(\vec p\leftrightarrow\vec q)\right]\nn\\
  &\times& \left(\frac{10}{7}E_2(\eta)+\frac{2}{21}E_3(\eta)-2E_{5/2}(\eta)+\frac{10}{21}E_{7/2}(\eta)\right)\nn\\
  &&  +{\cal}{O}(\epsilon^2,k^2)\,,
\eea
where $s_{pq}^2\equiv  1-(\vec p\cdot\vec q)^2/(p^2q^2)$. The second line scales linearly with $\vec k$, and we recall that $E_n={\cal O}(\epsilon)$.
Furthermore, one can check that the second line vanishes in the limit $\vec p+\vec q\to 0$, giving a $k^2$ scaling for the third-order kernel
entering the one-loop power spectrum, as stated above.

Note that, at first order in $\epsilon$, third and higher cumulants do not contribute, and therefore~\eqref{eq:F3thetakpq}
is independent of the truncation, such that it is identical for ({\bf cum2}) and ({\bf cum3+}), and would not change
when going to higher cumulant order. Furthermore, we find that tensor modes of the stress tensor do not contribute
to the linear term in $k$, such that it holds for both the ({\bf sv}) approximation as well as the full ({\bf svt}) case.
(In ({\bf sw}) approximation only the prefactor would change). From this we conclude that the linear scaling with $\vec k$
for the velocity divergence in the limit $\vec k\to 0$ cannot be an artifact of the approximation scheme,
but is present in the full Vlasov theory.

As a further check, we demonstrate that the linear term cancels when considering the momentum $P_i$. The divergence of the
momentum field can be written as
\be\label{eq:momentumdivergence}
  \nabla_i P_i = -{\cal H}f\left(\theta +\nabla_i\left(\delta\frac{\nabla_i\theta}{\nabla^2}\right)-\nabla_i\left(\delta\frac{\varepsilon_{ijk}\nabla_jw_k}{\nabla^2}\right)\right)\,.
\ee
Due to the momentum conservation equation~\eqref{eq:momentumevolution}, we know that the non-linear kernels obtained when perturbatively expanding $\nabla_i P_i$ have to
scale as $k^2$ in the limit where the sum of all wavenumbers goes to zero. In particular, at third order this requires that in the combination of kernels
\bea
  \lefteqn{ F_{3,\theta}(\vec k_1,\vec k_2,\vec k_3,\eta)+\frac13\Bigg[\Bigg( \frac{\vec k\cdot\vec k_3}{k_3^2}F_{2,\delta}(\vec k_1,\vec k_2,\eta)F_{1,\theta}(\vec k_3,\eta)}\nn\\
  &-& \frac{(\vec k\times(\vec k_2+\vec k_3))_i}{(\vec k_1+\vec k_3)^2} F_{1,\delta}(\vec k_1,\eta)F_{2,w_i}(\vec k_2,\vec k_3,\eta)\nn\\
  &+& \frac{\vec k\cdot(\vec k_2+\vec k_3)}{(\vec k_2+\vec k_3)^2}F_{1,\delta}(\vec k_1,\eta)F_{2,\theta}(\vec k_2,\vec k_3,\eta)\Bigg)  + 2\,\text{perm.}\Bigg]\,,\nn\\
\eea
the linear term in $\vec k=\sum_i\vec k_i$ for $\vec k\to 0$ cancels.  Using the analytical results for the density, velocity divergence and vorticity kernels at ${\cal O}(\epsilon)$
obtained in Sec.\,\ref{sec:analytical_results} and Sec.\,\ref{sec:vortF2} as well as Eq.~\eqref{eq:F3thetakpq} 
this can indeed be shown to be the case. This provides a non-trivial check of the linear scaling for the velocity divergence kernels.

Since the leading contribution to the power spectrum affected by the linear scaling in Eq.~\eqref{eq:F3thetakpq} is the two-loop velocity divergence power spectrum,
it is difficult to notice its impact in practice. In particular, note that the propagator-like terms (obtained from correlating one linear and one non-linear field)
contain kernels $F_{n,\theta}(\vec k,\vec q_1,-\vec q_1,\vec q_2,-\vec q_2,\dots,\eta)$ where the linear terms in $\vec k$ drop out for symmetry reasons, at all loop orders.
These propagator-like contributions therefore yield a contribution to the power spectrum $P_{\theta\theta}(k,\eta)$ that scales as $k^2P_0(k)$ for $k\to 0$, similarly as in SPT.
However, the linear scaling of $F_{3,\theta}(\vec k-\vec p-\vec q,\vec p,\vec q,\eta)$ 
produces an additional contribution arising from mode-coupling terms starting at two-loop order. It yields a contribution to $P_{\theta\theta}(k,\eta)$ that
scales as $k^2$ and should therefore dominate over $k^2P_0(k)$ on large enough scales. Nevertheless, within this regime, $P_{\theta\theta}(k,\eta)$ is dominated by the linear contribution to the power spectrum,  and thus 
the anomalous behavior is hard to verify in practice. However, as we will see below, a similar effects occurs for the vorticity, where it is potentially visible
given that the vorticity power spectrum has no linear contribution.

\section{Generation of vorticity, vector and tensor modes}
\label{sec:vorticity}

It is well known that within the perfect, pressureless fluid approximation vorticity decays as $(\nabla\times{\bf v})\propto 1/a$ within the linear regime~\cite{Pee80}, and is not generated by non-linear evolution due to the neglect of orbit crossing, in contradiction to the physical expectation.
While the former property essentially remains true when including velocity dispersion and higher cumulants, the latter does not~\cite{PueSco0908}.
In this section we discuss the generation of vorticity as well as vector and tensor modes of the stress tensor via non-linear effects within the framework underlying the present work.
The generation of vorticity and vector modes is coupled to each other, and we discuss them jointly in the following. The tensor mode is discussed in Sec.\,\ref{sec:tensor}.

Let us first recall the evolution equation~\eqref{eq:eom} for the vorticity ${\bf w}$, that is coupled to the vector mode ${\bf  \nu}$ of the velocity dispersion tensor,
\bea
  \partial_\eta w_{k,i}+ \left( \frac32\frac{\Omega_m}{f^2}-1\right)w_{k,i} + k^2\nu_{k,i} &=& \int_{pq}\gamma_{w_ibc}\psi_{p,b}\psi_{q,c}\,,\nn\\
  \partial_\eta \nu_{k,i}+ 2\left( \frac32\frac{\Omega_m}{f^2}-1\right)\nu_{k,i} -\epsilon w_{k,i} &=& \int_{pq}\gamma_{\nu_ibc}\psi_{p,b}\psi_{q,c}\,,\nn\\
\eea
with vertices listed below.

\subsection{Linear approximation}\label{sec:vortlinear}

Within the linear approximation, one obtains a coupled set of ordinary differential equations for each component of the vorticity and ${\bf \nu}$ vector, respectively.
In the limit $\epsilon\to 0$, and setting $\Omega_m/f^2\mapsto 1$ for simplicity, the second equation has the approximate solution $\nu_{k,i}(\eta)=\nu_{k,i}(\eta_0)e^{-(\eta-\eta_0)}$, being a purely decaying mode.
Inserting it in the linearized vorticity equation yields
\bea\label{eq:vorticitylinearearlytime}
  w_{k,i}(\eta) &=& (w_{k,i}(\eta_0)-2k^2\nu_{k,i}(\eta_0))e^{-\frac12(\eta-\eta_0)}\nn\\
  && {} +2k^2\nu_{k,i}(\eta_0)e^{-(\eta-\eta_0)}\,,
\eea
being a superposition of two decaying modes, that corresponds to a decay $\propto 1/a$ and $\propto a^{-3/2}$ for the unrescaled
vorticity $-f{\cal H}{\bf w}$. Therefore, the vector mode of the dispersion tensor leads to an additional decaying mode in the vorticity at linear level.
By inserting the solution for ${\bf w}$ from above into the linearized equation for the vector mode $\nu$, one can derive corrections at first order in $\epsilon$, that are relatively suppressed by
factors of $\epsilon k^2$ as long as this quantity is small, i.e. at early times. In the opposite limit the coupled set of linearized equations needs to be solved, and the background dispersion
leads to suppression as well. For example, for a power-law time dependence $\epsilon=\epsilon_0 e^{\alpha\eta}$, a closed-form
solution of the linear equations can be found in terms of hypergeometric functions (assuming $\alpha>1/2$), given by
\bea
  w_{k,i}(\eta) &=& A\, e^{-\frac12\eta}\ {}_0F_1\left(1+\frac{1}{2\alpha};-\frac{\epsilon(\eta)k^2}{\alpha^2}\right)\nn\\
  && {} + B\, e^{-\eta}\ {}_0F_1\left(1-\frac{1}{2\alpha};-\frac{\epsilon(\eta)k^2}{\alpha^2}\right)\,,
\eea
with coefficients $A, B$ related to the initial vorticity and vector mode. For $\epsilon k^2\ll 1$ one recovers Eq.~\eqref{eq:vorticitylinearearlytime} since ${}_0F_1\to 1$ in that limit.
For $\epsilon k^2\gg 1$ one finds oscillating behavior with a damped envelope,
\bea
  w_{k,i}(\eta) &\to& e^{-\frac{3+\alpha}{4}\eta}\,\Bigg[ A'\sin\left(2\sqrt{\epsilon} k+\frac{\alpha-1}{4\alpha}\pi\right)\nn\\
  && {} + B'\sin\left(2\sqrt{\epsilon} k+\frac{\alpha+1}{4\alpha}\pi\right)\Bigg]\,,
\eea
where 
\bea
  A'&\equiv& A \left(\frac{\epsilon_0 k^2}{\alpha^2}\right)^{-\frac14-\frac{1}{4\alpha}}\frac{\Gamma\left(1+\frac{1}{2\alpha}\right)}{\sqrt{\pi}}\,,\nn\\
  B'&\equiv& B \left(\frac{\epsilon_0 k^2}{\alpha^2}\right)^{-\frac14+\frac{1}{4\alpha}}\frac{\Gamma\left(1-\frac{1}{2\alpha}\right)}{\sqrt{\pi}}\,.
\eea
Since the linear solutions correspond to decaying modes, initial vorticity or $\nu$ modes can be neglected.
Indeed, as stated already above, this means that the first-order kernels vanish,
\be
  F_{1,w_i}=F_{1,\nu_i}=0\,.
\ee
This implies in particular that the linear contribution to the vorticity and vector power spectra is zero.

\subsection{Vorticity generation at second order}\label{sec:vortF2}

When including velocity dispersion and higher cumulants, vorticity is generated at the non-linear level~\cite{PueSco0908}.
The leading contribution arises at second order in perturbation theory. Two scalar modes can generate vorticity as well
as vector perturbations of the stress tensor.

At second order, the evolution equations for the ${\bf w}$ and $\nu$ kernels read
\bea\label{eq:F2wnu}
  \left(\partial_\eta+\frac52\right)&F_{2,w_i}(\vec p,\vec q,\eta) =& -(\vec p+\vec q)^2 F_{2,\nu_i}(\vec p,\vec q,\eta) \nn\\
  && +\gamma_{w_i bc}(\vec p,\vec q) F_{1,b}(p,\eta) F_{1,c}( q,\eta) \,,\nn\\\nn\\
  \left(\partial_\eta+3\right)&F_{2,\nu_i}(\vec p,\vec q,\eta) =& \epsilon(\eta) F_{2,w_i}(\vec p,\vec q,\eta) \nn\\
  &&+\gamma_{\nu_i bc}(\vec p,\vec q) F_{1,b}(p,\eta) F_{1,c}(q,\eta) \,.\nn\\
\eea
The list of potential vertices for vorticity is $\gamma_{w_iAg}$, $\gamma_{w_iA\epsilon}$, $\gamma_{w_iA\nu_j}$, $\gamma_{w_iAt_{jk}}$, $\gamma_{w_i\theta w_j}$, $\gamma_{w_iw_jw_k}$, and for the vector mode $\gamma_{\nu_i\theta g}$, $\gamma_{\nu_i\theta\epsilon}$, $\gamma_{\nu_iw_jg}$, $\gamma_{\nu_iw_j\epsilon}$, $\gamma_{\nu_i\theta\nu_j}$, $\gamma_{\nu_i\theta t_{jk}}$, $\gamma_{\nu_iw_j\nu_k}$, $\gamma_{\nu_iw_jt_{k\ell}}$.
Since only scalar modes have non-zero $F_{1,a}$, only the vertices $\gamma_{w_iAg}$, $\gamma_{w_iA\epsilon}$ and $\gamma_{\nu_i\theta g}$, $\gamma_{\nu_i\theta\epsilon}$ contribute at second order, and describe the generation of vorticity and
vector modes from a pair of scalar perturbations at non-linear level.
The leading contribution to the vorticity power spectrum is therefore given by
\bea\label{eq:P1Lww}
  P_{w_iw_i}^{1L}(k,\eta) &=& 2e^{4\eta}\int d^3q \, F_{2,w_i}(\vec k-\vec q,\vec q,\eta)\nn\\
  && {} \times F_{2,w_i}(\vec k-\vec q,\vec q,\eta)P_0(|\vec k-\vec q|)P_0(q)\,.\nn\\
\eea
The power spectra of the vector mode and their cross spectrum are given by analogous expressions.
Below we will see that it is actually necessary to include also the two-loop contribution to capture the correct scaling for $k\to 0$, see Sec.~\ref{sec:vortF3} and Sec.~\ref{sec:Pww}, but first discuss the one-loop part.

\subsubsection{Analytical results for $\epsilon\to 0$}

\begin{figure*}[t]
  \begin{center}
  \includegraphics[width=\columnwidth]{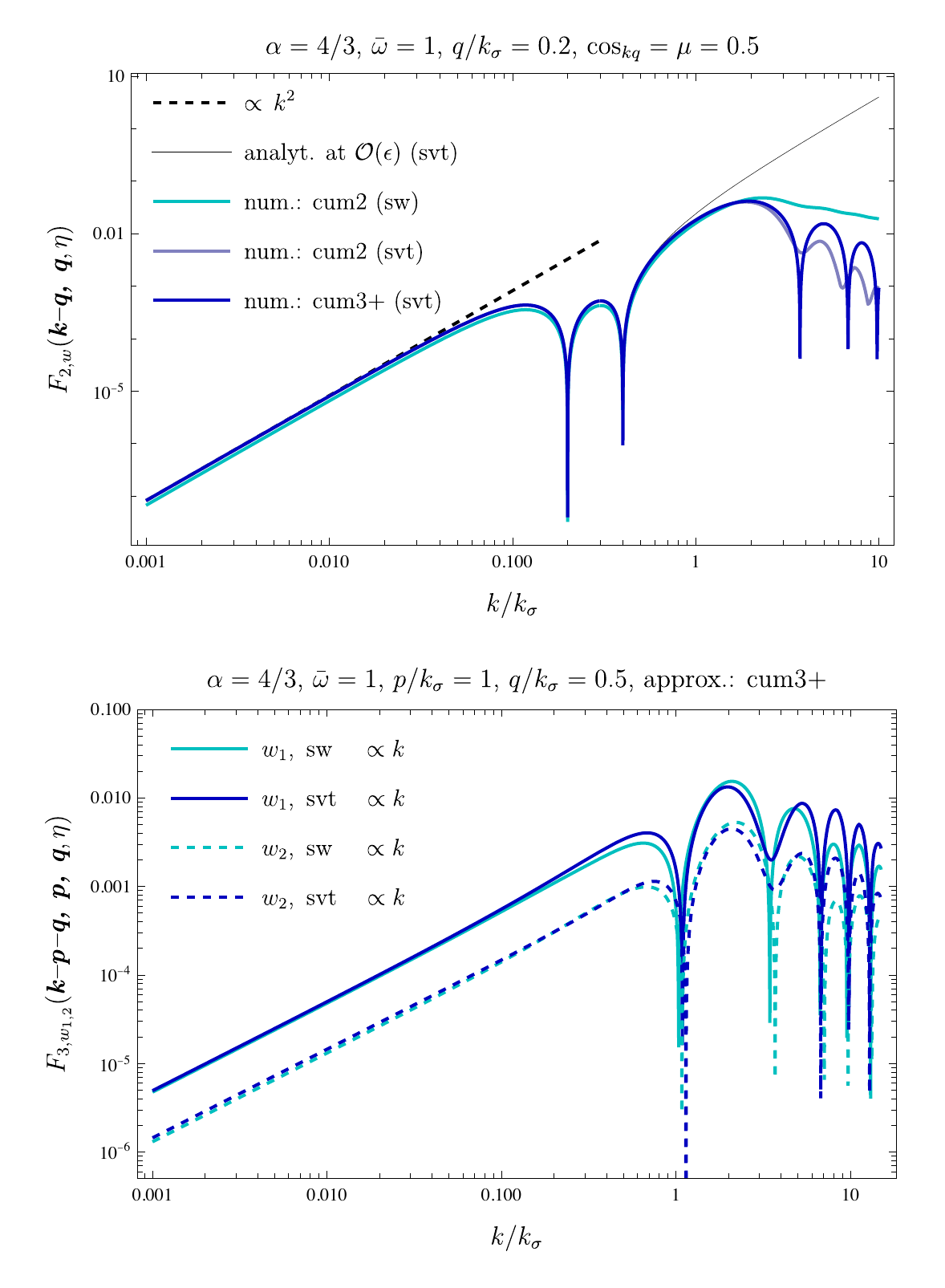}
  \includegraphics[width=\columnwidth]{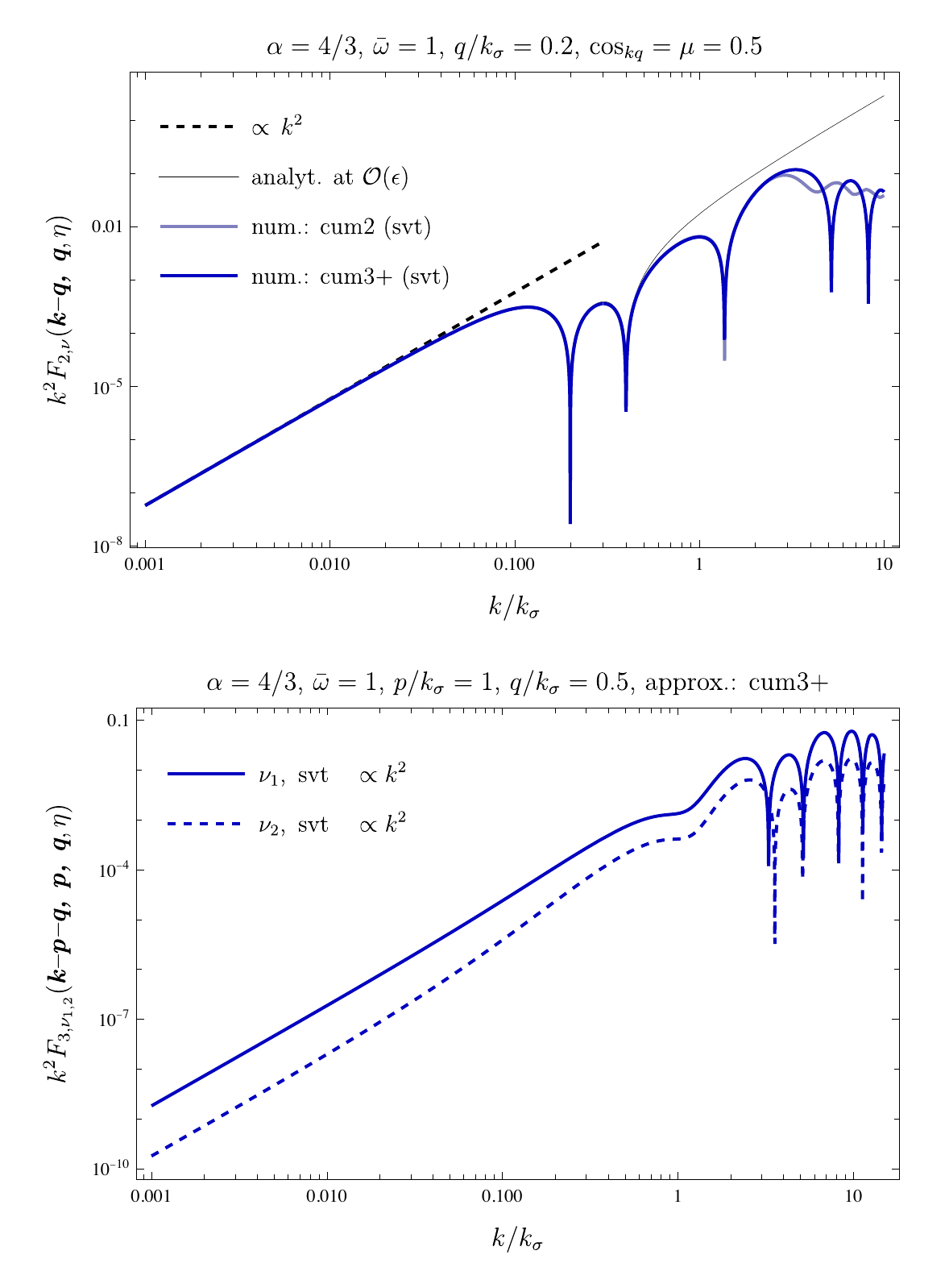}
  \end{center}
  \caption{\label{fig:F23w}
  Non-linear \vpt~kernels describing the generation of vorticity $F_{n,w_i}$ (left column) and the vector mode of the stress tensor $k^2F_{n,\nu_i}$ (right column), from non-linear coupling of two ($n=2$, upper row) or three ($n=3$, lower row) perturbation modes of the initial linear density field. The kernels are shown for various approximation
  schemes as indicated in the legend (see Table~\ref{tab:approxschemes}). For $n=2$ we also show the analytical results at first order in $\epsilon$ (black solid line), as well as a line $\propto k^2$ (black dashed) for comparison.
  Note that $F_{2,w_i}$ points in the direction perpendicular to the plane spanned by $\vec k$ and $\vec q$, and we show its projection on $\vec k\times\vec q/|\vec k\times\vec q|$.
  For $n=3$ we show two components $F_{3,w_i}$, $i=1,2$, perpendicular to $\vec k$ (see App.\,\ref{app:vorticity} for details),
  finding $F_{3,w_i}\propto k$ for $k\to 0$, similarly to $\theta$ (see Sec.~\ref{sec:ktozerolimit}).
  The impact of the vector mode on vorticity (({\bf sw}) vs ({\bf svt}) in the left panels) is noticeable at $k\gtrsim k_\sigma$,
  while the impact of third cumulant perturbations is relatively mild for vorticity as well as vector mode kernels (({\bf cum2}) vs ({\bf cum3+}) in the upper row).
  The analytical results Eqs.~(\ref{eq:F2nuanalyt},\,\ref{eq:F2wanalyt}) at first order in $\epsilon$ are shown as thin black lines, and agree well for $k\ll k_\sigma$. In SPT all these kernels are zero.
  For the figure we chose $\epsilon=\epsilon_0\,e^{\alpha\eta}$ with $\alpha=4/3$, and $c_{kp}=0.3, c_{kq}=0.5, c_{pq}=0.875, p=k_\sigma, q=0.5k_\sigma$, where $k_\sigma=1/\sqrt{\epsilon_0}$.} 
\end{figure*}

Let us first discuss an approximate analytical solution for the second-order kernels in the limit $\epsilon\to 0$, similar as above for the scalar kernels.
Both vorticity and the vector mode are generated starting at first order in $\epsilon$. Therefore, the $\epsilon F_{2,w_i}$ term in the equation for the vector mode
can be neglected at lowest order in $\epsilon$, and the equation for the vector kernel can be solved independently at this order. Using furthermore that $F_{1,\delta\epsilon}$
starts at order $\epsilon^2$, only the vertex $\gamma_{\nu_i\theta g}$ is relevant, and we find
\bea\label{eq:F2nuanalyt}
  F_{2,\nu_i}(\vec p,\vec q,\eta) &=& 2(\gamma_{\nu_i\theta g}(\vec p,\vec q)+\gamma_{\nu_i\theta g}(\vec q,\vec p))\nn\\
  && {} \times \int^\eta d\eta' e^{3(\eta'-\eta)}\int^{\eta'}d\eta''\,e^{2(\eta''-\eta')}\epsilon(\eta'')\nn\\
  && {} + {\cal O}(\epsilon^2)\,.
\eea
Inserting this result in the equation for vorticity, and using that only $\gamma_{w_i A g}$ contributes at first order in $\epsilon$, one finds
\bea
  F_{2,w_i}(\vec p,\vec q,\eta) &=& -(\vec p+\vec q)^2\int^\eta d\eta' e^{\frac52(\eta'-\eta)} \, F_{2,\nu_i}(\vec p,\vec q,\eta')\nn\\
  && {} + 2(\gamma_{w_i A g}(\vec p,\vec q)+\gamma_{w_i A g}(\vec q,\vec p))\nn\\
  && {} \times \int^\eta d\eta' e^{\frac52(\eta'-\eta)}\int^{\eta'}d\eta''\,e^{2(\eta''-\eta')}\epsilon(\eta'')\nn\\
  && {} + {\cal O}(\epsilon^2)\,.
\eea
Inserting the explicit form of the vertices~\cite{cumPT}, one finds
\be\label{eq:F2wanalyt}
  F_{2,w_i}(\vec p,\vec q,\eta) = (\vec p\times\vec q)_i(\vec p\cdot\vec q)\left(\frac{1}{p^2}-\frac{1}{q^2}\right)J^w(\eta)+ {\cal O}(\epsilon^2)\,,
\ee
with
\bea
  J^w(\eta) &=& \int^\eta d\eta' e^{\frac52(\eta'-\eta)} \Bigg[ \int^{\eta'}d\eta''\,e^{2(\eta''-\eta')}\epsilon(\eta'')\nn\\
  && + \int^{\eta'} d\eta'' e^{3(\eta''-\eta')}\int^{\eta''}d\eta'''\,e^{2(\eta'''-\eta'')}\epsilon(\eta''')\Bigg]\nn\\
  &=& 4E_2(\eta)-6E_{5/2}(\eta)+2E_3(\eta)\,.
\eea
Note that the contribution in the second line would be missed when neglecting the impact of the vector mode of the dispersion tensor.
As expected, the vorticity and vector mode are perpendicular to the plane spanned by the wavevectors ${\bf p}$ and ${\bf q}$.
Furthermore, $F_{2,w_i}(\vec k-\vec q,\vec q,\eta)$ scales as $k^2$ for $\vec k\to 0$, which can be readily checked
for the analytical result at order $\epsilon$ from above.
Inserting this first order result in $\epsilon$ into Eq.~\eqref{eq:P1Lww} yields a dependence on wavevectors identical to the integrand of Eq.~(70) in~\cite{CusTanDur1703}; in our numerical treatment (see below) we do not expand in $\epsilon$ and also fully include the impact of the vector mode on vorticity, while Eq.~\eqref{eq:F2wanalyt} captures only the lowest order in $\epsilon$.

\subsubsection{Numerical results}

As for the scalar kernels, we solve the differential equation~\eqref{eq:F2wnu}
for the vorticity and vector kernels numerically in our analysis, thereby resumming
terms of all powers in the background dispersion $\epsilon$. The numerical results for
a particular configuration of $F_{2,w_i}$ and $F_{2,\nu_i}$ are shown in the upper left and right panels of Fig.\,\ref{fig:F23w}, respectively.
The numerical and analytical results agree well in the limit $k\ll k_\sigma$, as expected. Furthermore, the scaling $F_{2,w_i}(\vec k-\vec q,\vec q,\eta) \propto k^2$
for $\vec k\to 0$ holds also for the numerical result. For the vector mode, we show the dimensionless quantity $k^2F_{2,\nu}$, which
also scales as $k^2$, meaning that $F_{2,\nu}$ approaches a constant, analogously as for the scalar modes of the dispersion tensor.
The simplest approximation capturing vorticity generation is ({\bf sw}), for which the vector mode $\nu$ is neglected. As can be seen in the left upper panel of Fig.\,\ref{fig:F23w},
this approximation leads to a small shift of $F_{2,w_i}$ for $k\lesssim k_\sigma$ as compared to the case ({\bf svt}) with vector mode, and a more shallow decline for $k\gg k_\sigma$.
Note that at second order the tensor mode is irrelevant for $F_{2,w_i}$ and $F_{2,\nu_i}$, such that there is no difference between ({\bf sv}) and ({\bf svt}), and only the latter is shown.
This is also the reason why only the ({\bf svt}) result is shown for $F_{2,\nu_i}$ in the right upper panel of Fig.\,\ref{fig:F23w}, noting in addition that $\nu$ is not taken into account within the ({\bf sw}) scheme.
Finally, the upper panels show the difference between ({\bf cum2}) and ({\bf cum3+}), both evaluated for ({\bf svt}). The impact of higher cumulants is negligibly small for $k\lesssim k_\sigma$,
while it somewhat impacts the asymptotic behaviour for $k\gg k_\sigma$.

\subsection{Vorticity generation at third order}\label{sec:vortF3}

Similarly as for the velocity divergence, we find that the non-linear kernels for the vorticity in general
scale only with the first power of the total wavenumber,
\be
  F_{n,w_i}(\vec k_1,\dots,\vec k_n,\eta)\propto k,\qquad \vec k=\sum_i\vec k_i\to 0\,.
\ee
As for $\theta$, this linear scaling appears first at third order, i.e. for $n\geq 3$, while
$F_{2,w_i}\propto k^2$ as discussed above.

In the lower left panel of Fig.~\ref{fig:F23w} we show the kernel $F_{3,w_i}(\vec k-\vec p-\vec q,\vec p,\vec q,\eta)$ versus $k$ in the ({\bf sw}) and ({\bf sv}) approximations,
and for two components of $w_i$ perpendicular to ${\bf k}$ (see App.\,\ref{app:vorticity} for details on the basis choice and projection). The linear scaling in $k$ can clearly
be seen for all cases. As for $\theta$, the scaling would be $\propto k^2$ for $\vec p+\vec q=0$. This implies that the vorticity power spectrum
scales as
\be
  P_{w_iw_i}(k,\eta)\propto k^2\,,
\ee
for $k\ll k_\sigma$. However, since $F_{2,w_i}\propto k^2$, \emph{this scaling can only be observed starting at two-loop order}, while $P_{w_iw_i}^{1L}(k,\eta)\propto k^4$ at one-loop.
We will confirm this behavior against measurements of the vorticity power spectrum in numerical simulations in Sec.\,\ref{sec:simulation}.

In contrast to the vorticity, the kernels $F_{n,\nu_i}$ approach constant values for $k\to 0$, such that the dimensionless combination $k^2F_{n,\nu_i}$ scales as $k^2$ also for
$n=3$. This is confirmed by our numerical results shown in the lower right panel of Fig.\,\ref{fig:F23w}.

\subsection{Generation of tensor modes}
\label{sec:tensor}

\begin{figure}[t]
  \begin{center}
  \includegraphics[width=\columnwidth]{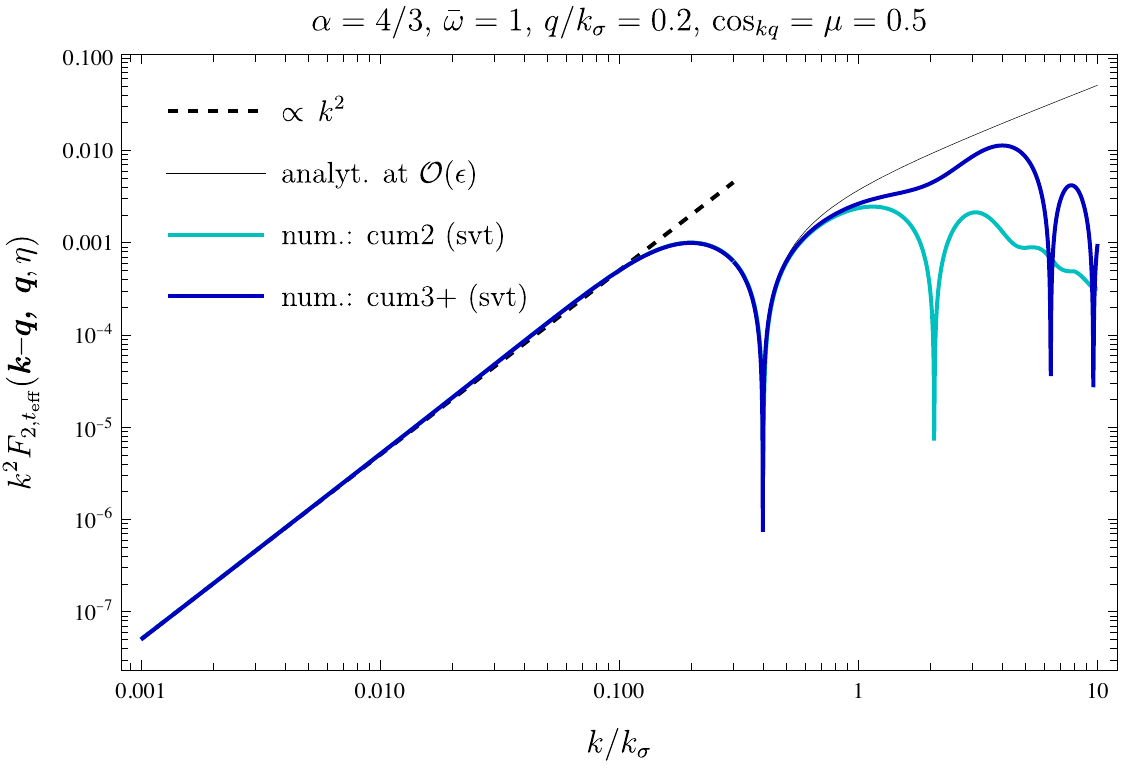}
  \end{center}
  \caption{\label{fig:F2T}
  Second order kernel of the tensor perturbation $t_{ij}$ of the velocity dispersion that describes the generation of tensor modes from coupling two scalar modes, captured by \vpt. We show the dimensionless kernel $k^2 F_{2,t_\text{eff}}(\vec k-\vec q,\vec q)$ versus $k$ in two approximations (see legend),
  as well as the analytical result at first order in $\epsilon$ (black solid line). The kernel scales as $k^2$ for $k\to 0$ (black dashed line).} 
\end{figure}

As discussed above, tensor perturbations of the stress tensor have only a very small impact on the non-linear kernels for the density contrast and velocity divergence (see Fig.\,\ref{fig:F3_delta_svt}), as opposed to impact of
vorticity and vector modes on the density.
Therefore, tensor modes could be neglected in practice when being interested in density and velocity power spectra (see Sec.\,\ref{sec:simulation}).
Nevertheless, as a consequence of non-linear mode coupling, a power spectrum of tensor modes is generated, analogously to the case of vorticity and vector modes discussed above.
The leading contribution arises from the second-order kernel,
\bea\label{eq:F2T}
  \left(\partial_\eta+3\right)&F_{2,t_{ij}}(\vec p,\vec q,\eta) = \gamma_{t_{ij} bc}(\vec p,\vec q) F_{1,b}(p,\eta) F_{1,c}(q,\eta) \,,\nn\\
\eea
giving rise to a tensor power spectrum at one-loop,
\bea\label{eq:P1LTT}
  P_{t_{ij}t_{ij}}^{1L}(k,\eta) &=& 2e^{4\eta}\int d^3q \, F_{2,t_{ij}}(\vec k-\vec q,\vec q,\eta)\nn\\
  && {} \times F_{2,t_{ij}}(\vec k-\vec q,\vec q,\eta)P_0(|\vec k-\vec q|)P_0(q)\,.\nn\\
\eea
Vertices that could potentially contribute (disregarding third and higher cumulant perturbations) are~\cite{cumPT} $\gamma_{t_{ij}\theta g}$, $\gamma_{t_{ij}\theta \epsilon}$, $\gamma_{t_{ij}w_k g}$, $\gamma_{t_{ij}w_k \epsilon}$, $\gamma_{t_{ij}\theta \nu_k}$, $\gamma_{t_{ij}\theta t_{k\ell}}$, $\gamma_{t_{ij}w_k \nu_\ell}$, $\gamma_{t_{ij}w_k t_{\ell m}}$.
Since $F_{1,b}$ and $F_{1,c}$ are non-zero only for scalar modes, only the vertices contributing to $F_{2,t_{ij}}$ are
\bea
  \gamma_{t_{ij} \theta g}(\vec p,\vec q) 
  &=& -\frac12\frac{(\vec p\times \vec q)^2\,\vec p\cdot \vec q}{2(\vec p+\vec q)^2p^2q^2} {\cal T}_{ij}\,,\nn\\
  \gamma_{t_{ij} \theta \epsilon}(\vec p,\vec q) 
  &=& \frac12\frac{(\vec p\times \vec q)^2}{(\vec p+\vec q)^2p^2} {\cal T}_{ij}\,,
\eea
where we defined
\be
 {\cal T}_{ij} \equiv \delta_{ij}-\frac{(\vec p+\vec q)_i(\vec p+\vec q)_j}{(\vec p+\vec q)^2} -2 \frac{(\vec p\times \vec q)_i(\vec p\times \vec q)_j}{(\vec p\times \vec q)^2}\,.
\ee
Note that $0={\cal T}_{ii}=(\vec p+\vec q)_i{\cal T}_{ij}={\cal T}_{ij}(\vec p+\vec q)_j$, as required for a tensor mode.

Let us first discuss the analytical result when expanding at first order in $\epsilon$.
In this case, only the vertex $\gamma_{t_{ij}\theta g}$ is relevant since
$F_{1,\delta\epsilon}\propto\epsilon^2$ (we remind the reader that the index $\epsilon$ in $\gamma_{abc}$ stands for the mode $\delta\epsilon$).
Furthermore, at ${\cal O}(\epsilon)$ we use Eq.~\eqref{eq:Tgsmalleps} for $F_{1,g}$ and can set $F_{1,\theta}\to 1$, giving
\bea
  F_{2,t_{ij}}(\vec p,\vec q,\eta) &=& 2(\gamma_{t_{ij}\theta g}(\vec p,\vec q)+\gamma_{t_{ij}\theta g}(\vec q,\vec p))\nn\\
  && {} \times \int^\eta d\eta' e^{3(\eta'-\eta)}\int^{\eta'}d\eta''\,e^{2(\eta''-\eta')}\epsilon(\eta'')\nn\\
  && {} + {\cal O}(\epsilon^2)\,.
\eea
This analytical result is compared to the full numerical kernel in Fig.\,\ref{fig:F2T}. Here we defined
\be
  F_{2,t_{ij}}(\vec p,\vec q,\eta)=F_{2,t_\text{eff}}(\vec p,\vec q,\eta) \, {\cal T}_{ij}\,,
\ee
factoring out the dependence on the indices captured by ${\cal T}_{ij}$ (see App.\,\ref{app:vorticity} for details).
The figure shows the dimensionless combination $k^2F_{2,t_\text{eff}}(\vec k-\vec q,\vec q)$, that scales as $k^2$ for $k\to 0$, analogously to the vector perturbation
of the stress tensor. As for the latter, the same scaling is expected at higher orders, and therefore no qualitative change is expected when including contributions beyond one-loop order.
Thus, $P_{t_{ij}t_{ij}}(k,\eta)$ approaches a constant for $k\to 0$ at all loop orders.
For the dimensionless tensor power spectrum, this means that $k^4P_{t_{ij}t_{ij}}(k,\eta)\propto k^4$ for $k\to 0$.

The tensor modes of the stress tensor generated by non-linearities constitute a source for a stochastic gravitational wave background. Since their frequency is related to the wavenumber $k$, its maximal amplitude is expected to
occur for ultra-low frequencies related to the dispersion scale $f_\text{gw}\sim c k_\sigma/(2\pi)\simeq 1.5\cdot 10^{-15}\, h$ Hz $\times (k_\sigma/(1h/\text{Mpc}))$.
In addition, due to the non-relativistic velocities of order ${\cal H}\times \sqrt{\epsilon}\simeq 100 \text{km}/\text{s} \times \sqrt{\epsilon}/(\text{Mpc}/h)$ the amplitude of the gravitational wave
background is expected to be tiny, although the small velocities can be somewhat compensated by the large masses that are involved. For a more detailed assessment, we refer to future work. 

\section{Background dispersion from simulations and perturbation theory}
\label{sec:powerlaw}

In this section we discuss various estimates of the background dispersion $\epsilon(\eta)$,
that plays a central role in \vpt.
We discuss two complementary approaches, one based on a self-consistent solution of the equation of motion 
for $\epsilon(\eta)$ within perturbation theory (Eq.~\ref{eq:epseom}), and one based on accounting for the dispersion generated in halos following paper~I~\cite{cumPT},
and taking the halo profile as well as the halo mass function measured in numerical N-body simulations into account.

For simplicity, we focus on a scaling universe with linear input power spectrum
\be
  P_0(k)=A k^{n_s}\equiv \frac{1}{4\pi k^3}\left(\frac{k}{k_\text{nl}}\right)^{n_s+3}\,,
\ee
with constant power law index $n_s$ and amplitude $A$ related to the non-linear scale $k_\text{nl}$, for an EdS background cosmology.
Scaling symmetry ensures that the dimensionless ratio of any dimensionfull quantity to an appropriate power of the time-dependent non-linear scale
$k_\text{nl}(\eta)=k_\text{nl}e^{-2\eta/(n_s+3)}$ is constant in time.
This implies~\cite{cumPT}
\be\label{eq:epspowerlaw}
  \epsilon(\eta)=\epsilon_0\,e^{\alpha\eta},\quad \alpha=\frac{4}{n_s+3}\,.
\ee
For given $n_s$, this leaves a single free parameter $\epsilon_0\equiv 1/k_\sigma^2$ for the
background dispersion, that we can conveniently parameterize by the ratio
\be
  k_\sigma/k_\text{nl} \equiv (\epsilon_0 k_\text{nl}^2)^{-1/2}\,.
\ee
We note that when defining a time-dependent dispersion scale by $\epsilon(\eta)\equiv1/k_\sigma^2(\eta)$, the time dependence given above
indeed implies that $k_\sigma(\eta)/k_\text{nl}(\eta)=k_\sigma/k_\text{nl}$ is constant.
Similarly, scaling symmetry implies a constant dimensionless ratio $\bar\omega\equiv \omega(\eta)/\epsilon(\eta)^2$ of the fourth cumulant expectation value to the dispersion squared.

\subsection{Self-consistent solution in perturbation theory}\label{sec:selfconsistent}

\begin{table}
  \centering
  \caption{Self-consistent solutions in linear and one-loop approximation for the velocity dispersion scale $k_\sigma=\epsilon_0^{-1/2}$ relative to the non-linear scale, for scaling universes with spectral indices $n_s=2,1,0$ and in various approximation schemes (see Table~\ref{tab:approxschemes}). Note that vorticity+vector ({\bf sv}) and tensor ({\bf svt}) modes enter starting at one-loop only.
  For ({\bf cum3+}), we show results for two choices $\bar\omega=\bar\omega_\text{fid}=0.579, 0.616, 0.668$ for $n_s=2,1,0$ and $\bar\omega=1$, respectively (see text for details).}
  \begin{ruledtabular}
    \begin{tabular}{c|ccc|ccccc} 
   &&&&&&&&\\[-2ex]
   $n_s$ & \multicolumn{3}{c|}{$k_\sigma/k_\text{nl}$ (linear)} &  \multicolumn{5}{c}{$k_\sigma/k_\text{nl}$ (one-loop)}   \\[1.5ex] \hline &&&&&&&&\\[-1.ex]
    & ({\bf cum2}) & \multicolumn{2}{c|}{({\bf cum3+})} & \multicolumn{3}{c}{({\bf cum2})} & \multicolumn{2}{c}{({\bf cum3+})}  \\[1.5ex]
   & - & - & - &  ({\bf s}) &  ({\bf sv})&  ({\bf svt})&  ({\bf svt})&  ({\bf svt}) \\[1.5ex] 
   & - & $\bar\omega_\text{fid}$ & $\bar\omega=1$ & - & - & - & $\bar\omega_\text{fid}$ & $\bar\omega=1$ \\[1.5ex] \hline &&&&&&&&\\[-1.ex]
  $2$ & $2.34$ & $2.00$ & $1.85$ & $1.80$ & $1.74$ & $1.74$ & $1.62$ & $1.51$ \\[1.5ex]
  $1$ & $2.55$ & $2.17$ & $2.00$ & $1.96$ & $1.86$ & $1.87$ & $1.69$ & $1.64$ \\[1.5ex]
  $0$ & $2.97$ & $2.50$ & $2.29$ & $2.28$ & $2.11$ & $2.11$ & $2.13$ & $1.97$ \\
    \end{tabular}
  \end{ruledtabular}
  \label{tab:selfconsistent}
\end{table}

The equation of motion~\eqref{eq:epseom} for the background dispersion can for a scaling universe and EdS background be turned
into an algebraic equation for $k_\sigma/k_\text{nl}$, that can be written as (see paper~I~\cite{cumPT})
\be\label{eq:epseompowerlaw}
  \frac{\epsilon'(\eta)}{\epsilon(\eta)}+1=\frac13\left( x I^\text{lin}(n_s)+x^2 I^{1L}(n_s)+\dots\right)\,,
\ee
where $x\equiv (k_\sigma/(\sqrt{3}k_\text{nl}))^{n_s+3}$ and the ellipsis denote two and higher loop contributions.
Using Eq.g~\eqref{eq:epspowerlaw}, the left-hand side evaluates to a constant, $\epsilon'/\epsilon+1=(n_s+7)/(n_s+3)$.
Furthermore, the coefficients $I^\text{lin}(n_s)$ and $I^{1L}(n_s)$ are independent of $x$ by virtue of the scaling symmetry,
and are given by integrals over the sum of the (suitably rescaled) power spectra $P_{\theta\tilde g}=P_{\theta g}-P_{\theta \delta\epsilon}$, $2P_{w_i\nu_i}$
and $P_{A\pi}$ evaluated in linear and one-loop approximation, respectively~\cite{cumPT}. These power spectra arise from the source term, Eq.~\eqref{eq:Q3rd}, 
entering the equation of motion~\eqref{eq:epseom} for $\epsilon(\eta)$. 
When including one-loop and linear terms on the right-hand side of Eq.~\eqref{eq:epseompowerlaw} one therefore obtains a quadratic equation for $x$. Its solution(s) $x_*$ determine the self-consistent value for
\be
  k_\sigma/k_\text{nl}\big|_\text{self-consistent}=\sqrt{3}x_*^{1/(n_s+3)}\,,
\ee
within perturbation theory up to one-loop order. Correspondingly, dropping also the term $x^2 I^{1L}(n_s)$ yields self-consistent values in linear approximation.
Note that since $I^\text{lin}(n_s)$ and $I^{1L}(n_s)$ are independent of $x$, and therefore of $k_\sigma$, the power spectra entering these expressions can be
computed at an arbitrary reference value $k_\sigma^\text{ref}$ (see App.\,\ref{app:rescaling} for details). We evaluate them in various approximation schemes (see Table~\ref{tab:approxschemes}).
At one-loop, we find that only one of the roots $x_*$ of the quadratic equation~\eqref{eq:epseompowerlaw} is positive, thus leading only to a single physical solution for $k_\sigma/k_\text{nl}$.

The results for the self-consistent value of $k_\sigma/k_\text{nl}$ are shown in Table~\ref{tab:selfconsistent}. Including the one-loop correction leads to an increase in the background dispersion $\epsilon(\eta)$ relative to the linear approximation, corresponding to a decrease in $k_\sigma$ by about $20-30\%$. While being significant, the shift is consistent with the one-loop
contribution being a perturbative correction. In addition, we observe that the self-consistent background dispersion is larger (i.e. $k_\sigma$ smaller) when including third cumulant perturbations ({\bf cum3+}) compared to the second cumulant approximation ({\bf cum2}). For the ({\bf cum3+}) case, we fix $\bar\omega$ to the value obtained from self-consistently solving the equation of motion of the fourth cumulant expectation value in linear approximation~\cite{cumPT} as fiducial value (see Table~\ref{tab:selfconsistent}), and consider the alternative value $\bar\omega=1$ to assess the dependence on this quantity. A self-consistent determination of $\bar\omega$ at one-loop is beyond the scope of this work, but would be an interesting extension. Nevertheless, the dependence of $k_\sigma/k_\text{nl}$ on $\bar\omega$ is relatively mild, such that our fiducial choice may be considered as a reasonable estimate. Note that vorticity, vector and tensor modes do not enter in the linear approximation. At one-loop, taking vorticity and vector modes into account ({\bf sv}) leads to a moderate increase in $\epsilon(\eta)$ compared to the case with scalar perturbations only ({\bf s}). On the contrary, adding also tensor modes ({\bf svt}) has a negligible impact. Finally, we note that we checked the convergence of the integrals over the power spectra entering
$I^{1L}(n_s)$, with negligible differences when using $30k_\sigma$ versus $40k_\sigma$ as cutoff for $n_s=2,1$. For $n_s=0$, the difference is at the percent level.

Overall, we find that the self-consistent prediction for the background dispersion $\epsilon(\eta)$ has the tendency to increase (such that $k_\sigma$ decreases) when enhancing the complexity of the approximation scheme, i.e. when successively including one-loop, vorticity/vector mode and higher cumulant effects. The self-consistent determination of $k_\sigma$ at one-loop therefore may be considered as indicative, with the actual value likely being smaller. In order to bracket the uncertainty on the background dispersion, we consider in addition an orthogonal approach based on a halo model in the following.

\subsection{Dispersion from simulated halos}\label{sec:halo}
Let us calculate the velocity dispersion contribution expected from dark matter halos. If the dispersion is only non-zero inside dark matter halos, we have
\be
\epsilon(z)=\lexp \epsilon_{ii}(\bm{x},z)/3 \rexp = \int \epsilon P(\epsilon)  d\epsilon =  \int \epsilon P(\epsilon/h) P(h) dh\,  d\epsilon
\label{epsH1}
\ee
where we used Bayes' theorem to introduce the conditional probability $P(\epsilon/h)$, i.e. the probability for $\epsilon(\bm{x},z)$ given that we are inside a dark matter halo. Since halos are characterized by their mass we can write, more precisely
\be
\epsilon(z)= \int_0^\infty\bar{\epsilon}_h(m,z)\, f(m)\, dm\,,
\label{epsFromH2}
\ee
where $f(m)$ is the fraction of mass in halos of mass $m$, which plays the role of $P(h)$, and the conditional average
\be
\int \epsilon P(\epsilon/h) \, d\epsilon = 
\bar{\epsilon}_h(m,z)\equiv \Big({3\over 4\pi r_\text{vir}^3}\Big) \int d^3x\  \epsilon_{h,m}(\bm{x},z)\,,
\label{epsbarh}
\eeq
where the integral on the RHS is done over the halo. The mass fraction $f(m)$ is related to the mass function $dn/dm$, i.e. the number density of halos of mass $m$ per unit mass, by $dn/dm \equiv \bar{\rho}f(m)/m$.
A convenient parametrization of the mass function is given by~\cite{SheTor9909,DesGioAng1603}
\bea
m\, f(m)&=& A \Big[1+ {1\over (a \nu)^p} \Big] \sqrt{a\nu \over 2\pi} {\rm e}^{-a\nu/2}\, {d\ln \nu \over d\ln m} \nonumber \\ & \equiv &\nu f(\nu) \ {d\ln \nu \over d\ln m} \,,
\label{MFst}
\eea
where $\sqrt{\nu}\equiv\delta_c/\sigma(R,z)$, with $\delta_c$  the critical density for collapse at the corresponding time and $\sigma^2(R,z)$ the variance of linear-SPT  fluctuations, with $R$ the Lagrangian radius of halos of mass $m$, i.e. $m=4\pi \bar{\rho}R^3/3$ where $\bar\rho$ is the total homogeneous matter density. The constants $a,p,A$ are fit to N-body simulations.   
The condition $\nu=1$ defines the non-linear scale $R_*$ and its associated non-linear mass scale $m_*$, with $R_*$ being the Lagrangian size of halos of mass $m_*$. 
This mass scale determines when the abundance of halos is exponentially suppressed from Eq.~(\ref{MFst}), i.e. essentially there is not enough time to build significant number of halos more massive than $m=m_*$.

For the application we are interested in this paper, a scale-free universe with $-1\leq n_s \leq 2$ and $\Omega_m=1$, $\delta_c(z)=1.686$. However, the usual calculation of $\sigma^2(R,z)$ becomes problematic for blue spectral indices as the standard choice of a top-hat filter gives a divergent answer for $\sigma^2(R,z)$ when $n_s\geq 1$. Replacing  the linear-SPT evolution by the linear evolution with dispersion found in paper~I~\cite{cumPT} cures this UV divergence, thus providing an interesting application of our results. In this case, the variance is given by,
\beq
\sigma^2= \Big({k_\sigma \over k_\text{nl}}\Big)^{n_s+3} \int_0^\infty {dx\over 2}\, W_{\rm TH}^2(\sqrt{x} k_\sigma R)\, x^{{n_s+1}\over 2} \, F_{1,\delta}^2(x,n_s)\,,
\label{sigmaTHvlasov}
\eeq
where $x\equiv k^2 \epsilon = (k/k_\sigma)^2$, $W_{\rm TH}$ is the top-hat window in Fourier space, the linear kernel is given by
\beq
F_{1,\delta}(x,n_s) \equiv {}_1  F_2\left(\frac{4+\alpha}{3\alpha};1+\frac{2}{\alpha},1+\frac{5}{2\alpha};-\frac{3x}{\alpha^2}\right)\,,
\label{1F2}
\eeq
with  ${}_1 F_2$ a hypergeometric function, $\alpha(n_s)=4/(n_s+3)$ as usual,  and $k_\sigma/k_\text{nl}$ is obtained from the self-consistent linear solution with dispersion (see Fig.~9 and Table~I in paper~I~\cite{cumPT}). For example, setting $\sigma=\delta_c$ gives $k_\text{nl}R_*=0.83,0.98,1.09,1.15$ for $n_s=-1,0,1,2$ respectively. For reference, using linear-SPT fluctuations gives $k_\text{nl}R_*=0.89,1.18$ for $n_s=-1,0$. 

It is instructive to first consider the contribution to $\bar{\epsilon}_h$ in Eq.~(\ref{epsFromH2}) from an isothermal halo, before we address the more realistic case of an NFW profile discussed in paper~I~\cite{cumPT}. In this case, the contribution to 1D velocity dispersion is simply $Gm/(2r_{\rm vir})$ independent of location, which in the normalization corresponding to $\bar{\epsilon}_h$ means  an isothermal halo of mass $m$ contributes to $\epsilon(z)$ as:
\beq
{G\, m \over 2\, r_{\rm vir}\, ({\cal H} f)^2} = {\widehat{m}^{2/3}\over 4f^2}\, \frac{\Omega_{m,0}^{2/3}\ \Delta_{\rm vir}^{1/3} }{\Omega_{m,0}+\Omega_{\Lambda,0} \,a^3}\ R_*^2\,,
\label{isothermal}
\eeq
where we have defined the dimensionless variable $\widehat{m}\equiv m/m_*$, and used that $m=(4\pi/3)\rho_{c,0} \Delta_{\rm vir} r_{\rm vir}^3/a^3$, the non-linear mass $m_*=(4\pi/3)\rho_{c,0}\Omega_{m,0} R_*^3$ and assumed a flat $\Lambda$CDM cosmology that goes to the self-similar case we are mostly interested in here when $\Omega_{\Lambda,0}=1-\Omega_{m,0}\to 0$. Here quantities with $0$ subindices correspond to $z=0$, and note that we assume halos are identified as having a mean physical density of $\Delta_{\rm vir}$ over the critical density $\rho_c$~\cite{DesGioAng1603}. A simple estimate is  $\Delta_{\rm vir}(z)=18 \pi^2 +82 (\Omega_m-1)-39 (\Omega_m-1)^2$~\cite{BryNor9803}. Since we are interested in predicting $k_\sigma/k_\text{nl}$ it is useful to consider
\beq
\Big({G\, m \, k_\text{nl}^2\over 2\, r_{\rm vir}\, ({\cal H} f)^2} \Big)^{-1/2}= 2 f \widehat{m}^{-1/3} \sqrt{\frac{\Omega_{m,0}+\Omega_{\Lambda,0} \,a^3}{\Omega_{m,0}^{2/3}\ \Delta_{\rm vir}^{1/3} }} \ {1\over k_\text{nl}R_*}\,,
\label{isothermal2}
\eeq
which is the contribution from an isothermal halo of mass $m$ to $k_\sigma/k_\text{nl}$. In the $\Omega_m\to 1$ limit, this becomes,
\beq
\Big({G\, m \, k_\text{nl}^2\over 2\, r_{\rm vir}\, {\cal H}^2} \Big)^{-1/2}= {2\,  \widehat{m}^{-1/3}\over (18\pi^2)^{1/6}} \ {1\over k_\text{nl}R_*} \simeq 0.84 \ {\widehat{m}^{-1/3}\over k_\text{nl}R_*}\,,
\label{isothermal3}
\eeq
which given the values of $ k_\text{nl}R_*$ quoted above says that halos of order $m\sim m_*$ ($\widehat{m}=1$) contribute $k_\sigma \sim k_\text{nl}$. More precisely, if we integrate over the mass function we obtain $(k_\sigma/k_\text{nl})\simeq 0.71, 0.81,0.91,1.04$ for $n_s=2,1,0,-1$ respectively, i.e. a dispersion scale larger than the non-linear scale. This gives a useful order of magnitude, but isothermal halos overestimate the velocity dispersion compared to realistic halo profiles. Therefore, let us now consider NFW halos. 

\begin{figure}[t]
  \begin{center}
  \includegraphics[width=\columnwidth]{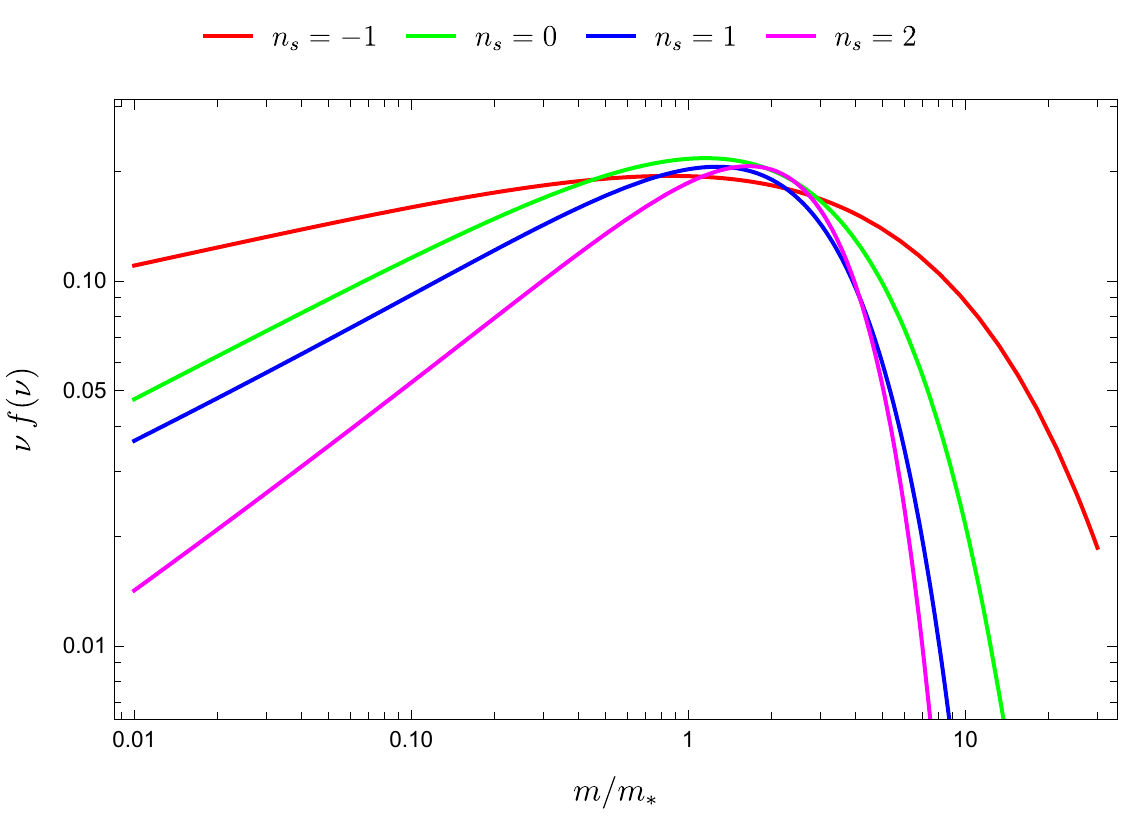}
  \end{center}
  \caption{\label{fig:nuFnu} 
    Mass function $\nu f(\nu)$ fits to simulation measurements (see App.~\ref{app:HMF}) as a function of $\widehat{m}=m/m_*$ for $n_s=-1,0,1,2$ (top to bottom from the low-mass end). 
  }
\end{figure}

From paper~I~\cite{cumPT}, we have that the 1D velocity dispersion $m_2(r)$ as a function of radius $r$ inside a halo with profile $\rho(r)$ and velocity dispersion tensor anisotropy parameter $\beta$ (assumed to be constant) is given by
\beq
m_2(r) =\frac{(1-2\beta/3)}{\rho\, r^{2\beta}} \int_r^\infty{d\Phi\over dr} \rho(r)\, r^{2\beta}\, dr\,,
\label{epsilonNFW}
\eeq
where for the case of NFW halos the potential $\Phi(r)$ can be written as
\beq
\Phi(r)= -{G m \over r_{\rm vir}}\, c\, g(c)\, \Big[{\ln (1+x) \over x} -{1\over 1+c}\Big]\,,
\label{PhiNFW}
\eeq
where $x\equiv c\,r/r_{\rm vir}$, $c$ is the concentration parameter and $g^{-1}(c) \equiv \ln(1+c)-c/(1+c)$. According to simulations, the concentration of a halo is a weak function of halo mass, we take $c(\widehat{m})=9~\widehat{m}^{-0.13}$~\cite{BulKolSig01}. A more recent study that includes scale-free simulations with red spectral indices $n_s\leq -1$  builds a more accurate model with six free parameters~\cite{DieKra1501}, but  their own comparison (see their Fig.~10) shows small differences at high-mass with~\cite{BulKolSig01} that  become smaller in the direction of increasing $n_s$, which is most relevant to our application. Given Eqs.~(\ref{epsilonNFW}-\ref{PhiNFW}), we can write the contribution to $\epsilon$ in analogy to the results above for an isothermal halo. We have,
\beq
 \bar{\epsilon}_h(m,z)= 
{ \widehat{m}^{2/3}\over 4f^2}\, \frac{\Omega_{m,0}^{2/3}\ \Delta_{\rm vir}^{1/3} \ R_*^2}{\Omega_{m,0}+\Omega_{\Lambda,0} \,a^3}\ I(\beta,c)\,,
\label{epshNFW}
\eeq
where we have factorized the isothermal halo contribution (see Eq.~\ref{isothermal}) and the integral $I$ is given by
\beqa
 I(\beta,c) &\equiv &
 \Big(6- 4\beta\Big)  {g(c)\over c^2}
\int_0^c dx\, x^{3-2\beta}(1+x)^2  \\ & & \times \int_x^\infty dy {y^{2\beta-1}\over (1+y)^2}\Big[{\ln (1+y) \over y^2} -{1\over y(1+y)}\Big]\,, \nonumber 
\label{Ibetac}
\eeqa
and is a function of $\widehat{m}$ through the concentration parameter. It can be computed analytically, see Eqs.~(\ref{Ibeta0}-\ref{Ibetahalf}) in App.~\ref{app:HMF} where we present results for the cases of interest, namely $\beta=0$ (isotropic case) and $\beta=1/2$ (radially biased case, i.e. radial dispersion larger than tangential dispersion). Simulations show $\beta$ to be scale dependent, with $\beta\approx 0$ near halo centers and $\beta$ approaching 1/2 at the halo outskirts. Since the profile is volume weighted we consider $\beta= 1/2$ to be more realistic for our purposes. 

\begin{table}
  \centering
  \caption{Velocity dispersion scale (in terms of the non-linear scale) from NFW  halos  as a function of spectral index.}
  \begin{ruledtabular}
    \begin{tabular}{rcc} 
   $n_s$ & $(k_\sigma/k_\text{nl})_{\beta=0}$  & $(k_\sigma/k_\text{nl})_{\beta=1/2}$    \\[1.5ex] \hline & &  \\[-1.ex]
  $2$ & $0.845$ & $0.863$   \\[1.5ex]
  $1$ & $0.957$ & $0.977$  \\[1.5ex]
  $0$ & $1.084$ & $1.105$  \\[1.5ex]
  $-1$ & $1.219$ & $1.241$  \\
    \end{tabular}
  \end{ruledtabular}
  \label{tab:ksigma}
\end{table}

For the case of a scale-free universe we have, integrating over the mass function and using $d\ln\nu/d\ln m=(n_s+3)/3$,

\beq
 \epsilon= {{n_s+3}\over 12} 
 (18\pi^2)^{1/3}  R_*^2 \int_0^\infty {d\widehat{m}\over \widehat{m}^{1/3}}\, [\nu f(\nu)]_{\widehat{m}}\  I(\beta,c)\,,
\label{epsNFW}
\eeq
which gives us the desired $k_\text{nl}/k_\sigma$ as a function of spectral index $n_s$
\beq
{k_\sigma \over k_\text{nl}} = \Big(\frac{96}{\pi^2}\Big)^{1/6} \frac{k_\text{nl}R_*}{{\sqrt(n_s+3)  \int d\widehat{m}\, \widehat{m}^{-1/3} [\nu f(\nu)]_{\widehat{m}}\  I(\beta,c) }}\,.
\label{ksigmaNFW}
\eeq

The mass function is measured in our simulations for different spectral indices as discussed in App.~\ref{app:HMF}, where we check our mass function measurements obey the expected self-similarity. As a summary of these measurements, Fig.~\ref{fig:nuFnu} presents the resulting fits for $[\nu f(\nu)]$ as a function of $\widehat{m}$.  The parameters $A,a,p$ (see Eq.~\ref{MFst}) for each spectral index are presented in Table~\ref{tab:HMF} in App.~\ref{app:HMF}. The behavior of the mass function at the high-mass end agrees with standard expectations, i.e. red spectra with more large-scale power have enhanced exponential tails. At the small-mass limit, however, blue spectra with more initial small-scale power show a suppressed halo abundance, highlighting the effects of increased velocity dispersion which is reflected already by linear evolution in \vpt~(see paper~I~\cite{cumPT}) and confirmed by our non-linear analysis. 

The results of Eq.~(\ref{ksigmaNFW}) after integration over the mass function are given in Table~\ref{tab:ksigma}. Overall we see that the dispersion values are smaller (higher $k_\sigma$) than for the corresponding isothermal halos quoted earlier, as expected. We also see that the dispersion scale changes weakly (by about 2\%) when $\beta$ is varied within reasonable limits. As mentioned earlier (see Eq.~\ref{1F2}) our results (as well as the mass function fits) are obtained using the self-consistent linear solution with dispersion only. Including higher cumulants in the linear self-consistent solution  (for both the computation of $R_*$ and the corresponding $m_*$ that appears in the mass function fits) changes the values in Table~\ref{tab:ksigma} by at most $0.1\%$, which is reassuring. 

It is worth noting comparing to Table~\ref{tab:selfconsistent} that these halo-based calculations yield stronger velocity dispersion than the self-consistent one-loop calculation done in the previous section, but by less than a factor of two. This is  remarkable, given the wide differences in the assumptions: perturbative dynamics with non-zero dispersion everywhere on the one hand, and dispersion only inside halos (modeled as spherical NFW profiles) on the other. Interestingly, we shall see that fitting for $k_\sigma$ from the density power spectrum measured in N-body simulations yields values that are in between the one-loop perturbative and halo calculations. We discuss such determinations next.

\section{{\vpt}~predictions vs Simulations}
\label{sec:simulation}

In this section we present results for density and velocity power spectra as well as the bispectrum computed within the framework of \vpt~and compare to N-body simulations. We focus on a scaling universe with linear input power spectrum $P_0\propto k^{n_s}$ and consider the values $n_s=2,1,0,-1$, with the very blue cases included to expose the strong screening of UV modes that occurs when taking velocity dispersion and higher cumulants into account. Note that within SPT the one-loop integrals become UV divergent for $n_s\geq -1$, therefore there are no SPT predictions for these spectra beyond tree-level. As we shall see, this shortcoming is remedied in \vpt.

Our perturbative results are based on the framework described above. In particular, we use ({\bf cum3+}, {\bf svt}) as our baseline scheme, and compare the impact of various approximation schemes  (see Table~\ref{tab:approxschemes}) against this fiducial choice. It includes apart from $\delta$ and $\theta$ the vorticity $w_i$, all scalar ($g,\delta\epsilon$) vector ($\nu_i$) and tensor ($t_{ij}$) modes of the second cumulant as well as its background value $\epsilon(\eta)$, and higher cumulants as detailed in Sec.\,\ref{sec:kernels} (scalar perturbations $\pi,\chi$ of the third cumulant and background value $\omega(\eta)$ of the fourth cumulant).

\subsection{Density power spectrum}\label{sec:Pdd}

\begin{figure}[t]
  \begin{center}
  \includegraphics[width=\columnwidth]{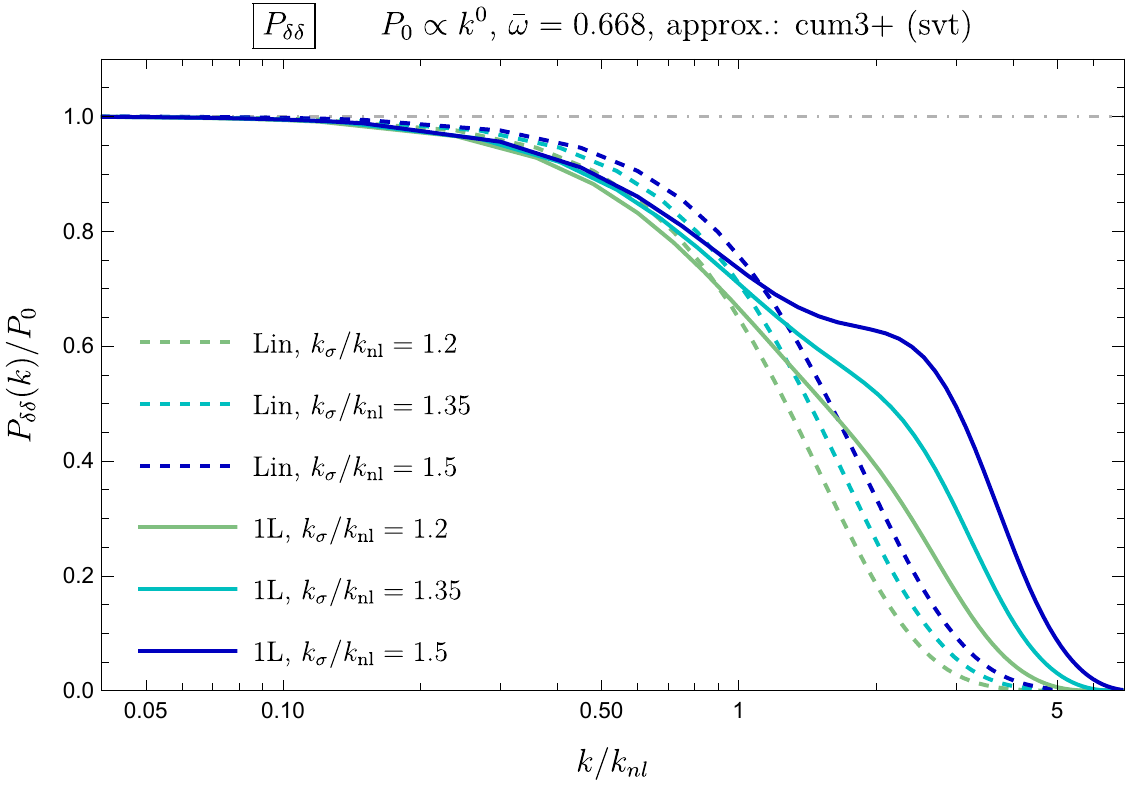}
  \end{center}
  \caption{\label{fig:Pdd_various_ksig} 
    Density power spectrum $P_{\delta\delta}(k,\eta)$ in one-loop (solid) and linear (dashed) approximation for various values of the background dispersion
    $\epsilon_0=1/k_\sigma^2$ for a power law input spectrum $P_0\propto k^{n_s}$ with $n_s=0$ and $\epsilon(\eta)=\epsilon_0e^{4\eta/(n_s+3)}$. The linear and one-loop result contains the numerical kernels
    $F_{1,\delta}(k,\eta)$, $F_{2,\delta}(\vec k-\vec q,\vec q,\eta)$ and $F_{3,\delta}(\vec k,\vec q,-\vec q,\eta)$ computed taking up to third cumulant scalar perturbations as well as vorticity, vector and tensor modes of the stress tensor into account ({\bf cum3+}, {\bf svt}). We set $\bar\omega=0.668$ (see text and~\cite{cumPT}) for the fourth cumulant expectation value $\omega(\eta)=\bar\omega\times\epsilon(\eta)^2$.
  }
\end{figure}

\begin{figure*}[t]
  \begin{center}
  \includegraphics[width=\textwidth]{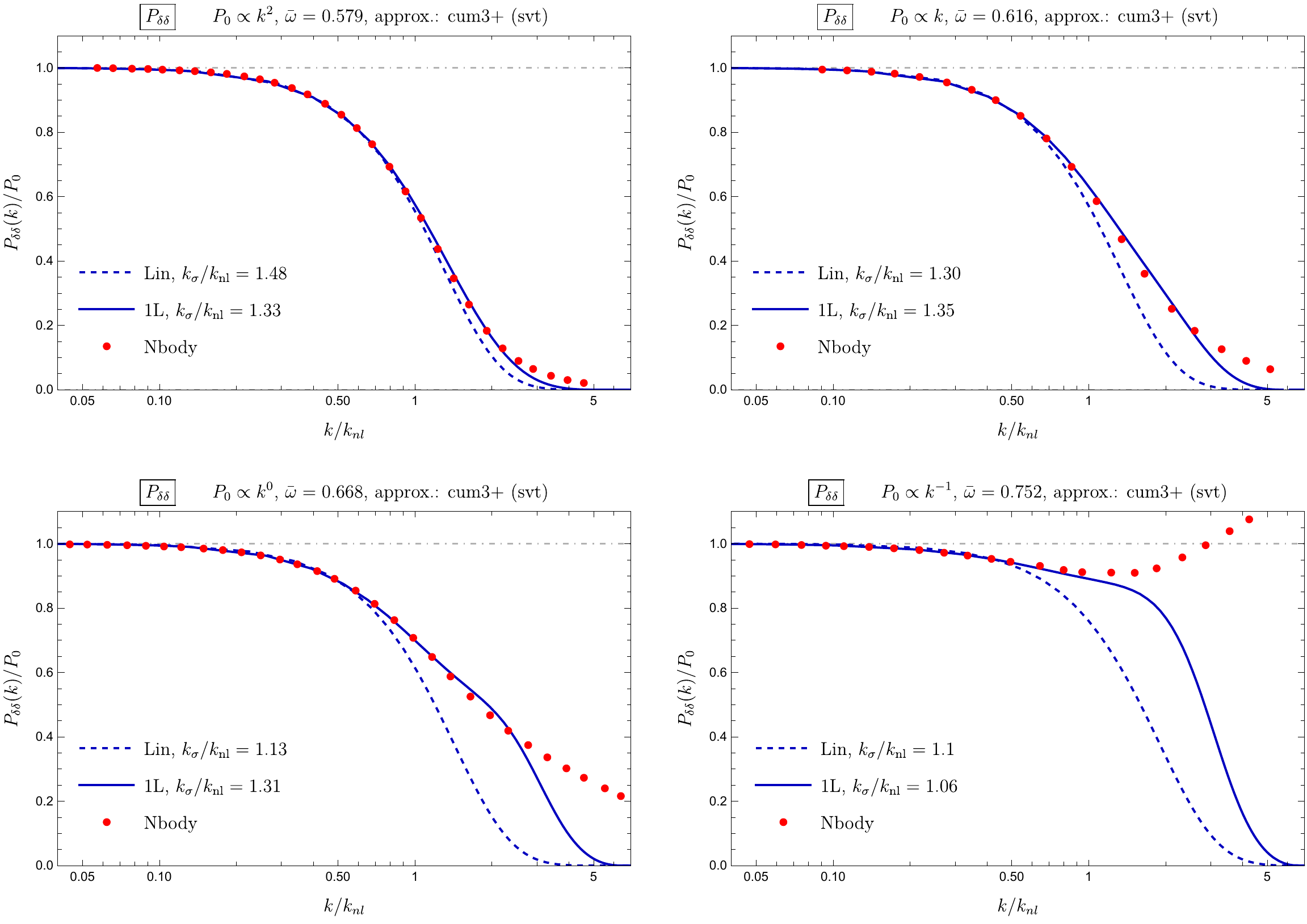}
  \end{center}
  \caption{\label{fig:Pdd_cumPT_Nbody} 
  Comparison of $P_{\delta\delta}$ in linear (dashed) and one-loop (solid) approximation in \vpt~within the ({\bf cum3+}, {\bf svt}) scheme to N-body results (red circles), for $n_s=2,1,0,-1$, as noted. In absence of a precise knowledge of the background dispersion, we adjusted $k_\sigma/k_\text{nl}$ by a one-parameter fit up to $k\leq k_\text{max}=0.6k_\text{nl}$, with best-fit values given in the legend for each case. Notably, these values of $k_\sigma/k_\text{nl}$ are consistent with our theoretical expectations given in Sec.\,\ref{sec:powerlaw}. We use the \emph{same fixed} values of $k_\sigma/k_\text{nl}$ as given here when comparing \vpt~and N-body results for velocity spectra and bispectra below (for each approximation and $n_s$, respectively) such that for all these statistics there is only a single free parameter that is adjusted once to a common value.
  }
\end{figure*}

\begin{figure*}[t]
  \begin{center}
  \includegraphics[width=0.49\textwidth]{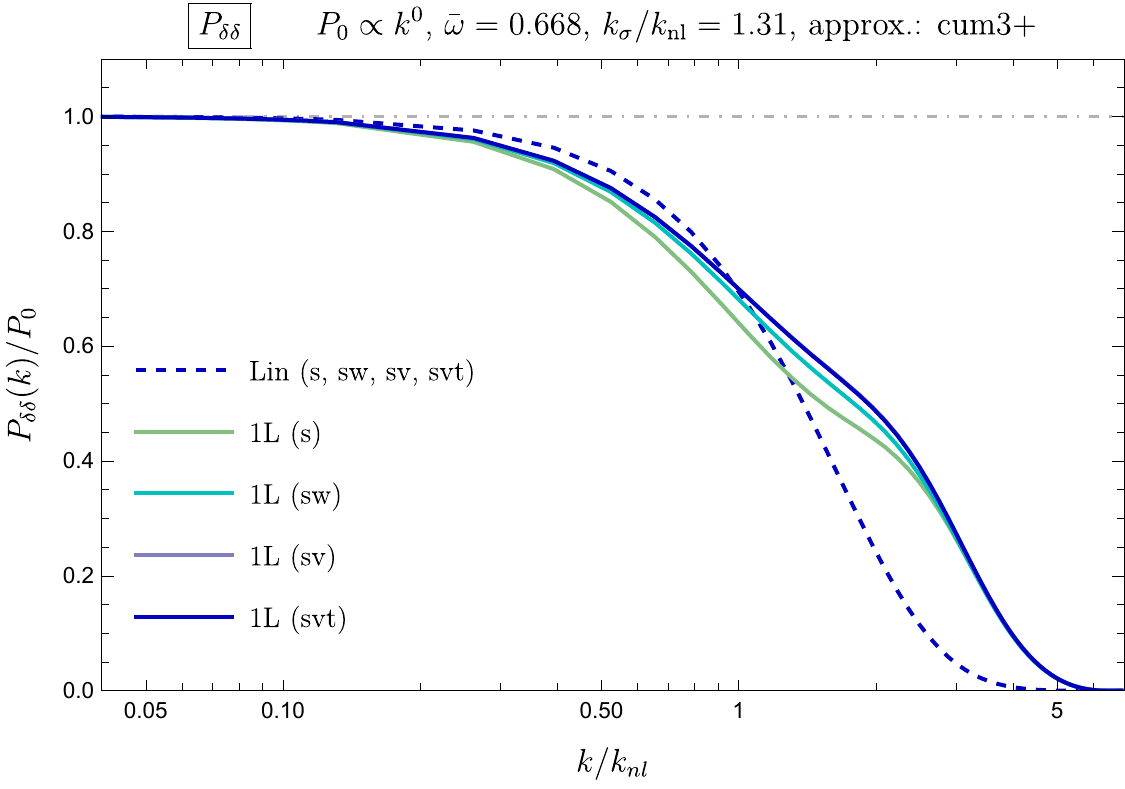}
  \includegraphics[width=0.49\textwidth]{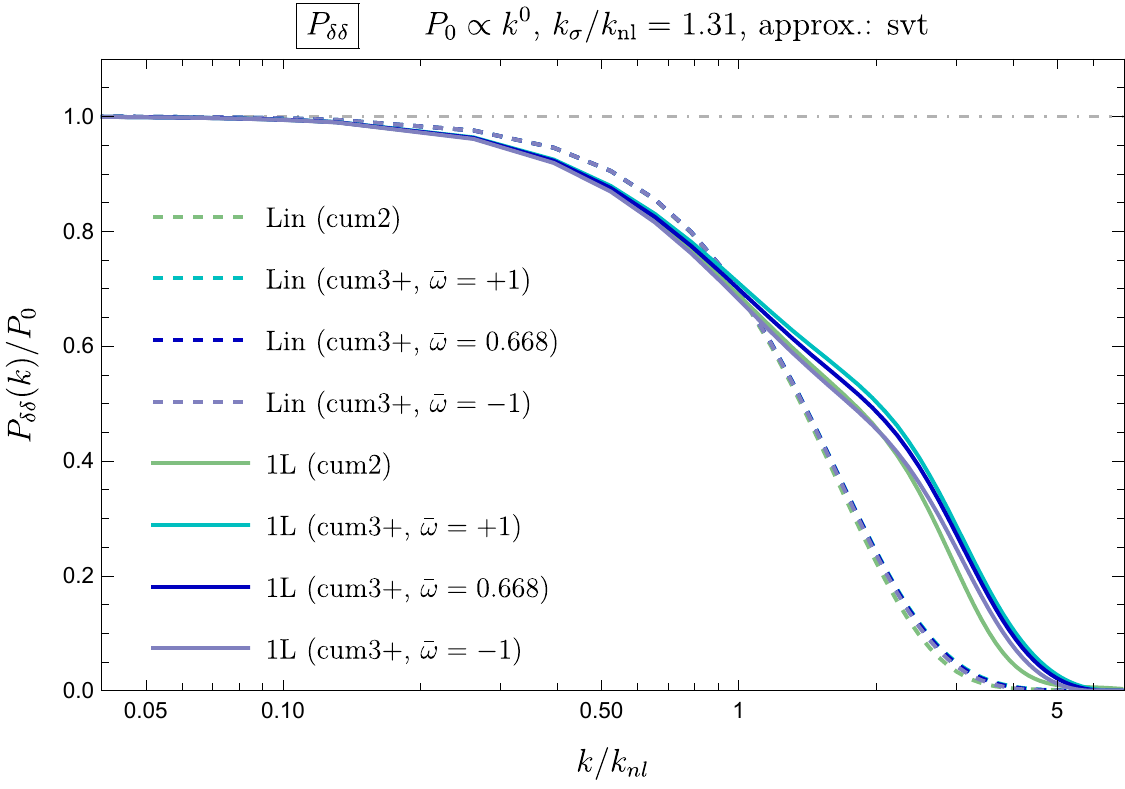}
  \end{center}
  \caption{\label{fig:Pdd_cumPT} 
    {\em Left panel:} dependence of the density power spectrum $P_{\delta\delta}(k,\eta)$ in one-loop (solid) and linear (dashed) order on the  approximation schemes ({\bf s}), ({\bf sw}), ({\bf sv}), ({\bf svt}) quantifying the impact of successively including vorticity  ({\bf sw}), vector ({\bf sv}) and tensor ({\bf svt}) modes. Note that ({\bf sv}) and ({\bf svt}) are essentially indistinguishable.  {\em Right panel:} second versus third cumulant approximations ({\bf cum2}) vs ({\bf cum3+}), with three choices of $\bar\omega$ for the latter    and a common value $k_\sigma=1.31 k_\text{nl}$ (see Fig.\,\ref{fig:Pdd_cumPT_Nbody})  in all cases.
  }
\end{figure*}

We start with a discussion of the perturbative prediction of the density power spectrum within \vpt, and then compare to N-body results.
We use Eq.~\eqref{eq:P1Ldef} to compute the linear and one-loop contribution, with kernels $F_{1,\delta}(k,\eta)$, $F_{2,\delta}(\vec k-\vec q,\vec q,\eta)$ and $F_{3,\delta}(\vec k,\vec q,-\vec q,\eta)$ determined numerically, as detailed in Sec.\,\ref{sec:numericalkernels}, for an input power spectrum $P_0\propto k^{n_s}$. Scaling symmetry implies 
\be
\epsilon(\eta)=\epsilon_0\, e^{4\eta/(n_s+3)}, \ \ \ \ \ \bar\omega=\omega(\eta)/\epsilon(\eta)^2={\rm const.}
\ee
Therefore, apart from $n_s$, the result depends only on the two parameters $\epsilon_0$ and $\bar\omega$. As we shall see, the latter has only a minor impact, and (as already mentioned in Sec.\,\ref{sec:selfconsistent}) we use the value obtained by self-consistently solving the equation of motion for the fourth cumulant in linear approximation~\cite{cumPT} as fiducial value, given by 
\be
\bar\omega=0.579, \, 0.616, \, 0.668, \, 0.752\,,
\label{wbarfiducial}
\ee
for $n_s=2,1,0,-1$, respectively (we assess the dependence on $\bar\omega$ below). This leaves $\epsilon_0\equiv 1/k_\sigma^2$ as remaining dimensionful parameter, which we quantify by the ratio $k_\sigma/k_\text{nl}$ of the dispersion to the non-linear scale. We discussed estimates for this quantity via two different approaches, perturbation theory and halo modeling, in Sec.\,\ref{sec:powerlaw}, indicating a value for $k_\sigma/k_\text{nl}$ in the range $\sim 1 - 2$. We shall see that this is indeed consistent with direct determinations from fitting the density power spectrum.

In Fig.~\ref{fig:Pdd_various_ksig} we show the density power spectrum in linear and one-loop approximation obtained in \vpt~based on the ({\bf cum3+}, {\bf svt}) scheme (see Table~\ref{tab:approxschemes}), for three different values of $k_\sigma/k_\text{nl}$\footnote{Note that scaling symmetry can be used to rescale numerical perturbative results for a given value of $k_\sigma/k_\text{nl}$ to any other value, see App.\,\ref{app:rescaling}.} and $n_s=0$. The result is normalized to the linear power spectrum in SPT, i.e. for the perfect pressureless fluid approximation, given by $P_0(k,\eta)=e^{2\eta}P_0(k)$. By showing the result for $P_{\delta\delta}(k,\eta)/P_0(k,\eta)$ versus $k/k_\text{nl}(\eta)$, the dependence on time $\eta$ drops out, and we therefore omit the time arguments for brevity.

From the dashed lines in Fig.\,\ref{fig:Pdd_various_ksig} one can see that the linear \vpt~result is suppressed compared to the linear SPT spectrum, with the suppression being  stronger for larger background dispersion $\epsilon$, i.e.  smaller $k_\sigma$, as expected. Adding the one-loop correction (solid lines) leads to additional suppression for $k\lesssim k_\text{nl}$, but enhances the power spectrum relative to the linear \vpt~result for $k\gtrsim k_\text{nl}$. Note that the linear and one-loop results are close to each other within the expected regime of perturbativity, i.e. for $k\lesssim {\cal O}(k_\text{nl})$. This is an important property, indicating perturbative stability. Note that within SPT the one-loop integrals are UV divergent, and we therefore do not show any one-loop SPT result. In contrast, we checked that the \vpt~one-loop integrals are finite; this is  a consequence of the strong suppression of the non-linear kernels for large loop wavenumber as discussed in Sec.\,\ref{sec:numericalkernels}.

For comparison, we performed N-body simulations for scaling universes with spectral indices $n_s=2,1,0,-1$. Our main suite of N-body simulations consists of  two sets of fixed amplitude initial conditions with opposite phases to cancel Gaussian cosmic variance,  each with $512^3$ particles. See App.~\ref{app:Nbody} for more details, including tests of self-similarity, initial conditions, and computation of error bars. 

Figure~\ref{fig:Pdd_cumPT_Nbody} shows the \vpt~power spectrum $P_{\delta\delta}$ in linear (dashed) and one-loop (solid) approximation along with the N-body measurements (red circles). Note that error bars are too small to be noticed in $P_{\delta\delta}$. In order to assess the level of agreement, and in absence of a precise independent determination of the background dispersion, we adjusted the value of $k_\sigma/k_\text{nl}$ by fitting the perturbative to the
N-body result at low wavenumber, specifically for $k\leq k_\text{max}=0.6k_\text{nl}$. Nevertheless, we stress that as a matter of principle this fitting procedure is not necessary to obtain predictions within \vpt.
It hinges on our ability to accurately determine the spatial average of the stress tensor, being a well-defined quantity with immediate physical meaning.

To further test the viability of this procedure, we use the \emph{same} value of $k_\sigma/k_\text{nl}$ as obtained from the density power spectrum (for a given approximation scheme and $n_s$) when comparing the velocity power spectrum and bispectrum to N-body results further below. In addition, we checked that our results are stable under changing $k_\text{max}$ as long
as it is below the non-linear scale. Finally, we note that the best-fit values for $k_\sigma/k_\text{nl}$ (as quoted in Fig.\,\ref{fig:Pdd_cumPT_Nbody}) lie within the expected range as obtained in Sec.\,\ref{sec:powerlaw}, giving further support. Notwithstanding, the fitting procedure likely slightly exaggerates the level of agreement, and we comment on the impact of missing two-loop contributions below.

With these caveats in mind, we observe that the one-loop \vpt~density power spectrum can reproduce the N-body result up to $k\lesssim {\cal O}(k_\text{nl})$. Furthermore, including the one-loop correction significantly improves the agreement compared to the linear result. To be conservative, we show the result for the optimal value of $k_\sigma/k_\text{nl}$ for the linear approximation as well as for the sum of the linear and one-loop contributions, respectively. This shows the actual improvement stemming from one-loop corrections while ``marginalizing'' over  $k_\sigma/k_\text{nl}$. Importantly, and in line with the previous observations, the difference between the two values of $k_\sigma/k_\text{nl}$ (see legends in Fig.\,\ref{fig:Pdd_cumPT_Nbody}) as well as the difference between linear and one-loop results for $P_{\delta\delta}$ (dashed and solid lines in Fig.\,\ref{fig:Pdd_cumPT_Nbody}) up to wavenumbers $k\lesssim k_\text{nl}$ are reasonably small, as required for a perturbative expansion.

\subsubsection{Impact of vorticity, vector and tensor modes}

\begin{figure*}[t]
  \begin{center}
  \includegraphics[width=\textwidth]{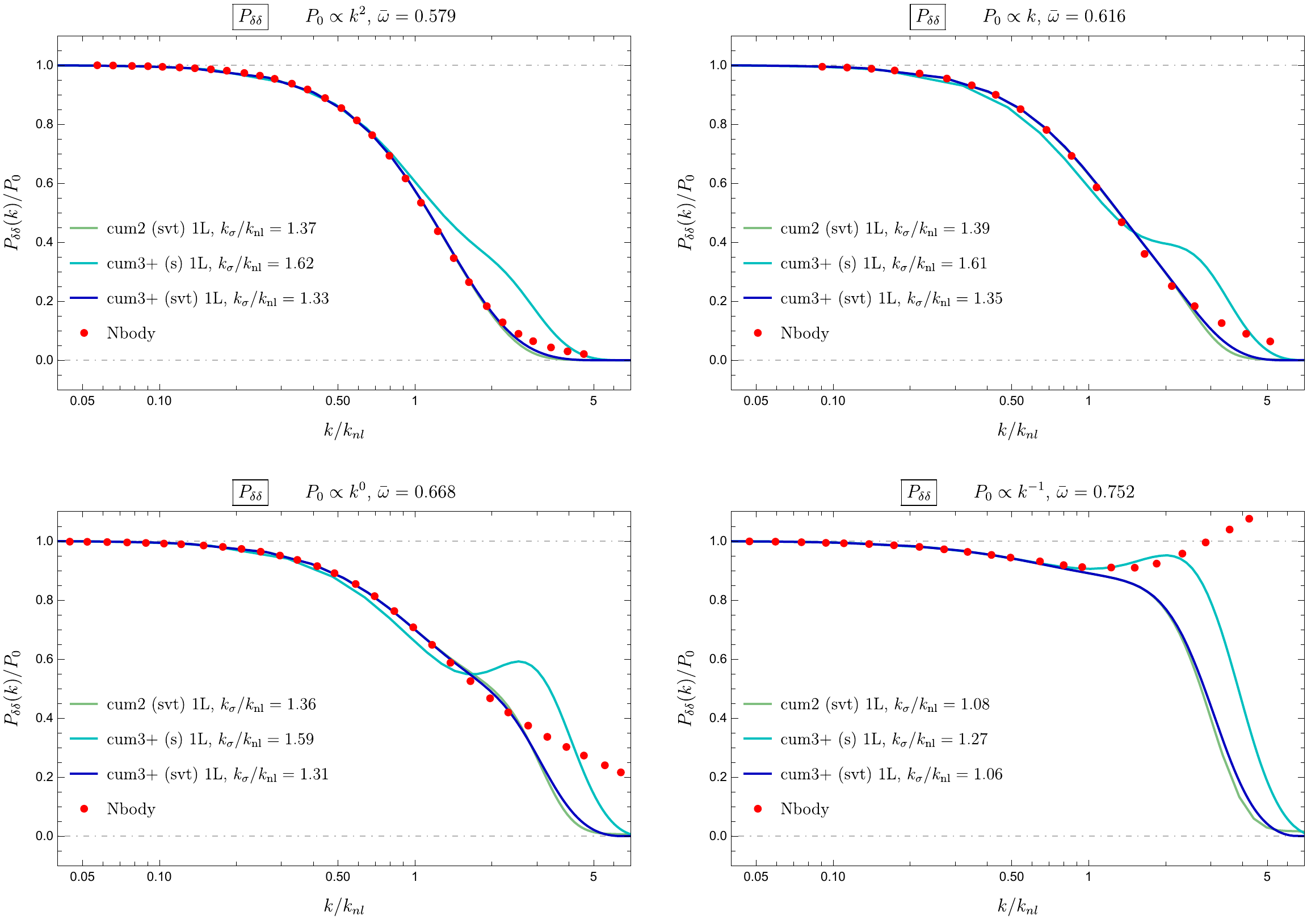}
  \end{center}
  \caption{\label{fig:Pdd_cumPT_Nbody_comp} 
  One-loop density power spectrum $P_{\delta\delta}(k,\eta)$ compared to N-body results for $n_s=2,1,0,-1$ and three approximation schemes. Taking only scalar perturbation modes into account ({\bf cum3+}, {\bf s}) leads to significant deviations from
  our fiducial scheme  ({\bf cum3+}, {\bf svt}) with vorticity, vector and tensor modes. On the contrary, the second cumulant approximation ({\bf cum2}, {\bf svt}) gives results close to ({\bf cum3+}, {\bf svt}). This means the impact of the third cumulant is mild whereas neglecting the backreaction of vorticity+vector(+tensor) modes would require a notably different $k_\sigma/k_\text{nl}$ which generally makes the agreement worse for $k\lesssim k_\text{nl}$ (grey lines). It is therefore important to take this backreaction into account, i.e.~use the ({\bf cum2/3+, svt}) (or ({\bf cum2/3+, sv}), see Fig.~\ref{fig:Pdd_cumPT}) schemes.
  }
\end{figure*}

Let us now turn to discussing the impact of various approximation schemes within \vpt~(see Table~\ref{tab:approxschemes}) on the density power spectrum.
We start with backreaction of vorticity, as well as vector and tensor modes of the velocity dispersion tensor.
They affect the kernel $F_{3,\delta}(\vec k,\vec q,-\vec q,\eta)$ entering the one-loop density power spectrum.
To assess the size of this backreaction, we compare the density power spectrum in ({\bf s}), ({\bf sw}),  ({\bf sv}) and  ({\bf svt}) approximation, all of them within the ({\bf cum3+}) scheme.
In the left panel of Fig.\,\ref{fig:Pdd_cumPT} we show the comparison for a fixed common value of $k_\sigma$ for all cases, and $n_s=0$. The linear results are identical.

We observe a significant impact of vorticity backreaction on the third-order density contrast, i.e. a sizable shift
between the one-loop density spectrum in the ({\bf s}) and ({\bf sw}) approximation. For example, at the non-linear scale the impact of vorticity backreaction is about 10\% for $n_s=0$ (Fig.\,\ref{fig:Pdd_cumPT}). We find the size of this depends on $n_s$. For blue indices $n_s=1,2$ this grows to about 15\%, while for $n_s=-1$ it drops to 3\%. 
Including the vector mode of the dispersion tensor leads to another small shift, when going from ({\bf sw}) to ({\bf sv}).
In contrast to that, the tensor mode has practically no effect, with the ({\bf sv}) and ({\bf svt}) lines in Fig.\,\ref{fig:Pdd_cumPT} lying on top of each other.

As discussed in Sec.\,\ref{sec:ktozerolimit}, including vorticity is necessary to ensure momentum conservation. Even though the scaling $F_{3,\delta}(\vec k,\vec q,-\vec q,\eta) \propto k^2$
for $k\to 0$ occurs in all schemes (and even in the momentum violating ({\bf s}) scheme for the particular set of arguments entering in the one-loop power spectrum), such that $P_{\delta\delta}/P_0\to 1$ for $k\to 0$, we find that vorticity backreaction has an important effect for $k\gtrsim 0.3k_\text{nl}$.

Note also that the impact of vorticity plus vector and tensor modes on the density sets in when $q\gtrsim k$ or when $q\gtrsim k_\sigma$ (if $k_\sigma>k$). This is the reason why in the left panel of Fig.~\ref{fig:Pdd_cumPT} for $k\gtrsim 3k_\text{nl}$ the relevance of them becomes negligible. The contribution to the one-loop result from loop wavenumbers for which $q\gtrsim k$ decreases when $k$ increases due to the UV screening in~\vpt. Hence vorticity backreaction eventually peters out for very high $k$ for the one-loop contribution. Note that this is also in line with vorticity backreaction shown in Fig.~\ref{fig:F3_delta_svt} and Fig.~\ref{fig:F3_d_t_kmpmq_vs_k}. Nevertheless, at these large wavenumbers, two- and higher loops have to become relevant (see below), which should themselves be affected by vorticity backreaction as well. Altogether, our results show that within the mildly non-linear regime that is most relevant for perturbative approaches, vorticity backreaction is significant.

The impact of vorticity backreaction can be seen even more clearly in Fig.\,\ref{fig:Pdd_cumPT_Nbody_comp}, by comparing the one-loop results in the ({\bf cum3+}, {\bf s}) and ({\bf cum3+}, {\bf svt}) scheme, respectively. In this figure we
adjusted $k_\sigma/k_\text{nl}$ to optimally match the N-body result for $k\ll k_\text{nl}$ in each scheme. The best-fit values as well as the power spectra differ significantly, and the ({\bf cum3+}, {\bf s}) scheme shows a markedly worse agreement with N-body results compared to ({\bf cum3+}, {\bf svt}) (except for $n_s=-1$, for which the difference however occurs at wavenumbers where two-loop corrections are expected to matter, see Sec.~\ref{sec:highercumandloops}).

Also note that in particular for $n_s=0,1$ the ({\bf cum3+}, {\bf s}) result presents a poor match to N-body data even at rather small $k$. The reason is that when varying $k_\sigma$ the value of $P_{\delta\delta}$ at a given fixed scale $k$ features a maximum for some particular $k_\sigma$, i.e. $P_{\delta\delta}$ at one-loop is constrained to be below a certain maximal possible value for any choice of $k_\sigma$. Since for ({\bf cum3+}, {\bf s}) and $n_s=0,1$ this maximum is below the N-body result, they cannot be brought into agreement even at low $k$ when varying $k_\sigma$\footnote{The reason for this maximum of $P_{\delta\delta}$ when varying $k_\sigma$ (for $k$ fixed) is that the linear \vpt~result decreases when decreasing $k_\sigma$ (corresponding to higher background dispersion), while the one-loop increases. The increase of the one-loop piece can be understood from UV screening within~\vpt: the most important role is played be the contribution to the one-loop from $F_{3,\delta}(\vec k,\vec q,-\vec q,\eta)$ for $k\lesssim q$, which is negative in SPT. The screening built into \vpt~makes it less negative or
even positive (see Fig.\,\ref{fig:F23_delta}). This explains the opposite behaviour of the linear and one-loop contributions when changing $k_\sigma$. Their interplay then creates the aforementioned maximum in the sum of linear and one-loop pieces.}.
This feature emphasizes that the agreement between \vpt~and N-body results when taking vorticity backreaction into account (as opposed to when not) is a non-trivial result.

We conclude that it is vital to take vorticity backreaction into account, both for physical reasons (momentum conservation, see Sec.\,\ref{sec:ktozerolimit}), and due to its quantitative impact on the density contrast. In addition, vector modes of the stress tensor have a certain effect on $P_{\delta\delta}$, while the backreaction of tensor modes is negligible at one-loop.

\begin{figure*}[t]
  \begin{center}
  \includegraphics[width=\textwidth]{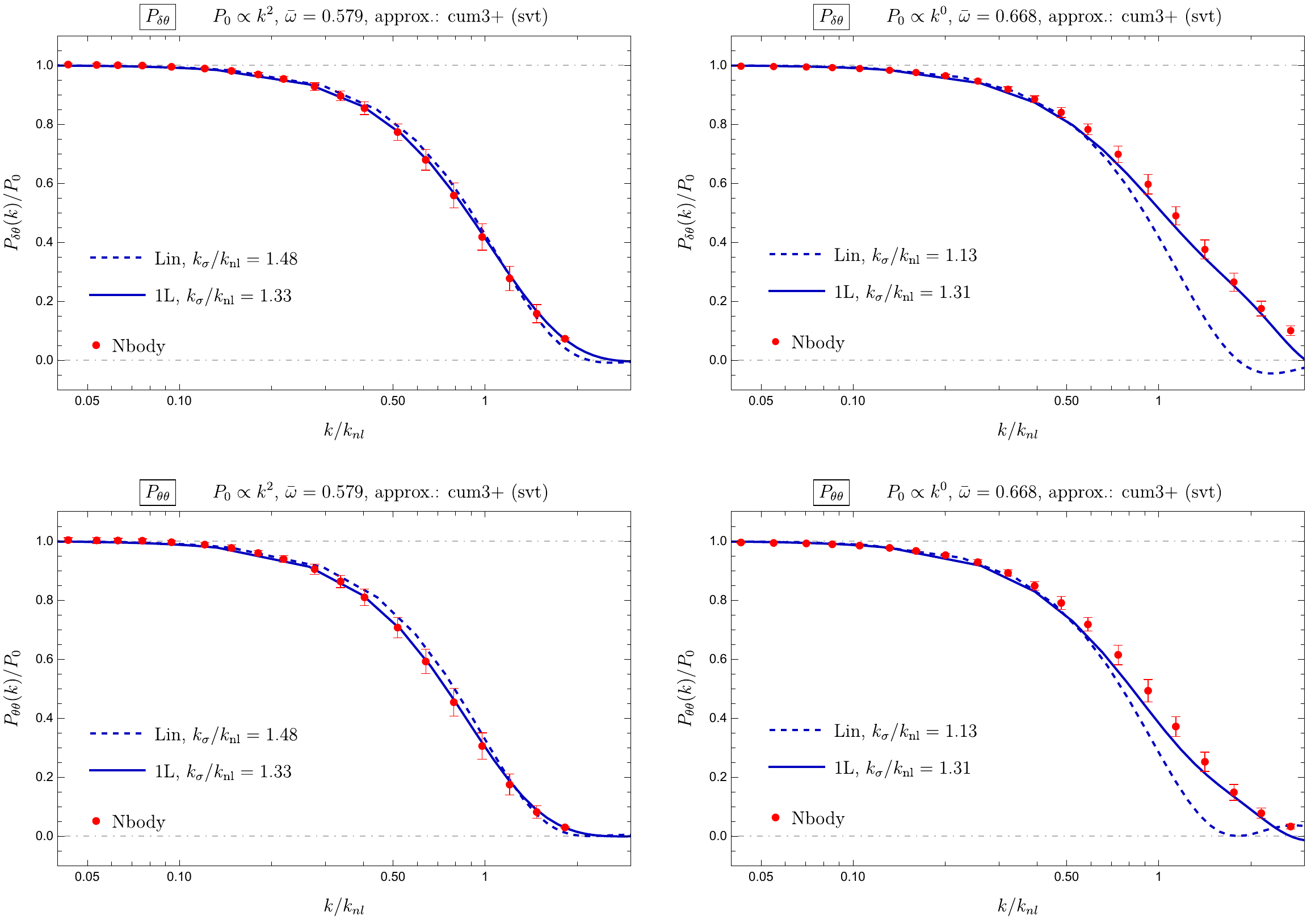}
  \end{center}
  \caption{\label{fig:Pdt_tt_cumPT_Nbody} 
  Velocity divergence and cross power spectra $P_{\delta\theta}$ (first row) and $P_{\theta\theta}$ (second row) for $n_s=2$ (left column) and $n_s=0$ (right column).
  Red circles with error bars show the corresponding N-body results. The linear (one-loop) approximation within the ({\bf cum3+, svt}) scheme of \vpt~is shown with dashed (solid) lines, with $k_\sigma/k_\text{nl}$  fixed to the \emph{same} value as for $P_{\delta\delta}$  (see Fig.\,\ref{fig:Pdd_cumPT_Nbody}) for each case, respectively. Therefore, these can be regarded as unique predictions of \vpt~as there is no free parameter being fit to the measurements. 
  }
\end{figure*}

\subsubsection{Impact of higher cumulants and loops}\label{sec:highercumandloops}

To assess the relevance of cumulants beyond second order we compare the
({\bf cum2}) and ({\bf cum3+}) approximation schemes, adopting ({\bf svt}) in both cases. For the latter, we also
consider three choices for the fourth cumulant expectation value, given by $\bar\omega=\pm 1$
in addition to our fiducial choice (see Eq.~\ref{wbarfiducial}). A comparison for
fixed $k_\sigma$ and $n_s=0$ is shown in the right panel of Fig.\,\ref{fig:Pdd_cumPT}.
The differences between the linear results for $P_{\delta\delta}$ are hardly visible (dashed lines).
At one-loop, the second cumulant approximation ({\bf cum2}) leads to a slightly smaller result compared to ({\bf cum3+}) with fiducial choice $\bar\omega=0.668$,
while ({\bf cum3+}) with $\bar\omega=1$ is slightly larger. The ({\bf cum3+}) case with $\bar\omega=-1$, motivated by the values found from halos in paper~I~\cite{cumPT},  shows a smaller amplitude of the one-loop correction, close to the second cumulant approximation. 
However, the differences are very small for the relevant range $k\lesssim k_\text{nl}$.
In addition, and in stark contrast to vorticity as discussed above, they can largely be compensated by a small shift in $k_\sigma$. This can be seen by
comparing the results for ({\bf cum2}, {\bf svt}) and ({\bf cum3+}, {\bf svt}) in Fig.\,\ref{fig:Pdd_cumPT_Nbody_comp}, with both lines lying on top of each other for $k\lesssim k_\text{nl}$.

We thus find no indication that higher cumulants beyond the velocity dispersion tensor invalidate the \vpt~approach for $k\lesssim {\cal O}(k_\text{nl})$. On the contrary, the good agreement between ({\bf cum2}) and ({\bf cum3+}) as well as small sensitivity on the fourth cumulant expectation value $\bar\omega$ can be taken as an indication that the truncation of higher cumulants leads to an acceptable uncertainty for the one-loop density power spectrum within the mildly non-linear regime and for the considered values of $n_s$. This finding is remarkable since e.g. for a single shell crossing, higher cumulants of all orders are generated at once. Nevertheless, the gradual build-up of an average velocity dispersion due to the superposition of many shell crossings, as well as the question of what impact do higher cumulants have on the large-scale density contrast  complicate this simple argument, rendering the relevance of third and higher cumulants unclear a priori. It's also worth pointing out that the picture that emerges from this one-loop calculation is consistent with that of the linear approximation detailed in paper~I~\cite{cumPT}, i.e. while higher cumulant perturbations are important to capture the suppression of the linear kernel for $k\gg k_\sigma$, they only mildly affect the transition region between the low-$k$ regime and the onset of suppression describing the screening of UV modes (see Fig.~6 in \cite{cumPT}).  For a more detailed discussion of the convergence of the cumulant expansion in the linear approximation we refer the reader to paper I~\cite{cumPT}.

However, while the ({\bf cum2}) scheme is apparently a reasonable approximation for $P_{\delta\delta}$ at one-loop, higher cumulants likely have a larger impact at higher order in perturbation theory, where backreaction effects can be more pronounced, and deserve further study in the future. Still, in order to explore the perturbative stability we checked in an explorative study that adding the {\em two-loop correction} for ({\bf cum2}, {\bf sv}) leads to only minor changes in $k_\sigma/k_\text{nl}$ and an agreement with N-body results that is comparable or slightly improved compared to the one-loop ({\bf cum3+}, {\bf svt}) case. Interestingly, by fitting the two-loop ({\bf cum2}, {\bf sv}) approximation to the N-body measurements we obtain the values:  
\be
k_\sigma/k_\text{nl}=1.32, \, 1.36, \, 1.39, \, 1.40\,,
\label{ksig2loop}
\ee
for $n_s=2,1,0,-1$, respectively. Note that here $k_\sigma/k_\text{nl}$  \emph{increases} when decreasing $n_s$. This is the same trend  observed in Sec.\,\ref{sec:powerlaw} both for the self-consistent and the halo determinations of $k_\sigma/k_\text{nl}$. We therefore conclude that the lack of a similar trend for the one-loop fits is most likely a limitation of the one-loop approximation. In addition, note that the values in Eq.~(\ref{ksig2loop}) lie in between the independent estimates from self-consistent perturbation theory and halo calculations (compare to Tables~\ref{tab:selfconsistent} and~\ref{tab:ksigma}). Overall, this is compatible with the fact that through our fitting procedure we are determining the spatial average of the dispersion tensor,  a well-defined quantity with clear physical meaning.

\begin{figure*}[t]
  \begin{center}
  \includegraphics[width=\textwidth]{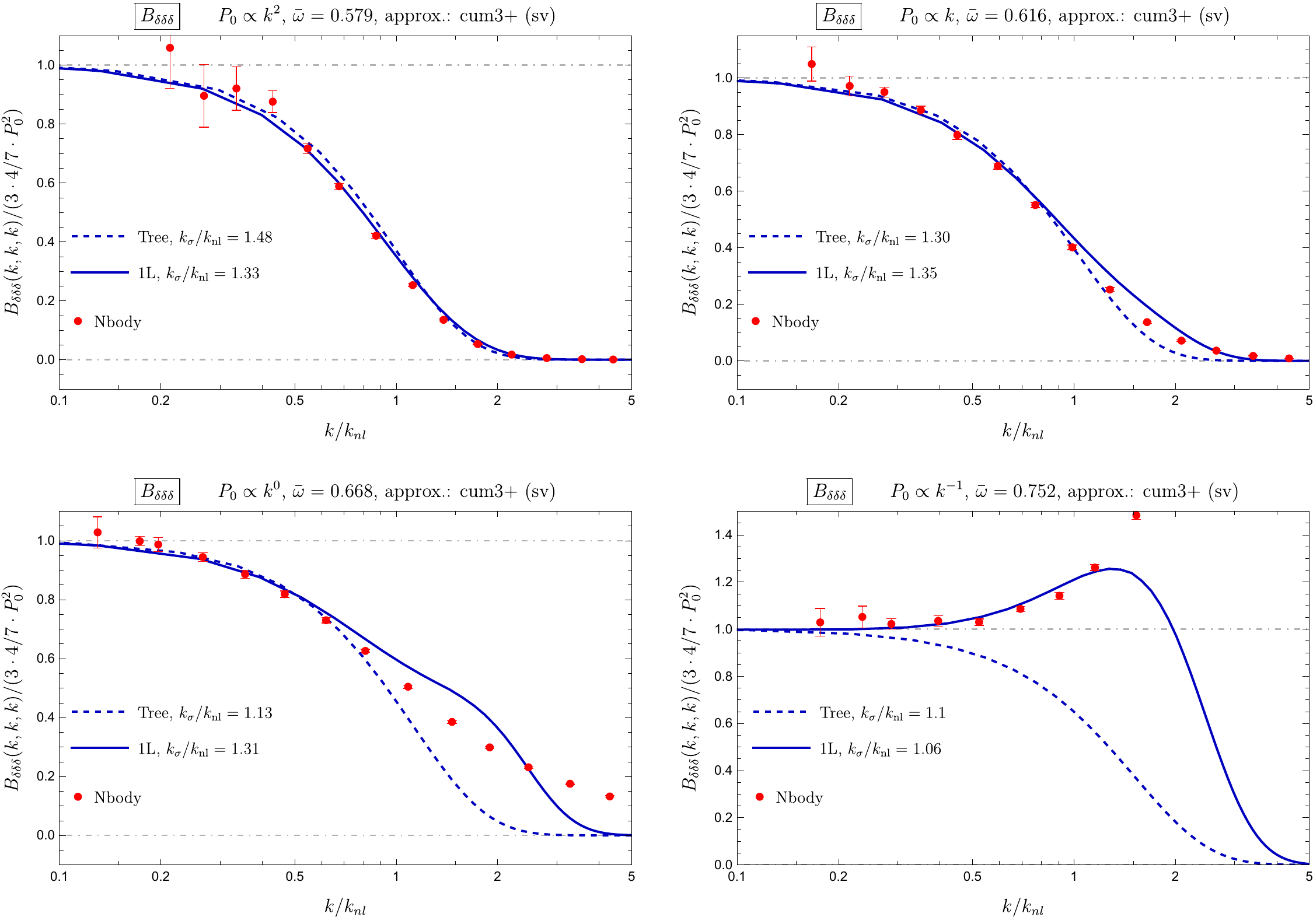}
  \end{center}
  \caption{\label{fig:Pddd}
  Equilateral bispectrum at tree-level (dashed) and one-loop (solid) \vpt~compared to N-body results for $n_s=2,1,0,-1$, respectively. The bispectra are normalized to the SPT tree-level bispectrum.  In each case $k_\sigma/k_\text{nl}$  is fixed to the \emph{same} value as for $P_{\delta\delta}$  (see Fig.\,\ref{fig:Pdd_cumPT_Nbody}).  Therefore, these can be regarded as unique predictions of \vpt~as there is no free parameter being fit to the measurements. 
  }
\end{figure*}

\subsection{Velocity divergence and cross power spectrum}\label{sec:Ptt}

Measuring the power spectrum of the velocity divergence and its cross correlation with the density contrast from N-body simulations requires a reconstruction of the velocity field. We describe our procedure to accomplish this in App.~\ref{app:DivVort}.

Within \vpt, $P_{\theta\theta}$ and $P_{\delta\theta}$ can be computed analogously to $P_{\delta\delta}$, using Eq.~\eqref{eq:P1Ldef} at one-loop with $a=b=\theta$ or $a=\delta, b=\theta$, respectively.
We use the \emph{same} values of $k_\sigma/k_\text{nl}$ as determined in Sec.\,\ref{sec:Pdd} by matching \vpt~and N-body results for $P_{\delta\delta}$, for each approximation scheme and $n_s$, respectively.
Therefore, there are no free parameters entering $P_{\theta\theta}$ and $P_{\delta\theta}$, and they are predicted uniquely. The corresponding linear and one-loop results are shown in Fig.\,\ref{fig:Pdt_tt_cumPT_Nbody} 
for $n_s=2$ as well as $n_s=0$. We find that the one-loop and N-body results are very close to each other for $k\lesssim k_\text{nl}$, with the agreement being somewhat better for $n_s=2$. Taking the uncertainty in the velocity
reconstruction into account, the level of agreement shows that \vpt~yields relevant results also for the velocity divergence. This is an important consistency check of the formalism. We stress again that no free parameters were
adjusted to obtain the velocity and cross power spectra, and the same is also true  for the next consistency check that we discuss, the bispectrum.

\begin{figure*}[t]
  \begin{center}
  \includegraphics[width=\textwidth]{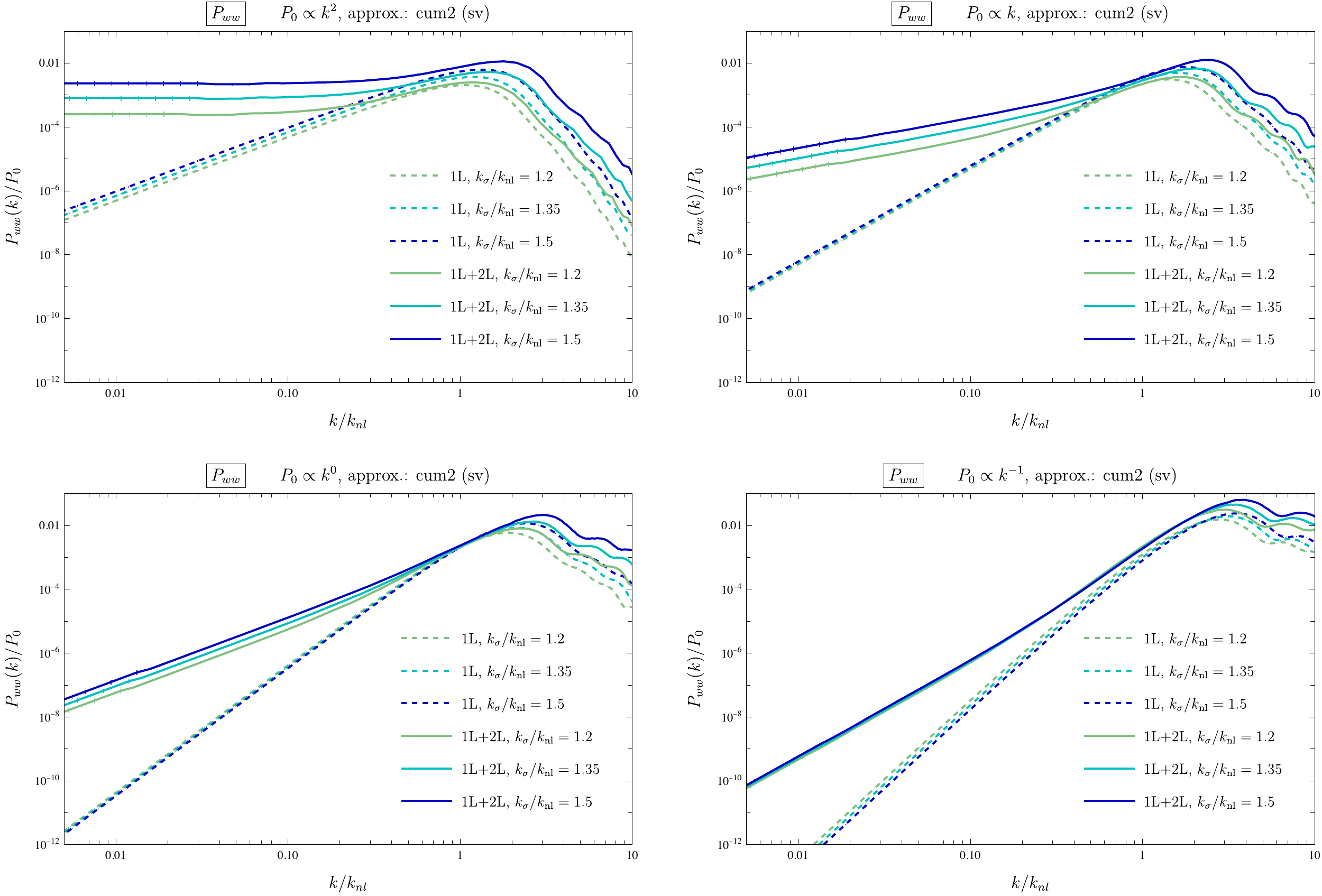}
  \end{center}
  \caption{\label{fig:Pww_cumPT_ksig} 
  Vorticity power spectrum $P_{w_iw_i}$ within \vpt~in one-loop (dashed) and two-loop (solid) approximation, for  $n_s=2,1,0,-1$. The scaling $P_{w_iw_i}\propto k^2$ occurs only starting  at two-loop order, while the one-loop result scales as $k^4$ for small $k$. The figure shows the dependence of the one- and two-loop pieces on $k_\sigma/k_\text{nl}$, see Eq.~\eqref{eq:PwwScaling}. The error bars correspond to the Monte Carlo integration error (visible only at low $k$).
  }
\end{figure*}

\subsection{Bispectrum}\label{sec:Pddd}

To further test the viability of \vpt, we compare perturbative and N-body results for the bispectrum.
The tree-level and one-loop bispectrum can be computed using the formal expressions as in SPT, but replacing the SPT kernels $F_n$ by the numerically computed kernels $F_{n,\delta}$ within \vpt. A subtlety is that also the first order kernel $F_{1,\delta}$ is non-trivial and therefore needs to be taken into account, giving the expressions
\bea
  \lefteqn{ B^\text{tree}(k_1,k_2,k_3,\eta) = 2e^{4\eta}F_{2,\delta}(\vec k_1,\vec k_2,\eta)F_{1,\delta}(k_1,\eta) }\nn\\
  &\times& F_{1,\delta}(k_2,\eta) P_0(k_1)P_0(k_2)  +2\ \text{perm.}\,,\nn\\
  \lefteqn{ B^{1L}(k_1,k_2,k_3,\eta) = e^{6\eta}\int d^3q P_0(q) \Big\{ 8F_{2,\delta}(\vec q+\vec k_1,-\vec q,\eta) }\nn\\
  &\times& F_{2,\delta}(\vec q-\vec k_2,-\vec q-\vec k_1,\eta)F_{2,\delta}(\vec q,\vec k_2-\vec q,\eta)P_0(|\vec q+\vec k_1|)\nn\\
  &\times& P_0(|\vec q-\vec k_2|) 
  + 6 \big[ F_{1,\delta}(k_3,\eta)F_{3,\delta}(\vec q-\vec k_2,-\vec q,-\vec k_3,\eta)\nn\\
  &\times& F_{2,\delta}(\vec q,\vec k_2-\vec q,\eta)P_0(|\vec q-\vec k_2|)P_0(k_3) 
  + F_{2,\delta}(\vec k_2,\vec k_3,\eta)\nn\\
  &\times& F_{1,\delta}(k_2,\eta)F_{3,\delta}(\vec k_3,\vec q,-\vec q,\eta)P_0(k_2)P_0(k_3) +5\ \text{perm.}\big] \nn\\
  &+& 12\big[F_{1,\delta}(k_2,\eta)F_{1,\delta}(k_3,\eta)F_{4,\delta}(\vec k_2,\vec k_3,\vec q,-\vec q,\eta)\nn\\
  &\times& P_0(k_2)P_0(k_3) +2\ \text{perm.}\big]\Big\}\,.
\eea
For the non-linear kernels entering the one-loop bispectrum, the inclusion
of vorticity and the vector mode of the dispersion tensor requires to use the advanced algorithm described in App.\,\ref{app:vorticity_full}. We do not include
the tensor mode of the dispersion for the bispectrum for simplicity, since its impact on the power spectrum was found to be tiny. This corresponds to the  ({\bf cum3+, sv})
scheme. Furthermore, we use the algorithm described in App.\,B of~\cite{Floerchinger:2019eoj} to rewrite the integrand such that contributions
that are enhanced for $q\to 0$ cancel in the sum of all terms at the integrand level. 

We normalize our results to the tree-level SPT
bispectrum $B^\text{tree}_\text{SPT}(k_1,k_2,k_3,\eta)=2e^{4\eta}F_2(\vec k_1,\vec k_2)P_0(k_1)P_0(k_2)+2\ $perm.,
with EdS-SPT kernel $F_2(\vec k_1,\vec k_2)$. In the equilateral configuration $k_1=k_2=k_3$, the relevant SPT kernels are given by
$F_2(\vec k_1,\vec k_2)=F_2(\vec k_1,\vec k_3)=F_2(\vec k_2,\vec k_3)=2/7$.
Note that for the considered initial power spectra with $n_s\geq -1$ the SPT one-loop integrals would be UV divergent, while the \vpt~result is finite, as for the power spectrum.

In Fig.\,\ref{fig:Pddd} we compare the tree-level and one-loop equilateral bispectrum within \vpt~to N-body results.
As above, we use the \emph{same} value for $k_\sigma/k_\text{nl}$ as in Sec.\,\ref{sec:Pdd} for each case, such that the bispectrum is predicted
without any additional free parameters. We find that even the tree-level bispectrum within \vpt~(dashed lines) yields a rather reasonable result,
accounting for the suppression relative to the SPT tree-level prediction (the latter corresponding to the line at $1.0$ in Fig.\,\ref{fig:Pddd}). Adding the \vpt~one-loop
contribution (solid lines) further improves the agreement with the N-body results.

Importantly, we find that despite of the large values for $n_s$ the one-loop correction remains small
in the perturbative regime, indicating perturbative stability of \vpt~also for the bispectrum. Indeed, the one-loop correction is smaller for larger $n_s$. This can be explained
by the stronger damping of the non-linear kernels for larger $n_s$, which overcompensates the larger weight of large wavenumbers due to $P_0(q)\propto q^{n_s}$. The same trend can also be
observed regarding the agreement between one-loop and N-body results, which reaches out to rather large wavenumbers for $n_s=2$.
We checked that the results for the bispectrum are very similar for the ({\bf cum2}) scheme, with differences to ({\bf cum3+}) comparable in size as for the power spectrum. 

Thus, we find that the \vpt~formalism developed in this work yields an accurate description not only of the density and velocity divergence power spectra, but also of the bispectrum. This adds another piece of evidence that \vpt~is able to capture the relevant dynamics. It further demonstrates its predictivity, since, as for the velocity divergence spectra, \emph{no} free parameters were adjusted for obtaining the bispectrum.

\begin{figure}[t]
  \begin{center}
  \includegraphics[width=0.49\textwidth]{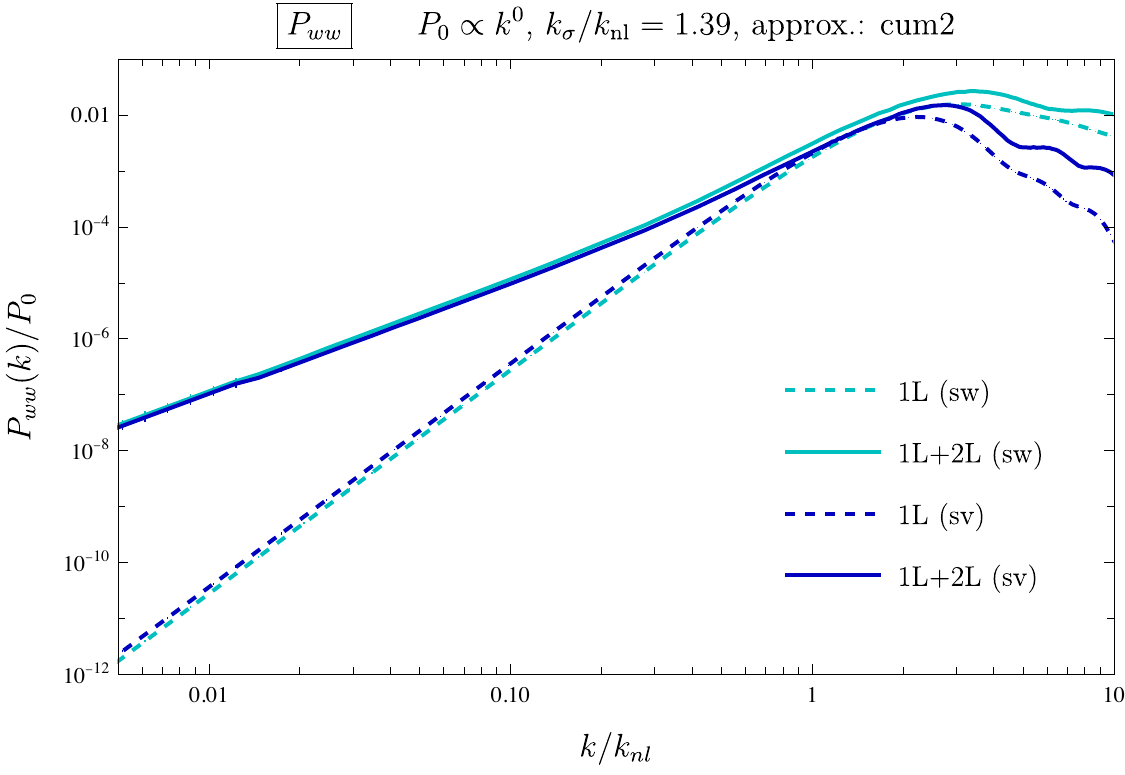}
  \end{center}
  \caption{\label{fig:Pww_cumPT_sw_sv} 
  Vorticity power spectrum $P_{w_iw_i}$ within \vpt~in one-loop (dashed) and two-loop (solid) approximation, for $n_s=0$. The generic scaling $P_{w_iw_i}\propto k^2$ for low $k$ is obtained starting at two-loop only, since the one-loop piece is suppressed as $k^4$. The figure shows the difference between the ({\bf cum2}, {\bf sw}) and ({\bf cum2}, {\bf sv}) scheme  (see Table~\ref{tab:approxschemes}), i.e. the backreaction of the vector modes of the velocity dispersion on vorticity.
   }
\end{figure}

\begin{figure}[t]
  \begin{center}
  \includegraphics[width=0.49\textwidth]{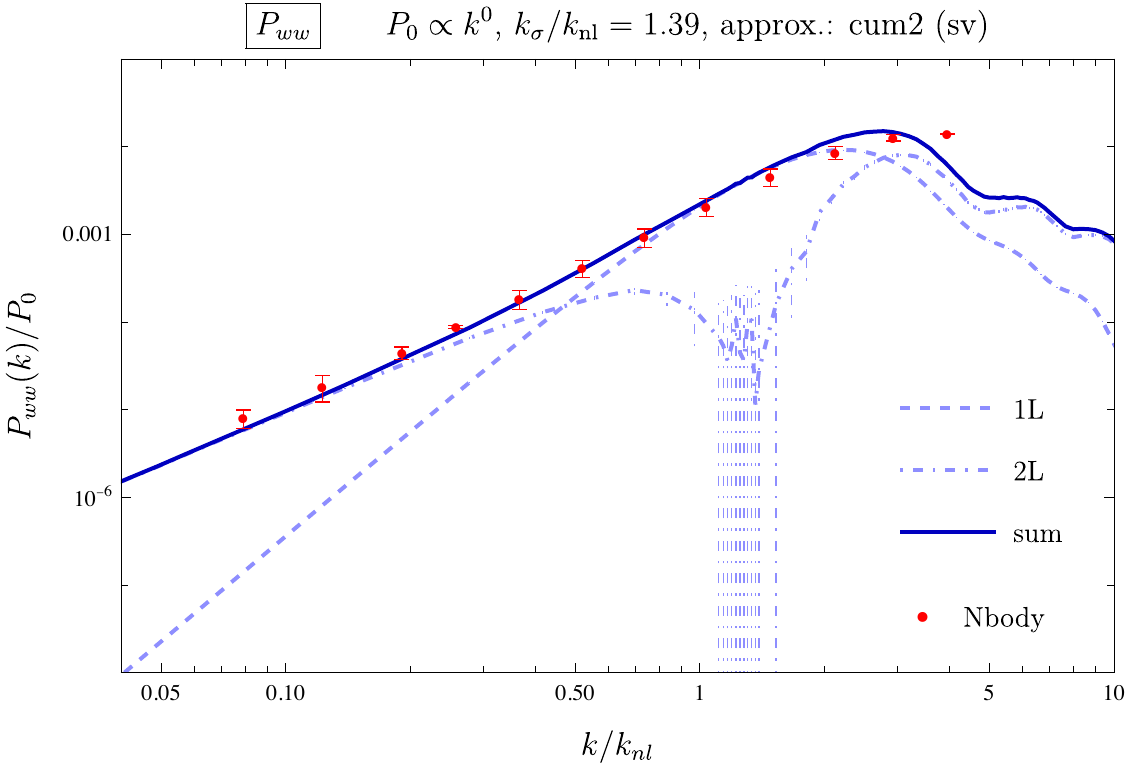}\\
  \includegraphics[width=0.49\textwidth]{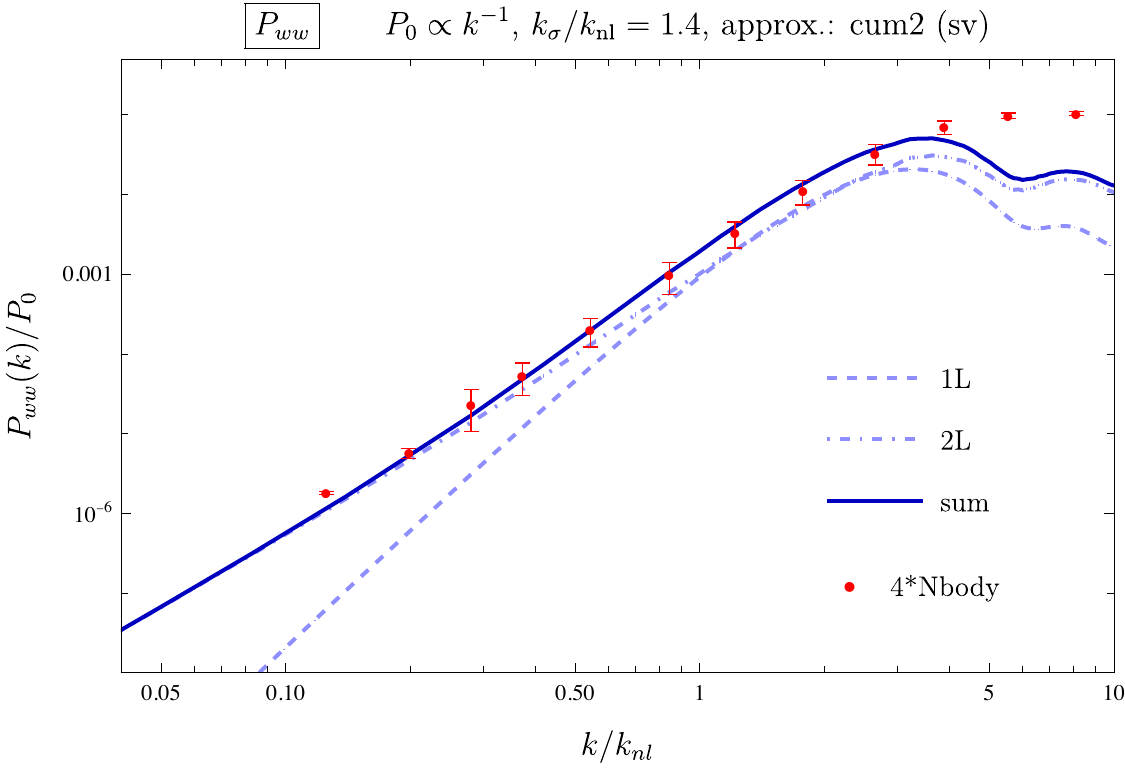}
  \end{center}
  \caption{\label{fig:Pww_cumPT_Nbody} 
  Comparison of the vorticity power spectrum $P_{w_iw_i}$ in \vpt~at two-loop order (solid line) to N-body results (red circles) for $n_s=0$ (top) and $n_s=-1$ (bottom).
  The cross-over from $k^4$ to $k^2$ scaling for low $k$ is clearly observed in both the N-body and two-loop \vpt~results.
  Note that the absolute normalization of the N-body vorticity power spectrum is sensitive to mass resolution (see App.\,\ref{app:DivVort}). For $n_s=-1$
  we rescaled the N-body result by a factor $4$ in order to ease  comparison of the shape.
  The dashed and dot-dashed lines show the one- and two-loop contributions, respectively. The value of $k_\sigma/k_\text{nl}$ is
  fixed to the one obtained from $P_{\delta\delta}$ in the same \vpt~approximation, two-loop ({\bf cum2}, {\bf sv}), see Eq.~(\ref{ksig2loop}). 
  }
\end{figure}

\subsection{Vorticity power spectrum}\label{sec:Pww}

The vorticity power spectrum  $P_{w_iw_i}(k,\eta)$ is a probe that is particularly sensitive to velocity dispersion, since no vorticity would be generated when neglecting second and higher cumulants, as is the case in SPT.  Within cumPT, the leading contribution arises at one-loop, since vorticity vanishes at the linear level.
Apart from the overall size,  the scaling with $k$ for small wavenumber is also an interesting issue. As discussed in Sec.\,\ref{sec:vortF3}, we find
a scaling 
\be
  P_{w_iw_i}(k,\eta)\propto k^2\ \text{for}\ k\to 0\,.
\ee
However, this scaling occurs only \emph{starting at two-loop order}, while the one-loop contribution scales as $k^4$. Therefore, the two-loop is expected to eventually dominate in the limit $k\to 0$, since the one-loop is  ``accidentally'' suppressed there. Note that, in contrast to the opposite regime, this behaviour on large scales is definitely not a sign that perturbation theory breaks down. It merely arises due to the strong suppression of the one-loop piece. Starting at two loops all higher loop corrections generically scale as $k^2$ for $k\to 0$. 

Nevertheless, we expect that the $k^2$ scaling can only be observed at small enough $k$, since also the one-loop piece contributes on intermediate scales.
To test this expectation, we compute the vorticity power spectrum up to two loops. Given the mild differences between ({\bf cum2}) and ({\bf cum3+}) as well as the tiny impact of tensor modes, we restrict ourselves to the second cumulant approximation including vorticity as well as scalar and vector modes of the dispersion tensor for this analysis, i.e. the ({\bf cum2}, {\bf sv}) scheme (see Table\,\ref{tab:approxschemes}). As for the one-loop bispectrum, computing the two-loop vorticity power spectrum requires the algorithm described in App.\,\ref{app:vorticity_full}, where
both modes of the vorticity in the plane perpendicular to the corresponding Fourier wavevector are included explicitly, and similarly for the vector mode.

In Fig.\,\ref{fig:Pww_cumPT_ksig} we show the sum of the one- and two-loop contributions to the vorticity power spectrum (solid lines), compared to the one-loop result only (dashed lines), for  $n_s=2,1,0,-1$. The different scaling of the one- ($\propto k^4$) and two-loop ($\propto k^2$) results for $k\ll k_\text{nl}$ is clearly visible. As expected, the power spectrum is dominated by the two-loop piece at low wavenumber, specifically for $k\lesssim 0.5 k_\text{nl}$. The $k^2$ scaling is approached only gradually, occurring  for $k\lesssim 0.1-0.2 k_\text{nl}$.
The panels of Fig.\,\ref{fig:Pww_cumPT_ksig} show the dependence on $k_\sigma/k_\text{nl}$ for each $n_s$. The different scaling with $k_\sigma$ of the one- and two-loop results can also be understood analytically. The dimensionless power spectrum $k^3 P_{w_iw_i}$ (as well as $P_{ww}/P_0$ shown in Fig.~\ref{fig:Pww_cumPT_ksig})  can only depend on dimensionless ratios, that we can take as $k/k_\sigma$ and $k_\sigma/k_\text{nl}$. Furthermore, the power of $k_\text{nl}$ is fixed at a given order in perturbation theory, since the $L$-loop contribution contains $L+1$ factors of $P_0=Ak^{n_s}$, and $A\propto 1/k_\text{nl}^{n_s+3}$. Using furthermore that $P_{w_iw_i}^{1L}\propto k^4$ and $P_{w_iw_i}^{2L}\propto k^2$ for small $k$ implies
\bea
  k^3P_{w_iw_i}^{1L} &\propto& \left(\frac{k}{k_\sigma}\right)^{4+3}\left(\frac{k_\sigma}{k_\text{nl}}\right)^{2(n_s+3)} \propto k_\sigma^{2n_s-1}\,,\nn\\
  k^3P_{w_iw_i}^{2L} &\propto& \left(\frac{k}{k_\sigma}\right)^{2+3}\left(\frac{k_\sigma}{k_\text{nl}}\right)^{3(n_s+3)} \propto k_\sigma^{3n_s+4}\,,\nn\\
  && \qquad\qquad\qquad\qquad\text{for}\ k\ll k_\text{nl},k_\sigma\,.
\label{eq:PwwScaling}
\eea
For $n_s=2,1$ ($n_s=0,-1$) this explains the increase (decrease) of $P_{w_iw_i}^{1L}$ when increasing $k_\sigma$ for some fixed $k/k_\text{nl}\ll 1$ (dashed lines in each panel of Fig.\,\ref{fig:Pww_cumPT_ksig}).
In addition, the second line in Eq.~\eqref{eq:PwwScaling} explains the increase of $P_{w_iw_i}^{2L}$ for all $n_s=2,1,0,-1$, which is least pronounced for $n_s=-1$ (solid lines, note that the sum of one- and two-loop is dominated by the latter for small $k$ as discussed above).

The impact of vector modes of the stress tensor is shown in Fig.\,\ref{fig:Pww_cumPT_sw_sv}, where we compare the ({\bf sw}) and ({\bf sv}) schemes. While both include vorticity, the vector mode is taken into account only in the latter case.
We find a very mild difference within the perturbative regime $k\lesssim {\cal O}(k_\text{nl})$.

Finally, a comparison to N-body results for the vorticity power spectrum is presented in Fig.\,\ref{fig:Pww_cumPT_Nbody}. The extraction of vorticity from N-body data requires a careful treatment, see discussion in App.\,\ref{app:DivVort} for details. In particular, as is well-known~\cite{PueSco0908,HahAngAbe1512,JelLepAda1809}, the overall amplitude of the vorticity power spectrum is rather sensitive to the mass resolution in the simulations, but it's \emph{shape} is rather robust. We verify this as well for our scale-free simulations in App.\,\ref{app:DivVort}, showing in particular that our $n_s=1,2$ simulations do not have enough resolution to capture the amplitude reliably. We therefore show results here only for $n_s=0,-1$, cautioning that in particular the former case may not be fully converged yet (see discussion in App.\,\ref{app:DivVort}). Similarly, our two-loop calculation of the vorticity power spectrum has some uncertainty as well, as it only includes the second cumulant. A full two-loop calculation would include higher cumulants, including their vector modes as well. This is well beyond the scope of the present paper. We therefore primarily base our comparison on the $k$-dependence of the vorticity power spectrum. To aid this comparison note that we have multiplied the N-body measurements for $n_s=-1$ by a factor of four, while the $n_s=0$ measurement is unchanged.
 
 Given these limitations, we find a good agreement of the shape with our perturbative \vpt~result. The solid line in Fig.\,\ref{fig:Pww_cumPT_Nbody} shows the sum of one- and two-loop contributions, while the dashed and dot-dashed lines show both of them individually. The dip of the two-loop correction at $k\sim 1.5k_\text{nl}$ for $n_s=0$ is due to an (almost) cancellation among the contributions $P^{2L}_{w_1w_1}$ and $P^{2L}_{w_2w_2}$ of the two vorticity degrees of freedom (see App.\,\ref{app:vorticity_full}) at this point, while the sum is positive below and above, as is the one-loop result. Note that in Fig.\,\ref{fig:Pww_cumPT_Nbody} we fixed $k_\sigma/k_\text{nl}$ to the value
obtained from $P_{\delta\delta}$ within the same approximation scheme, and going up to two-loop order (see Eq.~\ref{ksig2loop}), but our conclusion is independent of the precise value. We observe in particular that adding the two-loop contribution significantly improves the agreement of the shape of $P_{w_iw_i}$ with N-body results at small wavenumbers, where the transition from $k^4$ to $k^2$ scaling occurs, as captured by the two-loop part.
It is remarkable that this transition is seen both in two-loop \vpt~and N-body results. 

Thus, even though the overall normalization of the vorticity power spectrum is subject of further study, the \emph{shape} of the perturbative and N-body results agrees very well in the weakly non-linear regime. This constitutes yet another non-trivial consistency check of our approach, and demonstrates the versatility and predictivity of \vpt. 

The low-$k$ behavior of the vorticity power spectrum has been the subject of some study in the literature. In the context of $\Lambda$CDM simulations at $z=0$, \cite{JelLepAda1809} measure a low-$k$ asymptotic behavior $P_{ww}\propto k^{n_w}$ with $n_w=2.55\pm0.02$ (for $k \la 0.4 \iMpc$) in agreement with~\cite{HahAngAbe1512} which quote $n_w\simeq 5/2$ for $k \simeq 0.1 \iMpc$. Note that these works use small simulation boxes, $256 \Mpc$ and $100-300 \Mpc$, respectively, thus most likely they have not reached the regime where \vpt~predicts a transition to  $k^2$ scaling, i.e. $k \simeq (0.1-0.2)\, k_{\rm nl}$.  On the other hand, from the perturbative side, there have been two predictions based on EFT and one based on Lagrangian perturbation theory (LPT). From EFT, \cite{CarForGre1407}  predict a ``leading order" contribution at the one-loop level with $n_w=7+3n_s$,  and a stochastic component with $n_w=4$. On the other hand, ~\cite{MerPaj1403}  predict $n_w=7+2n_s$  and $n_w=4$. From LPT,  \cite{CusTanDur1703} also predict $n_w=4$. For our simulations with $n_s\geq -1$, all these three predictions would lead to a low-$k$ slope of $n_w=4$. As discussed above, the collisionless Vlasov dynamics captured by \vpt~yields $n_w=2$, in agreement with our N-body results. This result is independent of the input power spectrum, and thus also holds for $\Lambda$CDM power spectra.

\section{Conclusions}
\label{sec:conclusions}

In this paper  we presented non-linear perturbative solutions to the Vlasov hierarchy, a cumulant expansion of the Vlasov-Poisson equations of motion describing collisionless matter. This  Vlasov perturbation theory (\vpt) represents a systematic and predictive extension of the well-known framework of standard perturbation theory (SPT).

Formally, this approach is similar to SPT, in that the calculation of density and velocity divergence fields has the same expressions but with different non-linear kernels $F_n/G_n\to F_{n,a}$ and an enlarged set of perturbation modes $a=\delta,\theta,w_i,...$. One key modification is that the linear kernel ($n=1$) is not unity, except in the $k\to 0$ limit, with a scale-dependence describing how linear  growth is suppressed at small scales once $k$ crosses the (time-dependent) dispersion scale $k_\sigma$ set by the expectation value of the velocity dispersion tensor.  In contrast with SPT, there are additional degrees of freedom corresponding to vorticity $w_i$, the velocity dispersion tensor $\epsilon_{ij}$ and higher cumulants ${\cal C}_{ijk\dots}$ of the DF, which provide an important new window to testing the theory against simulations. 

For the non-linear kernels appearing in loop corrections, the suppression compared to the equivalent SPT kernels is present even when their total momentum $k < k_\sigma$ as long as the loop momenta $q_i$ cross the dispersion scale. Indeed, when such crossing takes place the linear modes cease to grow (as described by the linear kernel) and therefore they  cease to source the non-linear kernels. This reduces the UV sensitivity of \vpt~loops compared to SPT, capturing the expected decoupling of UV modes. This allows us to compute non-linear corrections to power spectra even for cosmologies with very blue power-law input spectra, for which SPT does not converge. Furthermore, it realizes the screening of UV modes for any input spectrum, considerably reducing the sensitivity of non-linear corrections to highly non-linear scales.

Along the way, we obtained a number of noteworthy results, namely: 

\begin{itemize}

\item[i)] We  provide analytical results when expanding in powers of the average background value of the velocity dispersion tensor (Sec.~\ref{sec:analytical_results}). These solutions illustrate the type of corrections to SPT, but are not sufficient to capture the UV screening.

\item[ii)] We characterize the impact of higher cumulants of the DF on the \vpt~non-linear kernels, and in addition we study the impact of scalar, vector and tensor modes of the second cumulant, i.e. the velocity dispersion tensor (Sec.~\ref{sec:numericalkernels}). 

\item[iii)] We demonstrate screening of UV modes, finding a universality that can be explained analytically. We derive an asymptotic scaling relation Eq.~\eqref{eq:kernelscalingvpt} showing how non-linear \vpt~density and velocity divergence kernels are suppressed relative to their SPT counterparts in the limit $k=|\sum_i\vec k_i|\ll k_\sigma\ll |\vec k_i|$,
\bea
\ \ \ \ \  F_{n,\delta/\theta}(\vec k_1,\dots,\vec k_n,\eta)\sim \frac{k^2}{q^2}\,\left(\frac{k_\sigma}{k_1}\right)^{2/\alpha}\!\!\!\cdots\left(\frac{k_\sigma}{k_n}\right)^{2/\alpha}\!\!\!\,.\nn\\
\eea
This makes the UV screening manifest for large internal wavenumbers $k_i$,  with an extra suppression on top of the $k^2/q^2$ scaling  imposed by momentum conservation well-known from SPT kernels, $F_n/G_n\sim k^2/q^2$ (where $q\sim\text{max}_i|\vec k_i|$). 
The exponent is related to the growth rate of the background dispersion $\alpha\equiv \partial_\eta\ln\epsilon$, and given by $2/\alpha=(n_s+3)/2$ for a scaling universe with spectral index $n_s$. This result is universal (e.g. valid in all truncation schemes we considered) and only hinges on the stalling of usual linear growth once a perturbation mode enters the ``dispersion horizon'', i.e. for $k_i>k_\sigma(\eta)=1/\epsilon(\eta)^{1/2}$. This scaling implies that the one-loop integral Eq.~\eqref{eq:P1Ldef} is UV finite for any spectral index $n_s$ and explains Eq.~\eqref{eq:vptintegrandscaling}.

\item[iv)] We derive constraints on the \vpt~non-linear kernels from underlying symmetries, such as Galilean invariance and momentum conservation (Sec.~\ref{sec:limits}). These generalize well-known relations to the case when including dispersion and higher cumulants, and provide non-trivial checks and implications for viable approximation schemes that truncate the hierarchy. In particular, we show that neglecting vorticity when including dispersion leads to a violation of momentum conservation. However, when including vorticity the truncations we explore do respect momentum conservation and Galilean invariance.

\item[v)] We discuss the generation of vorticity as well as vector and tensor modes of the velocity dispersion at second and higher order in perturbation theory (Sec.~\ref{sec:vorticity}). We find that the vorticity power spectrum generated non-linearly scales as $k^2$ for low wavenumber in general. However, the leading (one-loop) contribution is accidentally suppressed as $k^4$, such that the $k^2$ scaling appears only starting at two-loop order.

\item[vi)] We provide estimates of the dispersion scale within two complementary frameworks. First, we use a purely perturbative approach based on self-consistent solutions of the coupled equations for the perturbations and the average dispersion (Sec.~\ref{sec:selfconsistent}). Second, we use the halo mass function measured in simulations together with the calculations of the velocity dispersion profile in NFW halos to compute the dispersion scale from virialized regions (Sec.~\ref{sec:halo}). Remarkably, both of these approaches yield estimates that are close to each other, while not allowing for a precise determination.

\item[vii)] The \vpt~linear theory gives for the first time a way to calculate  the non-linear halo mass for blue spectra (Sec.~\ref{sec:halo}), avoiding UV divergences that appear for spectral indices $n_s\geq 1$. This is relevant as well for CDM spectra when considering biased tracers, to avoid UV divergences that appear when dealing with derivatives of the smoothed density field in some bias schemes,  see e.g.~\cite{ParSheDes1305}. 

\item[viii)] We compare predictions for the density and velocity divergence power spectra as well as the bispectrum at one-loop order to N-body results in a scaling universe with spectral indices  $-1\leq n_s\leq +2$ (Sec.~\ref{sec:simulation}). We use the density power spectrum to fix the precise value of the background dispersion, which then completely determines all other power- and bispectra. We find a good agreement up to the non-linear scale for all cases, with a reach that increases with the spectral index $n_s$.  Even though, naively (e.g. within SPT), the loop integration would be UV dominated for large $n_s$, we find that this is in fact not the case when including higher cumulants due to the screening of small-scale modes that becomes  more efficient with larger $n_s$. As a result of this, \vpt~reconciles perturbation theory behavior with the observation that blue spectra in simulations show most suppression of non-linear power, explaining the dependence of non-linear power on linear power with spectral index.

\item[ix)]  Interestingly, we find the values for the dispersion scale fixed from the two-loop density power spectrum (Eq.~\ref{ksig2loop}) is in between the perturbative estimate and the halo estimate (compare to Tables~\ref{tab:selfconsistent} and~\ref{tab:ksigma}, respectively). Furthermore, the trend of these values with spectral index is also consistent. This  is compatible with the fact that by matching the density power spectrum we are determining the spatial average of the dispersion tensor, a well-defined quantity with clear physical meaning.

\item[x)]   We observe a significant impact of vorticity backreaction on the (third- and higher order) density contrast. At the non-linear scale the impact of vorticity backreaction on the density power spectrum is $\sim 10\%$ for $n_s=0$ (Fig.\,\ref{fig:Pdd_cumPT}). For blue indices $n_s=1,2$ this grows to about 15\%, while for $n_s=-1$ it drops to 3\%. The backreaction of the vector mode of the dispersion tensor also has a certain impact, while its tensor modes yield a negligible backreaction up to one-loop.

\item[xi)]   A nontrivial prediction of \vpt~is the vorticity power spectrum, which we compute up to two-loop order in order to recover the correct large-scale limit. Comparing to our N-body measurements confirms the cross-over from $k^4$ to $k^2$ scaling on large scales (Sec.~\ref{sec:Ptt}).

\end{itemize}
Our results provide a proof-of-principle that perturbative techniques for dark matter clustering can be systematically improved based on the known underlying collisionless dynamics. In particular, this deterministic approach features a screening of UV modes and thereby abandons one of the main shortcomings of the ubiquitous SPT approximation. This motivates to develop the framework further and study also its application in the context of $\Lambda$CDM cosmologies. Another obvious extension is to carry out a systematic study of two-loop corrections, and check the convergence of \vpt, which given the UV screening may be expected to hold best for blue spectra, the opposite of what happens for SPT. Finally, along the same lines, we plan to  study  power spectra response functions, i.e. the dependence of the non-linear power spectrum on the initial spectrum. The suppression of this reponse seen in N-body simulations~\cite{NisBerTar1611,NisBerTar1712} compared to SPT is another manifestation of the screening of UV modes (see Fig.~\ref{fig:integrand}), and a quantitative comparison of this provides yet another test of \vpt.

\vspace*{2em}
\acknowledgments

We thank Michael Buehlmann, ChangHoon Hahn, Oliver Hahn and Lam Hui for useful discussions, and E.~Sefusatti for useful exchange on fixed-amplitude initial conditions and sharing his unpublished results with A.~Oddo~\cite{OddSef22}. 
We acknowledge support by the Excellence Cluster ORIGINS, which is funded by the Deutsche Forschungsgemeinschaft (DFG, German Research
Foundation) under Germany's Excellence Strategy - EXC-2094 - 390783311.

\begin{widetext}

\appendix

\section{Tadpoles}
\label{app:tadpoles}

In this appendix we discuss subtleties related to the splitting between background value and perturbation. For example, for the
velocity dispersion the equation of motion for the perturbations $\delta\epsilon_{ij}=\epsilon_{ij}-\langle\epsilon_{ij}\rangle$
are obtained by subtracting the Eq.~\eqref{eq:epseom} for the average value $\langle\epsilon_{ij}\rangle=\epsilon(\eta)\delta_{ij}$
from the evolution equation for $\epsilon_{ij}=\sigma_{ij}/(f{\cal H})^2$ obtained from the second moment of the Vlasov equation Eq.~\eqref{eq:eomsigmaij}.
The same holds for the perturbation of $A=\ln(1+\delta)$ around its average ${\cal A}=\langle A\rangle$, that we denote by $\delta A\equiv A-\langle A\rangle$
in this section for clarity (in the rest of the work we use the symbol $A$ also for the perturbation part in a slight abuse of notation). The equation of motion
for its background value reads ${\cal A}' = Q_A(\eta)$ with source term $Q_A(\eta)\equiv -\int d^3 k P_{\theta A}(k,\eta)$~\cite{cumPT}.

The evolution equation for the vector Eq.~\eqref{eq:psidef} of perturbation modes obtained in this way reads
\be\label{eq:eomtadpole}
  \psi_{k,a}'(\eta)+\Omega_{ab}(k,\eta)\psi_{k,b}(\eta) = \int_{pq} \hat\gamma_{abc}(\vec p,\vec q)\psi_{p,b}(\eta)\psi_{q,c}(\eta) - Q_a(\eta)\delta^{(3)}(\vec k)\,,
\ee
which differs from Eq.~\eqref{eq:eom} given in the main text by the last term, and contains vertices $\hat\gamma_{abc}(\vec p,\vec q)$, that, within this section, denote the vertices as derived from the Vlasov equation (we use the hatted notation to discriminate them from modified vertices to be defined below).
The last term arises from subtracting the equations for the average values from the original evolution equations for
$\delta\epsilon$ and $\delta A$, respectively. We use a vector notation with $Q_{a}\equiv Q$ (see Eq.~\eqref{eq:Q3rd}) for $a=\delta\epsilon$,
$Q_{a}\equiv Q_A$ for $a=\delta A$, and $Q_a=0$ for all other perturbation modes considered here, that do not possess average values.
The last term in Eq.~\eqref{eq:eomtadpole} ensures that $\langle\psi_{k,a}(\eta)\rangle=0$, i.e. that $\delta\epsilon$ and $\delta A$ indeed correspond to the perturbation around the average.

In this appendix we show that the term $- Q_a(\eta)\delta^{(3)}(\vec k)$ can be skipped in the evolution equations provided one replaces the vertices $\hat\gamma_{abc}(\vec p,\vec q)$ as derived from the Vlasov equation
by vertices $\gamma_{abc}(\vec p,\vec q)$ given by
\be\label{eq:gammared}
  \gamma_{abc}(\vec p,\vec q) \equiv \left\{\begin{array}{ll}
    \hat\gamma_{abc}(\vec p,\vec q) & \vec p+\vec q\not=0\,,\\
    0 & \vec p+\vec q=0\,.
  \end{array}\right.
\ee
Note that, apart from this section, we do not discriminate between $\gamma_{abc}$ and $\hat\gamma_{abc}$, but it is implicitly understood that vertices are set to zero
for $\vec p+\vec q=0$.

The statement from above can be most easily understood using a diagrammatic representation of the perturbative expansion, see e.g.~\cite{BerColGaz02}.
We note that it can be extended in a straightforward way to the generalized case of a vector of perturbations considered here. Starting point is the integral
representation of the equation of motion
\be\label{eq:eomtadpoleintegral}
  \psi_{k,a}(\eta) = \psi^\text{lin}_{k,a}(\eta) + \int^\eta d\eta' g_{aa'}(\eta,\eta',k) \left[ \int_{pq} \hat\gamma_{a'bc}(\vec p,\vec q)\psi_{p,b}(\eta')\psi_{q,c}(\eta') - Q_{a'}(\eta)\delta^{(3)}(\vec k)\right]\,,
\ee
where $g_{aa'}(\eta,\eta',k)$ is the retarded propagator, defined by the solution of the linear equation
$\partial_\eta g_{aa'}(\eta,\eta',k)+\Omega_{ab}(k,\eta)g_{ba'}(\eta,\eta',k)=0$ with boundary condition $g_{aa'}(\eta',\eta',k)=\delta_{aa'}$ for $\eta=\eta'$, and $g_{aa'}(\eta,\eta',k)=0$ for $\eta<\eta'$.
The perturbative expansion Eq.~\eqref{eq:psiexpansion} can be generated by iteratively inserting the linear solution $\psi^\text{lin}_{k,a}(\eta)=e^\eta F_{1,a}(k,\eta)\delta_{k0}$ inside the non-linear terms in Eq.~\eqref{eq:eomtadpoleintegral},
und using a lower integration boundary formally shifted to $-\infty$.

The propagator and vertices lead to a diagrammatic expansion that is formally analogous to the SPT case~\cite{BerColGaz02}, except that the propagator depends on both time arguments as well as on wavenumber, has more components, and more vertices contribute. In addition, the source terms $Q_a$ can formally be represented by insertions given by $- Q_{a'}(\eta)\delta^{(3)}(\vec k)$ from which a single propagator line with wavenumber $\vec k=0$ emerges. An example will be shown below. As we will see, their role is to precisely cancel ``tadpoles diagrams''. Similar cancellations are well-known in the context of effective actions considered in quantum field theory.

Tadpole diagrams contain some loops
that are connected only via a \emph{single} line to the rest of the diagram. By momentum conservation, this line
is restricted to carry vanishing wavenumber. Such diagrams can arise for perturbations that also
possess a homogeneous background value, i.e. $\delta A$ and $\delta \epsilon$ in the present case.

An example for a tadpole contribution to the one-loop power spectrum is shown in Fig.\,\ref{fig:tadpole} (left).
The tadpole loop can be non-zero if the vertex $\gamma_{abc}(p,q)$ belonging to the loop has a non-vanishing limit for $\vec p+\vec q\to 0$.
In the following we restrict the discussion to scalar perturbations up to the second cumulant for simplicity, but note that the arguments below can be extended to the general case.
Within this restriction, the vertices with a non-zero limit for $\vec p+\vec q\to 0$ are
\be
  \gamma_{A\theta A}(p,q) \to -\frac12\,, 
  \gamma_{g \theta g}(p,q) \to \frac14(3c_p^2-1)\,,
  \gamma_{g \theta \epsilon}(p,q) \to \frac12(3c_p^2-1)\,,
  \gamma_{\epsilon \theta g}(p,q) \to \frac14 s_p^2\,,
  \gamma_{\epsilon \theta \epsilon}(p,q) \to -\frac12 c_p^2\,.
\ee
Here $c_p (s_p)$ denotes the cosine (sine) of the angle between between $\vec p$ and $\vec k\equiv\vec p+\vec q$.
The tadpole loop is given by
\be\label{eq:tadpoleloop}
  T_a(\vec k,\eta) \equiv \delta^{(3)}(\vec k)\int d^3p \, \gamma_{abc}(\vec p,\vec q)\Big|_{\vec p+\vec q\to 0}\,P_{bc}(p,\eta)\,,
\ee
where $P_{bc}(p,\eta)$ denotes the linear power spectrum at time $\eta$ (see below for a generalization to the non-linear case).
Since $P_{bc}(p,\eta)$ does not depend on the direction of $\vec p$, we can consider the angle-averaged vertices
\be
  \int \frac{d\Omega}{4\pi} \gamma_{A\theta A}(\vec p,\vec q) \to -\frac12\,, 
  \int \frac{d\Omega}{4\pi} \gamma_{g \theta g}(\vec p,\vec q) \to 0\,,
  \int \frac{d\Omega}{4\pi} \gamma_{g \theta \epsilon}(\vec p,\vec q) \to 0\,,
  \int \frac{d\Omega}{4\pi} \gamma_{\epsilon \theta g}(\vec p,\vec q) \to \frac16\,,
  \int \frac{d\Omega}{4\pi} \gamma_{\epsilon \theta \epsilon}(\vec p,\vec q) \to -\frac16\,.
\ee
The contribution from tadpoles containing
$\gamma_{g \theta g}$ and $\gamma_{g \theta \epsilon}$ averages to zero, consistent with the
fact that $g$ does not have a homogeneous background value.

\begin{figure}[t]
\begin{center}
\includegraphics[width=0.65\textwidth]{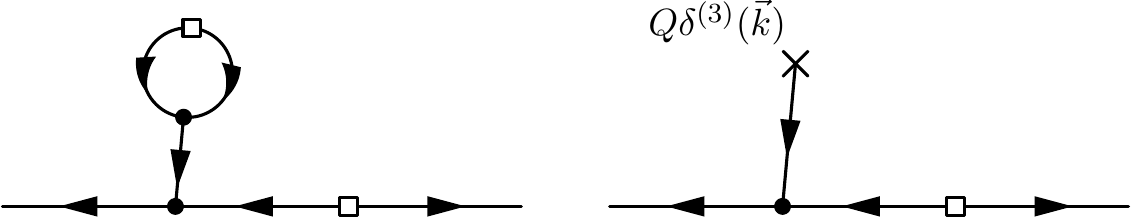}
\end{center}
\caption{\label{fig:tadpole}
Tadpole diagram contributing to the one-loop power spectrum (left), and a corresponding diagram with an insertion of the source term $Q(\eta)\delta^{(3)}(\vec k)$
contributing to the equation of motion for the perturbations (right). Both diagrams cancel each other, to ensure that the perturbation around the background maintains
a vanishing mean value. It is equivalent to skip the right diagram, and use the prescription to set the vertices to zero when the ingoing lines \emph{exactly} add up to zero, thereby eliminating also the left diagram.
}
\end{figure}

Let us consider as an example the tadpole $T_a$ for $a=\delta\epsilon$.
Using the angle-averaged vertices, the tadpole Eq.~\eqref{eq:tadpoleloop} is given by
\be
  T_{\delta\epsilon}(\vec k,\eta) = \delta^{(3)}(\vec k)\,\frac13\int d^3p \,(P_{\theta g}(p,\eta)-P_{\theta\, \delta\epsilon}(p,\eta))\,.
\ee
The tadpole is \emph{exactly} cancelled by the contribution $-Q(\eta)\delta^{(3)}(\vec k)$ from the source term
to the perturbation equation for $\delta\epsilon$ Eq.~\eqref{eq:Q3rd} (considering the restriction to scalar modes up to the second cumulant for the purpose of this discussion), as expected. 
The corresponding contribution to the power spectrum is shown in the right part of Fig.\,\ref{fig:tadpole}.

Note that this cancellation also works at higher loop order. We can decompose any loop diagram in tadpole contributions,
that are attached to some internal or external lines of the full diagram. The tadpole itself can also contain loops,
that precisely yield the full non-linear power spectrum $P_{bc}(p,\eta)$ in the integrand in Eq.~\eqref{eq:tadpoleloop}.
On the other hand, the source term $Q$ also contains the integral over the corresponding full non-linear power spectrum, such that the cancellation occurs at any
order in perturbation theory. The same argument applies to the tadpole Eq.~\eqref{eq:tadpoleloop} for $a=\delta A$. It can also be generalized to perturbations of the fourth, sixth, eighth, or any even cumulant, that may possess an average value (see paper~I~\cite{cumPT}), but are not considered in this work.

One may wonder whether the vertex connecting the line coming from the tadpole to the rest of the diagram can potentially be singular for vanishing ingoing wavenumber. One can check that either
this singularity is removed by a factor of $k^2$ coming from the internal propagating line connecting with the tadpole, or that the vertex is not singular.

Therefore, we conclude that with the prescription of setting vertices to zero when the ingoing lines \emph{exactly} add up to zero
effectively eliminates all tadpoles, such that we can also ignore diagrams with insertions of the sources $Q$ or $Q_A$.
This is equivalent to using the equation of motion Eq.~\eqref{eq:eom} instead of Eq.~\eqref{eq:eomtadpole}, with vertices Eq.~\eqref{eq:gammared}, as claimed above.

\section{Time integrals}
\label{app:timeintegrals}

The integrals appearing in the analytic results Eq.~\eqref{eq:F2analytRaw} for the non-linear kernels $F_{2,\delta}$ and $F_{2,\theta}$
when expanding to first order in the power of the background dispersion $\epsilon(\eta)$ are given by
\bea
  J^\delta_1 &\equiv & \frac45\int^\eta d\eta' \,\left(e^{\eta'-\eta}- e^{7(\eta'-\eta)/2}\right)E_3(\eta')
  = \frac25 (E_1(\eta)  - 5 E_3(\eta) + 4 E_{7/2}(\eta))\,,\nn\\
  J^\delta_2 &\equiv & \frac45\int^\eta d\eta' \,\left(e^{\eta'-\eta}- e^{7(\eta'-\eta)/2}\right)(E_2(\eta')-E_3(\eta'))
  = \frac{2}{15} (3 E_{1}(\eta) - 10 E_{2}(\eta) + 15 E_3(\eta) - 8 E_{7/2}(\eta)) \,,\nn\\
  J^\delta_3 &\equiv & \frac25\int^\eta d\eta' \,\left(e^{\eta'-\eta}- e^{7(\eta'-\eta)/2}\right)\epsilon(\eta')
  = \frac25 (E_{1}(\eta) - E_{7/2}(\eta)) \,,\nn\\
  J^\delta_4 &\equiv & \frac25\int^\eta d\eta' \,\left(e^{\eta'-\eta}- e^{7(\eta'-\eta)/2}\right)I_\theta(\eta') 
  = \frac{2}{105} (12 E_{0}(\eta) - 70 E_{2}(\eta) + 63 E_{5/2}(\eta) - 5 E_{7/2}(\eta)) \,,\nn\\
  J^\delta_5 &\equiv & J^\delta_1+ J^\delta_2
  = \frac{4}{15} (3 E_{1}(\eta) - 5 E_{2}(\eta) + 2 E_{7/2}(\eta))\,,\nn\\
  J^\delta_6 &\equiv & \frac25\int^\eta d\eta'\,\left(\frac32 e^{\eta'-\eta}+ e^{7(\eta'-\eta)/2}\right) I_\theta(\eta')
  = \frac{2}{21} (6 E_{0}(\eta) - 7 E_{2}(\eta) + E_{7/2}(\eta)) \,,\nn\\
  J^\delta_7 &\equiv & \frac25\int^\eta d\eta'\,\left(\frac32 e^{\eta'-\eta}+ e^{7(\eta'-\eta)/2}\right) I_\delta(\eta')
  = \frac{2}{105} (30 E_{0}(\eta) - 63 E_{1}(\eta) + 35 E_{2}(\eta) - 2 E_{7/2}(\eta)) \,,
\eea
and
\bea
  J^\theta_1 &\equiv & \frac45\int^\eta d\eta' \,\left(e^{\eta'-\eta}+\frac32 e^{7(\eta'-\eta)/2}\right)E_3(\eta')
  = \frac25 (E_{1}(\eta) + 5 E_3(\eta) - 6 E_{7/2}(\eta)) \,,\nn\\
  J^\theta_2 &\equiv & \frac45\int^\eta d\eta' \,\left(e^{\eta'-\eta}+\frac32 e^{7(\eta'-\eta)/2}\right)(E_2(\eta')-E_3(\eta'))
  =  \frac25 (E_{1}(\eta) - 5 E_3(\eta) + 4 E_{7/2}(\eta)) \,,\nn\\
  J^\theta_3 &\equiv & \frac25\int^\eta d\eta' \,\left(e^{\eta'-\eta}+\frac32 e^{7(\eta'-\eta)/2}\right)\epsilon(\eta')
  =  \frac15 (2 E_{1}(\eta) + 3 E_{7/2}(\eta)) \,,\nn\\
  J^\theta_4 &\equiv & \frac25\int^\eta d\eta' \,\left(e^{\eta'-\eta}+\frac32 e^{7(\eta'-\eta)/2}\right)I_\theta(\eta') 
  =  \frac{1}{35} (16 E_{0}(\eta) - 21 E_{5/2}(\eta) + 5 E_{7/2}(\eta))\,,\nn\\
  J^\theta_5 &\equiv & J^\theta_1+ J^\theta_2
  = \frac45 (E_{1}(\eta) - E_{7/2}(\eta))\,,\nn\\
  J^\theta_6 &\equiv & \frac25\int^\eta d\eta'\,\left(\frac32 e^{\eta'-\eta}-\frac32 e^{7(\eta'-\eta)/2}\right) I_\theta(\eta')
  =  \frac{1}{35} (12 E_{0}(\eta) - 70 E_{2}(\eta) + 63 E_{5/2}(\eta) - 5 E_{7/2}(\eta)) \,,\nn\\
  J^\theta_7 &\equiv & \frac25\int^\eta d\eta'\,\left(\frac32 e^{\eta'-\eta}-\frac32 e^{7(\eta'-\eta)/2}\right) I_\delta(\eta')
  = \frac{2}{35} (6 E_{0}(\eta) - 21 E_{1}(\eta) + 35 E_{2}(\eta) - 21 E_{5/2}(\eta) +    E_{7/2}(\eta)) \,.
\eea

\section{Recursion relation and Galilean invariance}
\label{app:galilei}

In this appendix we prove the relation Eq.~\eqref{eq:Fnsoft}, by generalizing the proof within EdS-SPT presented in~\cite{Sugiyama:2013pwa}.
The crucial observation is that the non-linear vertices $\gamma_{abc}(\vec p,\vec q)$ satisfy the
property Eq.~\eqref{eq:gammasoft} in the limit $\vec p\to 0$.

We proceed by induction and assume that Eq.~\eqref{eq:Fnsoft} is satisfied for all $m\leq n$. For the moment we also assume that it is satisfied for the starting point $n=1$, and show that this is indeed the case further below.
Now we want to proove Eq.~\eqref{eq:Fnsoft} for $F_{n+1,a}$.
The differential equation Eq.~\eqref{eq:kernelODE} for the kernel can be written as
\bea\label{eq:Fnsoftproof}
 \lefteqn{ (\partial_{\eta}\delta_{ab}+(n+1)\delta_{ab}+\Omega_{ab}(|\vec k+\vec p|,\eta))F_{n+1,b}(\vec p,\vec k_1,\dots ,\vec k_n,\eta) }\nn\\
  &=& \sum_{m=1}^{n-1} \frac{m+1}{n+1}\Big\{\gamma_{abc}(\vec p+\vec q_1+\cdots+\vec q_m,\vec q_{m+1}+\cdots+\vec q_n) F_{m+1,b}(\vec p,\vec q_1,\dots,\vec q_m,\eta)F_{n-m,c}(\vec q_{m+1},\dots,\vec q_n,\eta) \Big\}^s \nn\\
  &+& \sum_{m=1}^{n-1} \frac{n-m+1}{n+1}\Big\{\gamma_{abc}(\vec q_1+\cdots+\vec q_m,\vec p+\vec q_{m+1}+\cdots+\vec q_n) F_{m,b}(\vec p,\vec q_1,\dots,\vec q_m,\eta)F_{n-m+1,c}(\vec p,\vec q_{m+1},\dots,\vec q_n,\eta) \Big\}^s \nn\\
  &+& \frac{1}{n+1}\gamma_{abc}(\vec p,\vec k_1+\cdots+\vec k_n) F_{1,b}(\vec p,\eta)F_{n,c}(\vec k_{1},\dots,\vec k_n,\eta)
  + \frac{1}{n+1}\gamma_{abc}(\vec k_1+\cdots+\vec k_n,\vec p) F_{n,b}(\vec k_{1},\dots,\vec k_n,\eta)F_{n,c}(\vec p,\eta)\,,\nn\\
\eea
where $\{\cdots\}^s=\sum_\text{perm}\{\cdots\}/|\text{perm}|$ denotes an average over all $|\text{perm}|=n!/m!/(n-m)!$ possibilities to
choose the subset of wavevectors $\{\vec q_1,\dots,\vec q_m\}$ from $\{\vec k_1,\dots,\vec k_n\}$. The symmetrization with respect to the
vector $\vec p$ is \emph{not} included in $\{\cdots\}^s$ in the equation above, but explicitly accounted for by the two sums over $m$ in
the second and third line and the two contributions in the last line, respectively, as well as their prefactors.

In this form, the limit $\vec p\to 0$ can be performed, using Eq.~\eqref{eq:Fnsoft} for the kernels $F_{m+1,b}(\vec p,\vec q_1,\dots,\vec q_m,\eta)$ and $F_{n-m+1,c}(\vec p,\vec q_{m+1},\dots,\vec q_n,\eta)$, respectively,
as well as
\bea
  \gamma_{abc}(\vec p+\vec q_1+\cdots+\vec q_m,\vec q_{m+1}+\cdots+\vec q_n)&\to& \gamma_{abc}(\vec q_1+\cdots+\vec q_m,\vec q_{m+1}+\cdots+\vec q_n)\,,\nn\\
  \gamma_{abc}(\vec q_1+\cdots+\vec q_m,\vec p+\vec q_{m+1}+\cdots+\vec q_n)&\to& \gamma_{abc}(\vec q_1+\cdots+\vec q_m,\vec q_{m+1}+\cdots+\vec q_n)\,.
\eea
For the first term in the last line of Eq.~\eqref{eq:Fnsoftproof} we use the property Eq.~\eqref{eq:gammasoft} of the vertices, as well as $F_{1,\theta}(\vec p,\eta)\to 1$ for $\vec p\to 0$
and $F_{1,w_i}(\vec p,\eta)=0$. At this point we have to assume that $|\vec p|\ll k_\sigma$ in addition to $|\vec p|\ll|\vec k_i|$.
Using that the vertices are symmetrized, see Eq.~\eqref{eq:gammasymm}, the second term in the last line  of Eq.~\eqref{eq:Fnsoftproof} is identical to the first one. Finally, we use that $\Omega_{ab}(|\vec k+\vec p|,\eta))\to \Omega_{ab}(\vec k,\eta))$.
Altogether, we obtain
\bea
 \lefteqn{ (\partial_{\eta}\delta_{ab}+(n+1)\delta_{ab}+\Omega_{ab}(k,\eta))F_{n+1,b}(\vec p,\vec k_1,\dots ,\vec k_n,\eta)   }\nn\\
  &=&  \frac{1}{n+1} \frac{\vec k\cdot\vec p}{p^2}\sum_{m=1}^{n-1}\Big\{\gamma_{abc}(\vec q_1+\cdots+\vec q_m,\vec q_{m+1}+\cdots+\vec q_n)  F_{m,b}(\vec q_1,\dots,\vec q_m,\eta)F_{n-m,c}(\vec q_{m+1},\dots,\vec q_n,\eta) \Big\}^s \nn\\
  &+& \frac{1}{n+1}\frac{\vec k\cdot\vec p}{p^2} F_{n,a}(\vec k_{1},\dots,\vec k_n,\eta) \ + {\cal O}(p^0)\,.
\eea
The sum in the middle line is precisely the right-hand side of the differential equation Eq.~\eqref{eq:Fnsoft} for the kernel $F_{n,a}(\vec k_{1},\dots,\vec k_n,\eta)$.
Therefore, taking also the last line into account, we can also write this equation as
\be\label{eq:}
   (\partial_{\eta}\delta_{ab}+(n+1)\delta_{ab}+\Omega_{ab}(k,\eta))\left[F_{n+1,b}(\vec p,\vec k_1,\dots ,\vec k_n,\eta) -\frac{1}{n+1}\frac{\vec k\cdot\vec p}{p^2} F_{n,b}(\vec k_{1},\dots,\vec k_n,\eta)\right]={\cal O}(p^0)\,.
\ee
For $a=\delta,\theta$, the kernels $F_{n,a}$ approach the EdS-SPT kernels at early times, which are known to satisfy Eq.~\eqref{eq:Fnsoft}~\cite{Sugiyama:2013pwa}. For all other perturbation modes, the kernels can be assumed to
vanish at early times, and are generated only by time evolution. Therefore, the square bracket can be taken to be zero in the limit $\eta\to-\infty$ for all kernels, up to corrections of order $p^0$.
Consequently, the unique solution of the differential equation is that the square bracket is zero (up to $p^0$ terms) at \emph{all} times. This completes the proof for all $n\geq 2$. What is still missing is to show Eq.~\eqref{eq:Fnsoft}
for the starting point of the induction, $n=1$. This can be achieved by realizing that when proceeding analogously as above for the kernels $F_{2,a}(\vec p,\vec k,\eta)$, the second and third line in Eq.~\eqref{eq:Fnsoftproof} is
absent, and only the fourth line contributes. Therefore, the desired relation Eq.~\eqref{eq:Fnsoft} follows directly from the vertex property Eq.~\eqref{eq:gammasoft} in that case, when proceeding analogously as in the general case.

\section{Treatment of vorticity, vector and tensor modes}
\label{app:vorticity}

In Fourier space, the vorticity $w_i$ satisfies the transversality condition $\vec k\cdot\vec w=0$, such that only two of its three components are independent.
The same holds for the vector mode $\nu_i$ of the second cumulant, while its tensor mode $t_{ij}$ satisfies $k_it_{ij}=0, t_{ij}=t_{ji}, t_{ii}=0$, corresponding
to $3\times 3 - (3+3+1)=2$ independent degrees of freedom. In our implementation we use a scheme taking advantage of these constraints. For example, for the vorticity
perturbation $w_i$ at some wavenumber $\vec k$, we can choose the two independent degrees of freedom in the plane perpendicular to $\vec k$. There is a remaining freedom
to select a basis in this plane. It turns out that for the one-loop power spectrum a suitable choice of this basis allows us to track only
a single perturbation mode for the vorticity. The same applies to the vector and tensor mode, respectively. For the two-loop power spectrum or one-loop bispectrum, we need to include all
independent degrees of freedom. We describe both algorithms below, and checked their exact equivalence whenever the simpler one is applicable.

Before that, we discuss some general properties. In this appendix we refer to scalar modes with the letter $s$ (with $s\in\{\delta,\theta,g,\delta\epsilon,A,\pi,\chi\}$ up to ({\bf cum3+})), vector modes with $v_i$ (with $v_i\in\{w_i,\nu_i\}$ encompassing \emph{both} vorticity and the vector mode of the velocity dispersion tensor), and tensor modes with $t_{ij}$.
Vector and tensor modes can appear in two different roles:
\begin{itemize}
\item[($i$)] when computing power spectra of vector or tensor modes themselves (involving non-linear kernels $F_{n,v_i}$ or $F_{n,t_{ij}}$), 
\item[($ii$)] as \emph{backreaction} contributions to the power- (and bi-)spectra of scalar modes via non-linear kernels $F_{n,s}$. 
\end{itemize}
In a diagrammatic representation of the perturbative solution of the equations of motion (see e.g.~\cite{BerColGaz02} and App.\,\ref{app:tadpoles}), vector and
tensor modes appear as \emph{external, outgoing lines} in case ($i$) and as \emph{internal} lines in both ($i$) and ($ii$), while ``ingoing'' lines (that emerge from insertions of the initial power spectrum $P_0$ in loop diagrams) are always (growing) scalar modes.

Furthermore, statistical isotropy implies rotational invariance of power spectra involving vector and tensor modes, which implies that all
cross spectra $P_{v_is}=P_{t_{ij}s}=P_{v_it_{jk}}=0$ vanish. In addition, rotational invariance requires
\be\label{eq:Pvivj}
  P_{v_iv_j}(k,\eta)=\frac12\left(\delta_{ij}-\frac{k_ik_j}{k^2}\right)\sum_{n}P_{v_nv_n}(k,\eta)\,,\qquad
  P_{t_{ij}t_{ls}}(k,\eta)=\frac12 P^T_{ij,ls}(k)\sum_{m,n}P_{t_{nm}t_{nm}}(k,\eta)\,,
\ee 
where we explicitly indicated the summation over $n,m$ here for clarity (while being implicitly understood in the rest of this work), and $i,j,l,s$ are not summed over.
The tensor projector $P^T_{ij,ls}(k)$ can be found in~\cite{cumPT}. These results can be derived by making the ansatz that e.g. $P_{v_iv_j}(k,\eta)$ is a general linear
combination of all rotationally covariant objects with two indices built from the vector $\vec k$, being $\delta_{ij}$ and $k_ik_j$, and using the transversality condition $\vec k\cdot\vec v=0$.
An analogous derivation holds for the tensor mode and the cross spectra. Note that the structure of $P_{v_iv_j}$ applies correspondingly to the three cases $P_{w_iw_j}, P_{w_i\nu_j}, P_{\nu_i\nu_j}$. Altogether, this implies that it is sufficient to compute the vector and tensor power spectra on the right-hand sides of Eq.~\eqref{eq:Pvivj}, where a summation over all indices is performed.
Our numerical implementations take advantage of this observation.

\begin{figure}[t]
\begin{center}
\includegraphics[width=0.75\textwidth]{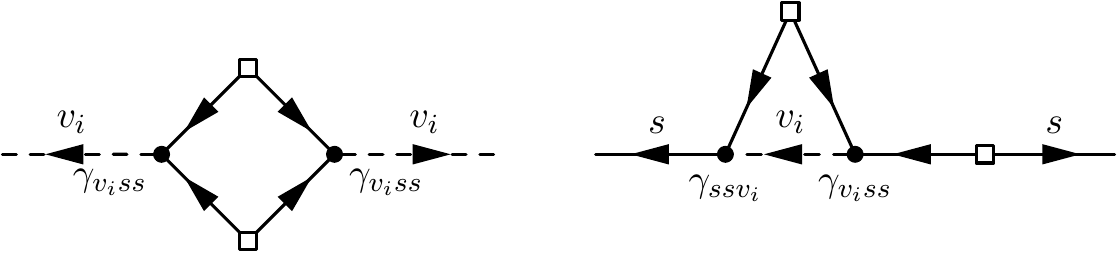}
\end{center}
\caption{\label{fig:vort}
Generation of vorticity/vector perturbations at one-loop (left diagram) and backreaction on scalar power spectra (right diagram).
Here $v_i\in\{w_i,\nu_i\}$ corresponds to dashed lines, and solid lines represent any scalar perturbations $s\in\{\delta,\theta,g,\delta\epsilon,A,\pi,\chi\}$ 
(possibly of different type for each line). The open square denotes the initial power spectrum, and the filled circles are non-linear vertices.
}
\end{figure}

\subsection{Simplified algorithm for the one-loop power spectrum}\label{app:vorticity_simple}

The algorithm described here can be used for all scalar, vector and tensor power spectra up to one-loop order, including backreaction of vector and tensor
modes for the former. It exploits that, at one-loop, it is sufficient to track only one of the two independent degrees of freedom for each vector perturbation $v_i$
and tensor perturbation $t_{ij}$, respectively. The general case is discussed in Sec.~\ref{app:vorticity_full}.

When being interested in the one-loop power spectrum, vector and tensor modes can appear only in two places:
($i$) in the kernel $F_{2,v_i}(\vec q,\vec k-\vec q,\eta)$ or $F_{2,t_{ij}}(\vec q,\vec k-\vec q,\eta)$ for the
vector- or tensor power spectra $P_{v_iv_i}$ or $P_{t_{ij}t_{ij}}$, respectively, and
($ii$) in the kernels $F_{3,s}(\vec k,\vec q,-\vec q,\eta)$ contributing to power spectra of scalar modes.
These two possibilities are specific examples for the generic cases ($i$) and ($ii$) from above.

In the following we discuss an algorithm that is sufficient to compute these kernels. Let us start with the vector modes $v_i\in\{w_i,\nu_i\}$.
We generically refer to non-linear vertices $\gamma_{abc}(\vec k_1,\vec k_2)$ with e.g. $a$ representing a vector mode and $b,c$ any two
scalar modes $s,s'$ as $\gamma_{v_iss'}(\vec k_1,\vec k_2)$, and analogously for other cases. The only relevant vertex combinations containing vector modes when computing 
one-loop power spectra are
\be\label{eq:vector_eff}
  \gamma_{v_iss'}(\vec q,\vec k-\vec q)\gamma_{v_i's''s'''}(\vec q,\vec k-\vec q)\,,\qquad
   \gamma_{ss'v_i}(\vec q,\vec k-\vec q)\gamma_{v_i's''s'''}(-\vec q,\vec k)\,,
\ee
which contribute to case ($i$) and ($ii$), respectively (see Fig.\,\ref{fig:vort}), and with summation over the spatial index $i=x,y,z$ in both cases.
Here $\vec k$ is the external wavenumber, and $\vec q$ the loop integration variable.
Note that there are only those two independent directions. In addition, we observe that all vertices of the form
$\gamma_{v_iss'}(\vec k_1,\vec k_2)$ as well as $\gamma_{sv_is'}(\vec k_1,\vec k_2)=\gamma_{ss'v_i}(\vec k_2,\vec k_1)$ are proportional to $\vec k_1\times \vec k_2$.
Therefore, the vorticity has to lie along the axis $\vec k\times \vec q$. Thus, it is sufficient to track a single degree of freedom for each vector mode. In practice, we can replace the three kartesian by a single component, $w_i\mapsto w_\text{eff}$ and $\nu_i\mapsto \nu_\text{eff}$, and replace the factor $(\vec k_1\times \vec k_2)_i$ contained in the relevant vertices
by $\sigma_{\vec k_1\vec k_2}|\vec k_1\times \vec k_2|$. Here $\sigma_{\vec k_1\vec k_2}\in\{\pm1,0\}$ keeps track of the correct sign, and is chosen such that
$(\vec k_1\times \vec k_2)_i=\sigma_{\vec k_1\vec k_2}|\vec k_1\times \vec k_2|(\vec k\times \vec q)_i/|\vec k\times \vec q|$ for all relevant vectors $\vec k_{1,2}\in\{\pm\vec q,\pm\vec k,\pm(\vec k-\vec q),\pm(\vec k+\vec q)\}$
that occur when evaluating the integrand of the one-loop power spectrum, including the symmetrization $\vec q\leftrightarrow -\vec q$ to ensure cancellation of infrared singularities at the integrand level. We use $\sigma_{\vec k_1\vec k_2}=\text{sgn}((\vec k_1\times \vec k_2)\cdot(\vec k\times \vec q))$.

Altogether, the effective single-component vorticity and vector modes correctly take into account the backreaction on the scalar kernels $F_{3,s}(\vec k,\vec q,-\vec q,\eta)$, and allow for the
computation of the vector power spectra according to $P_{w_iw_i}=P_{w_\text{eff}w_\text{eff}}, P_{w_i\nu_i}=P_{w_\text{eff}\nu_\text{eff}}, P_{\nu_i\nu_i}=P_{\nu_\text{eff}\nu_\text{eff}}$.

For the tensor mode, the relevant combinations of vertices are analogous to Eq.~\eqref{eq:vector_eff} with $v_i,v_i'$ replaced by $t_{ij},t_{ij}'$. We observe that for vertices of the form
$\gamma_{t_{ij}ss'}(\vec k_1,\vec k_2)$ the dependence on the indices $i,j$ is universally given by the factor 
\be
  f_{ij}(\vec k_1,\vec k_2)\equiv \delta_{ij}-(\vec k_1+\vec k_2)_i(\vec k_1+\vec k_2)_j/(\vec k_1+\vec k_2)^2-2(\vec k_1\times\vec k_2)_i(\vec k_1\times\vec k_2)_j/(\vec k_1\times\vec k_2)^2\,,
\ee
which satisfies the transversality and trace conditions $0=(\vec k_1+\vec k_2)_if_{ij}(\vec k_1,\vec k_2), 0=f_{ii}$. Furthermore, for all vertices of the form $\gamma_{st_{ij}s'}(\vec k_1,\vec k_2)=\gamma_{ss't_{ij}}(\vec k_2,\vec k_1)$ the dependence on $i,j$ is captured by $g_{ij}(\vec k_1,\vec k_2)\equiv (\vec k_2)_i (\vec k_2)_j$. We find that for the one-loop power spectrum we can again represent the tensor mode by a single degree of freedom, $t_{ij}\mapsto t_\text{eff}$, with vertices related to the full ones via $\gamma_{t_{ij}ss'}(\vec k_1,\vec k_2)=f_{ij}(\vec k_1,\vec k_2)\gamma_{t_\text{eff}ss'}(\vec k_1,\vec k_2)$ and $k_2^2\sin^2(\vec k_1,\vec k_2)\gamma_{st_{ij}s'}(\vec k_1,\vec k_2)=g_{ij}(\vec k_1,\vec k_2)\gamma_{st_\text{eff}s'}(\vec k_1,\vec k_2)$,
where $\sin^2(\vec k_1,\vec k_2)=(\vec k_1\times \vec k_2)^2/(k_1^2k_2^2)$. With these definitions one has
\bea\label{eq:tensor_eff}
  \gamma_{t_{ij}ss'}(\vec q,\vec k-\vec q)\gamma_{t_{ij}'s''s'''}(\vec q,\vec k-\vec q) &=& 2\gamma_{t_\text{eff}ss'}(\vec q,\vec k-\vec q)\gamma_{t_\text{eff}'s''s'''}(\vec q,\vec k-\vec q)\,,\nn\\
   \gamma_{ss't_{ij}}(\vec q,\vec k-\vec q)\gamma_{t_{ij}'s''s'''}(-\vec q,\vec k) &=& \gamma_{ss't_\text{eff}}(\vec q,\vec k-\vec q)\gamma_{t_\text{eff}'s''s'''}(-\vec q,\vec k)\,.
\eea
The second line implies that the effective tensor mode correctly takes into account the backreaction of tensor perturbations on the scalar kernels $F_{3,s}(\vec k,\vec q,-\vec q,\eta)$.
Using the first line we find that the tensor power spectrum at one-loop order can be computed according to $P_{t_{ij}t_{ij}}(k,\eta)=2P_{t_\text{eff}t_\text{eff}}(k,\eta)$.

\subsection{General case}\label{app:vorticity_full}

For situations where the simplified treatment described above is not applicable, we also implemented a full treatment of vorticity and vector mode perturbations taking both
of the two independent degrees of freedom for each of them into account. This is relevant in particular for the two-loop vorticity power spectrum (see Sec.\,\ref{sec:Pww}) and the one-loop matter density bispectrum (see Sec.\,\ref{sec:Pddd}).
In practice, to make use of the constraint $\vec p\cdot \vec v=0$ for a vector mode $v_i\in\{w_i,\nu_i\}$ of wavenumber $\vec p$, we project $v_i$ on a basis that depends on $\vec p$. In particular, we use the $\vec p$-dependent orthogonal basis vectors
\be
  \vec b_{\vec p1} \equiv {\cal N}_{\vec p1}(\vec p\times\vec Q_{\vec p}),\ 
  \vec b_{\vec p2} \equiv {\cal N}_{\vec p2}(\vec p\times(\vec p\times\vec Q_{\vec p})) = {\cal N}_{\vec p2}\left(\vec p \,(\vec p\cdot\vec Q_{\vec p})-\vec Q_{\vec p}\,p^2\right),\ 
  \vec b_{\vec p3} \equiv {\cal N}_{\vec p3}\,\vec p\,,
\ee
where $\vec Q_{\vec p}$ is an a priori arbitrary reference vector that can be chosen by convenience for each $\vec p$ with the only condition that $\vec p\times\vec Q_{\vec p}\not=0$.
Furthermore, ${\cal N}_{\vec p1,2,3}$ are normalization factors. A natural choice is such that the basis vectors have norm equal to unity, but this is not required in our implementation.
We refer to this basis as the \emph{transverse basis}, and decompose the vorticity for wavevector $\vec p$ as
\be
  \vec w = w_{\vec p1} \vec b_{\vec p1} + w_{\vec p2}\vec b_{\vec p2}\,,
\ee
where $w_{\vec p1,2}$ are the two tranverse vorticity components, and the projection along $\vec b_{\vec p3}$ vanishes by construction. The decomposition of the vector mode is analogous.
In the list of perturbation variables, we correspondingly include $w_{\vec p\alpha}$ and $\nu_{\vec p\alpha}$ for $\alpha=1,2$. The linear
part of their evolution is not affected by the basis choice. For the non-linear part, the vertices in the transverse basis can be
obtained from the generic kartesian ones via
\be
  \gamma_{aw_{\vec p\alpha} c}(\vec p,\vec q) = \gamma_{aw_ic}(\vec p,\vec q)\vec b_{\vec p\alpha,i},\
  \gamma_{abw_{\vec q\alpha}}(\vec p,\vec q) = \gamma_{abw_i}(\vec p,\vec q)\vec b_{\vec q\alpha,i},\
  \gamma_{w_{\vec k\alpha}bc}(\vec p,\vec q) = \vec b_{\vec p+\vec q\alpha,i}/|\vec b_{\vec p+\vec q\alpha}|^2 \gamma_{w_ibc}(\vec p,\vec q)\,,
\ee
where summation over $i=x,y,z$ is implied. Note that the appropriate transverse basis for each of the wavevectors $\vec p$, $\vec q$ or $\vec p+\vec q$ is used in the three cases.
For vertices involving the vector mode $\nu_i$ the projection is analogous,
and for vertices where more than one of the indices $a,b,c$ corresponds to a vorticity or vector mode, the projection is applied for each of them correspondingly. For example,
\be
  \gamma_{w_{\vec p+\vec q\alpha}w_{\vec p\beta}w_{\vec q\gamma}}(p,q) = b_{\vec p+\vec q\alpha,i}/|\vec b_{\vec p+\vec q\alpha}|^2 \gamma_{w_iw_jw_k}(\vec p,\vec q)\vec b_{\vec p\beta,j}\vec b_{\vec q\gamma,k}\,,
\ee
with summation over $i,j,k=x,y,z$ and for $\alpha,\beta,\gamma=1,2$. Therefore the $www$ vertex corresponds in the transverse basis to $2^3=8$ distinct vertices, while in the kartesian basis
it would have in general $3^3=27$ components. This illustrates the advantage of using the transverse basis.

We note that in the transverse basis, and using well-known relations from vector algebra, the projected vertices can be reduced to expressions that depend solely on scalar products $\vec p\cdot\vec q$, $\vec p\cdot \vec Q_{\vec r}$
and $\vec q\cdot \vec Q_{\vec r}$, as well as on
\be
  D_{\vec r\vec s} \equiv \text{det}(\vec s,\vec r,\vec Q_r)=(\vec r\times\vec Q_r) \cdot \vec s\,,
\ee
with $\vec r,\vec s \in\{\vec p,\vec q,\vec p+\vec q\}$. This is advantageous when following the algorithm outlined in~\cite{BlaGarKon1309}, and we precompute these values for a given configuration of wavevectors corresponding to
a single evaluation of the integrand for the loop evaluation. In particular, the relevant elementary wavevectors that may appear are the external wavevector $\vec k$ for the power spectrum (or $\vec k_1,\vec k_2$ for the bispectrum)
as well as loop wavenumbers $\vec Q_1,\dots,\vec Q_L$ at $L$-loop order. The most general set of vectors that may appear as arguments of non-linear kernels as well as of vertices has then the form $\sum_{n=1}^L c_n Q_n+c_{L+1}\vec k$ (or $\sum_{n=1}^L c_n Q_n+c_{L+1}\vec k_1+c_{L+2}\vec k_2$ for the bispectrum), with some coefficients $c_n=0,\pm 1$. For each of those vectors $\vec p$, we define the associated vector $\vec Q_{\vec p}$ entering the definition of the corresponding transverse basis
as the vector within the set ${\cal B}=\{\vec k,\vec Q_1,\dots,\vec Q_L\}$ (or ${\cal B}=\{\vec k_1,\vec k_2,\vec Q_1,\dots,\vec Q_L\}$ for the bispectrum) that has the largest projection within the plane perpendicular to $\vec p$, i.e. for which $s_{\vec p\vec Q}^2 = 1 - \frac{(\vec p\cdot\vec Q)^2}{\vec p^2\vec Q^2}$ is maximal out of all $\vec Q\in {\cal B}$.
This allows for an efficient implementation of vertices involving vector modes within the algorithm used in this work, based on~\cite{BlaGarKon1309,Blas:2015tla,Floerchinger:2019eoj,Garny:2020ilv}.

\section{Rescaling for power law universe}
\label{app:rescaling}

For a scaling universe with linear input power spectrum $P_0(k)=A k^{n_s}$ and EdS background, the average value of the velocity dispersion
is given by $\epsilon(\eta)=\epsilon_0 e^{\alpha\eta}$ with $\alpha=4/(n_s+3)$, and the dimensionless ratio $\bar\omega=\omega(\eta)/\epsilon^2(\eta)$
involving the fourth cumulant expectation value is constant (as are dimensionless ratios of higher order cumulants~\cite{cumPT}). The only dimensionful
scales are therefore $k_\sigma\equiv 1/\sqrt{\epsilon_0}$ and $k_\text{nl}\equiv 1/(4\pi A)^{1/(n_s+3)}$. We indicate the dependence of
power spectra on $\epsilon_0$ in this appendix using the notation $P_{ab}(k,\eta;\epsilon_0)$, for any perturbation modes $a$ and $b$ (e.g. $a=b=\delta$ for the density spectrum), and taking $A$ as given.

Scaling symmetry can be used to rescale power pectra computed for a given reference value $\epsilon_0^\text{ref}$ to any other value $\epsilon_0$, while keeping all dimensionless ratios of higher cumulants fixed, specifically $\bar\omega$ (the dependence on which we suppress for brevity).
To see this, we consider for a moment power spectra $P_{\bar a\bar b}(k,\eta;\epsilon_0)=P_{ab}(k,\eta;\epsilon_0)/\epsilon^{d_a+d_b}(\eta)$ of dimensionless perturbation variables $\bar a=a/\epsilon^{d_a}(\eta)$ and $\bar b=b/\epsilon^{d_b}(\eta)$, with appropriate powers $d_a, d_b$. For example, $d_\delta=d_\theta=d_A=d_{w_i}=0$, $d_g=d_{\delta\epsilon}=d_{\nu_i}=d_{t_{ij}}=d_\pi=d_\chi=1$.
Then the dimensionless power spectrum $4\pi k^3 P_{\bar a\bar b}$ can depend only on dimensionless ratios of $k$, $k_\sigma$ and $k_\text{nl}$, which can be taken to be $k/k_\sigma$
and $k_\sigma/k_\text{nl}$. In addition, within perturbation theory, the $L$-loop contribution involves exactly $L+1$ powers of $P_0$, i.e. is proportional to $A^{L+1}$.
This implies that $4\pi k^3 P^{L-\text{loop}}_{\bar a\bar b}(k,\eta;\epsilon_0) = (k_\sigma/k_\text{nl})^{(n_s+3)(L+1)}\bar\Delta_{\bar a\bar b}^L(k/k_\sigma,\eta)$, with some dimensionless function $\bar\Delta_{\bar a\bar b}^L(k/k_\sigma,\eta)$ that depends on $k$ only via the ratio $k/k_\sigma$. From this we obtain the rescaling relation for the original power spectrum,
\be
  P_{ab}^{L-\text{loop}}(k,\eta;\epsilon_0) = \left(\frac{\epsilon_0^\text{ref}}{\epsilon_0}\right)^{\frac{(n_s+3)(L+1)}{2}-\frac32-d_a-d_b}  P_{ab}^{L-\text{loop}}(k\times (\epsilon_0/\epsilon_0^\text{ref})^{1/2},\eta;\epsilon_0^\text{ref})\,.
\ee
Similarly, for the bispectrum we find
\be
  B_{abc}^{L-\text{loop}}(k_1,k_2,k_3,\eta;\epsilon_0) = \left(\frac{\epsilon_0^\text{ref}}{\epsilon_0}\right)^{\frac{(n_s+3)(L+2)}{2}-\frac62-d_a-d_b-d_c}  B_{abc}^{L-\text{loop}}(k_1',k_2',k_3',\eta;\epsilon_0^\text{ref})\,,
\ee
where $k_i'=k_i\times (\epsilon_0/\epsilon_0^\text{ref})^{1/2}$ for $i=1,2,3$.

In addition, the dependence on time (or equivalently $\eta$) is also fixed by scaling symmetry. To see this it is useful to introduce generalized time-dependent scales $k_\sigma(\eta)\equiv 1/\sqrt{\epsilon(\eta)}$, $k_\text{nl}(\eta)\equiv 1/(4\pi A e^{2\eta})^{1/(n_s+3)}$. The dimensionless power spectrum $4\pi k^3 P_{\bar a\bar b}(k,\eta;\epsilon_0)$ can depend on time only via
the ratios $k/k_\sigma(\eta)$ and $k_\sigma(\eta)/k_\text{nl}(\eta)$. Observing that the latter ratio is constant in time implies that $\bar\Delta_{\bar a\bar b}^L(k/k_\sigma,\eta)=\bar\Delta_{\bar a\bar b}^L(k/k_\sigma(\eta))$ is a function of a single variable. This yields $P_{ab}(k,\eta;\epsilon_0)=(\epsilon(\eta')/\epsilon(\eta))^{-3/2-d_a-d_b} P_{ab}(k',\eta';\epsilon_0)$ and $B_{abc}(k_1,k_2,k_3,\eta;\epsilon_0)=(\epsilon(\eta')/\epsilon(\eta))^{-6/2-d_a-d_b-d_c} B_{abc}(k_1',k_2',k_3',\eta';\epsilon_0)$
where now $k'=k \times (\epsilon(\eta)/\epsilon(\eta'))^{1/2} =k e^{2(\eta-\eta')/(n_s+3)}$ and analogously for $k_i'$.

\section{N-body simulations}
\label{app:Nbody}

Our main suite of N-body simulations (see Tab.~\ref{tab:simulations}) consists of 4 realizations of $512^3$ particles for each power-law spectral index $n_s=-1,0,1,2$, run with $\Omega_m=1$ using the Gadget code~\cite{Spr0512}. The four realizations in each case correspond to two sets of fixed amplitude initial conditions with opposite phases~\cite{AngPon1610}, to cancel Gaussian cosmic variance. The initial conditions are set at scale factor $a_{\rm ic}=0.001$ and evolved until $a=1$ with outputs at $a=10^s$ with $s=-2$ to $s=0$ in steps of $\Delta s=0.2$. The setup of initial conditions necessarily induces transients in subsequent evolution, due to imperfect dynamics~\cite{Sco98,Val0204,CroPueSco0611,Jen1004,ReeSmiPot1305,MicHahRam2101} (the use of perturbation theory to compute initial displacements) and special configurations of discrete particles~\cite{MarBaeJoy0605,JoyMar0703,JoyMar0711,GarEisFer1610} (cartesian grid out of which the initial displacements are imposed) that are not exact solutions of the  equations of motion. 

For our application of interest, namely testing 3D gravitational clustering in the regime where orbit-crossing is most significant, we would like to focus on spectra with blue spectral indices, hence our choice of pushing $n_s$ to be as high as $n_s=+2$, which as far as we know has never before been considered in the literature of scale-free 3D cosmological simulations~\cite{ColBouHer96,PeaDod96,SmiPeaJen03,BagKhaKul0908,WidElaTha0908,JoyGarEis2103,LerGarEis2103}. This gives rise to some challenges from the numerical simulation point of view. In particular, since we use perturbation theory to set up initial conditions, we would like to have the maximum amplitude of fluctuations to be initially less than unity, which for such blue spectra makes the fluctuation amplitude at the scale of the box very small. This in turn makes following the evolution of fluctuations more difficult by the tree algorithm due to accumulation of numerical errors (e.g.~\cite{ReeSmiPot1305,AngHah2212}. To strike a balance, the amplitude of initial fluctuations at the Nyquist frequency of the particle grid is an increasing function of $n_s$, with $\Delta_{\rm Nyq}=(k_{\rm Nyq}/k_{\rm nl})^{n_s+3}=0.01,0.01,0.05,0.25$ for $n_s=-1,0,1,2$. This gives an initial amplitude at the scale of the box of $\Delta_{\rm box}=1.5\times 10^{-7}, 6\times 10^{-10},1.2\times 10^{-11}, 2.3\times10^{-13}$, respectively. Even then, we see some numerical noise in the evolution of the large-scale modes down to late times, particularly for the bispectrum when $n_s=2$ (see Fig.~\ref{fig:Pddd}).

This limitation imposes constraints on the order of perturbation theory that can be safely used in setting up initial conditions. Indeed, at the $\Delta_{\rm Nyq}$ values mentioned above, our {\em linear} solutions to the equations of motion with dispersion and higher cumulants tell us that already the corrections over the unperturbed initial spectrum $P_0 \propto k^{n_s}$ are about $0,3,15,30\%$ at the Nyquist frequency for $n_s=-1,0,1,2$ respectively {\em at the initial conditions} (see Fig.~\ref{fig:Pdd_cumPT_Nbody}). This means that for blue spectral indices $n_s=1,2$,  Lagrangian perturbation theory convergence is doubtful, as one would have to take into account dispersion and higher cumulants. For this reason in such cases we set initial conditions using the Zel'dovich approximation (ZA), rather than second-order Lagrangian perturbation theory (2LPT).

The use of ZA initial conditions introduces violations of self-similarity at early times due to transients, which we correct using tree-level standard perturbation theory (SPT), tested against $n_s=-1,0$ 2LPT initial conditions simulations for which we have also run their ZA counterparts. For the bispectrum for equilateral triangles, tree-level SPT gives~\cite{Sco98}
\beq
B_{\rm eq}(k,a) = \Big\{\frac{12}{7} - \frac{27}{20} \Big({a_{\rm ic}\over a}\Big) + \frac{27}{70} \Big({a_{\rm ic}\over a}\Big)^{7/2}\Big\} \Big[ P_{\rm lin}^{\rm SPT}(k,a) \Big]^2\ , 
\label{BeqTransients}
\eeq
which at the initial conditions ($a=a_{\rm ic}$) gives the ZA tree-level bispectrum, and for $a \gg a_{\rm ic}$ gives the tree-level equilateral bispectrum of the exact dynamics in the absence of dispersion (which is correctly predicted by 2LPT). In between,  transients (spurious $a/a_{\rm ic}$ dependence in Eq.~\ref{BeqTransients}) induce violations to self-similar evolution. At early times, when our simulations are weakly non-linear, we find that Eq.~(\ref{BeqTransients}) matches the differences between the ZA and 2LPT initial conditions simulations very well, as expected, and in agreement with previous tests of higher-order statistics in the literature~\cite{Sco98,CroPueSco0611}. In fact, we have checked this for all triangle shapes, not just equilateral. The use of Eq.~(\ref{BeqTransients}) at late times to compute the relative corrections due to transients is also justified in practice because, although tree-level SPT is not valid, the size of the transient corrections is rather small, so a larger error in estimating it is acceptable compared to the statistical errors. More specifically, we find that the {\em maximum} deviation between ZA initial conditions simulations corrected for transients by SPT compared to the 2LPT initial conditions simulations for all triangle shapes and simulation outputs is about $1.5\%$ for $n_s=-1$ and $0.5\%$ for $n_s=0$, suggesting our application to $n_s=1,2$ should yield sub-percent systematic errors. 

\begin{table}
  \centering
  \caption{N-Body simulations as a function of spectral index.}
  \begin{ruledtabular}
    \begin{tabular}{rccc} 
   $n_s$ & $N_{\rm particles}$ & $N_{\rm realizations}$  & Initial Conditions   \\[1.5ex] \hline & &  & \\[-1.ex]
  $2$ & $256^3, 512^3$ & 6 (3 paired fixed-amplitude) & ZA  \\[1.5ex]
   \hline & &  & \\[-1.ex]
  $1$ & $256^3, 512^3$ & 4 (2 paired fixed-amplitude) &ZA \\[1.5ex]
   \hline & &  & \\[-1.ex]
  $0$ & $256^3, 512^3$ & 4 (2 paired fixed-amplitude) & ZA \\[1.5ex]
  $0$ & $384^3, 768^3$ & 1 fixed amplitude  & ZA \\[1.5ex]
  $0$ & $512^3$ & 4 (2 paired fixed-amplitude)& 2LPT \\[1.5ex]
   \hline & &  & \\[-1.ex]
  $-1$ & $512^3$ & 4 (2 paired fixed-amplitude) & ZA \\[1.5ex]
  $-1$ & $512^3$ & 4 (2 paired fixed-amplitude) & 2LPT \\
    \end{tabular}
  \end{ruledtabular}
  \label{tab:simulations}
\end{table}

The situation for power spectrum measurements is  less challenging, since ZA initial conditions reproduce the correct linear spectrum in the absence of dispersion, with transients entering only trough loop corrections in the context of SPT~\cite{CroPueSco0611}. By comparing our ZA initial conditions simulations to those run with 2LPT for the cases $n_s=-1,0$, we see that correcting power spectrum measurements for transients is in practice not necessary, as deviations are at most $0.5\%$ for $n_s=-1$ and $0.3\%$ for $n_s=0$ once the power spectra from different outputs are binned into functions of $k/k_{\rm nl}$.

Apart from transients, there are other spurious effects that break self-similarity~\cite{ColBouHer96}. N-body simulations cannot follow modes with wavelength larger than the size of the box, and at small scales the gravitational force is softened. The former means in particular that at late times, when the non-linear scale becomes comparable to the box size (e.g. $k_{\rm nl} L \la 0.1 $), the simulation cannot follow the true dynamics as mode-coupling to relevant modes is missing. In addition, particles in simulations can be thought of as having an effective size equal to the softening  length, and to be in the fluid limit (avoid spurious discreteness) requires that the non-linear scale be larger than the mean interparticle separation, or equivalently that the amplitude of fluctuations at the Nyquist frequency of the particles be larger than unity. This is of course the opposite of the requirement that perturbation theory be valid (as required for initial conditions). In practice this means that between the initial conditions and the first output that can be analyzed  one must leave sufficient time for both conditions to be met. This also minimizes transients induced by simplified dynamics in initial conditions generation and associated initial grid effects. Therefore, to obtain measurements as a function of $k/k_{\rm nl}$ with a broad range of values, we use measurements starting at a scale factor 25 times larger than at the initial conditions. 

Finally, another issue that can potentially impact our simulations is the use of  fixed-amplitude initial conditions to reduce cosmic variance~\cite{AngPon1610}. Because the amplitude of the Fourier modes are fixed, these initial conditions are actually non-Gaussian. That's precisely the point when it comes to the initial four-point function, as its modification leads to Gaussian cosmic variance cancellation. However this also affects the bispectrum. At tree-level in SPT, the bispectrum gets modified by order unity only for triangles of the form $B(\vec k, \vec k, -2\vec k)$, which involves the initial four-point function of pairs of opposite modes (same that enters into the power spectrum cosmic variance). Since we present measurements of the bispectrum for equilateral triangles in this paper, this is of no concern, except for the fact that we use a wide bin in $k$ with $\Delta k=6 k_F$, with $k_F$ the fundamental model of the simulation box, and the first bin at $k=6 \pm \Delta k/2$ does have contributions from such fundamental triangles. Therefore, to avoid this issue we do not use the first two equilateral triangle bins  in any of our bispectrum results. 

At one-loop order, however, it's easy to see that these fixed-amplitude initial conditions induce bispectrum modifications to triangles of all shapes (as well as to the power spectrum). This is so because given any choice of external momenta (triangle shape) one can always match the configuration of initial four-point functions affected in the expectation values over linear fields for a suitable choice of the  loop momentum that is being integrated over. However, since only a special configuration is involved, the loop integral collapses to a particular case of discrete modes. By dimensional analysis the loop then gives a contribution proportional to $k_F^3 P_{\rm lin}(k) \propto (k/k_{\rm nl})^{n_s} (k_F/k_{\rm nl})^3$, which means that fixed-amplitude initial conditions induce violations of self-similarity through loop corrections. Following~\cite{OddSef22}, we computed these for the power spectrum and bispectrum, but found that these violations are too small (sub-percent even for the latest outputs used) to affect our measurements. 

To calculate error bars for our power spectrum and bispectrum measurements, we first build an independent realization from the two fixed-amplitude initial conditions simulations with opposite phases, and do the same for the second random seed for each spectral index. These two independent realizations contain several outputs (different scale factors), whose measurements (power spectrum and bispectrum) are then binned into functions of $k/k_{\rm nl}$ by self-similarity. The error bars are then computed from the scatter of these measurements at fixed $k/k_{\rm nl}$. Thus our error bars contain scatter from two independent random seeds, as well as scatter from several different outputs that land at the same $k/k_{\rm nl}$. To illustrate the extent to which self-similarity is satisfied for the density power spectrum, we show in Fig.~\ref{fig:PallnsSS} the density power spectrum in units of the initial spectrum for the four spectral indices ($n_s=-1$ to $n_s=2$) in terms of the self-similar variable $k/k_{\rm nl}$. Different outputs (each corresponding to a different color) lie on top of each other, as expected from self-similar scaling. In this plot, each output corresponds to the average of one fixed-amplitude realization over the two opposite phases, and we have removed the high-$k$ portion of the spectrum when violation of self-similarity occurs (e.g. due to the softening length of the simulations) for clarity.

\section{The halo mass function}
\label{app:HMF}

\subsection{Measuring the Mass Function}

To measure the mass function in our simulations, we run the Rockstar halo finder~\cite{BehWecWu1301}. We use the virial mass definition, and consider only halos with more than 50 particles. The halos are then binned in 200 logarithmically spaced bins, and the  fixed-amplitude realizations  of opposite phases are averaged together. This gives us 2 paired fixed-amplitude realizations for each output which are then combined using self-similarity. 

The scatter among these is used to compute the error bars, and a fit of the form given by Eq.~(\ref{MFst}) is used to find the best fit parameters for $A$, $a$ and $p$ which are given in Tab.~\ref{tab:HMF}. Figure~\ref{fig:nuFnu} shows the resulting fits as a function of the ratio of mass to the non-linear mass.  We have verified comparing the halo mass function with 2LPT initial conditions to those with ZA initial conditions for $n_s=-1,0$ (see Tab.~\ref{tab:simulations}) that transients from ZA initial conditions do not affect our mass function fits. Figure~\ref{fig:MFallnsSS} shows how well our simulations obey self-similarity for the mass function. For each spectral index, different colors denote different outputs. For this figure, we have binned halos into 50 logarithmically spaced bins and suppressed error bars for clarity. Each line shows the average over two paired fixed-amplitude realizations.

\subsection{From Isothermal to NFW halos}

The contribution to the expectation value of the stress tensor $\epsilon$ of an NFW halo can be written in terms of its isothermal halo counterpart as Eq.~(\ref{epshNFW}), where $I(\beta,c)$ is a 2D integral given by Eq.~(\ref{Ibetac}) that represents the correction over the isothermal value. Here we present the analytic result for this integral in the two cases of interest, $\beta=0$ (isotropic) and $\beta=1/2$ (radially biased dispersion). We have, as a function of halo concentration $c$, 

\beqa
I(\beta=0,c)&=& {g(c)\over 20c^2}
  \Big[c \{6 - c [6 + c (25 + 2 c (59 + 35 c))] + 
      c^3 [15 + 2 c (12 + 5 c)] \pi^2 \} - 
   c^4 [15 + 2 c (12 + 5 c)] \ln(1 + c)  \nonumber \\ & & + (1 + c) \{-6 + c (1 + c) [12 + c (3 + 2 c (-28 + 5 c))]\} \ln(1 + c) 
   + 3 \{-1 + c^4 [15 + 2 c (12 + 5 c)]\} [\ln(1 + c)]^2  \nonumber \\ & &+ 
   6 c^4 [15 + 2 c (12 + 5 c)] {\rm Li}_2( -c)\Big]
\label{Ibeta0}
\eeqa
and
\beqa
I(\beta=1/2,c)&=& {2 g(c)\over 45c^2}  \Big[c^2 \{-3 + 9 c [7 + 4 c] - c [10 + 3 c (5 + 2 c)] \pi^2\} +  6 c (1 + c) [1 + 6 c (1 + c)] \ln(1 + c) \nonumber \\ & & 
-3 (1 + c)^3 (1 - 3 c + 6 c^2) [\ln(1 + c)]^2 -   6 c^3 [10 + 3 c (5 + 2 c)] {\rm Li}_2( -c)\Big]
\label{Ibetahalf}
\eeqa
respectively. In these expressions ${\rm Li}_2$ represents the dilogarithm function.

\section{Measuring Velocity Divergence and Vorticity}
\label{app:DivVort}

Measuring the velocity field divergence and vorticity is somewhat challenging in N-body simulations, as the velocity field is sampled by particles (which are themselves a sampling of the phase-space distribution function $f$). This means that in low-density regions the sampling is sparser than in high-density regions. In particular, in voids there is little velocity information at scales smaller than the void size. One would like to construct a grid where the velocity field is defined, then Fourier transformed, and divergence/vorticity and their spectra  computed. But this can only be done at all grid points if there is always a contribution from nearby particles. In practical terms, the issue manifests itself when defining the peculiar velocity field, 
\be
  \bm{v}(\bm{x}) = \frac{1}{1+\delta}\int d^3p\, \frac{\bm{p}}{a} \, f(\bm{x},\bm{p}) \,.
  \label{vdef}
\ee
because at points $\bm{x}$ where $f$ vanishes (and $\delta= -1$), i.e. voids, this leads to a 0/0 indeterminacy. A number of strategies have been developed in the literature to overcome this limitation, among them Voronoi  and Delaunay tessellations~\cite{Bervan96,Romvan0711,PueSco0908}, phase-space constructions~\cite{HahAngAbe1512,BueHah1907} and rescaling earlier time velocities~\cite{JelLepAda1809}. In this work, we have developed an alternative approach based on a multigrid method. 
 
To avoid the indeterminacy problem, we start by interpolating the particle velocities to a coarse grid (e.g. $64^3$), where all grid points receive velocity information. To do this, we use the Piecewise Cubic Spline (PCS) algorithm discussed in~\cite{SefCroSco1512} which employs fourth-order interpolation in interlaced grids to reduce aliasing. Fourth-order interpolation has a broad kernel and thus helps in giving velocity information to all grid points. Then, we recursively make the grid finer by factors of two in each dimension ($128^3, 256^3$, etc). As the grid becomes finer, eventually there will be grid points (particularly at late times) for which the indeterminacy happens somewhere inside a void, even with the PCS algorithm. In that case, we look up the information from the coarser grid, and perform an interpolation from the nearby points in the coarser grid one level above (which has no indeterminacy problem) to the finer grid. In this way we recursively make the grid finer and finer to the desired level, i.e. enough to resolve the vorticity power spectrum peak at small scales (see below), ending up at a $1024^3$ grid which is then Fourier transformed. The divergence and vorticity are then calculated in Fourier space, and their power spectrum computed as usual. Note that as we make the grid finer, if no grid points have indeterminacy the coarser level grid information is never used, and gets discarded. At each stage in the algorithm there are thus only two concurrent grids, and the finer grid becomes the coarser grid in the next iteration of the recursive algorithm. 

A second issue is that the velocity field, as seen from Eq.~(\ref{vdef}), is the ratio of two interpolated quantities (momentum over density). Normally, it is easy to correct for the interpolation scheme used by dividing by the corresponding Fourier transform of the interpolation window in Fourier space. This corrects frequencies close to the Nyquist frequency of the grid, which get damped the most. However, since the velocity is a ratio of interpolated quantities it is difficult to reliably correct for the interpolation kernel (i.e. correction of the interpolated density is sensitive to noise), although being a ratio means that the damping caused by the interpolation window is not as severe. In practice we minimize this issue by using frequencies up to one tenth of the Nyquist frequency, comparison between results for different grid sizes show that this leads to no noticeable window interpolation effects. 

The most challenging of our measurements correspond to the vorticity power spectrum, which shows significant deviations of self-similarity except at late times, as we detail below. For the velocity divergence power spectrum, Fig.~\ref{fig:PTTallnsSS} shows a test of self-similarity by plotting each output for a given spectral index in a different color. We see comparing to Fig.~\ref{fig:PallnsSS} that deviations from self-similarity are  more significant than for the density power spectrum case, as expected given the challenges discussed above. However, in comparison to the vorticity power spectrum (see Fig.~\ref{fig:PwwConverge}), self-similarity of the velocity divergence power spectrum is fairly well satisfied. 

As has been known from CDM simulations, the vorticity power spectrum is sensitive to mass resolution, and poor mass resolution results in an artificial amplification of the vorticity power~\cite{PueSco0908,HahAngAbe1512,JelLepAda1809}.  For CDM initial conditions, the vorticity power spectrum typically has a peak at $k_{\rm peak} \sim 1 \iMpc$ at $z=0$, and at higher redshifts this peak shifts to higher $k_{\rm peak}$ as expected. A reasonable criterion for convergence is to characterize the mass resolution (in terms of the mean interparticle separation $\lambda \equiv L_{\rm box}/N_{\rm par}$) needed to ``resolve" this peak, i.e. the characteristic scale where most vorticity is produced, extracted from the convergence studies mentioned above. We obtain
\be
\lambda\, k_{\rm peak} \la 0.25, 0.27, 0.19,
\label{wConverge}
\ee
respectively from the $z=0$ vorticity power spectra convergence studies from Fig.~3 in~\cite{PueSco0908}, Fig.~13 in~\cite{HahAngAbe1512}  and Fig.~6 in~\cite{JelLepAda1809} and the mass resolutions of their simulations. 

For scale-free simulations, we have a similar situation to CDM, in that the vorticity power spectrum has a peak at small scales. It peaks at $k_{\rm peak}/k_\text{nl} \simeq 5, 4, 3.5, 2.75$ for $n_s=-1,0,1,2$ respectively.  Table~\ref{tab:LambdaKpeak} shows the resulting values of $\lambda\, k_{\rm peak}$ for the last four outputs of our $512^3$ simulations. We see comparing these values to the simple convergence criterion in Eq.~(\ref{wConverge}) that we would expect convergence to be achieved only in the case of $n_s=-1$ simulations for the last two or perhaps three outputs. The next best situation is for $n_s=0$ last output. 

Figure~\ref{fig:PwwConverge} shows that these estimates are quite reasonable, perhaps on the conservative side. The left panel shows the vorticity power spectrum for $n_s=-1$ for different outputs. Each line shows a single output (averaged over two pair-fixed realizations), starting from early outputs (highest amplitude at low-$k$) down to the last output. We see that the last three to four outputs show a reasonably converged answer to a self-similar result. As expected from CDM simulations, and results in Tab.~\ref{tab:LambdaKpeak}, earlier outputs increasingly overestimate the vorticity power spectrum amplitude, although its shape is fairly robust. The right panel shows the same convergence study for $n=0$. As expected, the convergence is less solid, although the last two outputs are fairly close. The situation is less convincing for $n_s=1,2$ (not shown), as expected from the values quoted in Tab.~\ref{tab:LambdaKpeak}, thus we won't present any vorticity results for these spectra in this work. 

As a result of this analysis, we proceed as follows to build the vorticity power spectrum that are compared to our predictions in the main text. For $n_s=-1$, we compute  the mean and scatter over the last four outputs (8 pair-fixed measurements). For $n_s=0$, we do the same for the last two outputs, with the caveat that the result may be not be fully converged. Still, since the shape of the power spectra are robust to resolution, it provides useful information in comparison to our theoretical predictions. 

For the velocity divergence power spectra the situation is  less delicate. The large-scale part of the spectrum, which is what tests our perturbative predictions is rather insensitive to mass resolution; there is more sensitivity to mass resolution at small scales ($k\ga k_\text{nl}$) in agreement with previous results in the literature~\cite{PueSco0908,HahAngAbe1512,JelLepAda1809}. Therefore, in this case we combine all outputs into a self-similar spectrum, with error bars describing the scatter between two pair-fixed realizations, and deviations from self-similarity.

\end{widetext}

\begin{figure}[t]
  \begin{center}
  \includegraphics[width=\columnwidth]{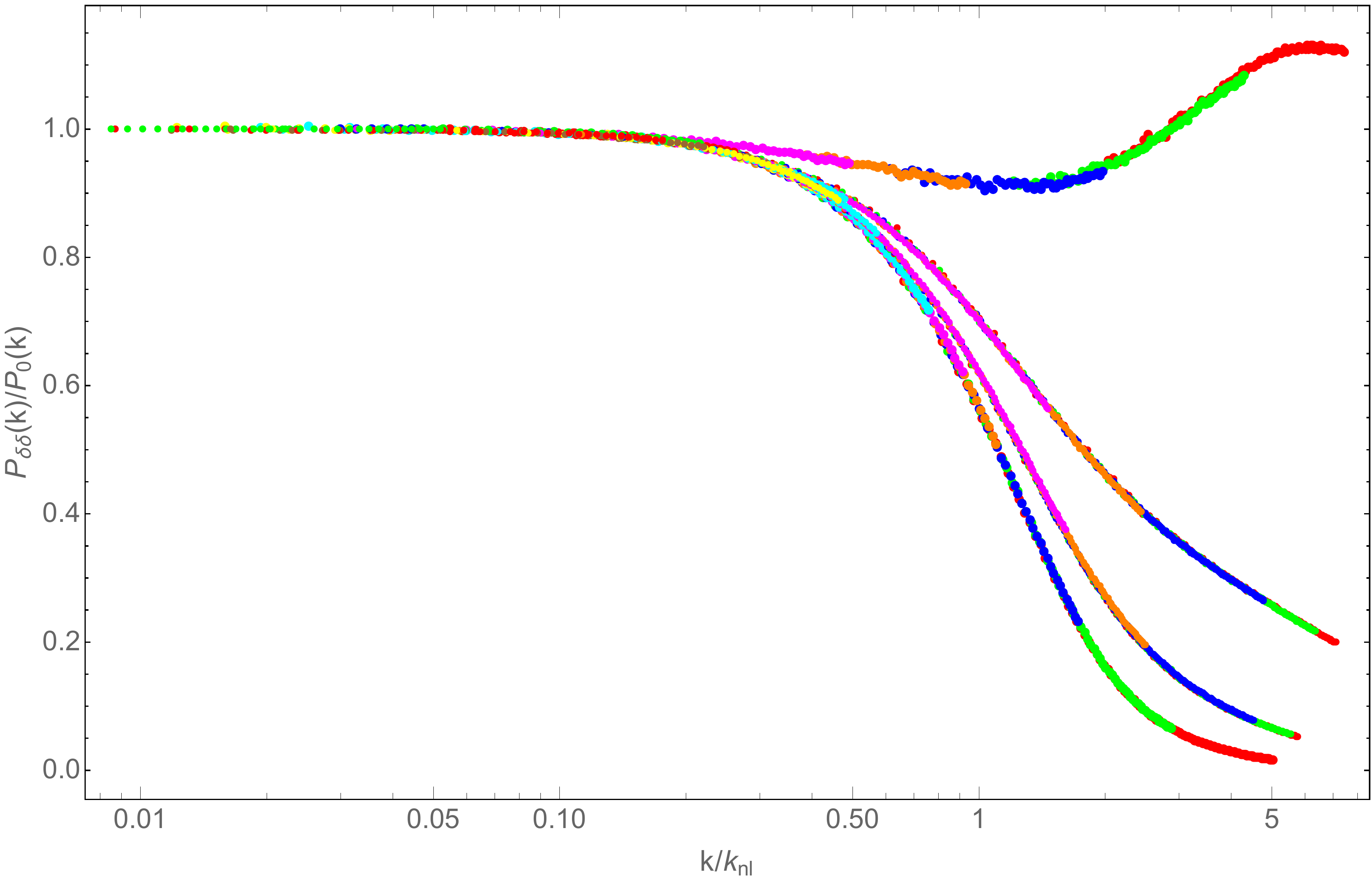}
  \end{center}
  \caption{\label{fig:PallnsSS} 
  Self-similarity of the density power spectrum for different spectral indices ($n_s=-1$ to $n_s=2$ from top to bottom at high-$k$). For each spectral index, different colors denote different outputs. 
  }
\end{figure}

\begin{figure}[t]
  \begin{center}
  \includegraphics[width=\columnwidth]{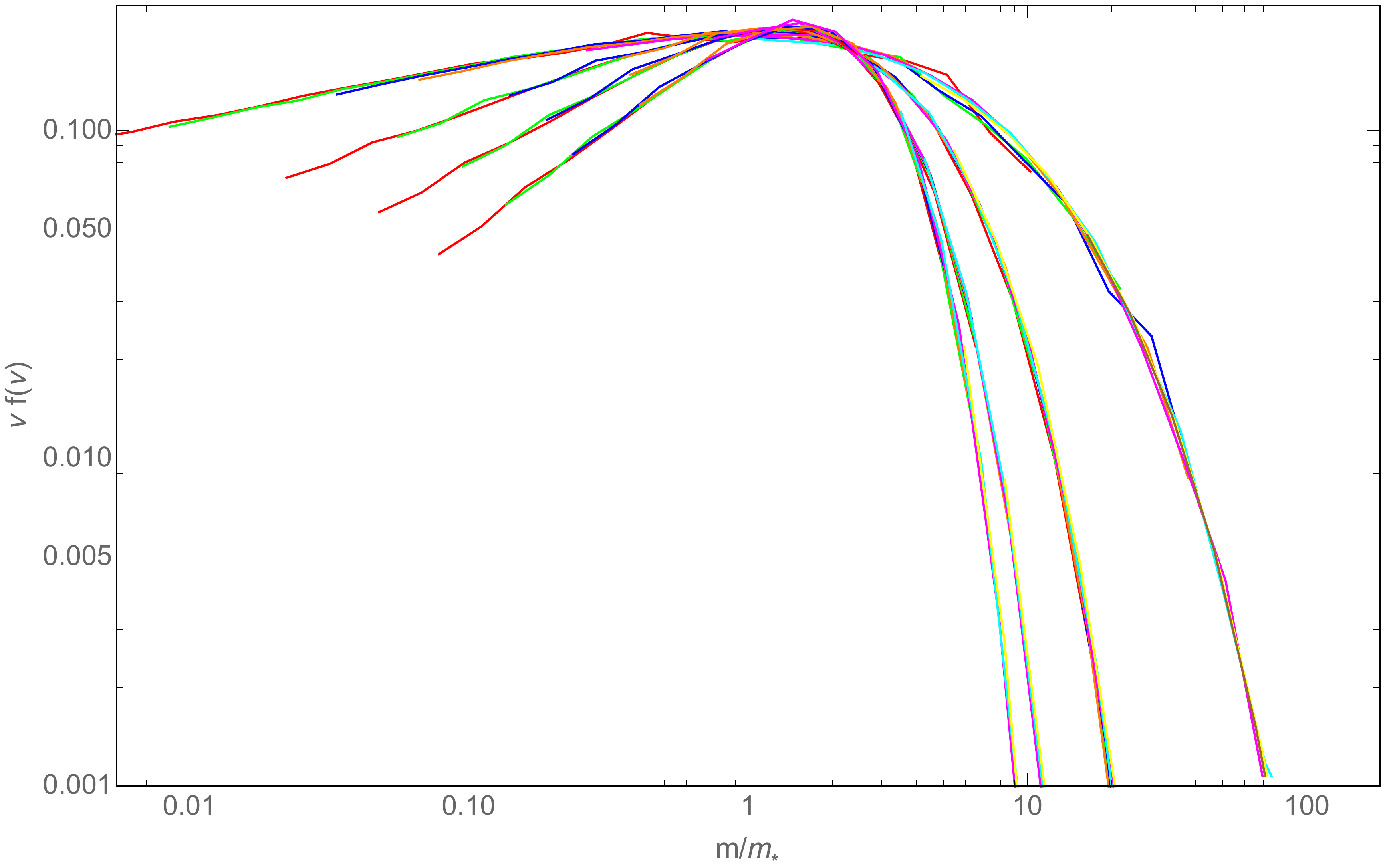}
  \end{center}
  \caption{\label{fig:MFallnsSS} 
  Self-similarity of the halo mass function for different spectral indices ($n_s=-1$ to $n_s=2$ from top to bottom at low $\hat{m}=m/m_*$). For each spectral index, different colors denote different outputs. 
  }
\end{figure}

\begin{table}
  \centering
  \caption{Mass function best fit parameters (see Eq.~\ref{MFst}) as a function of spectral index.}
  \begin{ruledtabular}
    \begin{tabular}{rccc} 
   $n_s$ & $A$  & $a$ & $p$   \\[1.5ex] \hline & & & \\[-1.ex]
  $2$ & $0.419$ & $0.344$ & $0.199$  \\[1.5ex]
  $1$ & $0.415$ & $0.538$ & $0.251$ \\[1.5ex]
  $0$ & $0.443$ & $0.732$ & $0.154$ \\[1.5ex]
  $-1$ & $0.387$ & $0.731$ & $0.317$ \\
    \end{tabular}
  \end{ruledtabular}
  \label{tab:HMF}
\end{table}

\begin{table}
  \centering
  \caption{Values for $\lambda\, k_{\rm peak}$ for the vorticity power spectrum for the last four outputs of our simulations as a function of spectral index.}
  \begin{ruledtabular}
    \begin{tabular}{rcccc} 
   $n_s$ & $a=0.25$  & $a=0.40$ & $a=0.63$ & $a=1$   \\[1.5ex] \hline & & & \\[-1.ex]
  $2$ & $1.25$ & $1.04$ & $0.86$ & $0.72$  \\[1.5ex]
  $1$ & $1.47$ & $1.16$ & $0.92$ & $0.73$\\[1.5ex]
  $0$ & $1.46$ & $1.08$ & $0.79$ & $0.58$\\[1.5ex]
  $-1$ & $0.62$ & $0.39$ & $0.24$ & $0.15$\\
    \end{tabular}
  \end{ruledtabular}
  \label{tab:LambdaKpeak}
\end{table}

\begin{figure}[t]
  \begin{center}
  \includegraphics[width=\columnwidth]{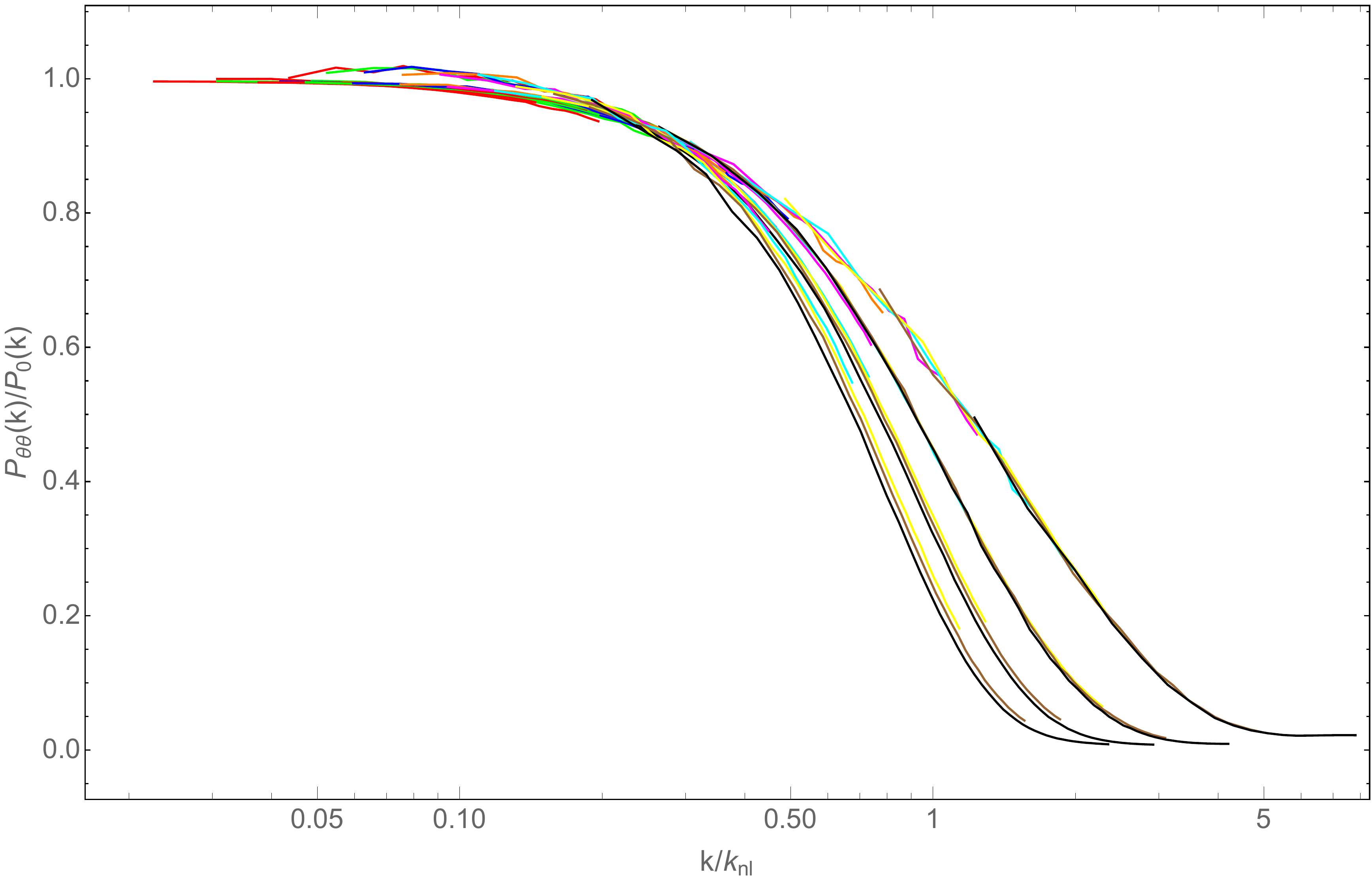}
  \end{center}
  \caption{\label{fig:PTTallnsSS} 
  Self-similarity of the velocity divergence power spectrum for different spectral indices ($n_s=-1$ to $n_s=2$ from top to bottom at high-$k$). For each spectral index, different colors denote different outputs. 
  }
\end{figure}

\begin{figure*}[t]
  \begin{center}
  \includegraphics[width=\columnwidth]{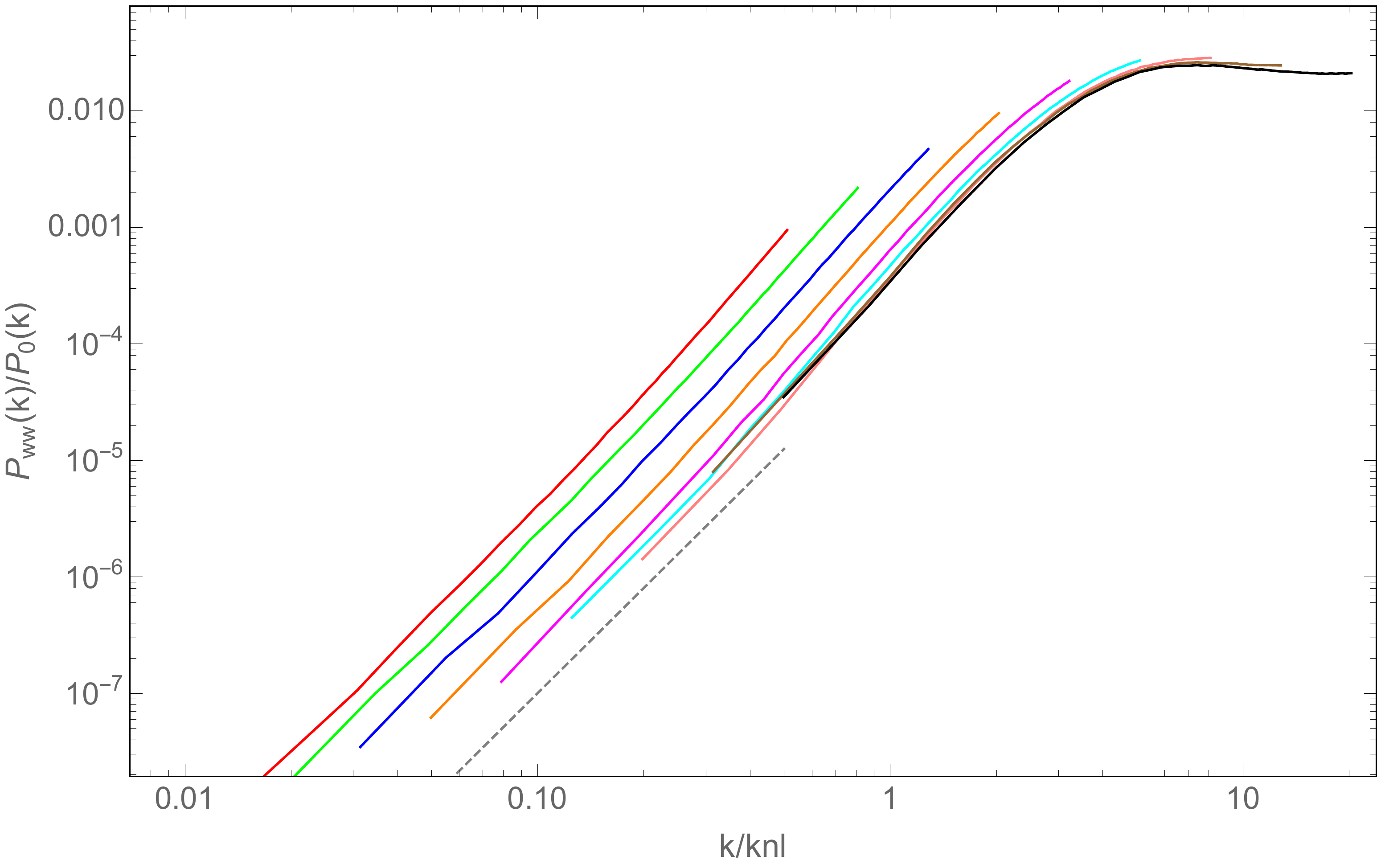}
  \includegraphics[width=\columnwidth]{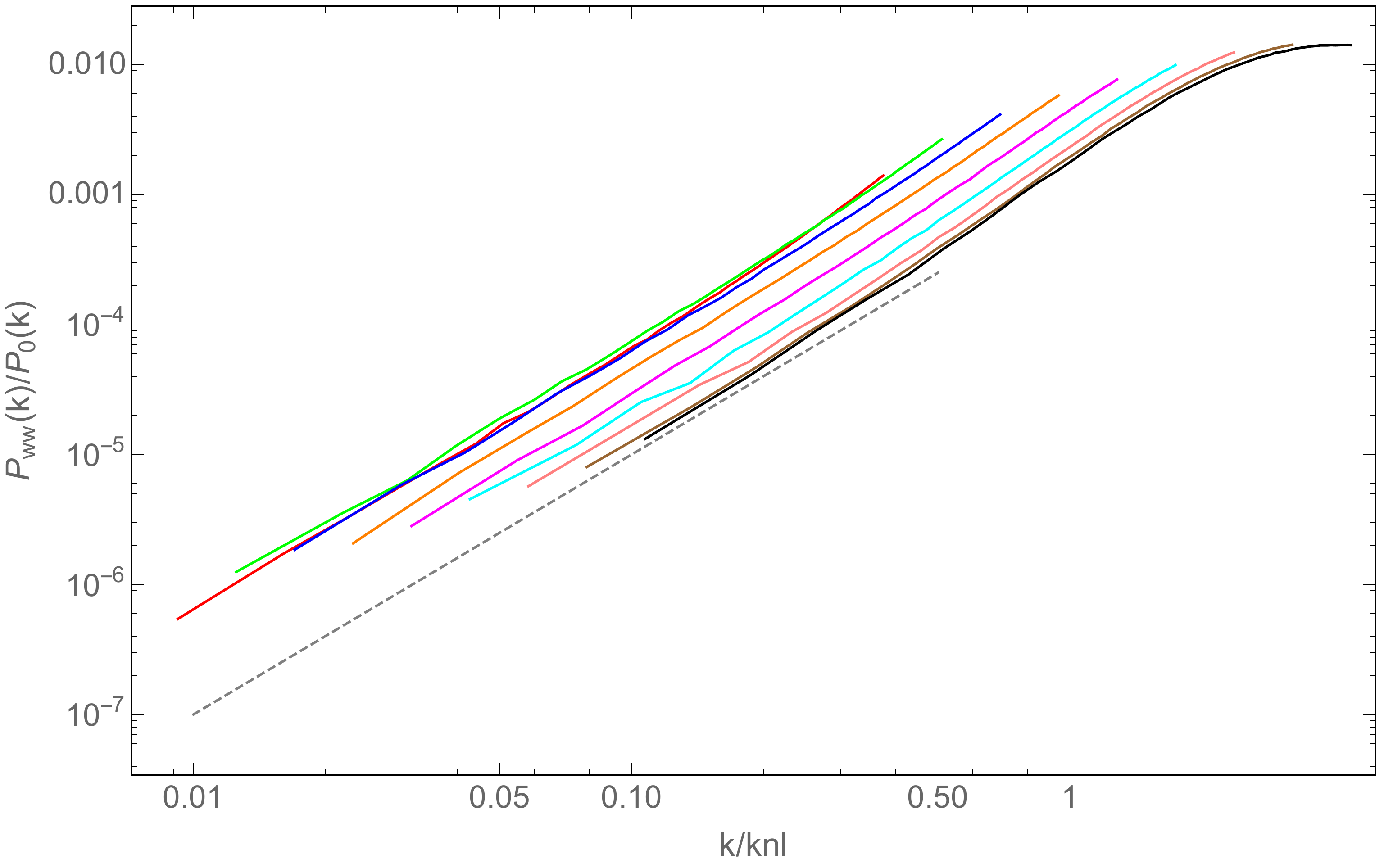}
  \end{center}
  \caption{\label{fig:PwwConverge} 
  Convergence study of the vorticity power spectrum measurements for $n_s=-1$ (left) and $n_s=0$ (right). Each line shows a single output (averaged over two pair-fixed realizations), starting from early outputs (highest amplitude at low-$k$) down to the last output. The dashed line denotes $P_{ww} \propto k^2$ expected as $k\to 0$. Convergence is achieved for the last 3-4 outputs for $n=-1$, which agree with each other at the 10\% level. For $n_s=0$ the situation is less clear, as expected from the higher values of $\lambda\, k_{\rm peak}$ for this spectral index, but the last two outputs are fairly consistent with each other. 
  }
\end{figure*}


\begin{thebibliography}{78}%
\makeatletter
\providecommand \@ifxundefined [1]{%
 \@ifx{#1\undefined}
}%
\providecommand \@ifnum [1]{%
 \ifnum #1\expandafter \@firstoftwo
 \else \expandafter \@secondoftwo
 \fi
}%
\providecommand \@ifx [1]{%
 \ifx #1\expandafter \@firstoftwo
 \else \expandafter \@secondoftwo
 \fi
}%
\providecommand \natexlab [1]{#1}%
\providecommand \enquote  [1]{``#1''}%
\providecommand \bibnamefont  [1]{#1}%
\providecommand \bibfnamefont [1]{#1}%
\providecommand \citenamefont [1]{#1}%
\providecommand \href@noop [0]{\@secondoftwo}%
\providecommand \href [0]{\begingroup \@sanitize@url \@href}%
\providecommand \@href[1]{\@@startlink{#1}\@@href}%
\providecommand \@@href[1]{\endgroup#1\@@endlink}%
\providecommand \@sanitize@url [0]{\catcode `\\12\catcode `\$12\catcode
  `\&12\catcode `\#12\catcode `\^12\catcode `\_12\catcode `\%12\relax}%
\providecommand \@@startlink[1]{}%
\providecommand \@@endlink[0]{}%
\providecommand \url  [0]{\begingroup\@sanitize@url \@url }%
\providecommand \@url [1]{\endgroup\@href {#1}{\urlprefix }}%
\providecommand \urlprefix  [0]{URL }%
\providecommand \Eprint [0]{\href }%
\providecommand \doibase [0]{http://dx.doi.org/}%
\providecommand \selectlanguage [0]{\@gobble}%
\providecommand \bibinfo  [0]{\@secondoftwo}%
\providecommand \bibfield  [0]{\@secondoftwo}%
\providecommand \translation [1]{[#1]}%
\providecommand \BibitemOpen [0]{}%
\providecommand \bibitemStop [0]{}%
\providecommand \bibitemNoStop [0]{.\EOS\space}%
\providecommand \EOS [0]{\spacefactor3000\relax}%
\providecommand \BibitemShut  [1]{\csname bibitem#1\endcsname}%
\let\auto@bib@innerbib\@empty
\bibitem [{\citenamefont {{Bernardeau}}\ \emph {et~al.}(2002)\citenamefont
  {{Bernardeau}}, \citenamefont {{Colombi}}, \citenamefont {{Gaztanaga}},\ and\
  \citenamefont {{Scoccimarro}}}]{BerColGaz02}%
  \BibitemOpen
  \bibfield  {author} {\bibinfo {author} {\bibfnamefont {F.}~\bibnamefont
  {{Bernardeau}}}, \bibinfo {author} {\bibfnamefont {S.}~\bibnamefont
  {{Colombi}}}, \bibinfo {author} {\bibfnamefont {E.}~\bibnamefont
  {{Gaztanaga}}}, \ and\ \bibinfo {author} {\bibfnamefont {R.}~\bibnamefont
  {{Scoccimarro}}},\ }\href@noop {} {\bibfield  {journal} {\bibinfo  {journal}
  {\physrep}\ }\textbf {\bibinfo {volume} {367}},\ \bibinfo {pages} {1}
  (\bibinfo {year} {2002})}\BibitemShut {NoStop}%
\bibitem [{\citenamefont {{Makino}}\ \emph {et~al.}(1992)\citenamefont
  {{Makino}}, \citenamefont {{Sasaki}},\ and\ \citenamefont
  {{Suto}}}]{MakSasSut92}%
  \BibitemOpen
  \bibfield  {author} {\bibinfo {author} {\bibfnamefont {N.}~\bibnamefont
  {{Makino}}}, \bibinfo {author} {\bibfnamefont {M.}~\bibnamefont {{Sasaki}}},
  \ and\ \bibinfo {author} {\bibfnamefont {Y.}~\bibnamefont {{Suto}}},\
  }\href@noop {} {\bibfield  {journal} {\bibinfo  {journal} {\prd}\ }\textbf
  {\bibinfo {volume} {46}},\ \bibinfo {pages} {585} (\bibinfo {year}
  {1992})}\BibitemShut {NoStop}%
\bibitem [{\citenamefont {{Scoccimarro}}\ and\ \citenamefont
  {{Frieman}}(1996{\natexlab{a}})}]{ScoFri9612}%
  \BibitemOpen
  \bibfield  {author} {\bibinfo {author} {\bibfnamefont {R.}~\bibnamefont
  {{Scoccimarro}}}\ and\ \bibinfo {author} {\bibfnamefont {J.~A.}\ \bibnamefont
  {{Frieman}}},\ }\href {\doibase 10.1086/178177} {\bibfield  {journal}
  {\bibinfo  {journal} {\apj}\ }\textbf {\bibinfo {volume} {473}},\ \bibinfo
  {pages} {620} (\bibinfo {year} {1996}{\natexlab{a}})},\ \Eprint
  {http://arxiv.org/abs/arXiv:astro-ph/9602070} {arXiv:astro-ph/9602070}
  \BibitemShut {NoStop}%
\bibitem [{\citenamefont {{Colombi}}\ \emph {et~al.}(1996)\citenamefont
  {{Colombi}}, \citenamefont {{Bouchet}},\ and\ \citenamefont
  {{Hernquist}}}]{ColBouHer96}%
  \BibitemOpen
  \bibfield  {author} {\bibinfo {author} {\bibfnamefont {S.}~\bibnamefont
  {{Colombi}}}, \bibinfo {author} {\bibfnamefont {F.~R.}\ \bibnamefont
  {{Bouchet}}}, \ and\ \bibinfo {author} {\bibfnamefont {L.}~\bibnamefont
  {{Hernquist}}},\ }\href {\doibase 10.1086/177398} {\bibfield  {journal}
  {\bibinfo  {journal} {\apj}\ }\textbf {\bibinfo {volume} {465}},\ \bibinfo
  {pages} {14} (\bibinfo {year} {1996})}\BibitemShut {NoStop}%
\bibitem [{\citenamefont {{Pueblas}}\ and\ \citenamefont
  {{Scoccimarro}}(2009)}]{PueSco0908}%
  \BibitemOpen
  \bibfield  {author} {\bibinfo {author} {\bibfnamefont {S.}~\bibnamefont
  {{Pueblas}}}\ and\ \bibinfo {author} {\bibfnamefont {R.}~\bibnamefont
  {{Scoccimarro}}},\ }\href {\doibase 10.1103/PhysRevD.80.043504} {\bibfield
  {journal} {\bibinfo  {journal} {\prd}\ }\textbf {\bibinfo {volume} {80}},\
  \bibinfo {pages} {043504} (\bibinfo {year} {2009})},\ \Eprint
  {http://arxiv.org/abs/0809.4606} {arXiv:0809.4606} \BibitemShut {NoStop}%
\bibitem [{\citenamefont {{Garny}}\ \emph {et~al.}(2022)\citenamefont
  {{Garny}}, \citenamefont {{Laxhuber}},\ and\ \citenamefont
  {{Scoccimarro}}}]{cumPT}%
  \BibitemOpen
  \bibfield  {author} {\bibinfo {author} {\bibfnamefont {M.}~\bibnamefont
  {{Garny}}}, \bibinfo {author} {\bibfnamefont {D.}~\bibnamefont {{Laxhuber}}},
  \ and\ \bibinfo {author} {\bibfnamefont {R.}~\bibnamefont {{Scoccimarro}}},\
  }\href@noop {} {\bibfield  {journal} {\bibinfo  {journal} {to appear}\ }
  (\bibinfo {year} {2022})}\BibitemShut {NoStop}%
\bibitem [{\citenamefont {{Davis}}\ and\ \citenamefont
  {{Peebles}}(1977)}]{DavPee7708}%
  \BibitemOpen
  \bibfield  {author} {\bibinfo {author} {\bibfnamefont {M.}~\bibnamefont
  {{Davis}}}\ and\ \bibinfo {author} {\bibfnamefont {P.~J.~E.}\ \bibnamefont
  {{Peebles}}},\ }\href {\doibase 10.1086/190456} {\bibfield  {journal}
  {\bibinfo  {journal} {\apjs}\ }\textbf {\bibinfo {volume} {34}},\ \bibinfo
  {pages} {425} (\bibinfo {year} {1977})}\BibitemShut {NoStop}%
\bibitem [{\citenamefont {{Peebles}}(1980)}]{Pee80}%
  \BibitemOpen
  \bibfield  {author} {\bibinfo {author} {\bibfnamefont {P.}~\bibnamefont
  {{Peebles}}},\ }\href@noop {} {\emph {\bibinfo {title} {{The large-scale
  structure of the universe}}}}\ (\bibinfo  {publisher} {Princeton University
  Press},\ \bibinfo {year} {1980})\BibitemShut {NoStop}%
\bibitem [{\citenamefont {{Scoccimarro}}(1997)}]{Sco97}%
  \BibitemOpen
  \bibfield  {author} {\bibinfo {author} {\bibfnamefont {R.}~\bibnamefont
  {{Scoccimarro}}},\ }\href@noop {} {\bibfield  {journal} {\bibinfo  {journal}
  {\apj}\ }\textbf {\bibinfo {volume} {487}},\ \bibinfo {pages} {1} (\bibinfo
  {year} {1997})}\BibitemShut {NoStop}%
\bibitem [{\citenamefont {{Pajer}}\ and\ \citenamefont
  {{Zaldarriaga}}(2013)}]{PajZal1308}%
  \BibitemOpen
  \bibfield  {author} {\bibinfo {author} {\bibfnamefont {E.}~\bibnamefont
  {{Pajer}}}\ and\ \bibinfo {author} {\bibfnamefont {M.}~\bibnamefont
  {{Zaldarriaga}}},\ }\href {\doibase 10.1088/1475-7516/2013/08/037} {\bibfield
   {journal} {\bibinfo  {journal} {\jcap}\ }\textbf {\bibinfo {volume} {8}},\
  \bibinfo {eid} {037} (\bibinfo {year} {2013})},\ \Eprint
  {http://arxiv.org/abs/1301.7182} {arXiv:1301.7182 [astro-ph.CO]} \BibitemShut
  {NoStop}%
\bibitem [{\citenamefont {{McDonald}}(2011)}]{McD1104}%
  \BibitemOpen
  \bibfield  {author} {\bibinfo {author} {\bibfnamefont {P.}~\bibnamefont
  {{McDonald}}},\ }\href {\doibase 10.1088/1475-7516/2011/04/032} {\bibfield
  {journal} {\bibinfo  {journal} {\jcap}\ }\textbf {\bibinfo {volume} {2011}},\
  \bibinfo {eid} {032} (\bibinfo {year} {2011})},\ \Eprint
  {http://arxiv.org/abs/0910.1002} {arXiv:0910.1002 [astro-ph.CO]} \BibitemShut
  {NoStop}%
\bibitem [{\citenamefont {{Aviles}}(2016)}]{Avi1603}%
  \BibitemOpen
  \bibfield  {author} {\bibinfo {author} {\bibfnamefont {A.}~\bibnamefont
  {{Aviles}}},\ }\href {\doibase 10.1103/PhysRevD.93.063517} {\bibfield
  {journal} {\bibinfo  {journal} {\prd}\ }\textbf {\bibinfo {volume} {93}},\
  \bibinfo {eid} {063517} (\bibinfo {year} {2016})},\ \Eprint
  {http://arxiv.org/abs/1512.07198} {arXiv:1512.07198 [astro-ph.CO]}
  \BibitemShut {NoStop}%
\bibitem [{\citenamefont {{Erschfeld}}\ and\ \citenamefont
  {{Floerchinger}}(2019)}]{ErsFlo1906}%
  \BibitemOpen
  \bibfield  {author} {\bibinfo {author} {\bibfnamefont {A.}~\bibnamefont
  {{Erschfeld}}}\ and\ \bibinfo {author} {\bibfnamefont {S.}~\bibnamefont
  {{Floerchinger}}},\ }\href {\doibase 10.1088/1475-7516/2019/06/039}
  {\bibfield  {journal} {\bibinfo  {journal} {\jcap}\ }\textbf {\bibinfo
  {volume} {2019}},\ \bibinfo {eid} {039} (\bibinfo {year} {2019})},\ \Eprint
  {http://arxiv.org/abs/1812.06891} {arXiv:1812.06891 [astro-ph.CO]}
  \BibitemShut {NoStop}%
\bibitem [{\citenamefont {Erschfeld}(2021)}]{Erschfeld:2021kem}%
  \BibitemOpen
  \bibfield  {author} {\bibinfo {author} {\bibfnamefont {A.~A.}\ \bibnamefont
  {Erschfeld}},\ }\emph {\bibinfo {title} {{Functional methods for cosmic
  structure formation.}}},\ \href {\doibase 10.11588/heidok.00030982} {Ph.D.
  thesis},\ \bibinfo  {school} {U. Heidelberg (main)} (\bibinfo {year}
  {2021})\BibitemShut {NoStop}%
\bibitem [{\citenamefont {{Uhlemann}}(2018)}]{Uhl1810}%
  \BibitemOpen
  \bibfield  {author} {\bibinfo {author} {\bibfnamefont {C.}~\bibnamefont
  {{Uhlemann}}},\ }\href {\doibase 10.1088/1475-7516/2018/10/030} {\bibfield
  {journal} {\bibinfo  {journal} {\jcap}\ }\textbf {\bibinfo {volume} {2018}},\
  \bibinfo {eid} {030} (\bibinfo {year} {2018})},\ \Eprint
  {http://arxiv.org/abs/1807.07274} {arXiv:1807.07274 [astro-ph.CO]}
  \BibitemShut {NoStop}%
\bibitem [{\citenamefont {{McDonald}}\ and\ \citenamefont
  {{Vlah}}(2018)}]{McDVla1801}%
  \BibitemOpen
  \bibfield  {author} {\bibinfo {author} {\bibfnamefont {P.}~\bibnamefont
  {{McDonald}}}\ and\ \bibinfo {author} {\bibfnamefont {Z.}~\bibnamefont
  {{Vlah}}},\ }\href {\doibase 10.1103/PhysRevD.97.023508} {\bibfield
  {journal} {\bibinfo  {journal} {\prd}\ }\textbf {\bibinfo {volume} {97}},\
  \bibinfo {eid} {023508} (\bibinfo {year} {2018})},\ \Eprint
  {http://arxiv.org/abs/1709.02834} {arXiv:1709.02834 [astro-ph.CO]}
  \BibitemShut {NoStop}%
\bibitem [{\citenamefont {{Cusin}}\ \emph {et~al.}(2017)\citenamefont
  {{Cusin}}, \citenamefont {{Tansella}},\ and\ \citenamefont
  {{Durrer}}}]{CusTanDur1703}%
  \BibitemOpen
  \bibfield  {author} {\bibinfo {author} {\bibfnamefont {G.}~\bibnamefont
  {{Cusin}}}, \bibinfo {author} {\bibfnamefont {V.}~\bibnamefont {{Tansella}}},
  \ and\ \bibinfo {author} {\bibfnamefont {R.}~\bibnamefont {{Durrer}}},\
  }\href {\doibase 10.1103/PhysRevD.95.063527} {\bibfield  {journal} {\bibinfo
  {journal} {\prd}\ }\textbf {\bibinfo {volume} {95}},\ \bibinfo {eid} {063527}
  (\bibinfo {year} {2017})},\ \Eprint {http://arxiv.org/abs/1612.00783}
  {arXiv:1612.00783 [astro-ph.CO]} \BibitemShut {NoStop}%
\bibitem [{\citenamefont {{Fry}}(1984)}]{Fry84}%
  \BibitemOpen
  \bibfield  {author} {\bibinfo {author} {\bibfnamefont {J.}~\bibnamefont
  {{Fry}}},\ }\href@noop {} {\bibfield  {journal} {\bibinfo  {journal} {\apj}\
  }\textbf {\bibinfo {volume} {279}},\ \bibinfo {pages} {499} (\bibinfo {year}
  {1984})}\BibitemShut {NoStop}%
\bibitem [{\citenamefont {{Goroff}}\ \emph {et~al.}(1986)\citenamefont
  {{Goroff}}, \citenamefont {{Grinstein}}, \citenamefont {{Rey}},\ and\
  \citenamefont {{Wise}}}]{GorGriRey86}%
  \BibitemOpen
  \bibfield  {author} {\bibinfo {author} {\bibfnamefont {M.}~\bibnamefont
  {{Goroff}}}, \bibinfo {author} {\bibfnamefont {B.}~\bibnamefont
  {{Grinstein}}}, \bibinfo {author} {\bibfnamefont {S.-J.}\ \bibnamefont
  {{Rey}}}, \ and\ \bibinfo {author} {\bibfnamefont {M.}~\bibnamefont
  {{Wise}}},\ }\href@noop {} {\bibfield  {journal} {\bibinfo  {journal} {\apj}\
  }\textbf {\bibinfo {volume} {311}},\ \bibinfo {pages} {6} (\bibinfo {year}
  {1986})}\BibitemShut {NoStop}%
\bibitem [{\citenamefont {Blas}\ \emph {et~al.}(2011)\citenamefont {Blas},
  \citenamefont {Lesgourgues},\ and\ \citenamefont {Tram}}]{Blas:2011rf}%
  \BibitemOpen
  \bibfield  {author} {\bibinfo {author} {\bibfnamefont {D.}~\bibnamefont
  {Blas}}, \bibinfo {author} {\bibfnamefont {J.}~\bibnamefont {Lesgourgues}}, \
  and\ \bibinfo {author} {\bibfnamefont {T.}~\bibnamefont {Tram}},\ }\href
  {\doibase 10.1088/1475-7516/2011/07/034} {\bibfield  {journal} {\bibinfo
  {journal} {JCAP}\ }\textbf {\bibinfo {volume} {07}},\ \bibinfo {pages} {034}
  (\bibinfo {year} {2011})},\ \Eprint {http://arxiv.org/abs/1104.2933}
  {arXiv:1104.2933 [astro-ph.CO]} \BibitemShut {NoStop}%
\bibitem [{\citenamefont {Lewis}\ \emph {et~al.}(2000)\citenamefont {Lewis},
  \citenamefont {Challinor},\ and\ \citenamefont {Lasenby}}]{Lewis:1999bs}%
  \BibitemOpen
  \bibfield  {author} {\bibinfo {author} {\bibfnamefont {A.}~\bibnamefont
  {Lewis}}, \bibinfo {author} {\bibfnamefont {A.}~\bibnamefont {Challinor}}, \
  and\ \bibinfo {author} {\bibfnamefont {A.}~\bibnamefont {Lasenby}},\ }\href
  {\doibase 10.1086/309179} {\bibfield  {journal} {\bibinfo  {journal}
  {Astrophys. J.}\ }\textbf {\bibinfo {volume} {538}},\ \bibinfo {pages} {473}
  (\bibinfo {year} {2000})},\ \Eprint {http://arxiv.org/abs/astro-ph/9911177}
  {arXiv:astro-ph/9911177} \BibitemShut {NoStop}%
\bibitem [{\citenamefont {Blas}\ \emph {et~al.}(2013)\citenamefont {Blas},
  \citenamefont {Garny},\ and\ \citenamefont {Konstandin}}]{Blas:2013bpa}%
  \BibitemOpen
  \bibfield  {author} {\bibinfo {author} {\bibfnamefont {D.}~\bibnamefont
  {Blas}}, \bibinfo {author} {\bibfnamefont {M.}~\bibnamefont {Garny}}, \ and\
  \bibinfo {author} {\bibfnamefont {T.}~\bibnamefont {Konstandin}},\ }\href
  {\doibase 10.1088/1475-7516/2013/09/024} {\bibfield  {journal} {\bibinfo
  {journal} {JCAP}\ }\textbf {\bibinfo {volume} {09}},\ \bibinfo {pages} {024}
  (\bibinfo {year} {2013})},\ \Eprint {http://arxiv.org/abs/1304.1546}
  {arXiv:1304.1546 [astro-ph.CO]} \BibitemShut {NoStop}%
\bibitem [{\citenamefont {{Blas}}\ \emph {et~al.}(2013)\citenamefont {{Blas}},
  \citenamefont {{Garny}},\ and\ \citenamefont {{Konstandin}}}]{BlaGarKon1309}%
  \BibitemOpen
  \bibfield  {author} {\bibinfo {author} {\bibfnamefont {D.}~\bibnamefont
  {{Blas}}}, \bibinfo {author} {\bibfnamefont {M.}~\bibnamefont {{Garny}}}, \
  and\ \bibinfo {author} {\bibfnamefont {T.}~\bibnamefont {{Konstandin}}},\
  }\href@noop {} {\bibfield  {journal} {\bibinfo  {journal} {ArXiv e-prints}\ }
  (\bibinfo {year} {2013})},\ \Eprint {http://arxiv.org/abs/1309.3308}
  {arXiv:1309.3308 [astro-ph.CO]} \BibitemShut {NoStop}%
\bibitem [{\citenamefont {Garny}\ and\ \citenamefont
  {Taule}(2021)}]{Garny:2020ilv}%
  \BibitemOpen
  \bibfield  {author} {\bibinfo {author} {\bibfnamefont {M.}~\bibnamefont
  {Garny}}\ and\ \bibinfo {author} {\bibfnamefont {P.}~\bibnamefont {Taule}},\
  }\href {\doibase 10.1088/1475-7516/2021/01/020} {\bibfield  {journal}
  {\bibinfo  {journal} {JCAP}\ }\textbf {\bibinfo {volume} {01}},\ \bibinfo
  {pages} {020} (\bibinfo {year} {2021})},\ \Eprint
  {http://arxiv.org/abs/2008.00013} {arXiv:2008.00013 [astro-ph.CO]}
  \BibitemShut {NoStop}%
\bibitem [{\citenamefont {Floerchinger}\ \emph {et~al.}(2019)\citenamefont
  {Floerchinger}, \citenamefont {Garny}, \citenamefont {Katsis}, \citenamefont
  {Tetradis},\ and\ \citenamefont {Wiedemann}}]{Floerchinger:2019eoj}%
  \BibitemOpen
  \bibfield  {author} {\bibinfo {author} {\bibfnamefont {S.}~\bibnamefont
  {Floerchinger}}, \bibinfo {author} {\bibfnamefont {M.}~\bibnamefont {Garny}},
  \bibinfo {author} {\bibfnamefont {A.}~\bibnamefont {Katsis}}, \bibinfo
  {author} {\bibfnamefont {N.}~\bibnamefont {Tetradis}}, \ and\ \bibinfo
  {author} {\bibfnamefont {U.~A.}\ \bibnamefont {Wiedemann}},\ }\href {\doibase
  10.1088/1475-7516/2019/09/047} {\bibfield  {journal} {\bibinfo  {journal}
  {JCAP}\ }\textbf {\bibinfo {volume} {09}},\ \bibinfo {pages} {047} (\bibinfo
  {year} {2019})},\ \Eprint {http://arxiv.org/abs/1907.10729} {arXiv:1907.10729
  [astro-ph.CO]} \BibitemShut {NoStop}%
\bibitem [{\citenamefont {Blas}\ \emph {et~al.}(2015)\citenamefont {Blas},
  \citenamefont {Floerchinger}, \citenamefont {Garny}, \citenamefont
  {Tetradis},\ and\ \citenamefont {Wiedemann}}]{Blas:2015tla}%
  \BibitemOpen
  \bibfield  {author} {\bibinfo {author} {\bibfnamefont {D.}~\bibnamefont
  {Blas}}, \bibinfo {author} {\bibfnamefont {S.}~\bibnamefont {Floerchinger}},
  \bibinfo {author} {\bibfnamefont {M.}~\bibnamefont {Garny}}, \bibinfo
  {author} {\bibfnamefont {N.}~\bibnamefont {Tetradis}}, \ and\ \bibinfo
  {author} {\bibfnamefont {U.~A.}\ \bibnamefont {Wiedemann}},\ }\href {\doibase
  10.1088/1475-7516/2015/11/049} {\bibfield  {journal} {\bibinfo  {journal}
  {JCAP}\ }\textbf {\bibinfo {volume} {11}},\ \bibinfo {pages} {049} (\bibinfo
  {year} {2015})},\ \Eprint {http://arxiv.org/abs/1507.06665} {arXiv:1507.06665
  [astro-ph.CO]} \BibitemShut {NoStop}%
\bibitem [{\citenamefont
  {{Scoccimarro}}(1998{\natexlab{a}})}]{Scoccimarro:1998b}%
  \BibitemOpen
  \bibfield  {author} {\bibinfo {author} {\bibfnamefont {R.}~\bibnamefont
  {{Scoccimarro}}},\ }\href {\doibase 10.1046/j.1365-8711.1998.01845.x}
  {\bibfield  {journal} {\bibinfo  {journal} {\mnras}\ }\textbf {\bibinfo
  {volume} {299}},\ \bibinfo {pages} {1097} (\bibinfo {year}
  {1998}{\natexlab{a}})},\ \Eprint {http://arxiv.org/abs/astro-ph/9711187}
  {astro-ph/9711187} \BibitemShut {NoStop}%
\bibitem [{\citenamefont {Eggemeier}\ \emph {et~al.}(2019)\citenamefont
  {Eggemeier}, \citenamefont {Scoccimarro},\ and\ \citenamefont
  {Smith}}]{Eggemeier:2018qae}%
  \BibitemOpen
  \bibfield  {author} {\bibinfo {author} {\bibfnamefont {A.}~\bibnamefont
  {Eggemeier}}, \bibinfo {author} {\bibfnamefont {R.}~\bibnamefont
  {Scoccimarro}}, \ and\ \bibinfo {author} {\bibfnamefont {R.~E.}\ \bibnamefont
  {Smith}},\ }\href {\doibase 10.1103/PhysRevD.99.123514} {\bibfield  {journal}
  {\bibinfo  {journal} {Phys. Rev. D}\ }\textbf {\bibinfo {volume} {99}},\
  \bibinfo {pages} {123514} (\bibinfo {year} {2019})},\ \Eprint
  {http://arxiv.org/abs/1812.03208} {arXiv:1812.03208 [astro-ph.CO]}
  \BibitemShut {NoStop}%
\bibitem [{\citenamefont {{Baldauf}}\ \emph {et~al.}(2015)\citenamefont
  {{Baldauf}}, \citenamefont {{Mercolli}}, \citenamefont {{Mirbabayi}},\ and\
  \citenamefont {{Pajer}}}]{Baldauf:2015}%
  \BibitemOpen
  \bibfield  {author} {\bibinfo {author} {\bibfnamefont {T.}~\bibnamefont
  {{Baldauf}}}, \bibinfo {author} {\bibfnamefont {L.}~\bibnamefont
  {{Mercolli}}}, \bibinfo {author} {\bibfnamefont {M.}~\bibnamefont
  {{Mirbabayi}}}, \ and\ \bibinfo {author} {\bibfnamefont {E.}~\bibnamefont
  {{Pajer}}},\ }\href {\doibase 10.1088/1475-7516/2015/05/007} {\bibfield
  {journal} {\bibinfo  {journal} {Journal of Cosmology and Astro-Particle
  Physics}\ }\textbf {\bibinfo {volume} {2015}},\ \bibinfo {eid} {007}
  (\bibinfo {year} {2015})},\ \Eprint {http://arxiv.org/abs/1406.4135}
  {arXiv:1406.4135 [astro-ph.CO]} \BibitemShut {NoStop}%
\bibitem [{\citenamefont {{Angulo}}\ \emph {et~al.}(2015)\citenamefont
  {{Angulo}}, \citenamefont {{Foreman}}, \citenamefont {{Schmittfull}},\ and\
  \citenamefont {{Senatore}}}]{Angulo:2015}%
  \BibitemOpen
  \bibfield  {author} {\bibinfo {author} {\bibfnamefont {R.~E.}\ \bibnamefont
  {{Angulo}}}, \bibinfo {author} {\bibfnamefont {S.}~\bibnamefont {{Foreman}}},
  \bibinfo {author} {\bibfnamefont {M.}~\bibnamefont {{Schmittfull}}}, \ and\
  \bibinfo {author} {\bibfnamefont {L.}~\bibnamefont {{Senatore}}},\ }\href
  {\doibase 10.1088/1475-7516/2015/10/039} {\bibfield  {journal} {\bibinfo
  {journal} {Journal of Cosmology and Astro-Particle Physics}\ }\textbf
  {\bibinfo {volume} {2015}},\ \bibinfo {eid} {039} (\bibinfo {year} {2015})},\
  \Eprint {http://arxiv.org/abs/1406.4143} {arXiv:1406.4143 [astro-ph.CO]}
  \BibitemShut {NoStop}%
\bibitem [{\citenamefont {Baldauf}\ \emph {et~al.}(2021)\citenamefont
  {Baldauf}, \citenamefont {Garny}, \citenamefont {Taule},\ and\ \citenamefont
  {Steele}}]{Baldauf:2021zlt}%
  \BibitemOpen
  \bibfield  {author} {\bibinfo {author} {\bibfnamefont {T.}~\bibnamefont
  {Baldauf}}, \bibinfo {author} {\bibfnamefont {M.}~\bibnamefont {Garny}},
  \bibinfo {author} {\bibfnamefont {P.}~\bibnamefont {Taule}}, \ and\ \bibinfo
  {author} {\bibfnamefont {T.}~\bibnamefont {Steele}},\ }\href {\doibase
  10.1103/PhysRevD.104.123551} {\bibfield  {journal} {\bibinfo  {journal}
  {Phys. Rev. D}\ }\textbf {\bibinfo {volume} {104}},\ \bibinfo {pages}
  {123551} (\bibinfo {year} {2021})},\ \Eprint
  {http://arxiv.org/abs/2110.13930} {arXiv:2110.13930 [astro-ph.CO]}
  \BibitemShut {NoStop}%
\bibitem [{\citenamefont {{Nishimichi}}\ \emph {et~al.}(2016)\citenamefont
  {{Nishimichi}}, \citenamefont {{Bernardeau}},\ and\ \citenamefont
  {{Taruya}}}]{NisBerTar1611}%
  \BibitemOpen
  \bibfield  {author} {\bibinfo {author} {\bibfnamefont {T.}~\bibnamefont
  {{Nishimichi}}}, \bibinfo {author} {\bibfnamefont {F.}~\bibnamefont
  {{Bernardeau}}}, \ and\ \bibinfo {author} {\bibfnamefont {A.}~\bibnamefont
  {{Taruya}}},\ }\href {\doibase 10.1016/j.physletb.2016.09.035} {\bibfield
  {journal} {\bibinfo  {journal} {Physics Letters B}\ }\textbf {\bibinfo
  {volume} {762}},\ \bibinfo {pages} {247} (\bibinfo {year} {2016})},\ \Eprint
  {http://arxiv.org/abs/1411.2970} {arXiv:1411.2970} \BibitemShut {NoStop}%
\bibitem [{\citenamefont {{Nishimichi}}\ \emph {et~al.}(2017)\citenamefont
  {{Nishimichi}}, \citenamefont {{Bernardeau}},\ and\ \citenamefont
  {{Taruya}}}]{NisBerTar1712}%
  \BibitemOpen
  \bibfield  {author} {\bibinfo {author} {\bibfnamefont {T.}~\bibnamefont
  {{Nishimichi}}}, \bibinfo {author} {\bibfnamefont {F.}~\bibnamefont
  {{Bernardeau}}}, \ and\ \bibinfo {author} {\bibfnamefont {A.}~\bibnamefont
  {{Taruya}}},\ }\href {\doibase 10.1103/PhysRevD.96.123515} {\bibfield
  {journal} {\bibinfo  {journal} {\prd}\ }\textbf {\bibinfo {volume} {96}},\
  \bibinfo {eid} {123515} (\bibinfo {year} {2017})},\ \Eprint
  {http://arxiv.org/abs/1708.08946} {arXiv:1708.08946 [astro-ph.CO]}
  \BibitemShut {NoStop}%
\bibitem [{\citenamefont {{Zeldovich}}(1965)}]{Zel6501}%
  \BibitemOpen
  \bibfield  {author} {\bibinfo {author} {\bibfnamefont {Y.~B.}\ \bibnamefont
  {{Zeldovich}}},\ }\href {\doibase 10.1016/B978-1-4831-9921-4.50011-9}
  {\bibfield  {journal} {\bibinfo  {journal} {Advances in Astronomy and
  Astrophysics}\ }\textbf {\bibinfo {volume} {3}},\ \bibinfo {pages} {241}
  (\bibinfo {year} {1965})}\BibitemShut {NoStop}%
\bibitem [{\citenamefont {Kehagias}\ and\ \citenamefont
  {Riotto}(2013)}]{Kehagias:2013yd}%
  \BibitemOpen
  \bibfield  {author} {\bibinfo {author} {\bibfnamefont {A.}~\bibnamefont
  {Kehagias}}\ and\ \bibinfo {author} {\bibfnamefont {A.}~\bibnamefont
  {Riotto}},\ }\href {\doibase 10.1016/j.nuclphysb.2013.05.009} {\bibfield
  {journal} {\bibinfo  {journal} {Nucl. Phys. B}\ }\textbf {\bibinfo {volume}
  {873}},\ \bibinfo {pages} {514} (\bibinfo {year} {2013})},\ \Eprint
  {http://arxiv.org/abs/1302.0130} {arXiv:1302.0130 [astro-ph.CO]} \BibitemShut
  {NoStop}%
\bibitem [{\citenamefont {Peloso}\ and\ \citenamefont
  {Pietroni}(2013)}]{Peloso:2013zw}%
  \BibitemOpen
  \bibfield  {author} {\bibinfo {author} {\bibfnamefont {M.}~\bibnamefont
  {Peloso}}\ and\ \bibinfo {author} {\bibfnamefont {M.}~\bibnamefont
  {Pietroni}},\ }\href {\doibase 10.1088/1475-7516/2013/05/031} {\bibfield
  {journal} {\bibinfo  {journal} {JCAP}\ }\textbf {\bibinfo {volume} {05}},\
  \bibinfo {pages} {031} (\bibinfo {year} {2013})},\ \Eprint
  {http://arxiv.org/abs/1302.0223} {arXiv:1302.0223 [astro-ph.CO]} \BibitemShut
  {NoStop}%
\bibitem [{\citenamefont {{Scoccimarro}}\ and\ \citenamefont
  {{Frieman}}(1996{\natexlab{b}})}]{ScoFri9607}%
  \BibitemOpen
  \bibfield  {author} {\bibinfo {author} {\bibfnamefont {R.}~\bibnamefont
  {{Scoccimarro}}}\ and\ \bibinfo {author} {\bibfnamefont {J.}~\bibnamefont
  {{Frieman}}},\ }\href {\doibase 10.1086/192306} {\bibfield  {journal}
  {\bibinfo  {journal} {\apjs}\ }\textbf {\bibinfo {volume} {105}},\ \bibinfo
  {pages} {37} (\bibinfo {year} {1996}{\natexlab{b}})},\ \Eprint
  {http://arxiv.org/abs/arXiv:astro-ph/9509047} {arXiv:astro-ph/9509047}
  \BibitemShut {NoStop}%
\bibitem [{\citenamefont {{Creminelli}}\ \emph {et~al.}(2013)\citenamefont
  {{Creminelli}}, \citenamefont {{Nore{\~n}a}}, \citenamefont
  {{Simonovi{\'c}}},\ and\ \citenamefont {{Vernizzi}}}]{CreNorSim1312}%
  \BibitemOpen
  \bibfield  {author} {\bibinfo {author} {\bibfnamefont {P.}~\bibnamefont
  {{Creminelli}}}, \bibinfo {author} {\bibfnamefont {J.}~\bibnamefont
  {{Nore{\~n}a}}}, \bibinfo {author} {\bibfnamefont {M.}~\bibnamefont
  {{Simonovi{\'c}}}}, \ and\ \bibinfo {author} {\bibfnamefont {F.}~\bibnamefont
  {{Vernizzi}}},\ }\href {\doibase 10.1088/1475-7516/2013/12/025} {\bibfield
  {journal} {\bibinfo  {journal} {\jcap}\ }\textbf {\bibinfo {volume} {2013}},\
  \bibinfo {eid} {025} (\bibinfo {year} {2013})},\ \Eprint
  {http://arxiv.org/abs/1309.3557} {arXiv:1309.3557 [astro-ph.CO]} \BibitemShut
  {NoStop}%
\bibitem [{\citenamefont {{Horn}}\ \emph {et~al.}(2014)\citenamefont {{Horn}},
  \citenamefont {{Hui}},\ and\ \citenamefont {{Xiao}}}]{HorHuiXia1409}%
  \BibitemOpen
  \bibfield  {author} {\bibinfo {author} {\bibfnamefont {B.}~\bibnamefont
  {{Horn}}}, \bibinfo {author} {\bibfnamefont {L.}~\bibnamefont {{Hui}}}, \
  and\ \bibinfo {author} {\bibfnamefont {X.}~\bibnamefont {{Xiao}}},\ }\href
  {\doibase 10.1088/1475-7516/2014/09/044} {\bibfield  {journal} {\bibinfo
  {journal} {\jcap}\ }\textbf {\bibinfo {volume} {2014}},\ \bibinfo {pages}
  {044} (\bibinfo {year} {2014})},\ \Eprint {http://arxiv.org/abs/1406.0842}
  {arXiv:1406.0842 [hep-th]} \BibitemShut {NoStop}%
\bibitem [{\citenamefont {{Crocce}}\ and\ \citenamefont
  {{Scoccimarro}}(2006)}]{CroSco0603b}%
  \BibitemOpen
  \bibfield  {author} {\bibinfo {author} {\bibfnamefont {M.}~\bibnamefont
  {{Crocce}}}\ and\ \bibinfo {author} {\bibfnamefont {R.}~\bibnamefont
  {{Scoccimarro}}},\ }\href {\doibase 10.1103/PhysRevD.73.063520} {\bibfield
  {journal} {\bibinfo  {journal} {\prd}\ }\textbf {\bibinfo {volume} {73}},\
  \bibinfo {pages} {063520} (\bibinfo {year} {2006})},\ \Eprint
  {http://arxiv.org/abs/arXiv:astro-ph/0509419} {arXiv:astro-ph/0509419}
  \BibitemShut {NoStop}%
\bibitem [{\citenamefont {Sugiyama}\ and\ \citenamefont
  {Futamase}(2013)}]{Sugiyama:2013pwa}%
  \BibitemOpen
  \bibfield  {author} {\bibinfo {author} {\bibfnamefont {N.~S.}\ \bibnamefont
  {Sugiyama}}\ and\ \bibinfo {author} {\bibfnamefont {T.}~\bibnamefont
  {Futamase}},\ }\href {\doibase 10.1088/0004-637X/769/2/106} {\bibfield
  {journal} {\bibinfo  {journal} {Astrophys. J.}\ }\textbf {\bibinfo {volume}
  {769}},\ \bibinfo {pages} {106} (\bibinfo {year} {2013})},\ \Eprint
  {http://arxiv.org/abs/1303.2748} {arXiv:1303.2748 [astro-ph.CO]} \BibitemShut
  {NoStop}%
\bibitem [{\citenamefont {{Jain}}\ and\ \citenamefont
  {{Bertschinger}}(1996)}]{JaiBer9601}%
  \BibitemOpen
  \bibfield  {author} {\bibinfo {author} {\bibfnamefont {B.}~\bibnamefont
  {{Jain}}}\ and\ \bibinfo {author} {\bibfnamefont {E.}~\bibnamefont
  {{Bertschinger}}},\ }\href {\doibase 10.1086/176625} {\bibfield  {journal}
  {\bibinfo  {journal} {\apj}\ }\textbf {\bibinfo {volume} {456}},\ \bibinfo
  {pages} {43} (\bibinfo {year} {1996})},\ \Eprint
  {http://arxiv.org/abs/astro-ph/9503025} {arXiv:astro-ph/9503025 [astro-ph]}
  \BibitemShut {NoStop}%
\bibitem [{\citenamefont {{Ma}}\ and\ \citenamefont
  {{Bertschinger}}(1995)}]{MaBer9512}%
  \BibitemOpen
  \bibfield  {author} {\bibinfo {author} {\bibfnamefont {C.-P.}\ \bibnamefont
  {{Ma}}}\ and\ \bibinfo {author} {\bibfnamefont {E.}~\bibnamefont
  {{Bertschinger}}},\ }\href {\doibase 10.1086/176550} {\bibfield  {journal}
  {\bibinfo  {journal} {\apj}\ }\textbf {\bibinfo {volume} {455}},\ \bibinfo
  {pages} {7} (\bibinfo {year} {1995})},\ \Eprint
  {http://arxiv.org/abs/astro-ph/9506072} {arXiv:astro-ph/9506072 [astro-ph]}
  \BibitemShut {NoStop}%
\bibitem [{\citenamefont {{Sheth}}\ and\ \citenamefont
  {{Tormen}}(1999)}]{SheTor9909}%
  \BibitemOpen
  \bibfield  {author} {\bibinfo {author} {\bibfnamefont {R.~K.}\ \bibnamefont
  {{Sheth}}}\ and\ \bibinfo {author} {\bibfnamefont {G.}~\bibnamefont
  {{Tormen}}},\ }\href {\doibase 10.1046/j.1365-8711.1999.02692.x} {\bibfield
  {journal} {\bibinfo  {journal} {\mnras}\ }\textbf {\bibinfo {volume} {308}},\
  \bibinfo {pages} {119} (\bibinfo {year} {1999})},\ \Eprint
  {http://arxiv.org/abs/arXiv:astro-ph/9901122} {arXiv:astro-ph/9901122}
  \BibitemShut {NoStop}%
\bibitem [{\citenamefont {{Despali}}\ \emph {et~al.}(2016)\citenamefont
  {{Despali}}, \citenamefont {{Giocoli}}, \citenamefont {{Angulo}},
  \citenamefont {{Tormen}}, \citenamefont {{Sheth}}, \citenamefont {{Baso}},\
  and\ \citenamefont {{Moscardini}}}]{DesGioAng1603}%
  \BibitemOpen
  \bibfield  {author} {\bibinfo {author} {\bibfnamefont {G.}~\bibnamefont
  {{Despali}}}, \bibinfo {author} {\bibfnamefont {C.}~\bibnamefont
  {{Giocoli}}}, \bibinfo {author} {\bibfnamefont {R.~E.}\ \bibnamefont
  {{Angulo}}}, \bibinfo {author} {\bibfnamefont {G.}~\bibnamefont {{Tormen}}},
  \bibinfo {author} {\bibfnamefont {R.~K.}\ \bibnamefont {{Sheth}}}, \bibinfo
  {author} {\bibfnamefont {G.}~\bibnamefont {{Baso}}}, \ and\ \bibinfo {author}
  {\bibfnamefont {L.}~\bibnamefont {{Moscardini}}},\ }\href {\doibase
  10.1093/mnras/stv2842} {\bibfield  {journal} {\bibinfo  {journal} {\mnras}\
  }\textbf {\bibinfo {volume} {456}},\ \bibinfo {pages} {2486} (\bibinfo {year}
  {2016})},\ \Eprint {http://arxiv.org/abs/1507.05627} {arXiv:1507.05627
  [astro-ph.CO]} \BibitemShut {NoStop}%
\bibitem [{\citenamefont {{Bryan}}\ and\ \citenamefont
  {{Norman}}(1998)}]{BryNor9803}%
  \BibitemOpen
  \bibfield  {author} {\bibinfo {author} {\bibfnamefont {G.~L.}\ \bibnamefont
  {{Bryan}}}\ and\ \bibinfo {author} {\bibfnamefont {M.~L.}\ \bibnamefont
  {{Norman}}},\ }\href {\doibase 10.1086/305262} {\bibfield  {journal}
  {\bibinfo  {journal} {\apj}\ }\textbf {\bibinfo {volume} {495}},\ \bibinfo
  {pages} {80} (\bibinfo {year} {1998})},\ \Eprint
  {http://arxiv.org/abs/astro-ph/9710107} {arXiv:astro-ph/9710107 [astro-ph]}
  \BibitemShut {NoStop}%
\bibitem [{\citenamefont {{Bullock}}\ \emph {et~al.}(2001)\citenamefont
  {{Bullock}}, \citenamefont {{Kolatt}}, \citenamefont {{Sigad}}, \citenamefont
  {{Somerville}}, \citenamefont {{Kravtsov}}, \citenamefont {{Klypin}},
  \citenamefont {{Primack}},\ and\ \citenamefont {{Dekel}}}]{BulKolSig01}%
  \BibitemOpen
  \bibfield  {author} {\bibinfo {author} {\bibfnamefont {J.}~\bibnamefont
  {{Bullock}}}, \bibinfo {author} {\bibfnamefont {T.}~\bibnamefont {{Kolatt}}},
  \bibinfo {author} {\bibfnamefont {Y.}~\bibnamefont {{Sigad}}}, \bibinfo
  {author} {\bibfnamefont {R.}~\bibnamefont {{Somerville}}}, \bibinfo {author}
  {\bibfnamefont {A.}~\bibnamefont {{Kravtsov}}}, \bibinfo {author}
  {\bibfnamefont {A.}~\bibnamefont {{Klypin}}}, \bibinfo {author}
  {\bibfnamefont {J.}~\bibnamefont {{Primack}}}, \ and\ \bibinfo {author}
  {\bibfnamefont {A.}~\bibnamefont {{Dekel}}},\ }\href@noop {} {\bibfield
  {journal} {\bibinfo  {journal} {\mnras}\ }\textbf {\bibinfo {volume} {321}},\
  \bibinfo {pages} {559} (\bibinfo {year} {2001})}\BibitemShut {NoStop}%
\bibitem [{\citenamefont {{Diemer}}\ and\ \citenamefont
  {{Kravtsov}}(2015)}]{DieKra1501}%
  \BibitemOpen
  \bibfield  {author} {\bibinfo {author} {\bibfnamefont {B.}~\bibnamefont
  {{Diemer}}}\ and\ \bibinfo {author} {\bibfnamefont {A.~V.}\ \bibnamefont
  {{Kravtsov}}},\ }\href {\doibase 10.1088/0004-637X/799/1/108} {\bibfield
  {journal} {\bibinfo  {journal} {\apj}\ }\textbf {\bibinfo {volume} {799}},\
  \bibinfo {eid} {108} (\bibinfo {year} {2015})},\ \Eprint
  {http://arxiv.org/abs/1407.4730} {arXiv:1407.4730 [astro-ph.CO]} \BibitemShut
  {NoStop}%
\bibitem [{\citenamefont {{Hahn}}\ \emph {et~al.}(2015)\citenamefont {{Hahn}},
  \citenamefont {{Angulo}},\ and\ \citenamefont {{Abel}}}]{HahAngAbe1512}%
  \BibitemOpen
  \bibfield  {author} {\bibinfo {author} {\bibfnamefont {O.}~\bibnamefont
  {{Hahn}}}, \bibinfo {author} {\bibfnamefont {R.~E.}\ \bibnamefont
  {{Angulo}}}, \ and\ \bibinfo {author} {\bibfnamefont {T.}~\bibnamefont
  {{Abel}}},\ }\href {\doibase 10.1093/mnras/stv2179} {\bibfield  {journal}
  {\bibinfo  {journal} {\mnras}\ }\textbf {\bibinfo {volume} {454}},\ \bibinfo
  {pages} {3920} (\bibinfo {year} {2015})},\ \Eprint
  {http://arxiv.org/abs/1404.2280} {arXiv:1404.2280 [astro-ph.CO]} \BibitemShut
  {NoStop}%
\bibitem [{\citenamefont {{Jelic-Cizmek}}\ \emph {et~al.}(2018)\citenamefont
  {{Jelic-Cizmek}}, \citenamefont {{Lepori}}, \citenamefont {{Adamek}},\ and\
  \citenamefont {{Durrer}}}]{JelLepAda1809}%
  \BibitemOpen
  \bibfield  {author} {\bibinfo {author} {\bibfnamefont {G.}~\bibnamefont
  {{Jelic-Cizmek}}}, \bibinfo {author} {\bibfnamefont {F.}~\bibnamefont
  {{Lepori}}}, \bibinfo {author} {\bibfnamefont {J.}~\bibnamefont {{Adamek}}},
  \ and\ \bibinfo {author} {\bibfnamefont {R.}~\bibnamefont {{Durrer}}},\
  }\href {\doibase 10.1088/1475-7516/2018/09/006} {\bibfield  {journal}
  {\bibinfo  {journal} {\jcap}\ }\textbf {\bibinfo {volume} {2018}},\ \bibinfo
  {eid} {006} (\bibinfo {year} {2018})},\ \Eprint
  {http://arxiv.org/abs/1806.05146} {arXiv:1806.05146 [astro-ph.CO]}
  \BibitemShut {NoStop}%
\bibitem [{\citenamefont {{Carrasco}}\ \emph {et~al.}(2014)\citenamefont
  {{Carrasco}}, \citenamefont {{Foreman}}, \citenamefont {{Green}},\ and\
  \citenamefont {{Senatore}}}]{CarForGre1407}%
  \BibitemOpen
  \bibfield  {author} {\bibinfo {author} {\bibfnamefont {J.~J.~M.}\
  \bibnamefont {{Carrasco}}}, \bibinfo {author} {\bibfnamefont
  {S.}~\bibnamefont {{Foreman}}}, \bibinfo {author} {\bibfnamefont
  {D.}~\bibnamefont {{Green}}}, \ and\ \bibinfo {author} {\bibfnamefont
  {L.}~\bibnamefont {{Senatore}}},\ }\href {\doibase
  10.1088/1475-7516/2014/07/057} {\bibfield  {journal} {\bibinfo  {journal}
  {\jcap}\ }\textbf {\bibinfo {volume} {2014}},\ \bibinfo {eid} {057} (\bibinfo
  {year} {2014})},\ \Eprint {http://arxiv.org/abs/1310.0464} {arXiv:1310.0464
  [astro-ph.CO]} \BibitemShut {NoStop}%
\bibitem [{\citenamefont {{Mercolli}}\ and\ \citenamefont
  {{Pajer}}(2014)}]{MerPaj1403}%
  \BibitemOpen
  \bibfield  {author} {\bibinfo {author} {\bibfnamefont {L.}~\bibnamefont
  {{Mercolli}}}\ and\ \bibinfo {author} {\bibfnamefont {E.}~\bibnamefont
  {{Pajer}}},\ }\href {\doibase 10.1088/1475-7516/2014/03/006} {\bibfield
  {journal} {\bibinfo  {journal} {\jcap}\ }\textbf {\bibinfo {volume} {2014}},\
  \bibinfo {eid} {006} (\bibinfo {year} {2014})},\ \Eprint
  {http://arxiv.org/abs/1307.3220} {arXiv:1307.3220 [astro-ph.CO]} \BibitemShut
  {NoStop}%
\bibitem [{\citenamefont {{Paranjape}}\ \emph {et~al.}(2013)\citenamefont
  {{Paranjape}}, \citenamefont {{Sheth}},\ and\ \citenamefont
  {{Desjacques}}}]{ParSheDes1305}%
  \BibitemOpen
  \bibfield  {author} {\bibinfo {author} {\bibfnamefont {A.}~\bibnamefont
  {{Paranjape}}}, \bibinfo {author} {\bibfnamefont {R.~K.}\ \bibnamefont
  {{Sheth}}}, \ and\ \bibinfo {author} {\bibfnamefont {V.}~\bibnamefont
  {{Desjacques}}},\ }\href {\doibase 10.1093/mnras/stt267} {\bibfield
  {journal} {\bibinfo  {journal} {\mnras}\ }\textbf {\bibinfo {volume} {431}},\
  \bibinfo {pages} {1503} (\bibinfo {year} {2013})},\ \Eprint
  {http://arxiv.org/abs/1210.1483} {arXiv:1210.1483 [astro-ph.CO]} \BibitemShut
  {NoStop}%
\bibitem [{\citenamefont {{Oddo}}\ and\ \citenamefont
  {{Sefusatti}}(2022)}]{OddSef22}%
  \BibitemOpen
  \bibfield  {author} {\bibinfo {author} {\bibfnamefont {A.}~\bibnamefont
  {{Oddo}}}\ and\ \bibinfo {author} {\bibfnamefont {E.}~\bibnamefont
  {{Sefusatti}}},\ }\href@noop {} {\bibfield  {journal} {\bibinfo  {journal}
  {unpublished}\ } (\bibinfo {year} {2022})}\BibitemShut {NoStop}%
\bibitem [{\citenamefont {{Springel}}(2005)}]{Spr0512}%
  \BibitemOpen
  \bibfield  {author} {\bibinfo {author} {\bibfnamefont {V.}~\bibnamefont
  {{Springel}}},\ }\href {\doibase 10.1111/j.1365-2966.2005.09655.x} {\bibfield
   {journal} {\bibinfo  {journal} {\mnras}\ }\textbf {\bibinfo {volume}
  {364}},\ \bibinfo {pages} {1105} (\bibinfo {year} {2005})},\ \Eprint
  {http://arxiv.org/abs/arXiv:astro-ph/0505010} {arXiv:astro-ph/0505010}
  \BibitemShut {NoStop}%
\bibitem [{\citenamefont {{Angulo}}\ and\ \citenamefont
  {{Pontzen}}(2016)}]{AngPon1610}%
  \BibitemOpen
  \bibfield  {author} {\bibinfo {author} {\bibfnamefont {R.~E.}\ \bibnamefont
  {{Angulo}}}\ and\ \bibinfo {author} {\bibfnamefont {A.}~\bibnamefont
  {{Pontzen}}},\ }\href {\doibase 10.1093/mnrasl/slw098} {\bibfield  {journal}
  {\bibinfo  {journal} {\mnras}\ }\textbf {\bibinfo {volume} {462}},\ \bibinfo
  {pages} {L1} (\bibinfo {year} {2016})},\ \Eprint
  {http://arxiv.org/abs/1603.05253} {arXiv:1603.05253} \BibitemShut {NoStop}%
\bibitem [{\citenamefont {{Scoccimarro}}(1998{\natexlab{b}})}]{Sco98}%
  \BibitemOpen
  \bibfield  {author} {\bibinfo {author} {\bibfnamefont {R.}~\bibnamefont
  {{Scoccimarro}}},\ }\href@noop {} {\bibfield  {journal} {\bibinfo  {journal}
  {\mnras}\ }\textbf {\bibinfo {volume} {299}},\ \bibinfo {pages} {1097}
  (\bibinfo {year} {1998}{\natexlab{b}})}\BibitemShut {NoStop}%
\bibitem [{\citenamefont {{Valageas}}(2002)}]{Val0204}%
  \BibitemOpen
  \bibfield  {author} {\bibinfo {author} {\bibfnamefont {P.}~\bibnamefont
  {{Valageas}}},\ }\href {\doibase 10.1051/0004-6361:20020187} {\bibfield
  {journal} {\bibinfo  {journal} {\aap}\ }\textbf {\bibinfo {volume} {385}},\
  \bibinfo {pages} {761} (\bibinfo {year} {2002})},\ \Eprint
  {http://arxiv.org/abs/arXiv:astro-ph/0112102} {arXiv:astro-ph/0112102}
  \BibitemShut {NoStop}%
\bibitem [{\citenamefont {{Crocce}}\ \emph {et~al.}(2006)\citenamefont
  {{Crocce}}, \citenamefont {{Pueblas}},\ and\ \citenamefont
  {{Scoccimarro}}}]{CroPueSco0611}%
  \BibitemOpen
  \bibfield  {author} {\bibinfo {author} {\bibfnamefont {M.}~\bibnamefont
  {{Crocce}}}, \bibinfo {author} {\bibfnamefont {S.}~\bibnamefont {{Pueblas}}},
  \ and\ \bibinfo {author} {\bibfnamefont {R.}~\bibnamefont {{Scoccimarro}}},\
  }\href {\doibase 10.1111/j.1365-2966.2006.11040.x} {\bibfield  {journal}
  {\bibinfo  {journal} {\mnras}\ }\textbf {\bibinfo {volume} {373}},\ \bibinfo
  {pages} {369} (\bibinfo {year} {2006})},\ \Eprint
  {http://arxiv.org/abs/arXiv:astro-ph/0606505} {arXiv:astro-ph/0606505}
  \BibitemShut {NoStop}%
\bibitem [{\citenamefont {{Jenkins}}(2010)}]{Jen1004}%
  \BibitemOpen
  \bibfield  {author} {\bibinfo {author} {\bibfnamefont {A.}~\bibnamefont
  {{Jenkins}}},\ }\href {\doibase 10.1111/j.1365-2966.2010.16259.x} {\bibfield
  {journal} {\bibinfo  {journal} {\mnras}\ }\textbf {\bibinfo {volume} {403}},\
  \bibinfo {pages} {1859} (\bibinfo {year} {2010})},\ \Eprint
  {http://arxiv.org/abs/0910.0258} {arXiv:0910.0258 [astro-ph.CO]} \BibitemShut
  {NoStop}%
\bibitem [{\citenamefont {{Reed}}\ \emph {et~al.}(2013)\citenamefont {{Reed}},
  \citenamefont {{Smith}}, \citenamefont {{Potter}}, \citenamefont
  {{Schneider}}, \citenamefont {{Stadel}},\ and\ \citenamefont
  {{Moore}}}]{ReeSmiPot1305}%
  \BibitemOpen
  \bibfield  {author} {\bibinfo {author} {\bibfnamefont {D.~S.}\ \bibnamefont
  {{Reed}}}, \bibinfo {author} {\bibfnamefont {R.~E.}\ \bibnamefont {{Smith}}},
  \bibinfo {author} {\bibfnamefont {D.}~\bibnamefont {{Potter}}}, \bibinfo
  {author} {\bibfnamefont {A.}~\bibnamefont {{Schneider}}}, \bibinfo {author}
  {\bibfnamefont {J.}~\bibnamefont {{Stadel}}}, \ and\ \bibinfo {author}
  {\bibfnamefont {B.}~\bibnamefont {{Moore}}},\ }\href {\doibase
  10.1093/mnras/stt301} {\bibfield  {journal} {\bibinfo  {journal} {\mnras}\
  }\textbf {\bibinfo {volume} {431}},\ \bibinfo {pages} {1866} (\bibinfo {year}
  {2013})},\ \Eprint {http://arxiv.org/abs/1206.5302} {arXiv:1206.5302
  [astro-ph.CO]} \BibitemShut {NoStop}%
\bibitem [{\citenamefont {{Michaux}}\ \emph {et~al.}(2021)\citenamefont
  {{Michaux}}, \citenamefont {{Hahn}}, \citenamefont {{Rampf}},\ and\
  \citenamefont {{Angulo}}}]{MicHahRam2101}%
  \BibitemOpen
  \bibfield  {author} {\bibinfo {author} {\bibfnamefont {M.}~\bibnamefont
  {{Michaux}}}, \bibinfo {author} {\bibfnamefont {O.}~\bibnamefont {{Hahn}}},
  \bibinfo {author} {\bibfnamefont {C.}~\bibnamefont {{Rampf}}}, \ and\
  \bibinfo {author} {\bibfnamefont {R.~E.}\ \bibnamefont {{Angulo}}},\ }\href
  {\doibase 10.1093/mnras/staa3149} {\bibfield  {journal} {\bibinfo  {journal}
  {\mnras}\ }\textbf {\bibinfo {volume} {500}},\ \bibinfo {pages} {663}
  (\bibinfo {year} {2021})},\ \Eprint {http://arxiv.org/abs/2008.09588}
  {arXiv:2008.09588 [astro-ph.CO]} \BibitemShut {NoStop}%
\bibitem [{\citenamefont {{Marcos}}\ \emph {et~al.}(2006)\citenamefont
  {{Marcos}}, \citenamefont {{Baertschiger}}, \citenamefont {{Joyce}},
  \citenamefont {{Gabrielli}},\ and\ \citenamefont {{Sylos
  Labini}}}]{MarBaeJoy0605}%
  \BibitemOpen
  \bibfield  {author} {\bibinfo {author} {\bibfnamefont {B.}~\bibnamefont
  {{Marcos}}}, \bibinfo {author} {\bibfnamefont {T.}~\bibnamefont
  {{Baertschiger}}}, \bibinfo {author} {\bibfnamefont {M.}~\bibnamefont
  {{Joyce}}}, \bibinfo {author} {\bibfnamefont {A.}~\bibnamefont
  {{Gabrielli}}}, \ and\ \bibinfo {author} {\bibfnamefont {F.}~\bibnamefont
  {{Sylos Labini}}},\ }\href {\doibase 10.1103/PhysRevD.73.103507} {\bibfield
  {journal} {\bibinfo  {journal} {\prd}\ }\textbf {\bibinfo {volume} {73}},\
  \bibinfo {eid} {103507} (\bibinfo {year} {2006})},\ \Eprint
  {http://arxiv.org/abs/arXiv:astro-ph/0601479} {arXiv:astro-ph/0601479}
  \BibitemShut {NoStop}%
\bibitem [{\citenamefont {{Joyce}}\ and\ \citenamefont
  {{Marcos}}(2007{\natexlab{a}})}]{JoyMar0703}%
  \BibitemOpen
  \bibfield  {author} {\bibinfo {author} {\bibfnamefont {M.}~\bibnamefont
  {{Joyce}}}\ and\ \bibinfo {author} {\bibfnamefont {B.}~\bibnamefont
  {{Marcos}}},\ }\href {\doibase 10.1103/PhysRevD.75.063516} {\bibfield
  {journal} {\bibinfo  {journal} {\prd}\ }\textbf {\bibinfo {volume} {75}},\
  \bibinfo {eid} {063516} (\bibinfo {year} {2007}{\natexlab{a}})},\ \Eprint
  {http://arxiv.org/abs/astro-ph/0410451} {astro-ph/0410451} \BibitemShut
  {NoStop}%
\bibitem [{\citenamefont {{Joyce}}\ and\ \citenamefont
  {{Marcos}}(2007{\natexlab{b}})}]{JoyMar0711}%
  \BibitemOpen
  \bibfield  {author} {\bibinfo {author} {\bibfnamefont {M.}~\bibnamefont
  {{Joyce}}}\ and\ \bibinfo {author} {\bibfnamefont {B.}~\bibnamefont
  {{Marcos}}},\ }\href {\doibase 10.1103/PhysRevD.76.103505} {\bibfield
  {journal} {\bibinfo  {journal} {\prd}\ }\textbf {\bibinfo {volume} {76}},\
  \bibinfo {eid} {103505} (\bibinfo {year} {2007}{\natexlab{b}})},\ \Eprint
  {http://arxiv.org/abs/0704.3697} {arXiv:0704.3697} \BibitemShut {NoStop}%
\bibitem [{\citenamefont {{Garrison}}\ \emph {et~al.}(2016)\citenamefont
  {{Garrison}}, \citenamefont {{Eisenstein}}, \citenamefont {{Ferrer}},
  \citenamefont {{Metchnik}},\ and\ \citenamefont {{Pinto}}}]{GarEisFer1610}%
  \BibitemOpen
  \bibfield  {author} {\bibinfo {author} {\bibfnamefont {L.~H.}\ \bibnamefont
  {{Garrison}}}, \bibinfo {author} {\bibfnamefont {D.~J.}\ \bibnamefont
  {{Eisenstein}}}, \bibinfo {author} {\bibfnamefont {D.}~\bibnamefont
  {{Ferrer}}}, \bibinfo {author} {\bibfnamefont {M.~V.}\ \bibnamefont
  {{Metchnik}}}, \ and\ \bibinfo {author} {\bibfnamefont {P.~A.}\ \bibnamefont
  {{Pinto}}},\ }\href {\doibase 10.1093/mnras/stw1594} {\bibfield  {journal}
  {\bibinfo  {journal} {\mnras}\ }\textbf {\bibinfo {volume} {461}},\ \bibinfo
  {pages} {4125} (\bibinfo {year} {2016})},\ \Eprint
  {http://arxiv.org/abs/1605.02333} {arXiv:1605.02333 [astro-ph.CO]}
  \BibitemShut {NoStop}%
\bibitem [{\citenamefont {{Peacock}}\ and\ \citenamefont
  {{Dodds}}(1996)}]{PeaDod96}%
  \BibitemOpen
  \bibfield  {author} {\bibinfo {author} {\bibfnamefont {J.}~\bibnamefont
  {{Peacock}}}\ and\ \bibinfo {author} {\bibfnamefont {S.}~\bibnamefont
  {{Dodds}}},\ }\href@noop {} {\bibfield  {journal} {\bibinfo  {journal}
  {\mnras}\ }\textbf {\bibinfo {volume} {280}},\ \bibinfo {pages} {L19}
  (\bibinfo {year} {1996})}\BibitemShut {NoStop}%
\bibitem [{\citenamefont {{Smith}}\ \emph {et~al.}(2003)\citenamefont
  {{Smith}}, \citenamefont {{Peacock}}, \citenamefont {{Jenkins}},
  \citenamefont {{White}}, \citenamefont {{Frenk}}, \citenamefont {{Pearce}},
  \citenamefont {{Thomas}}, \citenamefont {{Efstathiou}},\ and\ \citenamefont
  {{Couchman}}}]{SmiPeaJen03}%
  \BibitemOpen
  \bibfield  {author} {\bibinfo {author} {\bibfnamefont {R.}~\bibnamefont
  {{Smith}}}, \bibinfo {author} {\bibfnamefont {J.}~\bibnamefont {{Peacock}}},
  \bibinfo {author} {\bibfnamefont {A.}~\bibnamefont {{Jenkins}}}, \bibinfo
  {author} {\bibfnamefont {S.}~\bibnamefont {{White}}}, \bibinfo {author}
  {\bibfnamefont {C.}~\bibnamefont {{Frenk}}}, \bibinfo {author} {\bibfnamefont
  {F.}~\bibnamefont {{Pearce}}}, \bibinfo {author} {\bibfnamefont
  {P.}~\bibnamefont {{Thomas}}}, \bibinfo {author} {\bibfnamefont
  {G.}~\bibnamefont {{Efstathiou}}}, \ and\ \bibinfo {author} {\bibfnamefont
  {H.}~\bibnamefont {{Couchman}}},\ }\href@noop {} {\bibfield  {journal}
  {\bibinfo  {journal} {\mnras}\ }\textbf {\bibinfo {volume} {341}},\ \bibinfo
  {pages} {1311} (\bibinfo {year} {2003})}\BibitemShut {NoStop}%
\bibitem [{\citenamefont {{Bagla}}\ \emph {et~al.}(2009)\citenamefont
  {{Bagla}}, \citenamefont {{Khandai}},\ and\ \citenamefont
  {{Kulkarni}}}]{BagKhaKul0908}%
  \BibitemOpen
  \bibfield  {author} {\bibinfo {author} {\bibfnamefont {J.~S.}\ \bibnamefont
  {{Bagla}}}, \bibinfo {author} {\bibfnamefont {N.}~\bibnamefont {{Khandai}}},
  \ and\ \bibinfo {author} {\bibfnamefont {G.}~\bibnamefont {{Kulkarni}}},\
  }\href@noop {} {\bibfield  {journal} {\bibinfo  {journal} {arXiv e-prints}\
  ,\ \bibinfo {eid} {arXiv:0908.2702}} (\bibinfo {year} {2009})},\ \Eprint
  {http://arxiv.org/abs/0908.2702} {arXiv:0908.2702 [astro-ph.CO]} \BibitemShut
  {NoStop}%
\bibitem [{\citenamefont {{Widrow}}\ \emph {et~al.}(2009)\citenamefont
  {{Widrow}}, \citenamefont {{Elahi}}, \citenamefont {{Thacker}}, \citenamefont
  {{Richardson}},\ and\ \citenamefont {{Scannapieco}}}]{WidElaTha0908}%
  \BibitemOpen
  \bibfield  {author} {\bibinfo {author} {\bibfnamefont {L.~M.}\ \bibnamefont
  {{Widrow}}}, \bibinfo {author} {\bibfnamefont {P.~J.}\ \bibnamefont
  {{Elahi}}}, \bibinfo {author} {\bibfnamefont {R.~J.}\ \bibnamefont
  {{Thacker}}}, \bibinfo {author} {\bibfnamefont {M.}~\bibnamefont
  {{Richardson}}}, \ and\ \bibinfo {author} {\bibfnamefont {E.}~\bibnamefont
  {{Scannapieco}}},\ }\href {\doibase 10.1111/j.1365-2966.2009.15075.x}
  {\bibfield  {journal} {\bibinfo  {journal} {\mnras}\ }\textbf {\bibinfo
  {volume} {397}},\ \bibinfo {pages} {1275} (\bibinfo {year} {2009})},\ \Eprint
  {http://arxiv.org/abs/0901.4576} {arXiv:0901.4576 [astro-ph.CO]} \BibitemShut
  {NoStop}%
\bibitem [{\citenamefont {{Joyce}}\ \emph {et~al.}(2021)\citenamefont
  {{Joyce}}, \citenamefont {{Garrison}},\ and\ \citenamefont
  {{Eisenstein}}}]{JoyGarEis2103}%
  \BibitemOpen
  \bibfield  {author} {\bibinfo {author} {\bibfnamefont {M.}~\bibnamefont
  {{Joyce}}}, \bibinfo {author} {\bibfnamefont {L.}~\bibnamefont {{Garrison}}},
  \ and\ \bibinfo {author} {\bibfnamefont {D.}~\bibnamefont {{Eisenstein}}},\
  }\href {\doibase 10.1093/mnras/staa3434} {\bibfield  {journal} {\bibinfo
  {journal} {\mnras}\ }\textbf {\bibinfo {volume} {501}},\ \bibinfo {pages}
  {5051} (\bibinfo {year} {2021})},\ \Eprint {http://arxiv.org/abs/2004.07256}
  {arXiv:2004.07256 [astro-ph.CO]} \BibitemShut {NoStop}%
\bibitem [{\citenamefont {{Leroy}}\ \emph {et~al.}(2021)\citenamefont
  {{Leroy}}, \citenamefont {{Garrison}}, \citenamefont {{Eisenstein}},
  \citenamefont {{Joyce}},\ and\ \citenamefont {{Maleubre}}}]{LerGarEis2103}%
  \BibitemOpen
  \bibfield  {author} {\bibinfo {author} {\bibfnamefont {M.}~\bibnamefont
  {{Leroy}}}, \bibinfo {author} {\bibfnamefont {L.}~\bibnamefont {{Garrison}}},
  \bibinfo {author} {\bibfnamefont {D.}~\bibnamefont {{Eisenstein}}}, \bibinfo
  {author} {\bibfnamefont {M.}~\bibnamefont {{Joyce}}}, \ and\ \bibinfo
  {author} {\bibfnamefont {S.}~\bibnamefont {{Maleubre}}},\ }\href {\doibase
  10.1093/mnras/staa3435} {\bibfield  {journal} {\bibinfo  {journal} {\mnras}\
  }\textbf {\bibinfo {volume} {501}},\ \bibinfo {pages} {5064} (\bibinfo {year}
  {2021})},\ \Eprint {http://arxiv.org/abs/2004.08406} {arXiv:2004.08406
  [astro-ph.CO]} \BibitemShut {NoStop}%
\bibitem [{\citenamefont {{Angulo}}\ and\ \citenamefont
  {{Hahn}}(2022)}]{AngHah2212}%
  \BibitemOpen
  \bibfield  {author} {\bibinfo {author} {\bibfnamefont {R.~E.}\ \bibnamefont
  {{Angulo}}}\ and\ \bibinfo {author} {\bibfnamefont {O.}~\bibnamefont
  {{Hahn}}},\ }\href {\doibase 10.1007/s41115-021-00013-z} {\bibfield
  {journal} {\bibinfo  {journal} {Living Reviews in Computational
  Astrophysics}\ }\textbf {\bibinfo {volume} {8}},\ \bibinfo {eid} {1}
  (\bibinfo {year} {2022})},\ \Eprint {http://arxiv.org/abs/2112.05165}
  {arXiv:2112.05165 [astro-ph.CO]} \BibitemShut {NoStop}%
\bibitem [{\citenamefont {{Behroozi}}\ \emph {et~al.}(2013)\citenamefont
  {{Behroozi}}, \citenamefont {{Wechsler}},\ and\ \citenamefont
  {{Wu}}}]{BehWecWu1301}%
  \BibitemOpen
  \bibfield  {author} {\bibinfo {author} {\bibfnamefont {P.~S.}\ \bibnamefont
  {{Behroozi}}}, \bibinfo {author} {\bibfnamefont {R.~H.}\ \bibnamefont
  {{Wechsler}}}, \ and\ \bibinfo {author} {\bibfnamefont {H.-Y.}\ \bibnamefont
  {{Wu}}},\ }\href {\doibase 10.1088/0004-637X/762/2/109} {\bibfield  {journal}
  {\bibinfo  {journal} {\apj}\ }\textbf {\bibinfo {volume} {762}},\ \bibinfo
  {eid} {109} (\bibinfo {year} {2013})},\ \Eprint
  {http://arxiv.org/abs/1110.4372} {arXiv:1110.4372 [astro-ph.CO]} \BibitemShut
  {NoStop}%
\bibitem [{\citenamefont {{Bernardeau}}\ and\ \citenamefont {{van de
  Weygaert}}(1996)}]{Bervan96}%
  \BibitemOpen
  \bibfield  {author} {\bibinfo {author} {\bibfnamefont {F.}~\bibnamefont
  {{Bernardeau}}}\ and\ \bibinfo {author} {\bibfnamefont {R.}~\bibnamefont
  {{van de Weygaert}}},\ }\href@noop {} {\bibfield  {journal} {\bibinfo
  {journal} {\mnras}\ }\textbf {\bibinfo {volume} {279}},\ \bibinfo {pages}
  {693} (\bibinfo {year} {1996})}\BibitemShut {NoStop}%
\bibitem [{\citenamefont {{Romano-D{\'\i}az}}\ and\ \citenamefont {{van de
  Weygaert}}(2007)}]{Romvan0711}%
  \BibitemOpen
  \bibfield  {author} {\bibinfo {author} {\bibfnamefont {E.}~\bibnamefont
  {{Romano-D{\'\i}az}}}\ and\ \bibinfo {author} {\bibfnamefont
  {R.}~\bibnamefont {{van de Weygaert}}},\ }\href {\doibase
  10.1111/j.1365-2966.2007.12190.x} {\bibfield  {journal} {\bibinfo  {journal}
  {\mnras}\ }\textbf {\bibinfo {volume} {382}},\ \bibinfo {pages} {2} (\bibinfo
  {year} {2007})}\BibitemShut {NoStop}%
\bibitem [{\citenamefont {{Buehlmann}}\ and\ \citenamefont
  {{Hahn}}(2019)}]{BueHah1907}%
  \BibitemOpen
  \bibfield  {author} {\bibinfo {author} {\bibfnamefont {M.}~\bibnamefont
  {{Buehlmann}}}\ and\ \bibinfo {author} {\bibfnamefont {O.}~\bibnamefont
  {{Hahn}}},\ }\href {\doibase 10.1093/mnras/stz1243} {\bibfield  {journal}
  {\bibinfo  {journal} {\mnras}\ }\textbf {\bibinfo {volume} {487}},\ \bibinfo
  {pages} {228} (\bibinfo {year} {2019})},\ \Eprint
  {http://arxiv.org/abs/1812.07489} {arXiv:1812.07489 [astro-ph.CO]}
  \BibitemShut {NoStop}%
\bibitem [{\citenamefont {{Sefusatti}}\ \emph {et~al.}(2015)\citenamefont
  {{Sefusatti}}, \citenamefont {{Crocce}}, \citenamefont {{Scoccimarro}},\ and\
  \citenamefont {{Couchman}}}]{SefCroSco1512}%
  \BibitemOpen
  \bibfield  {author} {\bibinfo {author} {\bibfnamefont {E.}~\bibnamefont
  {{Sefusatti}}}, \bibinfo {author} {\bibfnamefont {M.}~\bibnamefont
  {{Crocce}}}, \bibinfo {author} {\bibfnamefont {R.}~\bibnamefont
  {{Scoccimarro}}}, \ and\ \bibinfo {author} {\bibfnamefont {H.}~\bibnamefont
  {{Couchman}}},\ }\href@noop {} {\bibfield  {journal} {\bibinfo  {journal}
  {ArXiv e-prints}\ } (\bibinfo {year} {2015})},\ \Eprint
  {http://arxiv.org/abs/1512.07295} {arXiv:1512.07295} \BibitemShut {NoStop}%
\end{thebibliography}
%

\end{document}